\documentclass[12pt, oneside]{Thesis} 

\usepackage{algorithm}
\usepackage{algpseudocode}
\usepackage[square, numbers, comma, sort&compress]{natbib}
\usepackage{pdfpages}
\usepackage[titletoc]{appendix}
\usepackage{wrapfig}
\usepackage{multirow}
\usepackage{adjustbox}
\usepackage{times}
\usepackage{textcomp}
\usepackage{array}

\graphicspath{{Pictures/}} 
\newcommand*\xbar[1]{%
 \hbox{%
 \vbox{%
 \hrule height 0.5pt 
 \kern0.5ex
 \hbox{%
 \kern+0em
 \ensuremath{#1}%
 \kern+0em
 }%
 }%
 }%
}

\hypersetup{urlcolor=blue, colorlinks=true} 
\title{\ttitle} 
\begin{document}

\frontmatter 
\pagestyle{empty} 
\setstretch{1.5} 



\newcommand{\HRule}{\rule{\linewidth}{0.5mm}} 

\hypersetup{pdftitle={\ttitle}}
\hypersetup{pdfsubject=\subjectname}
\hypersetup{pdfauthor=\authornames}
\hypersetup{pdfkeywords=\keywordnames}


\begin{center}
\includegraphics[scale=0.5]{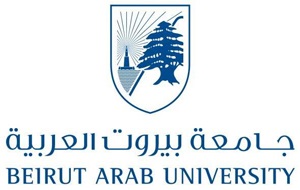}\\ [1cm] 
\setstretch{2}
\textbf{\Large \ttitle}\\[0.5cm] 
 \setstretch{1.5}
 by\\[0.5cm]
{\Large{RAMZI ALI JABER}}\\[0.5cm] 
 \textbf{Thesis}\\[1cm]
\textit{Submitted in Partial Fulfilment of the Requirements for the degree of }\\ 
PhD in Computer Engineering\\[3cm] 
Department of \deptname\\
Faculty of \facname\\[4cm]
{\large 2020}\\[1cm] 
\vfill
\end{center}


\clearpage 
\begin{center}
\includegraphics[scale=0.5]{Logo.jpg}\\ [1cm]
\setstretch{2}
\textbf{\Large \ttitle}\\[0.5cm] 
\setstretch{1.5}
 by\\[0.5cm]
{\Large{RAMZI ALI JABER}}\\[0.5cm] 
 \textbf{Thesis}\\[0.5cm]
\textit{Submitted in Partial Fulfilment of the Requirements for the degree of }\\ 
PhD in Computer Engineering\\[1cm] 
Department of \deptname\\
Faculty of \facname\\[1cm]
Supervised by\\
\vspace{5mm}
\begin{table}[h!]
\setlength{\tabcolsep}{1pt}
\centering
\begin{tabular}{ccc}
Dr. Ali Massoud \textsc{Haidar} &\hspace{1cm} &Dr. Lina \textsc{EL-Nimri} \\
Professor & &Associate Professor \\
Faculty of Engineering & &Faculty of Economics and Business Administration \\
Electrical \& Computer Eng. Dept. & &Computer Dept. \\
Beirut Arab University & &Lebanese University \\[1cm]
\end{tabular}
\end{table}

{\large 2020}\\ 
\vfill
\end{center}


\clearpage 
\begin{center}
\includegraphics[scale=0.5]{Logo.jpg}\\ [0.5cm]
The Thesis Defense Committee for \textbf{RAMZI ALI JABER}\\
Certifies that this is the approved version of the following thesis\\[0.5cm]
\setstretch{2}
\textbf{\Large \ttitle}\\[0.5cm] 
APPROVED BY:\\[0.5cm]\setstretch{2.5}
\textbf{Supervisor Signature:} \underline{\hspace{8cm}}\\ \hspace{4cm}\textbf{Prof. Ali Haidar}\\ 
\textbf{Examiner Signature:}\underline{\hspace{8cm}}\\ \hspace{4cm}\textbf{Prof. Ziad Osman}\\ 
\textbf{Examiner Signature:}\underline{\hspace{8cm}}\\ \hspace{4cm}\textbf{Associate Prof. Issam Damaj}\\ 
\textbf{Examiner Signature:}\underline{\hspace{8cm}}\\ \hspace{4cm}\textbf{Prof. Mohamed Dbouk}\\ 
\textbf{Examiner Signature:}\underline{\hspace{8cm}}\\ \hspace{4cm}\textbf{Prof. Mohamad Sawan}\\ 
\end{center}


\clearpage 
\pagestyle{empty} 
\setstretch{1.5} 
\begin{center} \textbf{\Large ACKNOWLEDGEMENT}\\[1.5cm] \end{center}
First of all, I thank my God (ALLAH) for protecting and giving me the ability to do this hard work.\\
Secondly, it is my pleasure to thank all the people who helped me throughout the achievement of this thesis.
Special thanks and appreciations to my supervisor Prof. Dr. Ali Haidar and Dr. Ahmad El-Hajj for granting me the opportunity to carry out this research and for their constant encouragement and valuable guidance during this thesis work.\\
Also, I would like to thank my co-supervisor Dr. Lina El-Nimri, Dr. Abdallah Kassem, and Dr. Ziad Osman for their trust and very kind support.\\
I would also like to say thanks to my friends and research colleagues for their constant support and encouragement: Dr. Bilal Owaidat, Eng. Imane Haidar, Eng. Hamza Damaj, and Eng. Abdallah Salem.\\
Most importantly, none of this would have been possible without the love and patience of my family. I am eternally grateful to them, 
for their careful considerations and high confidence in me through all these years.

Finally, I am grateful to the jury members:
\begin{enumerate}
  \item \textbf{Prof. Ali Haidar}: Committee Chair and Computer Eng. undergraduate \& graduate Program Coordinator at BAU
  \item \textbf{Prof. Ziad Osman}: Chair of Electrical \& Computer Eng. Dept. at BAU
  \item \textbf{Associate Prof. Issam Damaj}: Director of the Center for Quality Assurance at BAU
  \item \textbf{Prof. Mohamed Dbouk}: Director of L'ARiCoD: Lab of Advanced Research in Intelligent Computing \& Data and Master Coordinator at Lebanese Univ.
  \item \textbf{Prof. Mohamad Sawan}: Chair Professor at Westlake University, Hangzhou, China and Canada Research Chair at Polytechnique Montreal, Canada
 \end{enumerate}


\clearpage 
\pagestyle{plain} 
\begin{center} \textbf{\Large  Abstract}\\[1.5cm]\end{center}

Recently, the demand for portable electronics and embedded systems has increased. These devices need low-power circuit designs because they depend on batteries as an energy~resource. Moreover, Multi-Valued Logic (MVL) circuits provide notable improvements over binary circuits in terms of interconnect complexity, chip area, propagation delay, and~energy consumption.
Therefore, this thesis proposes novel ternary circuits aiming to reduce the energy (Power Delay Product (PDP)) to preserve battery consumption.

The proposed designs include eight ternary logic gates, three ternary combinational circuits, and six Ternary Arithmetic Logic Units (TALU).  
The ternary logic gates are seven unary operators of the ternary system ($A^1$, $A^2$, $\bar{A^2}$, $A_1$, $1.\bar{A_n}$, $1.\bar{A_p}$, and the Standard Ternary Inverter (STI) $\bar{A}$), and Ternary NAND based on Carbon Nanotube Field-Effect Transistor (CNFET). Ternary combinational circuits, two different designs for Ternary Decoders (TDecoder) and Ternary Multiplexer (TMUX): (1) TDecoder1 using CNFET-based proposed unary operators and TDecoder2 using Double-Pass Logic (DPL) binary gates. (2) TMUX using CNFET-based proposed unary operators. 
And Ternary Arithmetic Logic Units are three different designs for Ternary Half-Adders (THA) and Ternary Multipliers (TMUL): (1) The first design uses the proposed TDecoder1, STI, and TNAND. (2) While the second design uses the cascading proposed TMUX. (3) As for the third design, it uses the proposed unary operators and TMUX.
 
This thesis applies the best trade-off between reducing the number of used transistors,~utilizing energy-efficient transistor arrangement such as transmission gate and applying the dual supply voltages (Vdd, and Vdd/2) to achieve its objective.
The proposed designs are~compared to the latest ternary circuits using the HSPICE simulator for different  supply voltages, different temperatures, and different frequencies. Simulations are performed to prove the efficiency of the proposed designs. The results demonstrate the advantage of the proposed designs with a reduction over 73\% in terms of transistors count for the THA and over 88\%, 99\%, 98\%, 84\%, 98\%, and 99\% in energy consumption for the STI, TNAND, TDecoder, TMUX, THA, and TMUL, respectively. 

Moreover, the noise immunity curve (NIC) and Monte Carlo analysis for major process variations (TOX, CNT Diameter, CNT's Count, and Channel length) were studied. The results confirmed that the third proposed THA3 and TMUL3 had higher strength and higher noise tolerance, among other designs.

In addition, the second objective is using ternary data transmission to improve data communications between computer hosts. Also, this thesis proposes a bi-directional circuit that contains two converters: (1) A binary-to-ternary converter and (2) a ternary-to-binary converter.

Finally, logical analysis and simulation results prove the merits of the approaches compared to existing designs in terms of transistor count, reduced latency, and energy efficiency.


\clearpage 

\lhead{\emph{Contents}} 
\tableofcontents 

\lhead{\emph{List of Tables}} 
\listoftables 

\lhead{\emph{List of Figures}} 
\listoffigures 


\clearpage 

\setstretch{1.5} 

\lhead{\emph{Abbreviations}} 

\listofsymbols{ll} 
{
\textbf{BJT} & \textbf{B}ipolar \textbf{J}unction \textbf{T}ransistor\\
\textbf{CMOS} & \textbf{C}omplementary \textbf{MOSFET}\\
\textbf{CNT} & \textbf{C}arbon \textbf{N}anotube \textbf{T}ube\\
\textbf{CNFET} & \textbf{C}arbon \textbf{N}anotube \textbf{FET}\\
\textbf{CPL} & \textbf{C}omplementary \textbf{P}ass transistor \textbf{L}ogic\\
\textbf{CPU} & \textbf{C}entral \textbf{P}rocessing \textbf{U}nit \\
\textbf{DPL} & \textbf{D}ouble \textbf{P}ass transistor \textbf{L}ogic\\
\textbf{FET} & \textbf{F}ield \textbf{E}ffect \textbf{T}ransistor\\
\textbf{FinFET} & \textbf{F}in \textbf{FET}\\
\textbf{HDD} & \textbf{H}ard \textbf{D}isk \textbf{D}rive \\
\textbf{MOSFET} & \textbf{M}etal \textbf{O}xide \textbf{S}emiconductor \textbf{FET}\\
\textbf{MVL} & \textbf{M}ulti-\textbf{V}alued \textbf{L}ogic \\
\textbf{NIC} & \textbf{N}oise \textbf{I}mmunity \textbf{C}urve \\
\textbf{NTI} & \textbf{N}egative \textbf{T}ernary \textbf{I}nverter\\
\textbf{PTI} & \textbf{P}ositive \textbf{T}ernary \textbf{I}nverter\\
\textbf{RAM} & \textbf{R}andom \textbf{A}ccess \textbf{M}emory \\
\textbf{STI (TNOT)} & \textbf{S}tandard \textbf{T}ernary \textbf{I}nverter\\
\textbf{TAND} & \textbf{T}ernary \textbf{A}ND\\
\textbf{TNAND} & \textbf{T}ernary \textbf{N}AND\\
\textbf{TOR} & \textbf{T}ernary \textbf{O}R\\
\textbf{TNOR} & \textbf{T}ernary \textbf{N}OR\\
\textbf{TDecoder} & \textbf{T}ernary \textbf{D}ecoder\\
\textbf{TMUX} & \textbf{T}ernary \textbf{M}ultiplexer\\
\textbf{TALU} & \textbf{T}ernary \textbf{A}rithmetic \textbf{L}ogic \textbf{U}nit\\
\textbf{THA} & \textbf{T}ernary \textbf{H}alf \textbf{A}dder\\
\textbf{TMUL} & \textbf{T}ernary \textbf{M}ultiplier\\
\textbf{PDP} & \textbf{P}ower \textbf{D}elay \textbf{P}roduct (Energy Consumption)\\
}


\clearpage 

\lhead{\emph{Symbols}} 

\listofnomenclature{ll} 
{
$a_0$ &The inter-atomic distance between each carbon atom and its neighbor\\
$a$& The carbon to carbon atom distance\\
$b$& bit (Binary digit)\\
$d$ & The number of digits\\
$Dcnt$ &The CNT diameter\\
$e$&The electron charge unit\\
$R$& The base or radix of the numerical system \\
$t$& trit (Ternary digit)\\
$V\pi$ &The carbon bond energy

}




\clearpage 

\mainmatter 

\pagestyle{fancy} 



\chapter{Introduction} 

\label{Chapter1} 

\lhead{Chapter 1. \emph{Introduction}} 

``If it would have been possible to build reliable ternary architecture, everybody would be using it.” - Donald Ervin Knuth, a famous computer scientist at Stanford University~\cite{100}.

\section{Binary Logic System}
The two-valued logic system (Binary) dominates in circuit designs because the transistor has two states ``ON'' or ``OFF''. However, Binary systems have disadvantages in the design, which require a large number of interconnections around 70\% of the total area, 20\% for insulation, and only 10\% for components \cite{101}. Therefore, Binary systems suffer from high-power consumption. Whereas, Multi-Valued Logic (MVL) circuits have more than two-valued logic, which reduces the interconnections, chip area, and power consumption \cite{102}. 

\section{Multi-Valued Logic (MVL) System}
Multi-Valued Logic (MVL) returns to ancient Mesopotamian cultures. Five thousand years ago, the Babylonians and Sumerians used the sexagesimal system (base 60)~\cite{103}.

The first algebra of k-valued logic corresponding to the work of Post was known as Post Algebra of order k, where k $\geqslant$ 2 \cite{104}. Post algebra coincides with Boolean algebra for~k=~2. 

In 1958, Nikolai P. Brusenzov and his team constructed the world’s first and still unique ternary computer (called Setun) at the University of Moscow \cite{105}. Setun gained much interest among scientists. Unfortunately, the development of ternary architectures was not keeping up with the speed of the binary counterparts because a transistor has two stable states: ”ON” or ”OFF”.

\begin{figure}[!t]
\centering
\includegraphics[width=12cm]{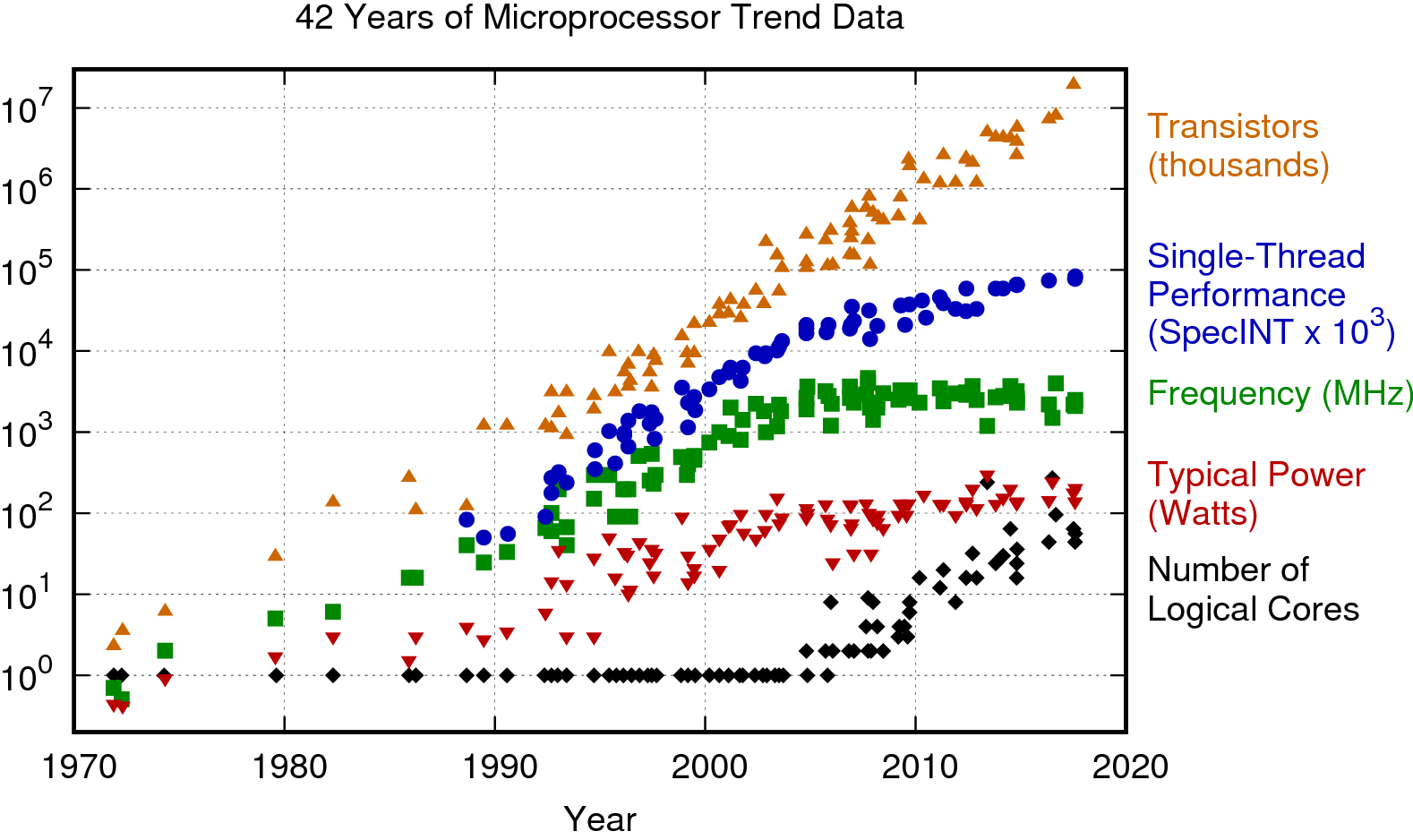}
\caption{Limitation of the CPU Clock Speed and its power consumption. }
\label{limitclock}
\end{figure}

Nowadays, many researchers believe that binary logic is doomed because of: 
\begin{enumerate}
  \item Moore's Law 1965 (transistors will double every two years) \cite{106} is stopped because the shrinkage sizes of transistors indefinitely is not possible
  \item Limitation of CPU Clock Speed and its power consumption are shown in Fig.~\ref{limitclock}
 \end{enumerate}

As shown in Fig. \ref{limitclock}, Since 2011, the transistor count inside the CPU core can not exceed 10 million transistors, and the CPU clock speed and power consumption remain constant since 2010.

Therefore recently, MVL has attracted researchers’ attention again over binary logic. MVL can be implemented almost in all topics. For example, in Algorithm (software), Artificial Intelligence, Big Data \cite{107}, MV-Quantum Logic \cite{108}, Wireless Sensor Networks \cite{109}, Cloud Vehicular Networks \cite{110}, Communication systems \cite{111}, Programmable Logic Arrays (PLAs), and circuit designs like Logic Gates, Arithmetic, Memory, and Memristor circuits \cite{112, 113,114}.

Example of MVL, Samsung Ternary Memory (1995), Intel Strata Flash Memory ( 2 bits per cell) (1997), Sony Ternary Flash Memory (2010), Samsung Multi-Level Cell NAND Flash Memory (2013), CISCO SF/SG 300 Series Managed Switches (2018), TOSHIBA hexa-valued logic memory (2018), Intel and Micron hexa-valued logic SSD 1 TB (2018), Samsung hexa-valued logic SSD 1 TB (2020).

\section{Ternary Logic System}
The number to represent the range $N$ from 0 to $R^{d-1}$ of any numerical system is given by the equation \eqref{1eq1}:

\begin{equation}
\label{1eq1}
N = R^d
\end{equation}
Where $R$ is the base or radix of the numerical system and $d$ is the number of digits.

The cost and complexity $C$ of the system is proportional to the digit. The capacity of the digits is ($R \cdot d$) then $C$ is given by the equation \eqref{1eq2}:
\begin{equation}
\label{1eq2}
C = k (R \cdot d) = k (R \cdot \frac{logN}{logR}) = k logN (\frac{R}{logR})
\end{equation}
Where $k$ and $logN$ are constant numbers.

To minimize the cost and complexity $C$, we must differentiate the equation \eqref{1eq2} with respect to $R$ and set it to zero, then:

\begin{align*}
\frac {\partial C}{\partial R}= \frac{\partial R}{\partial logR} = \frac{\frac {\partial R}{\partial R} logR - R \frac {\partial logR}{\partial R}}{(logR)^2} = \frac{logR - R (1/R)}{(logR)^2} = \frac{logR - 1}{(logR)^2} = 0 
\end {align*}
then $logR - 1 = 0 $ implies that $R=e=2.718$.\\
Since radix $R$ is an integer, then $R = 3$. Therefore, the ternary logic system is the most efficient in circuit complexity and cost compared to other bases \cite{115}.

Ternary digit, called trit, can be represented in two ways: balanced ternary logic~(-1,~0,~1) corresponding to~(-Vdd, 0, Vdd), and standard (unbalanced) ternary logic~(0, 1, 2) corresponding to~(0, Vdd/2, Vdd).

\subsection{Relation between Binary and Ternary Digit}
For the same amount of data, n-bits or m-trits are represented in the equation \eqref{r1}:
\begin{equation} \label{r1}
2^n=3^m
\end{equation}
This implies that
\begin{equation} \label{r2}
n=\frac{log3}{ log2} m=1.585  m
\end{equation}
Thus, the equation \eqref{r2} shows that a single trit can represent 58.5\% more data than a single bit.

For example, a decimal number 2186 is 1000 1000 1010 (12 bits) in binary whereas, in ternary equivalent is 2 222 222 (7 trits). Conclude that the ternary system has a notable reduction in wires around 41.7\% [(12-7)*100/12] compared to the binary system.

Therefore, Ternary systems can help to solve the problem of `` Cloud storage, big data, artificial intelligence, technical computing, and ever-richer digital content demand massive storage capacity and exceptional performance.'' 


\clearpage
\section{Transistor Technologies}
This section is summarizing the transistor technologies \cite{118}. 
The transistor, a semiconductor, is one of the most important devices used in electronic applications. It is mainly applied in two ways: (1) amplification, and (2) switching. In digital logic circuits, like in this thesis, it will be used as a switch.

Transistor has at least three terminals: The first one is for voltage supply or output [called Drain (D) or Collector (C)], the second one is for input voltage or controller [called Gate (G) or Base (B)], and the third one is for the output or ground [called Source (S) or Emitter (E)]. The terminals names depend on the transistor technology which will be discussed later.

Transistor is manufactured from a material like Silicon, Germanium, Carbon, or others.

\subsection{BJT}
The Bipolar Junction Transistor (BJT) invented in 1947 at the Bell Telephone Laboratories.
The BJT is a three-layer device. It is classified according to the types of the layers as NPN or PNP transistors.
NPN transistors are constructed by sandwiching P-type layer between two N-type layers, while PNP transistors are constructed by sandwiching N-type layer between two P-type layers, as shown in Fig. \ref{BJT}:

\begin{figure}[!b]
  \begin{center}
      \includegraphics [width=11.5cm]{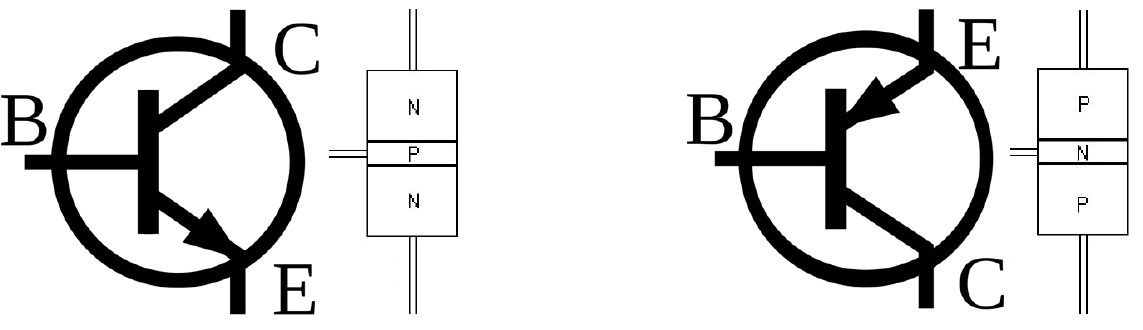}
  \end{center}
  \caption{BJT Transistor schematic symbols.}
\label{BJT}
\end{figure}  

\pagebreak  
\setstretch{1}
\textbf{Advantages:}
\begin{enumerate}
  \item BJT has a large gain bandwidth.
  \item It shows better performance at high frequency.
  \item It has a better voltage gain.
  \item It can be operated in low or high power applications.
  \item It has a high current density.
  \item There is a low forward voltage drop.
 \end{enumerate}
\textbf{Disadvantages:}
\begin{enumerate}
  \item BJT is sensitive of radiation.
  \item It has a very complex base control. So it may lead to confusion and requires skillful handling.
  \item The switching speed of a BJT is low.
  \item It produce more noise.
  \item It has low thermal stability.
 \end{enumerate}
\setstretch{1.5}

\subsection{FET}
Julius Edgar Lilienfeld first patented the concept of a Field-Effect Transistor (FET) in 1925. The first FET device to be successfully built was the Junction Field-Effect Transistor (JFET) by Heinrich Welker in 1945 \cite{118a}.

\subsubsection{MOSFET}
The Metal–Oxide–Semiconductor Field-Effect Transistor (MOSFET), unipolar transistor, was invented by Mohamed Atalla and Dawon Kahng in 1959. The MOSFET largely superseded both the BJT and the JFET. 
Apart from the terminals (Drain, Gate, Source), there is a substrate (Sub) or a body.

MOSFET utilizes an insulator (typically SiO2) between the gate and the body which provides high input impedance, thus capturing all the input signal.

\begin{figure}[!b]
  \begin{center}
    \includegraphics [width=12cm]{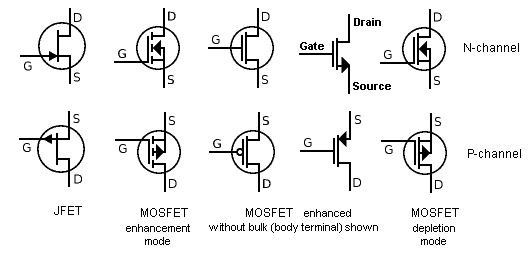}
  \end{center}
  \caption{MOSFET Transistor schematic symbols. }
\label{transistor3}
\end{figure}

MOSFET schematic symbols are shown in Fig. \ref{transistor3}.   

\subsubsection{CMOS}
CMOS (Complementary MOS) thechnology was invented at Fairchild Semiconductor in 1963 \cite{118c}. It is used in microprocessors, microcontrollers, memory chips, and other digital logic circuits. Also, It is used in several analog circuits such as data converters and image sensors (CMOS sensor).

\textbf{CMOS Advantages:}
\begin{enumerate}
  \item MOSFETs provide greater efficiency while operating at lower voltages.
  \item Absence of gate current results in high input impedance producing high switching speed.
  \item Lowest power dissipation of all gates (a few nW).
  \item Huge fan-out capability ($>$50).
  \item Very high noise-immunity and noise-margin (typically, VDD/2).
  \item Lower propagation delay than NMOS.
  \item Temperature stability is excellent.
   \end{enumerate}

\textbf{CMOS Disadvantages:}
\begin{enumerate}
  \item The thin oxide layer makes the MOSFETs vulnerable to permanent damage when evoked by electrostatic charges.
  \item Overload voltages make it unstable.
  \end{enumerate}

\subsubsection{FinFET}
Hitachi Central Research Laboratory in 1989 first fabricated a Fin Field-Effect Transistor (FinFET) \cite{118f}.
It is a multigate device and has faster switching times than the CMOS.

FinFET is a "3D" transistor or a type of non-planar transistor. It is the basis for nanoelectronic and becomes the dominant gate design at 14 nm, 10 nm, and 7 nm process nodes.

\subsubsection{CNFET}
A Carbon Nanotube Field-Effect Transistor (CNFET) was first demonstrated in 1998. It refers to a FET that uses a single carbon nanotube or more as the channel material instead of bulk silicon in the traditional MOSFET structure.

\textbf{CNFET Advantages:}
\begin{enumerate}
  \item Very fast switching time
  \item Better control over channel formation
  \item Customize the threshold voltage by changing CNT diameter
  \item High electron mobility
  \item High current density
  \item High transconductance
  \item High linearity
   \end{enumerate}
   
   \textbf{CNFET Disadvantages:}
\begin{enumerate}
  \item Lifetime (degradation): Carbon nanotubes degrade in a few months.
  \item Reliability issues when operated under high electric field or temperature gradients.
  \item Production cost.
  \end{enumerate}

Among the techniques mentioned above, CNFET provides the best trade-off in terms of energy efficiency and circuit speed \cite{119}.

Details about the Stanford CNFET model used in this thesis can be found in \cite{120,121,122}. However, it is worth mentioning that the CNFETs use a semiconducting single-walled CNT as a channel for conduction with high drive current 35 $\mu$ A when the supply voltage Vdd is equal to 0.9 V. 

\begin{figure}[!tb]
  \begin{center}
    \includegraphics [width=15cm]{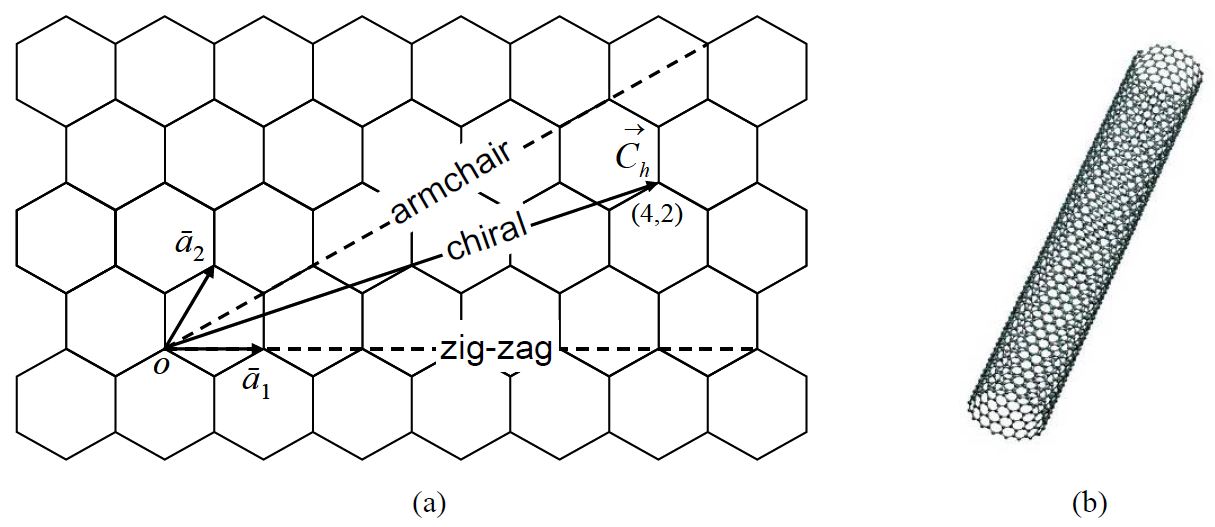}
  \end{center}
  \caption{(a) Unrolled graphite sheet and (b) the rolled carbon nanotube lattice structure.}
\label{chirality}
\end{figure}

A single-walled carbon nanotube (SWCNT) can be visualized as a sheet of graphite which is rolled up and joined together along a wrapping vector $C_h = i.\bar{a_1} + j.\bar{a_2}$, where $[\bar{a_1}, \bar{a_2}]$ are lattice unit vectors as shown by Fig. \ref{chirality}, and the indices $(i,j)$ are positive integers that specify the chirality of the tube.

This chirality vector determines if the CNT is metallic or semiconducting; if $i$  =  $j$ or $i$–$j$  =  3$t$, where $t$ is an integer, then the nanotube is metallic else it is semiconducting.

The CNFET diameter can be calculated from equation \eqref{eq1}:
\begin{equation}
\label{eq1}
  D_{\mathrm{cnt}}  =  \frac{\sqrt{3}\cdot  a_0}{\pi} \sqrt{i^2+j^2+ij}
 \end{equation}
Where $a_0$  =  0.142 nm is the inter-atomic distance between each carbon atom and its neighbor, and the integer pair ($i$, $j$) represents the chirality vector.

The characteristics of the CNFET model are similar to traditional MOSFETs. Except for the threshold voltage, which is calculated by the equation \eqref{eq2}:
\begin{equation}
\label{eq2}
  V_{\mathrm{th}}  = \frac{\sqrt{3}}{3} \frac{a\cdot V\pi}{e\cdot Dcnt}
 \end{equation}
Where $a$  =  2.49 Å is the carbon to carbon atom distance, $V\pi$  =  3.033 eV is the carbon bond energy in the tight binding model, $e$ is the electron charge unit, and $Dcnt$ is the CNT diameter.

In general, three chiralities can be used in the ternary logic design. The relationship between the chirality, diameter, and threshold voltage are shown in Table~\ref{t1}.

\renewcommand{\arraystretch}{1.2}
\begin{table} [!t]
\caption{The relation between the chirality, diameter, and threshold voltage.}
\label{t1}
\setlength{\tabcolsep}{3pt}
\centering
\begin{tabular}{cccc}
\hline \hline
&&\multicolumn{2}{c}{\textbf{Threshold voltage}}\\
\cline{3-4}
Chirality & CNT diameter & N-CNFET & P-CNFET\\
\hline
(10,0)	&0.783nm&	0.559V&	- 0.559V\\
(13,0)	&1.018nm&	0.428V&	- 0.428V\\
(19,0)	&1.487nm&	0.289V&	- 0.289V\\
\hline \hline
\end{tabular}
\end{table}
\renewcommand{\arraystretch}{1}

Table~\ref{1t6} shows some essential parameters of the CNFET model used in all proposed designs for embedded systems with brief descriptions.

\renewcommand{\arraystretch}{1.3}
\begin{table}[h!]
\caption{CNFET model parameters}
\label{1t6}
\setlength{\tabcolsep}{2pt}
\centering
\begin{tabular}{p{40pt}|p{250pt}|p{40pt}}
\hline \hline
	&	Description																					&	Value\\
\hline
$L_{ch}$	&			Physical channel length																	&	32 nm\\
$L_{geff}$&			The mean free path in the intrinsic CNT													&	100 nm\\
$L_{ss}$ ($L_{dd}$)&		The length of doped CNT source-side (drain-side) extension region 						&		10 nm\\
				
$Efi$			&	The Fermi level of the doped tube													&		0.6 eV\\
$K_{gate}$		&	The dielectric constant of high-k top gate dielectric material (planer gate) 				&			4\\
				
$T_{ox}$			&	The thickness of the high-k top gate dielectric material (planer gate) 						&		1 nm\\
					
$C_{sub}$		&	The coupling capacitance between the channel region and the substrate  &	10 pF/m\\
					
$C_{csd}$		&	The coupling capacitance between the channel region and the source/drain region 				&0 pF/m\\
					
$Pitch$		&	The distance between two adjacent CNTs within the same device 								&20 nm\\
					
$Tubes$		&	The number of tubes																&1\\
\hline \hline
\end{tabular}
\end{table}
\renewcommand{\arraystretch}{1}
\clearpage
\section{Power Consumption}
The power consumption in a transistor is composed of two types: Dynamic and Static.

Dynamic power is the power consumed while charging and discharging the capacitive nodes in the process of switching. 

Static power essentially consists of the power used when the transistor is not in the process of switching. If Diode-Connected transistors exist, then the transistors will act as Resistors. Thus, Joule effect power is created and generated heat in the circuit.

Total Power consumption	: $P= P_s + P_d + P_r$

Static Power				: $P_s= N* Vdd * Id$ 

Dynamic Power			: $P_d= N* Vdd^2* f * CL$

Joule effect Power			: $P_r= N*R*Id^2$ 

Where,\\
 $N$ 	: number of transistors in the circuit\\
$Vdd$: Power Supply\\
$Id$	: Current in transistors\\
$f$ 	: Frequency of the Vinput\\
$CL$	: Load Capacitor\\ 
$R$ 	: Resistor value of the Diode-Connected Transistor\\

\clearpage
\section{Logic Gates}
Logic gates are the main blocks of digital Integrated Circuits (IC) and they execute basic logical functions. They have one or more inputs and provide only one output \cite{201}. 

A one-input and one-output logic gate is also called a Unary operator.

\subsection{Unary Operators of MVL Systems}
The literature about unary operators of MVL is limited. It is formally defined in \cite{117} as the number of one-place, or unary, functions of a $p$-valued logic system which is $p^p$. 

For binary systems where $p$  =  2, there are only four unary functions, as shown in Table \ref{U1}, the identity ``01'', negation ``10'', and the two constant functions ``00'' and ``11''. Whereas, for ternary systems where $p$  =  3, there are twenty-seven ($3^3$) unary functions, as shown in Table \ref{U2}. 

\renewcommand{\arraystretch}{1}
\begin{table} [!b]
\caption{Unary operators of binary logic system}
\label{U1}
\setlength{\tabcolsep}{5pt}
\centering
\begin{tabular}{c|cccc}
\hline \hline
Binary input $X$ & $f_1(X)$ & $f_2(X)$& $f_3(X)$& $f_4(X)$\\
\hline
Logic 0 (0V)	&0&	0& 1& 1\\
Logic 1 (Vdd)	&0&	1&	0& 1\\
\hline \hline
\end{tabular}
\end{table}
\renewcommand{\arraystretch}{1}

\renewcommand{\arraystretch}{1}
\begin{table} [!b]
\caption{Unary operators of ternary logic system}
\label{U2}
\setlength{\tabcolsep}{1pt}
\centering
\begin{tabular}{l|ccc|ccc|ccc|ccc|ccc|ccc|ccc|ccc|ccc}
\hline \hline
Ternary &&&&&&&&&&&&&&&&&&&&&&&&&&\\
 input $A$ & $f_1$ & $f_2$& $f_3$ & $f_4$ & $f_5$& $f_6$ & $f_7$ & $f_8$& $f_9$ & $f_{10}$ & $f_{11}$& $f_{12}$ & $f_{13}$ & $f_{14}$& $f_{15}$ & $f_{16}$ & $f_{17}$& $f_{18}$ & $f_{19}$ & $f_{20}$& $f_{21}$ & $f_{22}$ & $f_{23}$& $f_{24}$ & $f_{25}$ & $f_{26}$& $f_{27}$\\
\hline
0 (0V)	&0&0&0&0&0&0&0&0&0  &1&1&1&1&1&1&1&1&1  &2&2&2&2&2&2&2&2&2\\
1 (Vdd/2)&0&0&0& 1&1&1 &2&2&2 &0&0&0& 1&1&1 &2&2&2 &0&0&0& 1&1&1 &2&2&2\\
2 (Vdd)	&0&1&	2 &0&1&2 &0&1&2 &0&1&2 &0&1&2 &0&1&2 &0&1&2 &0&1&2 &0&1&2\\
\hline \hline
\end{tabular}
\end{table}
\renewcommand{\arraystretch}{1}
This thesis will use, in the next chapters, some of Unary Operators to design Ternary Combinational circuits like Ternary Decoder, and Ternary Multiplexer. Also,  Ternary Arithmatic Logic Unit  (TALU) like Ternary Half-Adder, and Ternary Multuplier. 

The use of unary operators in our model designs is one of our novelty techniques.

\subsection{Binary Logic Gates}
Binary digit, called bit, can be represented as (0, 1) corresponding to (0 V, Vdd), (Low, High), or (False, True).

\subsubsection{Inverter Gate}
A NOT gate performs a unary function and is often called an inverter or negation. It has a single input $A$ and a single output $\bar{A}$, which is the complement of $A$, as shown in~Fig.~\ref{Bgates}(a).
 
In general, the equation of $\bar{A}$ is equal to $\bar{A}= (R-1)-A$. In binary $R=2$, then $\bar{A}=1-A$. 

When the input is true (Vdd), then the output will be false (0). Similarly, a false input results in a true output.

Table \ref{ti} shows the truth table of a NOT gate.

\renewcommand{\arraystretch}{1.2}
\begin{table} [!h]
\caption{Binary NOT truth table}
\label{ti}
\setlength{\tabcolsep}{5pt}
\centering
\begin{tabular}{c|c}
\hline \hline
Input $A$ & $\bar{A}$\\
\hline
Logic 0 (0V)	&1\\
Logic 1 (Vdd)	&0\\
\hline \hline
\end{tabular}
\end{table}
\renewcommand{\arraystretch}{1}

\subsubsection{AND Gate (Minimum function)}

An AND gate has more than one input and one output. The output is equal to the minimum of all its inputs.
The output is true only when all of the inputs are true. Whereas, the output is false when at least one of its inputs is false, as shown in Fig.~\ref{Bgates}(b).

Table \ref{tbgates} shows the truth table for a two-input AND gate.

\subsubsection{NAND Gate}
A NAND gate has more than one input and one output. It is the complement of the AND gate.
The output is true when at least one of its inputs is false. Whereas, the output is false only when all inputs are true, as shown in Fig.~\ref{Bgates}(c).

Table \ref{tbgates} shows the truth table for a two-input NAND gate.

\subsubsection{OR Gate (Maximum function)}
An OR gate has more than one input and one output. The output is equal to the maximum of all its inputs.
The output is true when at least one of its inputs is true. Whereas, the output is false only when all of the inputs are false, as shown in Fig.~\ref{Bgates}(d).

Table \ref{tbgates} shows the truth table for a two-input OR gate.

\subsubsection{NOR Gate}
\begin{figure}[!tb]
  \begin{center}
    \includegraphics [height=20cm]{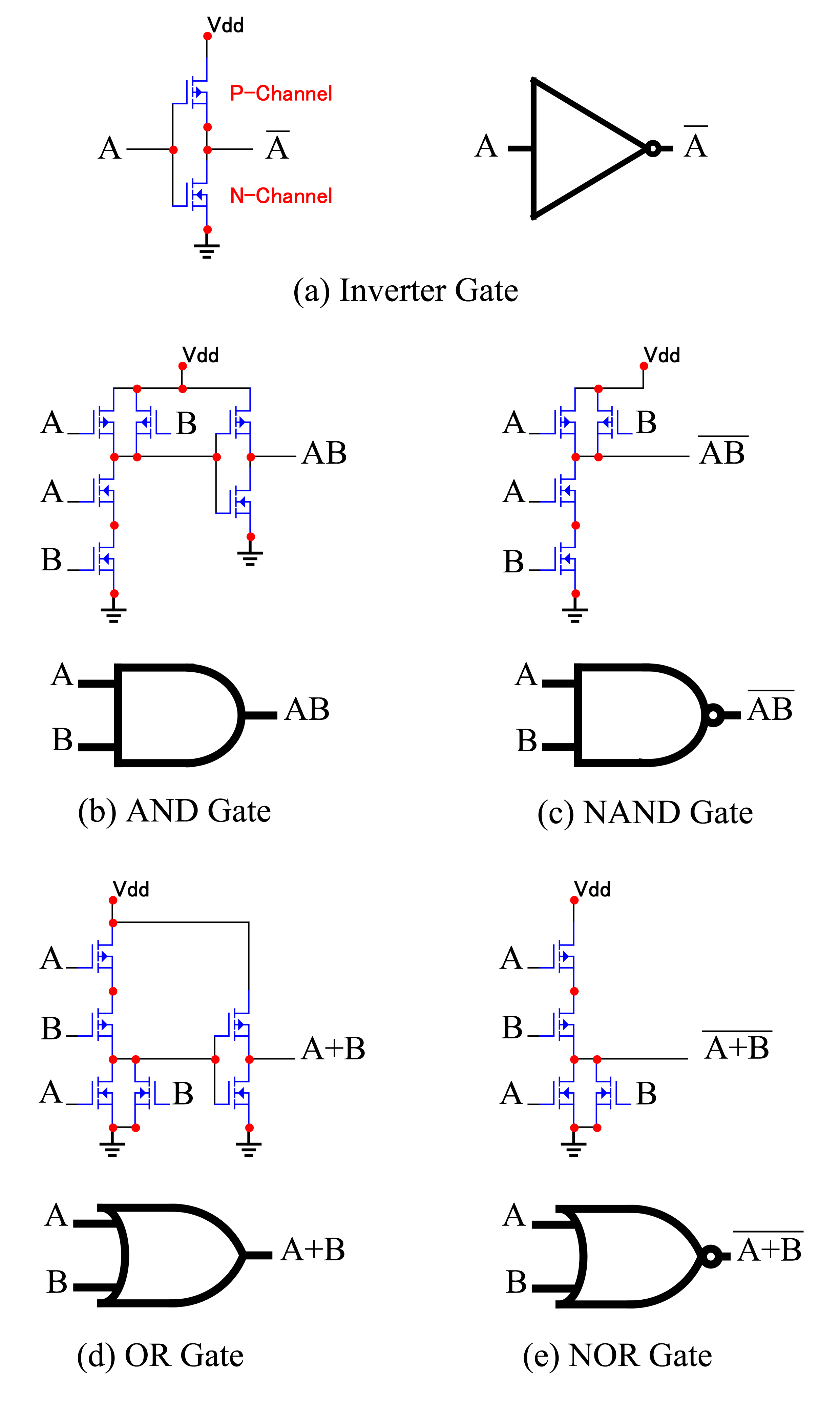}
  \end{center}
  \caption{Basic CMOS Binary Logic Gates.}
\label{Bgates}
\end{figure}

A NOR gate has more than one input and one output. It is the complement of the OR gate.
The output is true only when all of the inputs are false. Whereas, the output is false when at least one of its inputs is true, as shown in Fig.~\ref{Bgates}(e).

Table \ref{tbgates} shows the truth table for a two-input NOR gate.
\renewcommand{\arraystretch}{1.4}
\begin{table} [!h] 
\caption{Truth table of some basic binary gates}
\label{tbgates}
\setlength{\tabcolsep}{7pt}
\centering
\begin{tabular}{cc|cccc}
\hline \hline
\multicolumn{2}{c|}{Input}& \multicolumn{4}{c}{Output}\\
$A$ & $B$& AND & NAND & OR & NOR\\
\hline
0&0&0&1&0&1\\
0&1&0&1&1&0\\
1&0&0&1&1&0\\
1&1&1&0&1&0\\
\hline \hline
\end{tabular}
\end{table}
\renewcommand{\arraystretch}{1}

\clearpage
\subsection{Ternary Logic Gates}
Ternary digit, called trit, can be represented in two ways: balanced ternary logic~(-1,~0,~1) corresponding to~(-Vdd, 0, Vdd), and standard (unbalanced) ternary logic~(0, 1, 2) corresponding to~(0, Vdd/2, Vdd).

The basic ternary logic gates are TNOT, TAND, TOR, TNAND and TNOR. The ternary equations of the basic ternary logic gates are presented in \eqref{eq25}:

\begin{equation} 
\label{eq25}
\begin{aligned}
TNOT (STI) &: &\bar{A} &= 2 - A, \\
TAND &: &A \bullet B &= min\{A, B\}, \\
TNAND &: &\overline{A \bullet B} &= \overline{min\{A, B\}}, \\
TOR &: &A + B &= max\{A, B\}, \\
TNOR  &: &\overline{A + B} &= \overline{max\{A, B\}}, \\
\end{aligned}
\end{equation}  

Where $A$ and $B$ are ternary inputs.

\subsubsection{Ternary Inverter Gates}
Ternary Inverter Gates are unary functions (one input one output). There exist three types of ternary inverters, the first is a Positive Ternary Inverter (PTI), $A_p$, the second is a Negative Ternary Inverter (NTI), $A_n$, and the third one is a standard ternary inverter (STI), which is the complement of A, $\bar{A}$, as shown in Fig. \ref{Tgates}(a, b, c).

When the ternary input $A$ is equal to logic 0 (0V) or 2 (Vdd), then the output of the three ternary inverters is the same as the binary inverter: 0V input becomes Vdd as output and vice versa. 

However, when $A$ = logic 1 (Vdd/2), the PTI output will be logic 2, the NTI output will be logic 0, and the STI output will still logic 1.  

Table \ref{tit} shows the truth table of the three ternary inverters.

\renewcommand{\arraystretch}{1.2}
\begin{table} [!h]
\caption{Ternary inverters truth table}
\label{tit}
\setlength{\tabcolsep}{5pt}
\centering
\begin{tabular}{c|ccc}
\hline \hline
\multicolumn{1}{c|}{Ternary Input}& 	PTI &	NTI &	STI \\
\multicolumn{1}{c|}{$A$} 	& 	$A_p$ & $A_n$ &  $\bar{A}$  \\
\hline
Logic 0 (0V)	  &2&	2&	2  \\
Logic 1 (0.45V)	  &2&	0&	1  \\
Logic 2	(0.9V)    &0&	0&	0  \\
\hline \hline
\end{tabular}
\end{table}
\renewcommand{\arraystretch}{1}

\subsubsection{Other Ternary Logic Gates}
The Ternary AND (TAND), Ternary NAND (TNAND), Ternary OR (TOR), and Ternary NOR (TNOR) are described by their truth table in Table \ref{tgates}.

\begin{figure}[!tb]
  \begin{center}
    \includegraphics [height=20cm]{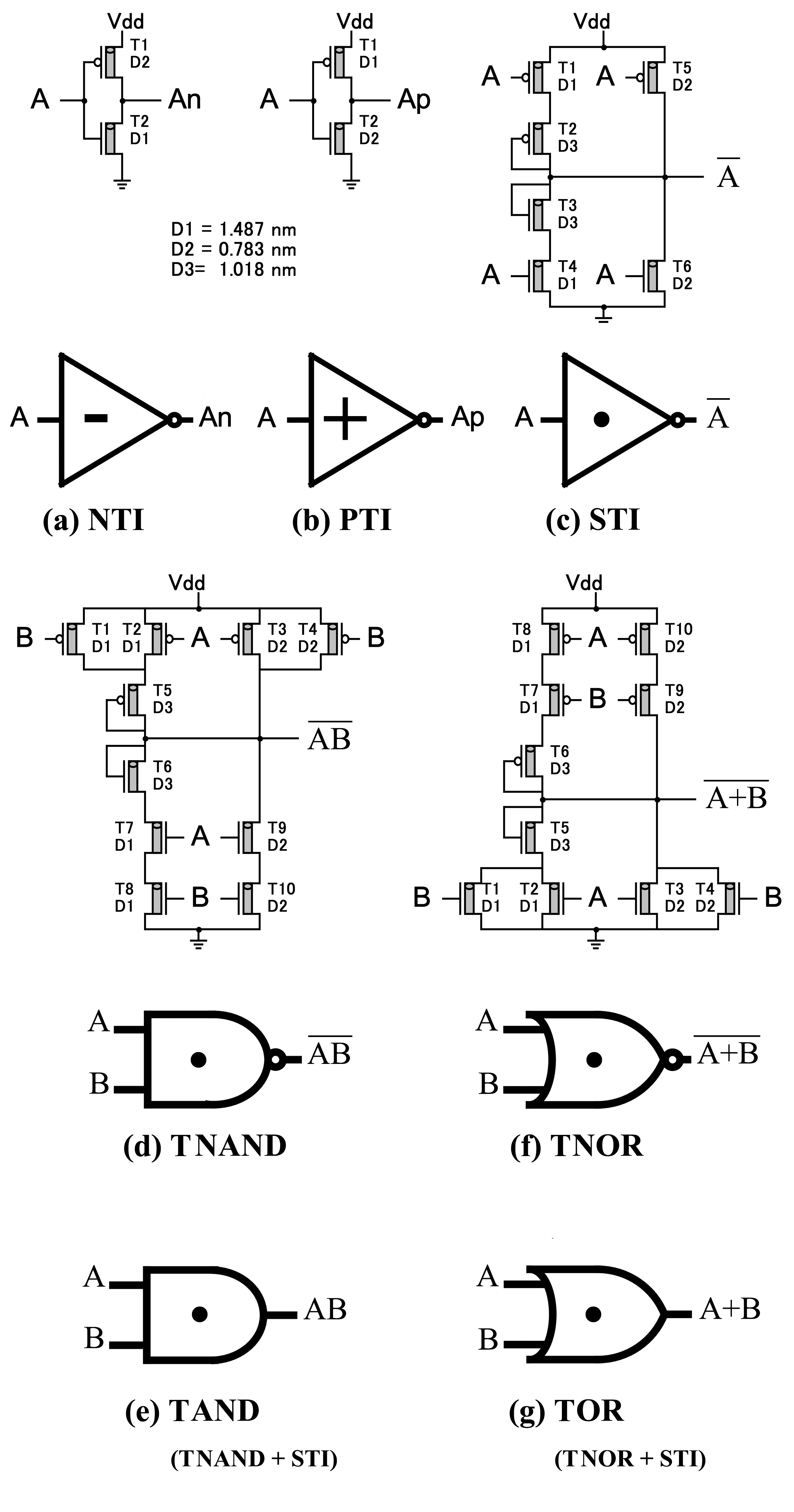}
  \end{center}
  \caption{Basic Ternary Logic Gates \protect\cite{123} using CNFET.}
\label{Tgates}
\end{figure}

The transistor level of basic Ternary Logic Gates are shown in Fig. \ref{Tgates}.

\renewcommand{\arraystretch}{1.4}
\begin{table}[h!]
\caption{Truth table of some basic ternary gates}
\label{tgates}
\setlength{\tabcolsep}{7pt}
\centering
\begin{tabular}{cc|cccc}
\hline \hline
\multicolumn{2}{c|}{Input}& \multicolumn{4}{c}{Output}\\
$A$ & $B$& TAND & TNAND & TOR & TNOR\\
\hline
0&0&0&2&0&2\\
0&1&0&2&1&1\\
0&2&0&2&2&0\\
1&0&0&2&1&1\\
1&1&1&1&1&1\\
1&2&1&1&2&0\\
2&0&0&2&2&0\\
2&1&1&1&2&0\\
2&2&2&0&2&0\\
\hline\hline
\end{tabular}
\end{table}
\renewcommand{\arraystretch}{1}

\clearpage
\subsection{Double Pass-transistor Logic (DPL)}
\begin{figure}[h!]
\centering
\includegraphics[width=12cm]{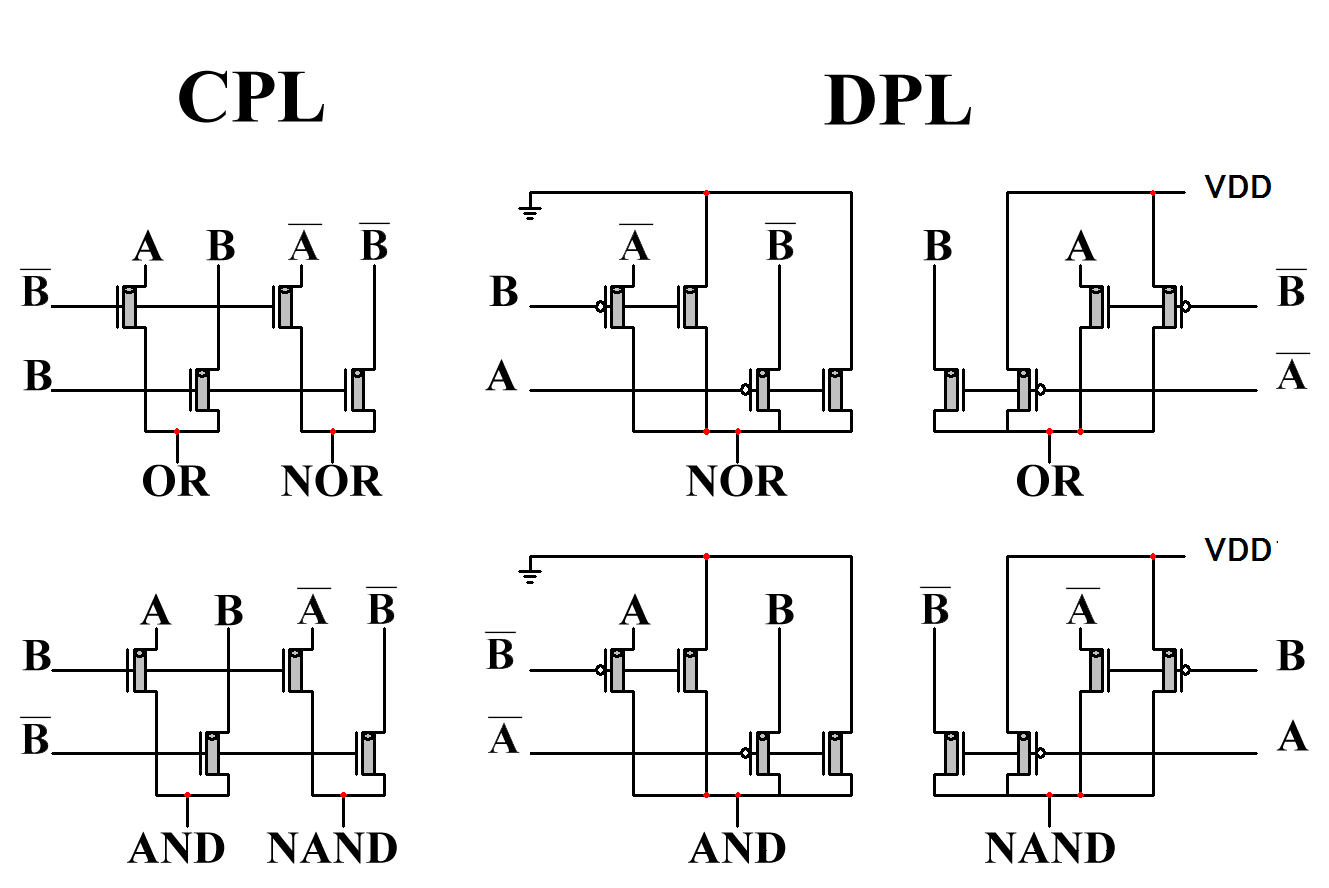}
\caption{DPL Gates Design \protect\cite{124}}
\label{DPL}
\end{figure}

Fig.\ref{DPL} shows the Double Pass-transistor Logic (DPL) \cite{124}, which is a modified version of Complementary Pass-transistor Logic (CPL) by adding P-MOSFET in parallel with the N-MOSFET to eliminate the problems of noise margin and speed degradation at reduced supply voltages associated in CPL circuits. Besides, basic logic gates using DPL ideally have identical propagation delay.
 
\clearpage
\section{Objectives}
This thesis has two Objectives to achieve:
 
\subsection{The First Objective}
The first Objective is to decrease the energy consumption of portable electronics and embedded systems to preserve their battery consumption. 

\subsection{The Second Objective}
The second objective is to improve Data Transmission between End devices.


\section{Methodology}
In order to decrease the energy consumption and to improve Data Transmission between End devices, there are five approaches considered for our designs.

\subsection{Choosing the High-Performance Transistor Technology}
Conduct a survey to choose the high-performance transistor technology and we found that Carbon Nanotube Field-Effect Transistor (CNFET) provides the best trade-off in terms of energy efficiency and circuit speed \cite{119} among other transistor technology like CMOS, FinFET, and others.

Therefore, this thesis uses the Stanford CNFET model in \cite{120,121,122} with Vdd is equal to~0.9~V.

\subsection{Utilizing Ternary Logic System}
As described before that the ternary logic system is the most efficient in circuit complexity and cost compared to other bases \cite{115}.

Therefore, this thesis uses ternary logic values (0, 1, 2) corresponding to (0, Vdd/2, Vdd) = (0 V, 0.45 V, 0.9 V).

\subsection{Applying the Dual Supply Voltages}
In general, to get logic 1 (0.45 V) in the ternary circuit from one power supply~Vdd~(0.9~V), two diode-connected transistors that act like resistors must be added. The use of two transistors is necessary to create a voltage divider in order to produce the ternary logic 1. The results show a notable increase in the static power dissipation due to the direct current from Vdd to the ground \cite{206}. 

Therefore, this thesis uses dual supply voltages (Vdd, Vdd/2) from external power supply to eliminate these two transistors that produce high power consumption and heat.

\subsection{Implementing Energy-Efficient Transistor Arrangement}
This thesis uses the best topology transistor arrangement according to paper \cite{tg} entitled ``Impact of Different Transistor Arrangements on Gate Variability'' published in 2018, to increase the robustness of the design regarding the process variability issues. 

In general, the best choice is to use the network topology that the transistor in series is as far as possible to the output node. Also, circuit designs with parallel transistors like transmission gates provide better performance.   

\subsection{Reducing the number of used transistors}
The minimization of transistors count is not the only factor that affects the performance of the proposed circuits, but it is a good factor.

As illustrated before that total Power consumption	: $P= P_s + P_d + P_r$

Static Power				: $P_s= N* Vdd * Id$ 

Dynamic Power			: $P_d= N* Vdd^2* f * CL$

Joule effect Power			: $P_r= N*R*Id^2$ 

Where,\\
 $N$ 	: number of transistors in the circuit\\
$Vdd$: Power Supply\\
$Id$	: Current in transistors\\
$f$ 	: Frequency of the Vinput\\
$CL$	: Load Capacitor\\ 
$R$ 	: Resistor value of the Diode-Connected Transistor\\

Therefore, this thesis reduces the number of used transistors, as follows:\\
a- Eliminating the diode-connected transistors by using Vdd/2\\
b- Applying De Morgan's Law in designing THA1 and TMUL1, to replace for example, (3-AND 1-OR) logic gates with 24 transistors to a 4-NAND logic gate with 16 transistors \\
c- Using Transmission Gate (TG) in designing TMUX with 15 transistors instead of using TDecoder, which has 28 transistors\\
d- Using unary operators as blocks in our designs

\section{Thesis Outline}
This thesis is organized as follow: Chapters 2, 3, and 4 will use CNFET in the designs and HSPICE simulator. Whereas, in chapter 5 will use CMOS DPL gates and Microcap simulator.

Chapter \ref{Chapter2} ``Ternary Logic Gates'': This chapter represents some of the existing 27 unary operators and TNAND design circuits and proposes seven unary operators and TNAND logic gate. 

Chapter \ref{Chapter3} ``Combinational Logic Circuits'': This chapter represents the existing TDecoders and TMUXs design circuits and proposes two different TDecoders and TMUX. 

Chapter \ref{Chapter4} ``Ternary Arithmetic Logic Units (TALUs)'': This chapter represents the existing THA and TMUL design circuits and proposes three different desings of THAs and three TMULs. 

Chapter \ref{Chapter5} ``Ternary Data Transmission'': This chapter represents the proposed two converters: a Binary-to-Ternary converter and a Ternary-to-Binary converter. 

Chapter \ref{Chapter6} ``Conclusion and Perspectives''.


\chapter{Ternary Logic Gates} 

\label{Chapter2} 

\lhead{Chapter 2. \emph{Ternary Logic Gates}} 

Chapter 1 described the definition and types of ternary logic gates.\\
This chapter will propose seven unary operators ($A^1$, $A^2$, $\bar{A^2}$, $A_1$, $1.\bar{A_n}$, $1.\bar{A_p}$, and STI $\bar{A}$)  and TNAND, which are published in (\cite{paper1}, \cite{paper2}) and can be found in the Appendices (A, B). 

\section{Literature Review}
In the 1970s, the Unary Operators (one input and one output logic gate) of Ternary~Systems were first used to design a universal CMOS circuit for implementation of all 27 unary operators \cite{202,203,204}.  

Recently, the researchers utilized the unary operators to design many applications in ternary circuits like TDecoder, TMUX, THA, and TMUL.

In \cite{123}, the authors proposed CNFET-based three types of ternary inverters (NTI, PTI, STI) which are three unary operators of the ternary system, TNAND \& TNOR logic gates, and TDecoder using the proposed unary operators and TNOR.
In \cite{205}, the authors presented new unary circuits which are subsequently used to design a ``decoder-less'' (3:1) TMUX, a THA, and a TMUL.
While in \cite{t16, t18}, they improved circuit designs for the STI, and TNAND.

In addition, this thesis uses a dual voltage supply.
In general, to get logic 1 (0.45V) in the ternary circuit from one power supply Vdd (0.9V), two transistors must be added. The use of two transistors is necessary to create a voltage divider in order to produce the ternary logic 1. The results showed a notable increase in the static power dissipation due to the direct current from Vdd to the ground \cite{206}. \textbf{\emph{Therefore, this thesis uses Vdd/2 from the same power supply to eliminate these two transistors that produce high power consumption}}.

The new designs provide a considerable gain in terms of system performance compared to the designs in \cite{123, 205}, as demonstrated in the HSPICE-based simulation results and get the lowest energy consumption (power-delay product (PDP)).  

\section{The Proposed Unary Operators of Ternary Systems}
\label{ch2.2}
This thesis will use nine unary operators to design the ternary circuits, which are shown in Table \ref{t211} and derived from the equation \eqref{eq23}.

\renewcommand{\arraystretch}{1.2}
\begin{table}[b!]
\caption{Truth table of the selected nine unary operators}
\label{t211}
\centering
\begin{tabular}{c|ccc|c|cc|ccc}
\hline \hline
\multicolumn{1}{c|}{Ternary Input}& 	PTI &	NTI &	STI &Decisive Literal &\multicolumn{2}{c|}{Cycle Operators}&&&\\
\multicolumn{1}{c|}{$A$} 	& 	$A_p$ & $A_n$ &  $\bar{A}$& $A_1$ & $A^1$ &  $A^2$ & $\bar{A^2}$  &$1\cdot \bar{A_n}$ & $1\cdot \bar{A_p}$  \\
\hline
0 (0V)	  &2&	2&	2 &	0	&1		&2	&0			&0			&0 \\
1 (0.45V)	  &2&	0&	1 &	2	&2		&0	&2		 	&1			&0 \\
2 (0.9V)    &0&	0&	0 &	0	&0		&1	&1			&1			&1 \\
\hline \hline
\end{tabular}
\end{table}
\renewcommand{\arraystretch}{1}

\begin{equation}
\label{eq23}
\begin{aligned}
A_p  & = &
 \begin{cases}
2,  &   if  A \neq 0\\
0,  &   if  A   =  0
\end{cases} \\
A_n & = &  
 \begin{cases}
2,  &   if  A  =  0\\
0,  &   if  A  \neq 0
\end{cases}\\
\bar{A} & = & 2 - A \\
A_1 & = & 
 \begin{cases}
2,  &   if  A  =  1\\
0,  &   if  A  \neq 1
\end{cases}\\
A^1 & = &  (A+1) mod (3) \\
A^2 & = &  (A+2) mod (3) \\
\bar{A^2} &=& 2 - A^2 \\
1\cdot \bar{A_n} & = &  min\{1, \bar{A_n}\} \\
1\cdot \bar{A_p} & = &  min\{1, \bar{A_p}\} \\
  \end{aligned} 
   \end{equation}
Where $A$ $\in$ \{0, 1, 2\} is the ternary input.  

The first three unary functions are three types of ternary inverters, the first is a positive ternary inverter (PTI), $A_p$, the second is a negative ternary inverter (NTI), $A_n$, and the third $\bar{A}$ is a standard ternary inverter (STI), which is the complement of A.
The fourth function is the Decisive literal, $A_1$. The fifth and the sixth functions are the cycle operators, $A^1$ and $A^2$. The seventh function $\bar{A^2}$ is the complement of $A^2$, and the last two unary functions are the two-inputs ternary AND Gate, $1\cdot \bar{A_n}$ and $1\cdot \bar{A_p}$.

\begin{figure}[!t]
\centering
\includegraphics[height=20cm]{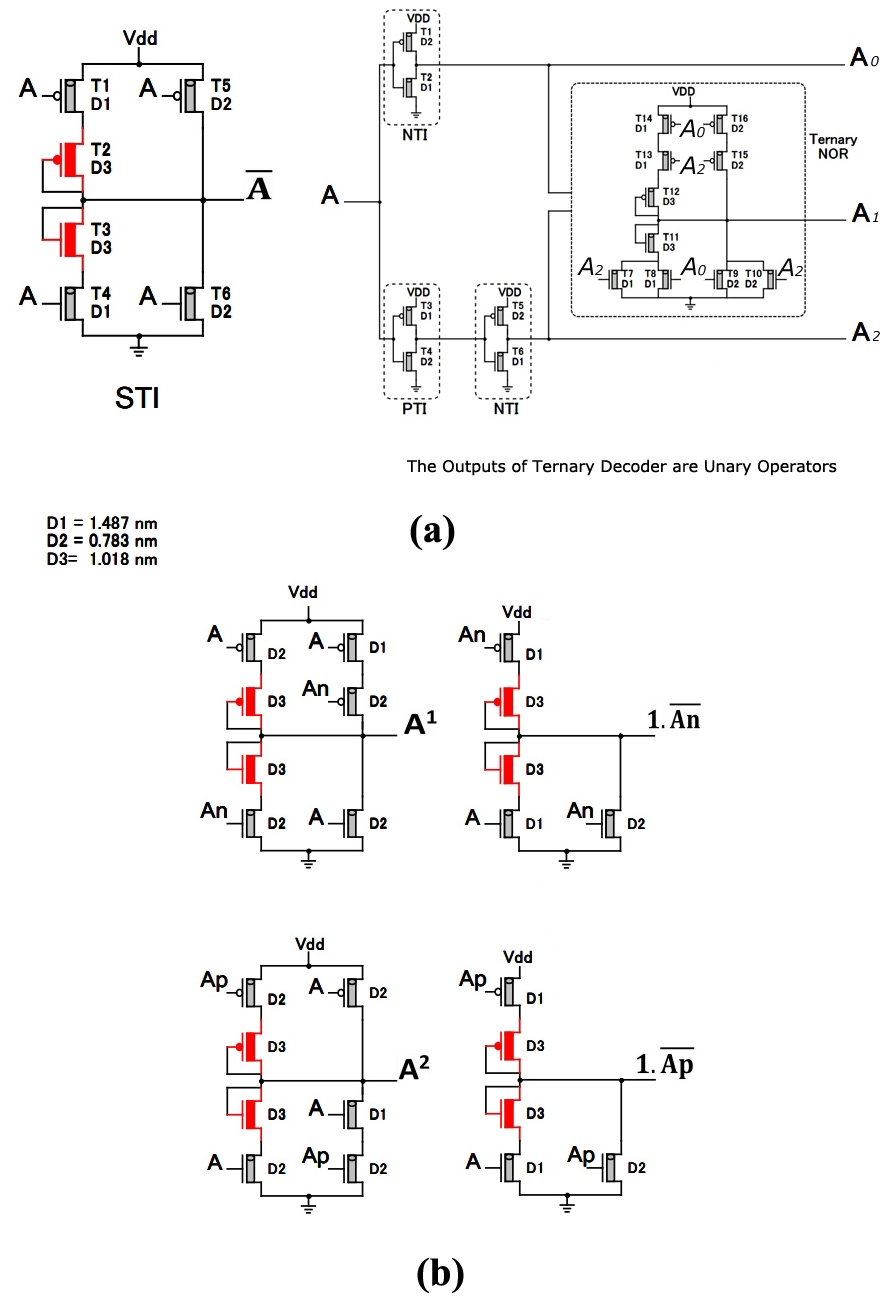}
\caption{The Existing unary operators (a) in \protect\cite{123}, $A_p$, $A_n$, $\bar{A}$, $A_0$, $A_1$, and $A_2$  (b) in \protect\cite{205}, $A^1$, $A^2$, $1\cdot \bar{A_n}$, and $1\cdot \bar{A_p}$.}
\label{figExistingUNARY}
\end{figure}

The existing Unary Operators' designs in \cite{123}, and \cite{205} are shown in Fig. \ref{figExistingUNARY};\\
Fig. \ref{figExistingUNARY} (a) shows three types of ternary inverters ($A_p$, $A_n$, $\bar{A}$) and three Decisive literals ($A_0$, $A_1$, $A_2$) of \cite{123}.\\
Fig. \ref{figExistingUNARY} (b) shows the existing four unary operators circuits of \cite{205} ($A^1$, $A^2$, $1\cdot \bar{A_n}$, and $1\cdot \bar{A_p}$).

\pagebreak
This chapter proposes new designs for seven unary operators using CNFET and dual supply voltage (Vdd, Vdd/2). This section will explain the six unary operator designs, whereas, the seventh unary operator (STI) will be described in the next section. 

\begin{figure}[t!]
\centering
\includegraphics[height=10cm]{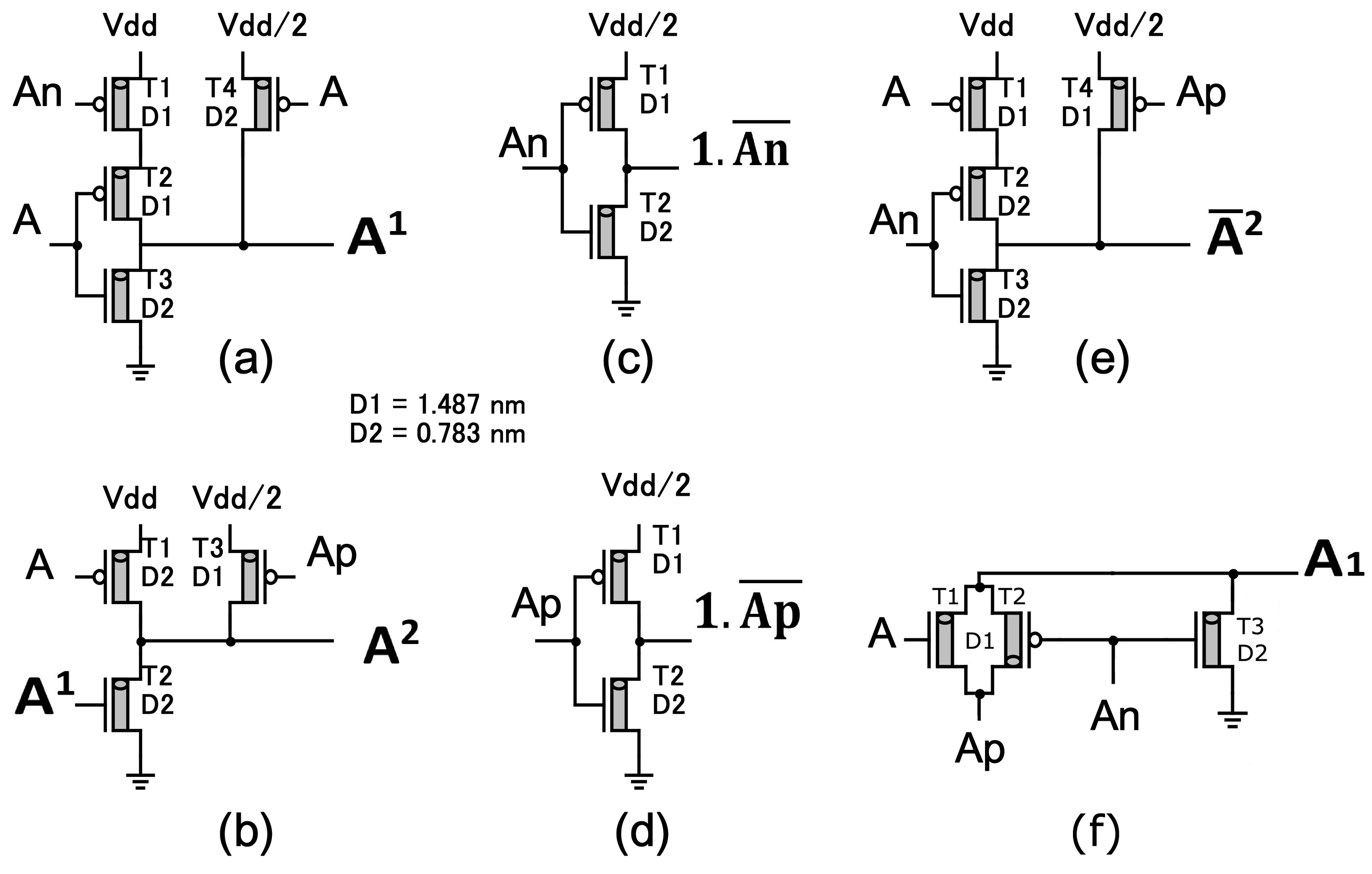}
\caption{The transistor level of the proposed six unary operators: (a) $A^1$, (b) $A^2$, (c) $1\cdot \bar{A_n}$, (d) $1\cdot \bar{A_p}$, (e) $\bar{A^2}$, and (f) $A_1$ }
\label{figunary}
\end{figure}

Fig. \ref{figunary} shows the transistor level design of the proposed six Unary Operators. (a) $A^1$, (b) $A^2$, (c) $1\cdot \bar{A_n}$, (d) $1\cdot \bar{A_p}$, (e) $\bar{A^2}$, and (f) $A_1$.

The chirality, diameter, and threshold voltage ($V_{th}$) of the CNFETs used in Fig. \ref{figunary} are shown in Table~\ref{t223}.

\renewcommand{\arraystretch}{1.3}
\begin{table}[!t]
\caption{The chirality, diameter, and threshold voltage of the CNTs used in the proposed six unary operators}
\label{t223}
\centering
\begin{tabular}{l|cccccc|ccc}
\hline\hline
&\multicolumn {6}{c|}{Fig.\ref{figunary}}&&&\\
CNFET Type & (a) & (b) & (c) & (d) & (e) &(f)	&Chirality&	Diameter&	Vth \\
\hline
P-CNFET&T4 				&T1 		&-  &-	&T2 		 			 			&-	&(10, 0)&	0.783nm&	- 0.559V\\
P-CNFET&T1, T2			 &T3 	&T1 									 		&T1 	&T1, T4 	 		&T2	&	(19, 0)&	1.487nm&	- 0.289V\\
N-CNFET&T3				&T2  &T2 											&T2  &T3 	 					&T3	&(10, 0)&	0.783nm&	 0.559V\\
N-CNFET&-				&-  &- 											&-  &- 	 					&T1	&(19, 0)&	1.487nm&	 0.289V\\
\hline \hline
\end{tabular}
\end{table}
\renewcommand{\arraystretch}{1}

\subsection{State of CNFETs with Different Diameters}
When the voltage of the gate varies (0 V, 0.45 V, 0.9 V), then the transistor will be turned ON or OFF, which depends on the type and the diameter of CNFET, as described in Table \ref{t236}.

The operations of the proposed unary operators are summarized in Table \ref{to}.

\renewcommand{\arraystretch}{1.3}
\begin{table}[!t]
\caption{The state of CNFETs with D1=1.487 nm and D2=0.783 nm}
\label{t236}
\setlength{\tabcolsep}{2pt}
\centering
\begin{tabular}{lc|ccc}
\hline \hline
&&\multicolumn{3}{c}{Voltage Gate }\\
Type &	Diameter & 0V & 0.45V & 0.9V\\
\hline

\multirow{2}{*}{P-CNFET}&	D2 & ON & OFF & OFF \\
							& D1 & ON & ON & OFF\\ \hline
\multirow{2}{*}{N-CNFET}& D2 & OFF & OFF & ON\\
							& D1 & OFF & ON & ON\\
\hline \hline
\end{tabular}
\end{table}

\renewcommand{\arraystretch}{1.3}
\begin{table}[!t]
\caption{The detailed operation of the proposed six unary operators with D1=1.487 nm and D2=0.783 nm}
\label{to}
\setlength{\tabcolsep}{2pt}
\begin{center}
\begin{tabular}{c|ccc|cc|l}
\hline \hline 
&&&&\multicolumn{2}{c|}{Transistors Turned}&\\
Circuit&$A$&$A_n$&$A_p$&ON& OFF & Output\\ \hline \hline
\multirow{5}{*}{\adjustbox{valign=c}{\includegraphics[width=2.5cm, height=2.5cm]{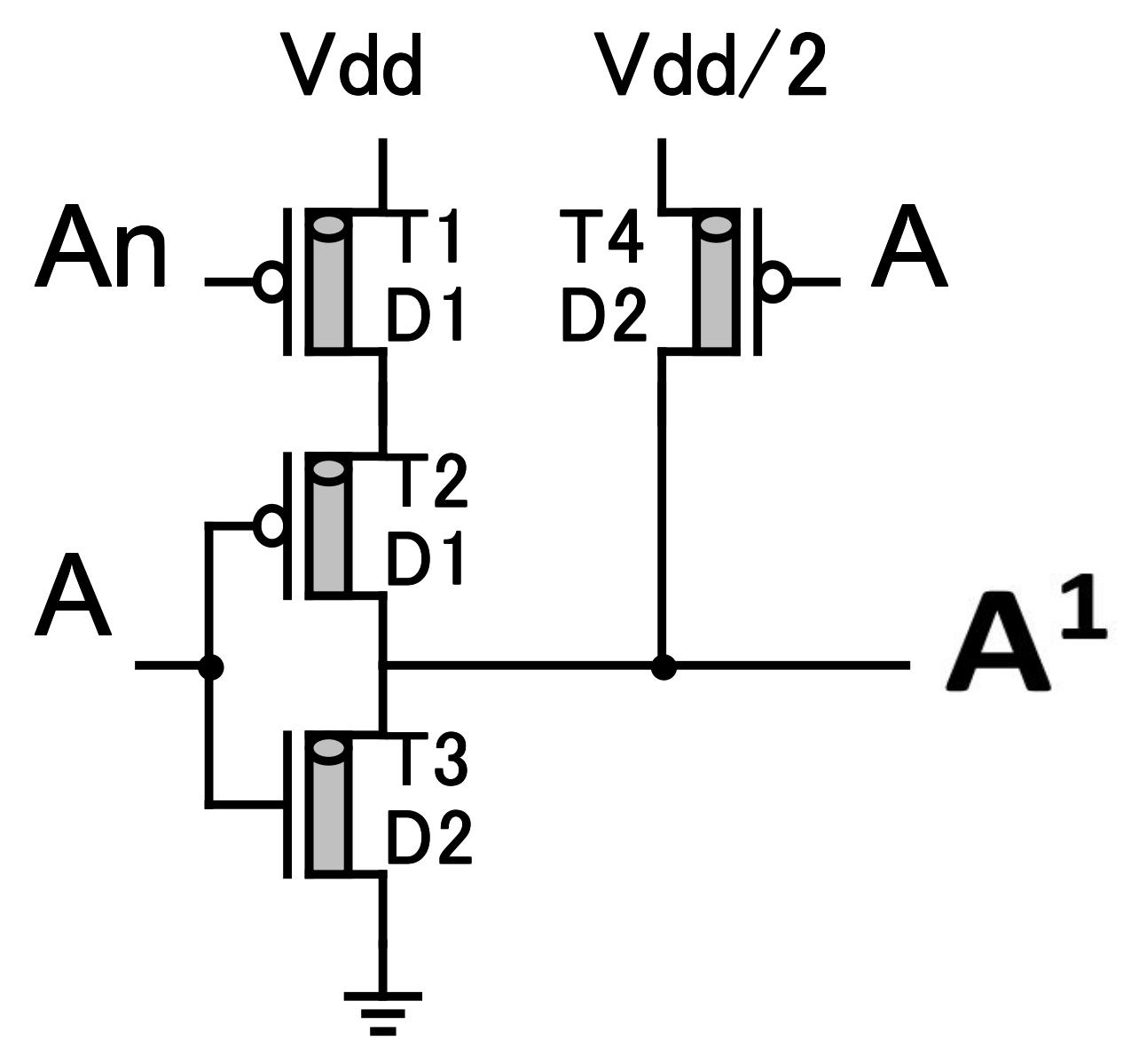}}} &&&&&&\\
&0 & 2 & &T2,T4 &T1,T3&1 \multirow{3}{*}{ \Bigg\}	$\mathbf{A^1}$ }\\
& 1 & 0 & &T1,T2 &T3,T4&2\\
& 2 & 0 & &T1,T3 &T2,T4&0\\
&&&&&&\\
\hline  

\multirow{5}{*}{\adjustbox{valign=c}{\includegraphics[width=2.5cm, height=2.5cm]{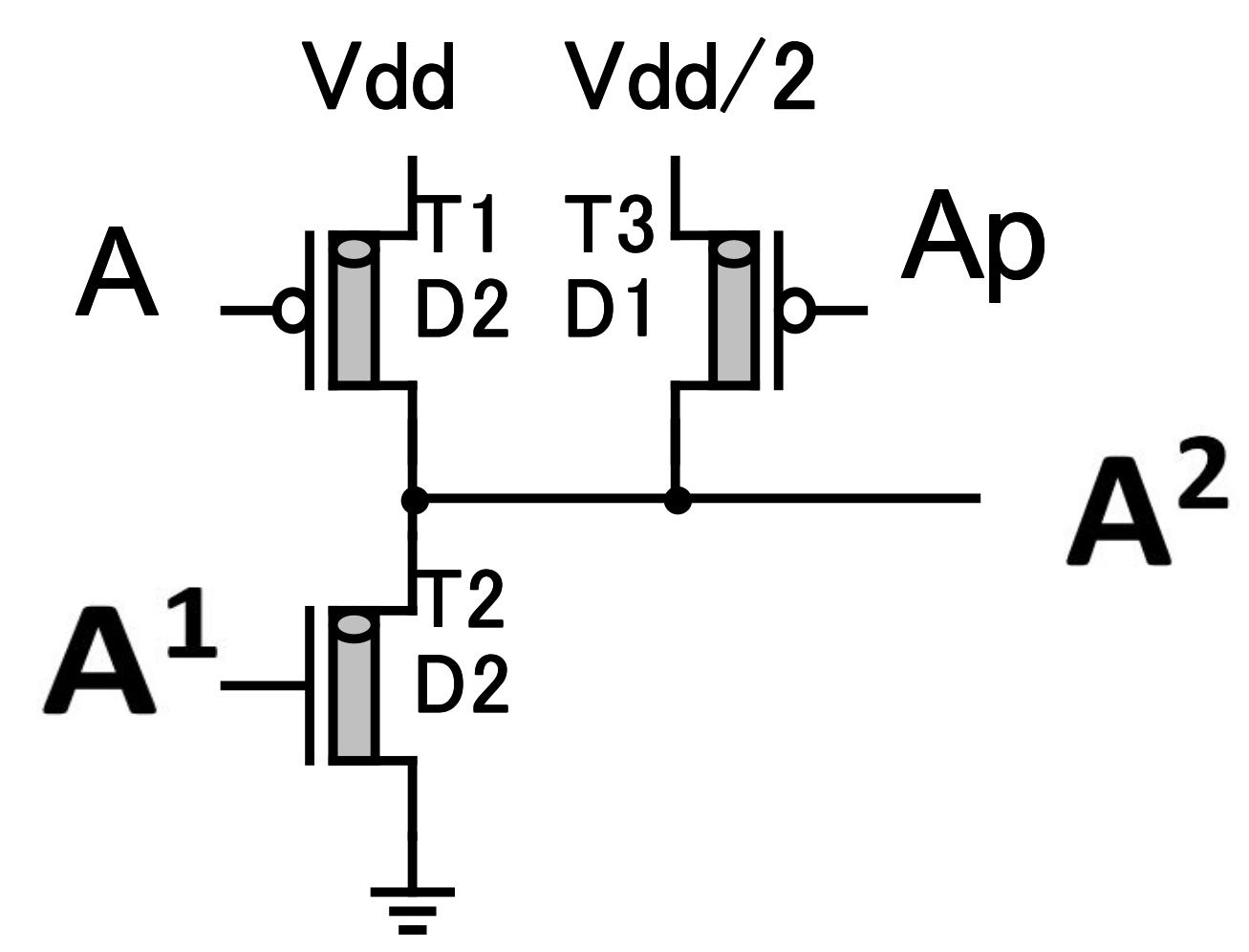}}} &&&&&&\\
& 0 &  &2 &T1 &T2,T3& 2\multirow{3}{*}{ \Bigg\} $\mathbf{A^2}$}\\
& 1 &  &2 &T2 &T1,T3& 0\\
& 2 &  &0 &T3 &T1,T2& 1\\
&&&&&&\\
\hline 

\multirow{5}{*}{\adjustbox{valign=c}{\includegraphics[width=2.5cm, height=2.5cm]{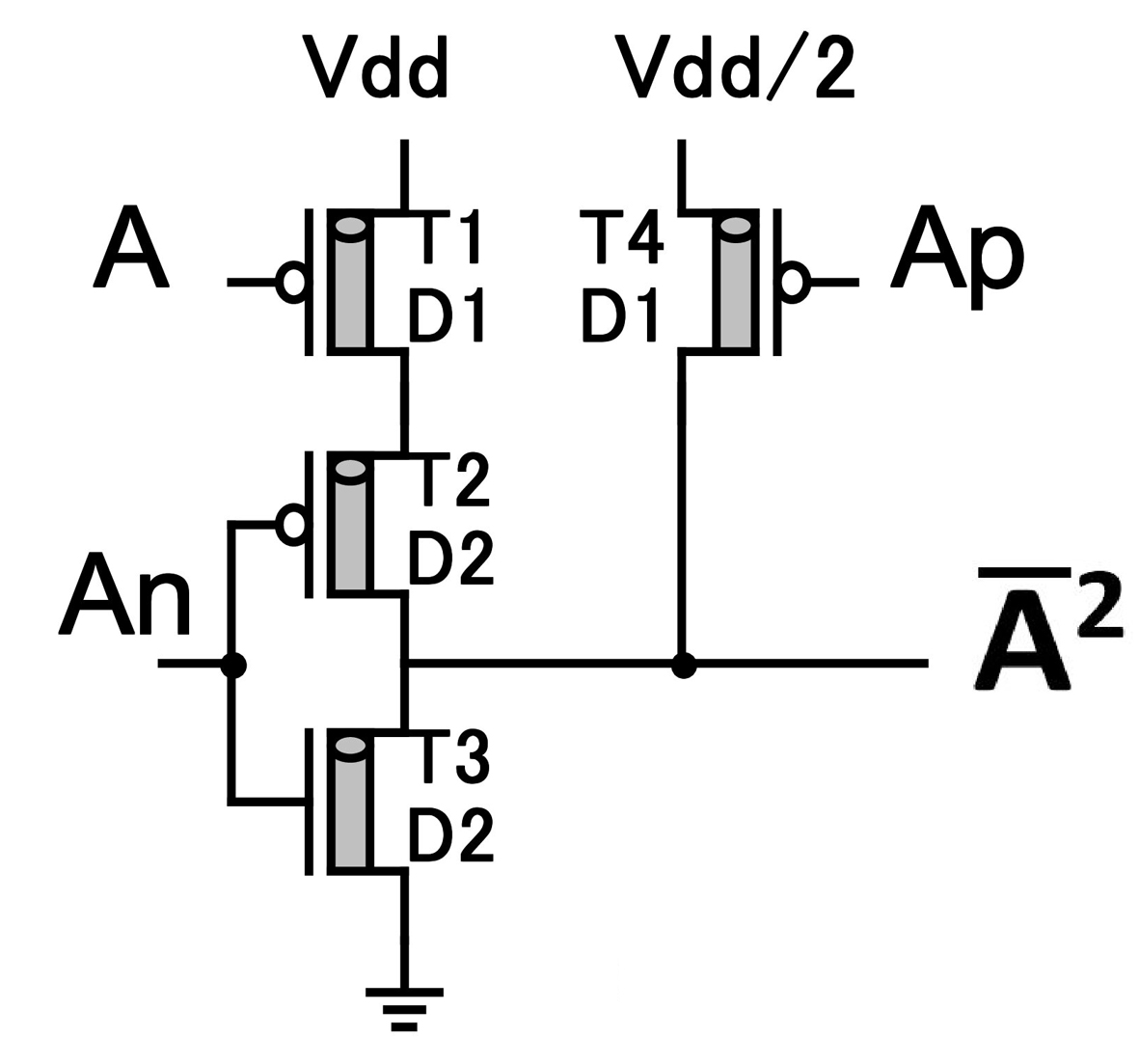}}} &&&&&&\\
 & 0 & 2 &2 &T1,T3 &T2,T4& 0\multirow{3}{*}{ \Bigg\}	$\mathbf{\overline{A^2}}$ } \\
& 1 & 0 & 2&T1,T2 &T3,T4& 2\\
& 2 &0  & 0&T2,T4 &T1,T3& 1\\
&&&&&&\\
\hline
\multirow{5}{*}{\adjustbox{valign=c}{\includegraphics[width=2.5cm, height=2.5cm]{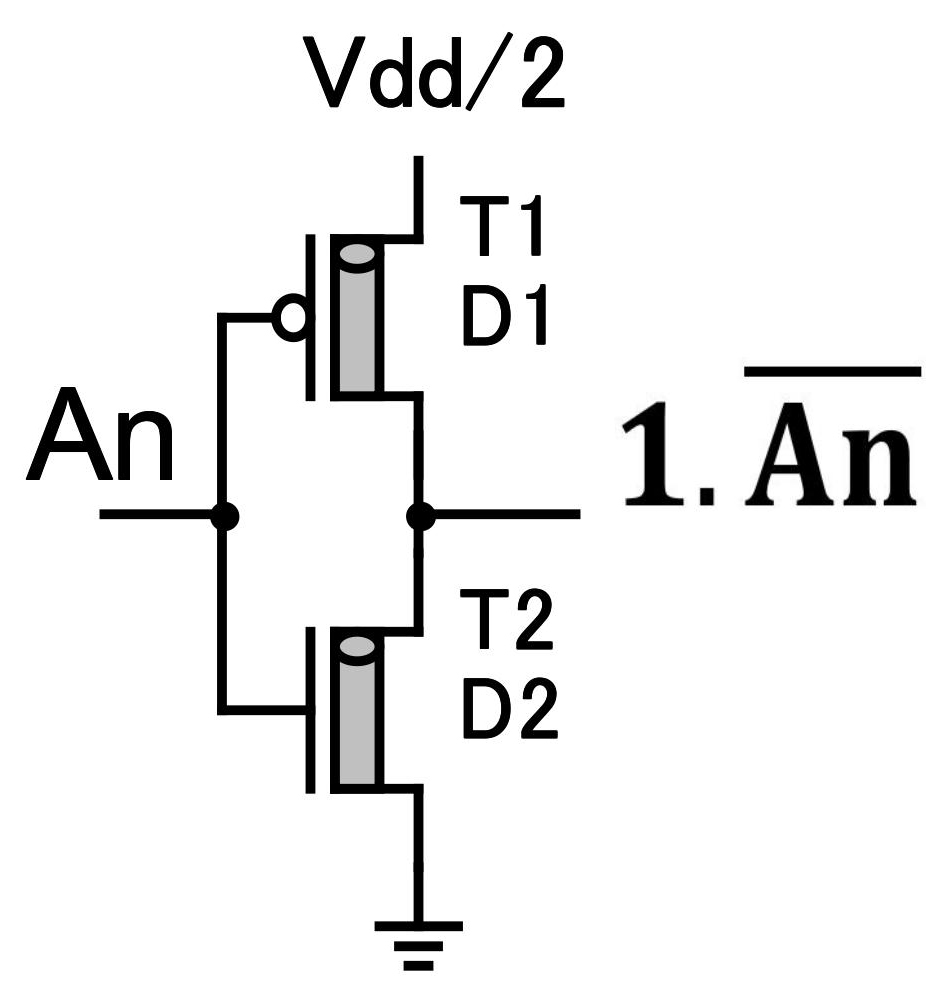}}} &&&&&&\\
& 0 & 2 & &T2 &T1& 0\multirow{3}{*}{ \Bigg\} $\mathbf{1\cdot \overline{A_n}}$ }\\
& 1 & 0 & &T1 &T2& 1\\
& 2 & 0 & &T1 &T2& 1\\
&&&&&&\\
\hline 

\multirow{5}{*}{\adjustbox{valign=c}{\includegraphics[width=2.5cm, height=2.5cm]{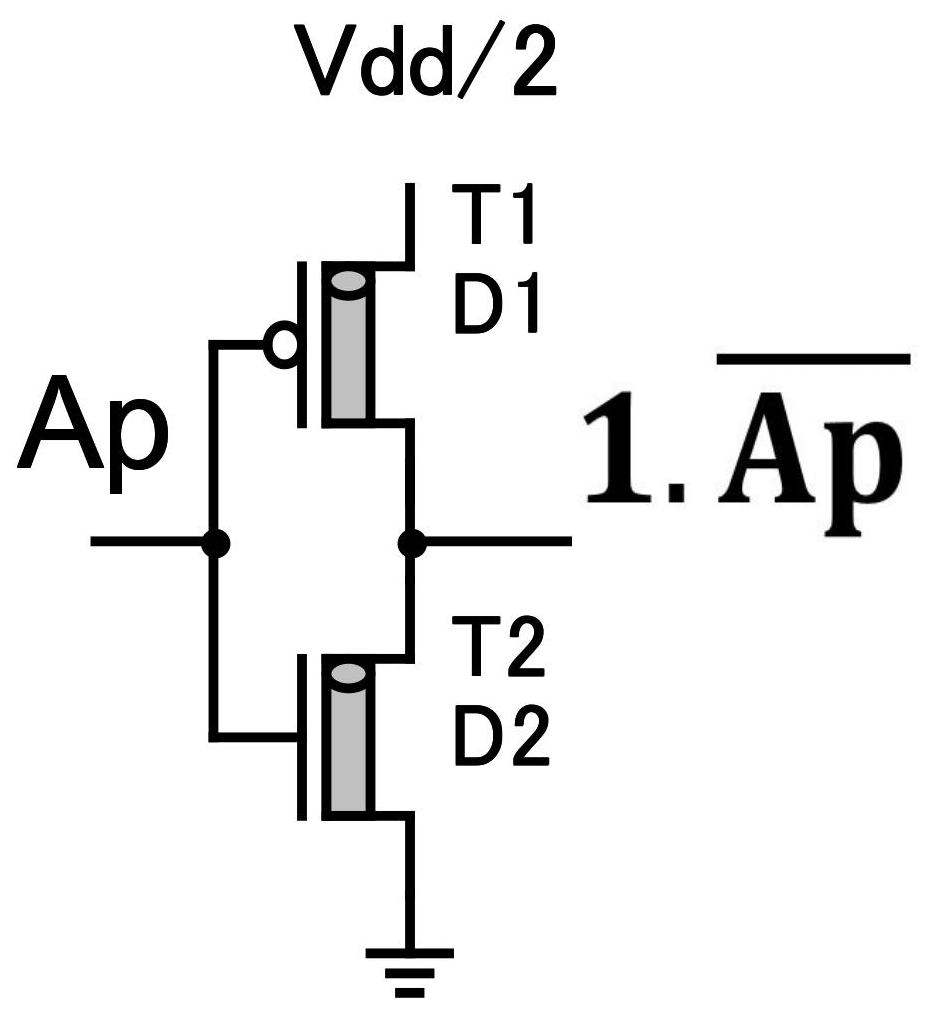}}} &&&&&&\\
 & 0 &  &2 &T2 &T1& 0\multirow{3}{*}{ \Bigg\}	$\mathbf{1\cdot \overline{A_p}}$ } \\
& 1 &  &2 &T2 &T1& 0\\
& 2 &  & 0&T1 &T2& 1\\
&&&&&&\\ \hline

\multirow{5}{*}{\adjustbox{valign=c}{\includegraphics[width=3cm, height=2cm]{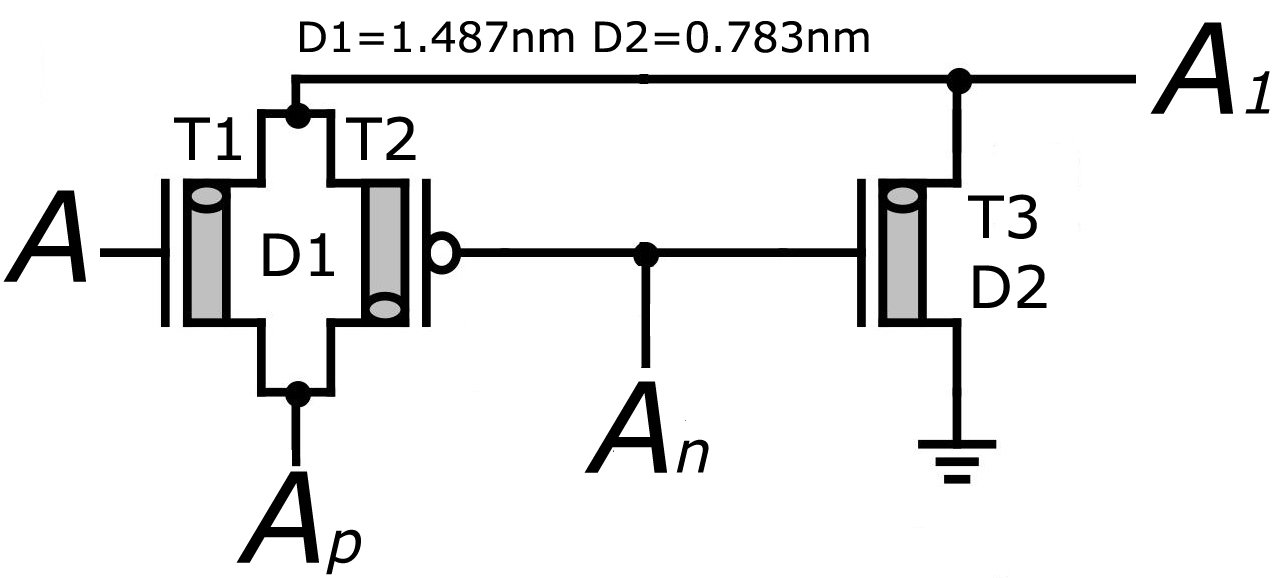}}} &&&&&&\\
 & 0 & 2 &2 &T3 &T1,T2& Ground \multirow{3}{*}{ \Bigg\}	$\mathbf{A_1}$ } \\
& 1 &  0&2 &T1,T2 &T3& $A_p$ = 2\\
& 2 &  0& 0&T1,T2 &T3& $A_p$ = 0\\
&&&&&&\\

\hline \hline 
\end{tabular}
\end{center}
\end{table}

The advantages of the proposed six Unary Operators are:
\begin{enumerate}
  \item They provide low power consumption due to eliminating the two transistors that act as resistors (T2, T3 in all circuits \cite{205}) by applying dual supply voltages (Vdd, Vdd/2)
  \item To get logic 1 (0.45 V): one or two transistors must be active (see Table \ref{to}) whereas, in \cite{205}, four transistors in series must be active (T1, T2, T3, T4)
  \item Reducing the Transistors Count
  \end{enumerate}
\clearpage
\section{The Proposed STI \& TNAND Logic Gates}
\begin{figure}[!b]
\centering
\includegraphics[width=\textwidth]{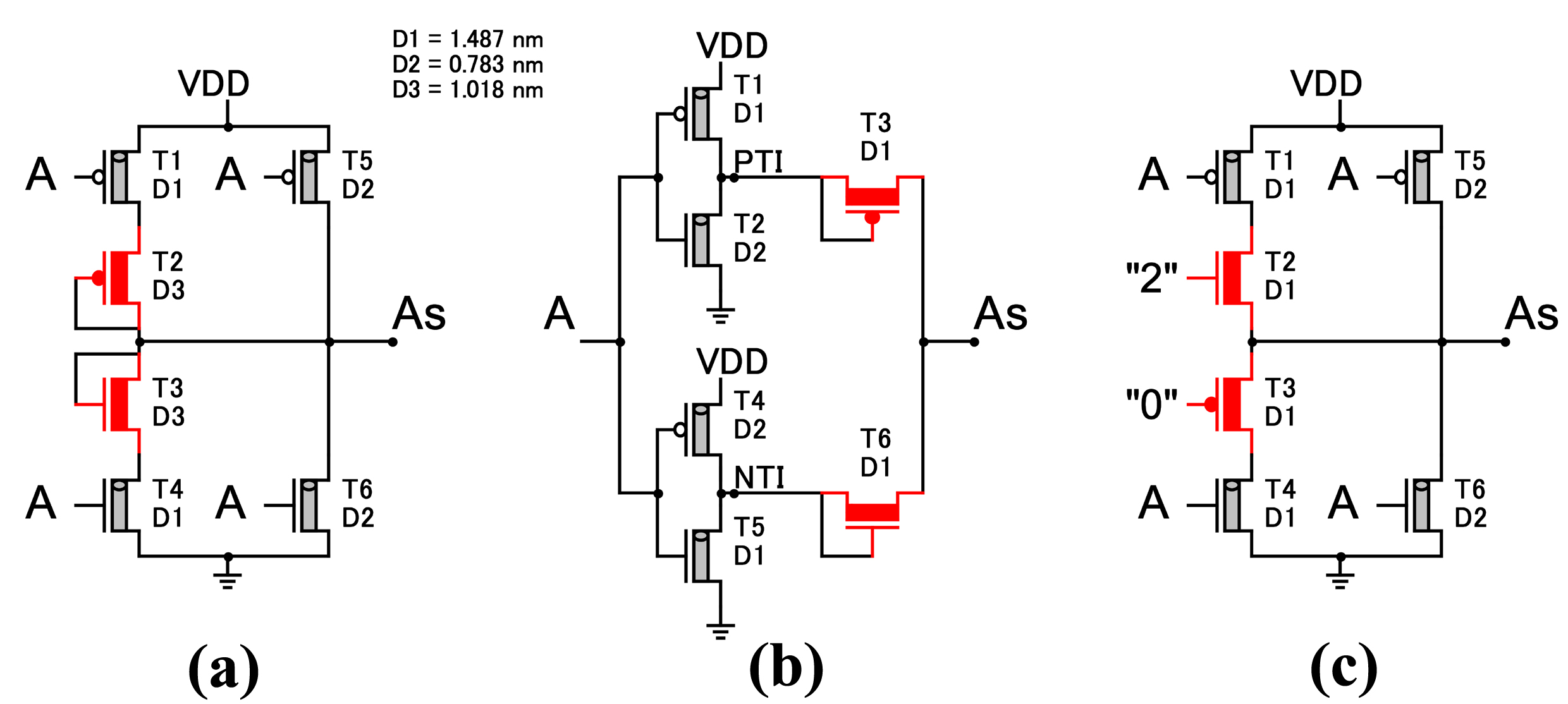}
\caption{The Existing STI in (a) \cite{123}, (b) \cite{t16}, and (c) \cite{t18}.}
\label{figstiexisting}
\end{figure}

\begin{figure}[!b]
\centering
\includegraphics[width=\textwidth]{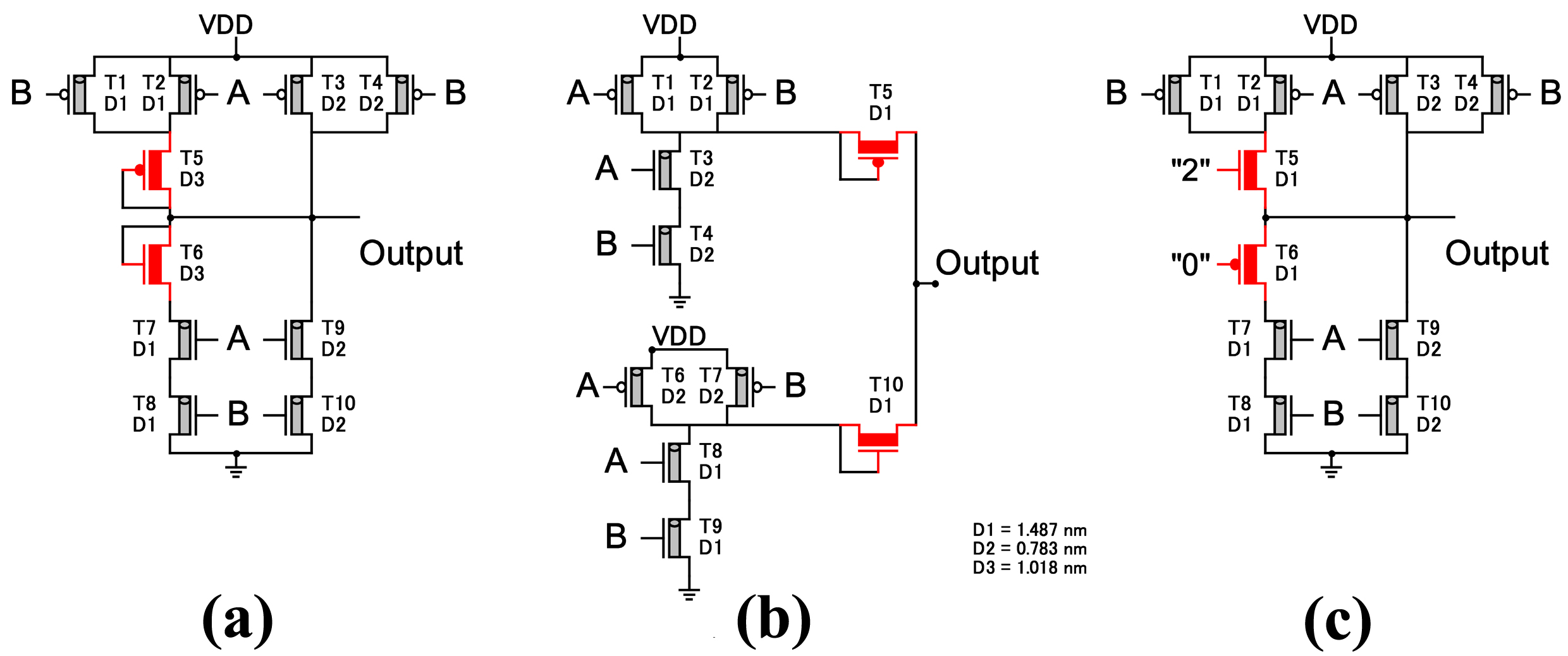}
\caption{The Existing TNAND in (a) \cite{123}, (b) \cite{t16}, and (c) \cite{t18}.}
\label{figTNANDexisting}
\end{figure}

The existing STI (TNOT) and TNAND logic gates designs in \cite{123,t16}, and \cite{t18} are shown in Fig. \ref{figstiexisting} and \ref{figTNANDexisting}.

This chapter proposes new designs of STI and TNAND logic gates, as shown in Fig.~\ref{figSTI} and Fig.~\ref{fignand}.

\subsection{The Proposed STI}
 
\begin{figure}[!b]
\centering
\includegraphics[width=12cm]{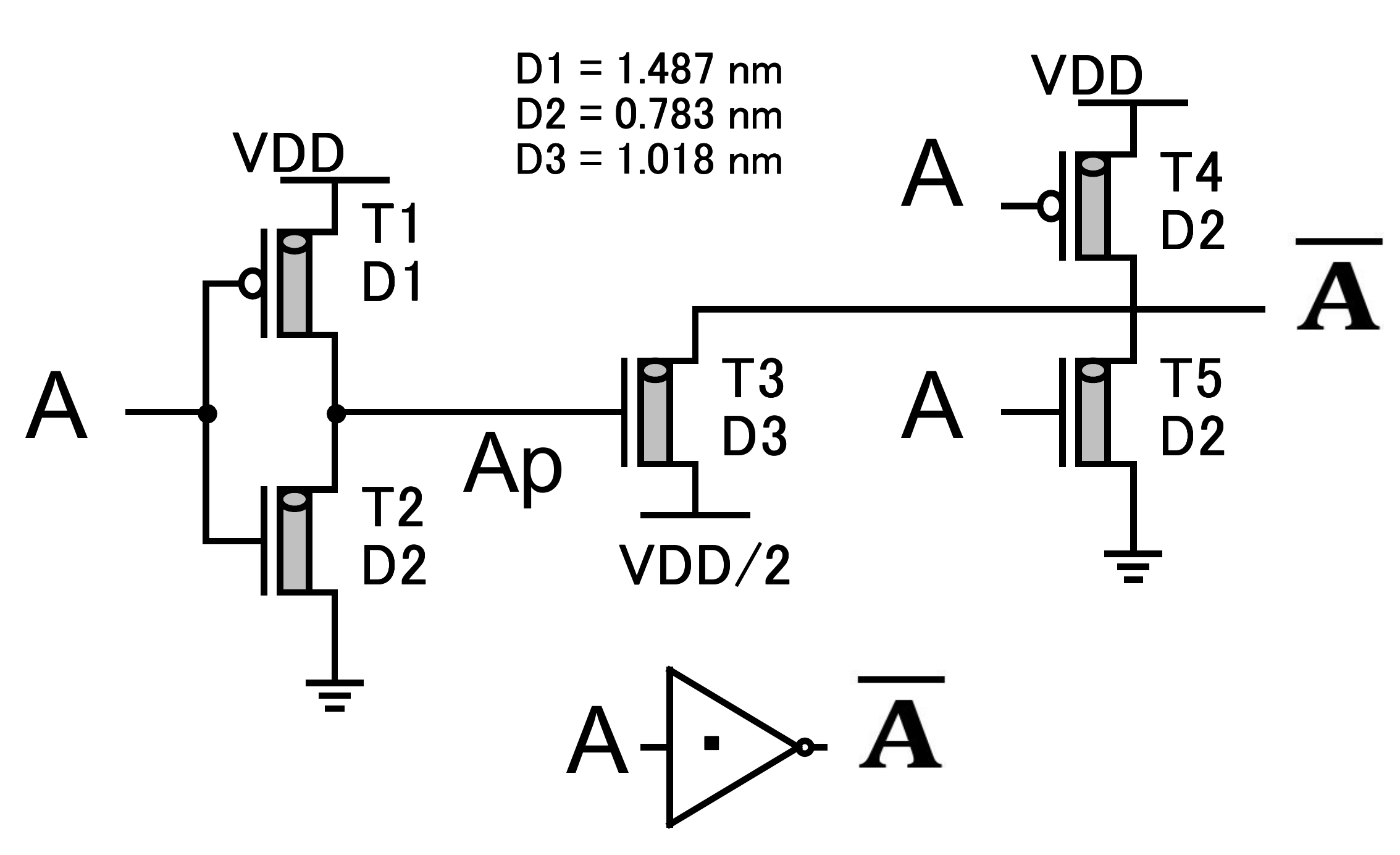}
\caption{Transistor level of the proposed STI.}
\label{figSTI}
\end{figure}

Fig.~\ref{figSTI} shows the transistor level design of the proposed STI where the chirality, diameter, and threshold voltage (Vth) of the CNFETs used are shown in Table~\ref{2t223}.

\renewcommand{\arraystretch}{1.3}
\begin{table}[!b]
\caption{The chirality, diameter, and threshold voltage of the CNTs used in the proposed STI}
\label{2t223}
\centering
\begin{tabular}{lccr}
\hline
CNFET Type&	Chirality&	Diameter&	Vth \\
\hline
\hline
P-CNFET (T4)&	(10, 0)&	0.783nm&	- 0.559V\\
P-CNFET (T1)&	(19, 0)&	1.487nm&	- 0.289V\\
N-CNFET (T2,T5)&	(10, 0)&	0.783nm&	 0.559V\\
N-CNFET (T3)&	(13, 0)&	1.018nm&	 0.428V\\
\hline
\end{tabular}
\end{table}

\renewcommand{\arraystretch}{1.2}

\begin{table}[t!]
\caption{The detailed operation of the STI}
\label{t224}
\centering
\begin{tabular}{lccc}
\hline
\textbf{Ternary Input $A$}&	\textbf{0 (0 V)}&\textbf{1 (0.45 V)}&\textbf{2 (0.9 V)} \\
\hline
\hline
P-CNFET T1&	ON	&ON	&OFF\\
N-CNFET T2&	OFF&	OFF&	ON\\
\hline
\textbf{$A_p$}&\textbf{2}&\textbf{2}&	\textbf{0}\\
\hline
\hline
N-CNFET T3&	ON&	ON&	OFF\\
P-CNFET T4&	ON&	OFF&	OFF\\
N-CNFET T5&	OFF&	OFF	&ON\\
\hline \hline
\textbf{Output $\overline{A}$}&\textbf{2}	&\textbf{1} &\textbf{0}\\
\hline
\end{tabular}
\end{table}
\renewcommand{\arraystretch}{1}

Table~\ref{t224} shows the detailed operation of the proposed STI circuit of Fig.~\ref{figSTI}.

When the input $A$ is logic 0, then transistors (T1, and T4) are turned ON and (T2, and T5) are turned OFF. The output $A_p$ is equal to logic 2, then transistor (T3) is turned ON. Therefore, the output $\bar{A}$ is equal to logic 2.

When the input $A$ is logic 1, then transistor (T1) is turned ON, and (T2, T4, and T5) are turned OFF. The output $A_p$ is equal to logic 2, then transistor (T3) is turned ON. Therefore, the output $\bar{A}$ is equal to logic 1.

Finally, when the input $A$ is logic 2, then transistors (T2, and T5) are turned ON and (T1, and T4) are turned OFF. The output $A_p$ is equal to logic 0, then transistor (T3) is turned OFF. Therefore, the output $\bar{A}$ is equal to logic 0.

\subsection{The Proposed TNAND}
 
\begin{figure}[!t]
\centering
\includegraphics[width=12cm]{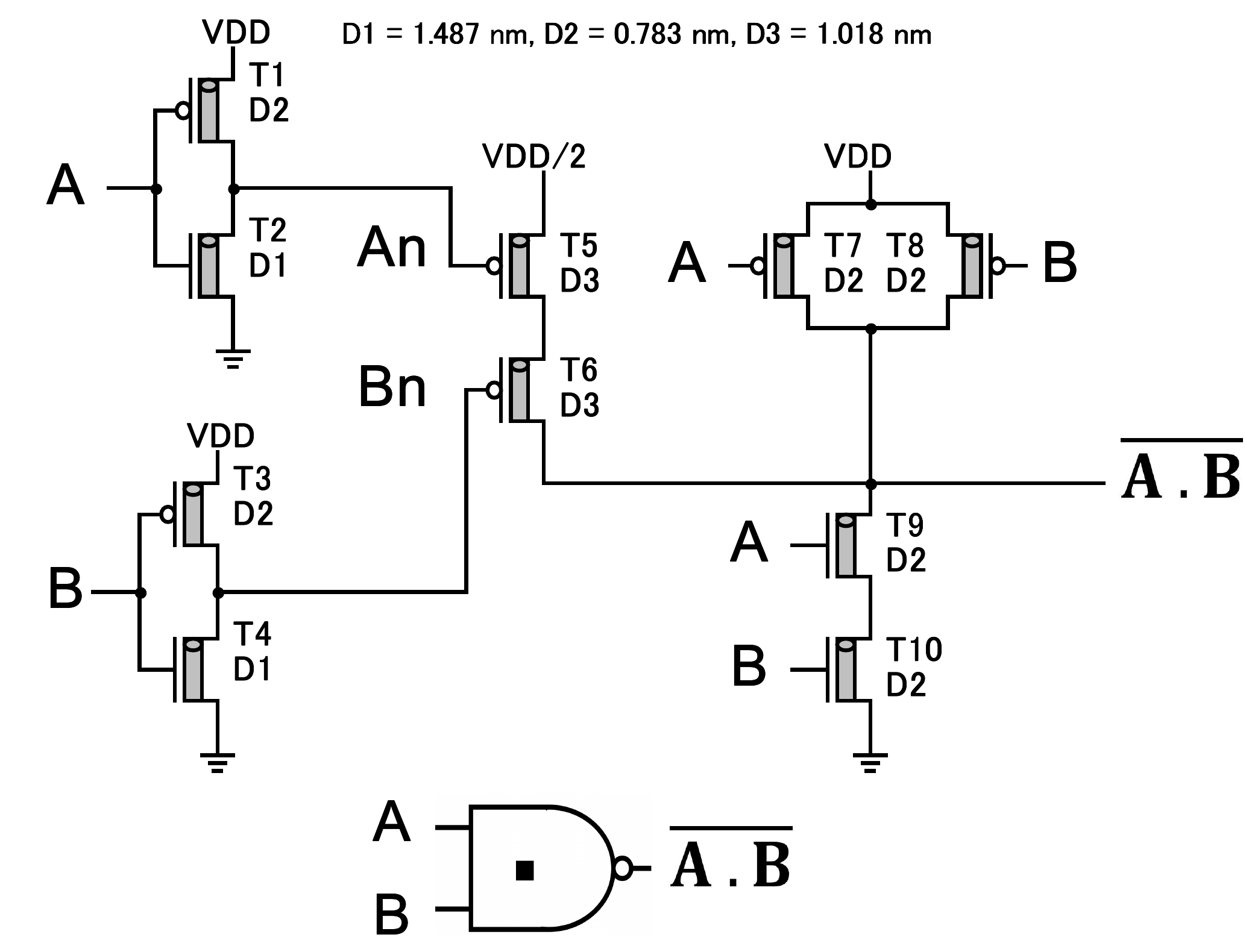}
\caption{Transistor level of the proposed TNAND.}
\label{fignand}
\end{figure}

Fig.~\ref{fignand} shows the transistor level design of the proposed two inputs TNAND where the chirality, diameter, and threshold voltage (Vth) of the CNFETs used are shown in Table~\ref{t225}.

\renewcommand{\arraystretch}{1.3}
\begin{table}[b!]
\caption{The chirality, diameter, and threshold voltage of the CNTs used in the proposed TNAND}
\label{t225}
\setlength{\tabcolsep}{5pt}
\centering
\begin{tabular}{lccr}
\hline
CNFET Type&	Chirality&	Diameter&	Vth \\
\hline
\hline
P-CNFET (T1,T3,T7,T8)&	(10, 0)&	0.783nm&	- 0.559V\\
P-CNFET (T5, T6)&	(13, 0)&	1.018nm&	- 0.428V\\
N-CNFET (T9, T10)&	(10, 0)&	0.783nm&	 0.559V\\
N-CNFET (T2, T4)&	(19, 0)&	1.487nm&	 0.289V\\
\hline
\end{tabular}
\end{table}

\begin{table}[t!]
\caption{The detailed operation of the TNAND with selected inputs}
\label{t226}
\centering
\begin{tabular}{lcccc}
\hline
\textbf{Ternary Inputs (A, B)}&	\textbf{(0, 0)}&\textbf{(0, 1)}&\textbf{(1, 1)} &\textbf{(1, 2)} \\
\hline
\hline
P-CNFET T1&	ON	&  ON& OFF	&OFF\\
N-CNFET T2&	OFF&	OFF& ON&	ON\\
\hline
\textbf{$A_n$}&\textbf{2}&\textbf{2}&\textbf{0}&	\textbf{0}\\
\hline
\hline
P-CNFET T3&	ON	&  OFF& OFF	&OFF\\
N-CNFET T4&	OFF&	ON& ON&	ON\\
\hline
\textbf{$B_n$}&\textbf{2}&\textbf{0}&\textbf{0}&	\textbf{0}\\
\hline
\hline
P-CNFET T5&	OFF&	OFF&ON&	ON\\
P-CNFET T6&	OFF&	ON&ON&	ON\\
\hline
P-CNFET T7&	ON&	ON	&OFF &OFF\\
P-CNFET T8&	ON&	OFF	&OFF&OFF\\
\hline
N-CNFET T9&	OFF&	OFF	&OFF &OFF\\
N-CNFET T10&	OFF&	OFF	&OFF&ON\\
\hline \hline
\textbf{Output TNAND }&\textbf{2}&\textbf{2}	&\textbf{1} &\textbf{1}\\
\hline
\end{tabular}
\end{table}
\renewcommand{\arraystretch}{1}

Table~\ref{t226} shows the selected combinations of two ternary inputs $A$ and $B$ to describe the operation of the proposed TNAND logic gate circuit of Fig.~\ref{fignand}.

When the inputs ($A$, $B$) are (0 V, 0 V), then transistors (T1, T3, T7, and T8) are turned ON and (T2, T4, T9, and T10) are turned OFF. The outputs ($A_n$, $B_n$) are equal to (0.9 V, 0.9 V), then transistors (T5, and T6) are turned OFF. Therefore, the output is equal to 0.9 V.

When the inputs ($A$, $B$) are (0 V, 0.45 V), then transistors (T1, T4, and T7) are turned ON and (T2, T3, T8, T9, and T10) are turned OFF. The outputs ($A_n$, $B_n$) are equal to (0.9 V, 0 V), then transistor (T6) is turned ON and (T5) is turned OFF. Therefore, the output is equal to 0.9 V.

When the inputs ($A$, $B$) are (0.45 V, 0.45 V), then transistors (T2, and T4) are turned ON and (T1, T3, T7, T8, T9, and T10) are turned OFF. The outputs ($A_n$, $B_n$) are equal to (0 V, 0 V), then transistors (T5, and T6) are turned ON. Therefore, the output is equal to 0.45 V.

Finally, when the inputs ($A$, $B$) are (0.45 V, 0.9 V), then transistors (T2, T4, and T10) are turned ON and (T1, T3, T7, T8, and T9) are turned OFF. The outputs ($A_n$, $B_n$) are equal to (0 V, 0 V), then transistors (T5, and T6) are turned ON. Therefore, the output is equal to 0.45 V.

\pagebreak
Table \ref{t235} shows the disadvantage of the existing STI and TNAND in \cite{123,t16} and \cite{t18}.

\renewcommand{\arraystretch}{1.5}
\begin{table}[t!]
\caption{The advantage of proposed STI \& TNAND logic gates}
\label{t235}
\setlength{\tabcolsep}{4pt}
\begin{center}
\begin{tabular}{p{5pt}p{225pt}|m{180pt}}
\hline
 &\multicolumn {1}{c|}{Disadvantage of Existing Logic gates in \cite{123,t16} and \cite{t18}}	& \multicolumn {1}{c}{Advantage} \\ \hline \hline
&\multicolumn {1}{c|}{Shown in Fig. \ref{figstiexisting} and \ref{figTNANDexisting}}&\multicolumn {1}{c}{Shown in Fig. \ref{figSTI} and \ref{fignand}}\\
 &\multicolumn {1}{c|}{The existing designs suffer from } 	  	& \multicolumn {1}{c}{The proposed designs provide }\\
 &\multicolumn {1}{c|}{High power consumption due to:} 	  	& \multicolumn {1}{c}{ Low power consumption due to:}\\
\hline
1- &Two transistors that act as resistors in STI (T2, T3) and in TNAND (T5, T6) \cite{123}.& \multirow{3}{180pt}{Eliminating these two transistors by applying dual supply voltages (Vdd and Vdd/2).} \\ 
2- &Two transistors that act as resistors in STI (T3, T6) and in TNAND (T5, T10) \cite{t16}.&  \\ 
3- &Two transistors that are always active in STI (T2, T3) and in TNAND (T5, T6) \cite{t18}.&  \\ \hline

4- &For STI \cite{123,t18}: Four transistors (T1, T2, T3, T4) in series must be active to get logic 1.& \multirow{2}{180pt}{For the proposed STI: Only one transistor (T3) must be active to get logic 1.}  \\
5- & For STI \cite{t16}: Four transistors (T1, T3, T6, T5) in series must be active to get logic 1. & \\ \hline

6- &For TNAND \cite{123,t18}: Five or Six transistors [(T1 or T2) or both, T5, T6, T7, T8)] in series must be active to get logic 1.& \multirow{2}{180pt}{For the proposed TNAND: Two transistors (T5, T6) in series must be active to get logic 1.} \\
7- & For TNAND \cite{t16}: Five or Six transistors [(T1 or T2) or both, T5, T10, T8, T9)] in series must be active to get logic 1.& \\
 
\hline
\end{tabular}
\end{center}
\end{table}
\renewcommand{\arraystretch}{1}

\clearpage
\section{Simulation Results and Comparisons}
 
The proposed six Unary Operators circuits are simulated, tested, and compared to \cite{205, 123} using the HSPICE simulator and CNFET with 32-nm channel length at a power supply~(0.9V), a room temperature~(27 \textdegree{}C), and a frequency~(1 GHz).

Whereas, the proposed STI and TNAND are simulated, tested, and compared to \cite{123,t16,t18} at different power supplies~(0.8V, 0.9V, 1V), different temperatures~(10\textdegree{}C, 27\textdegree{}C, 70\textdegree{}C ), and different frequencies~(0.5GHz, 1GHz, 2GHz).

All input signals have a fall and rise time equal to 10 ps. The average power consumption,~maximum propagation delay, and maximum power delay product (PDP) are obtained for all circuits.

The performance of the proposed circuits will be compared to other designs for the PDP.

\begin{figure}[!b]
\centering
\includegraphics[width=10cm]{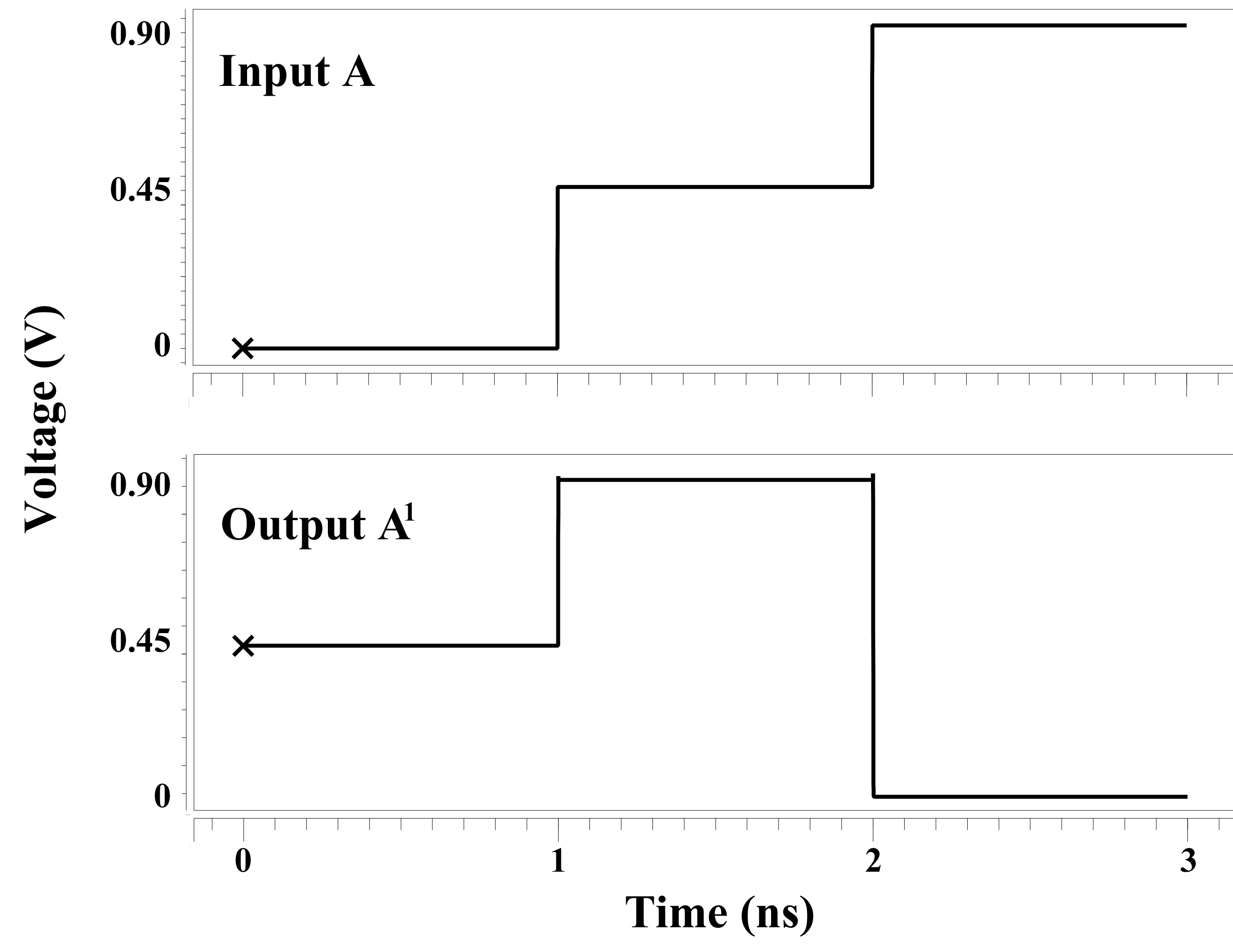}
\caption{Transient analysis of the proposed $A^1$.}
\label{simA1}
\end{figure}

\begin{figure}[!b]
\centering
\includegraphics[width=10cm]{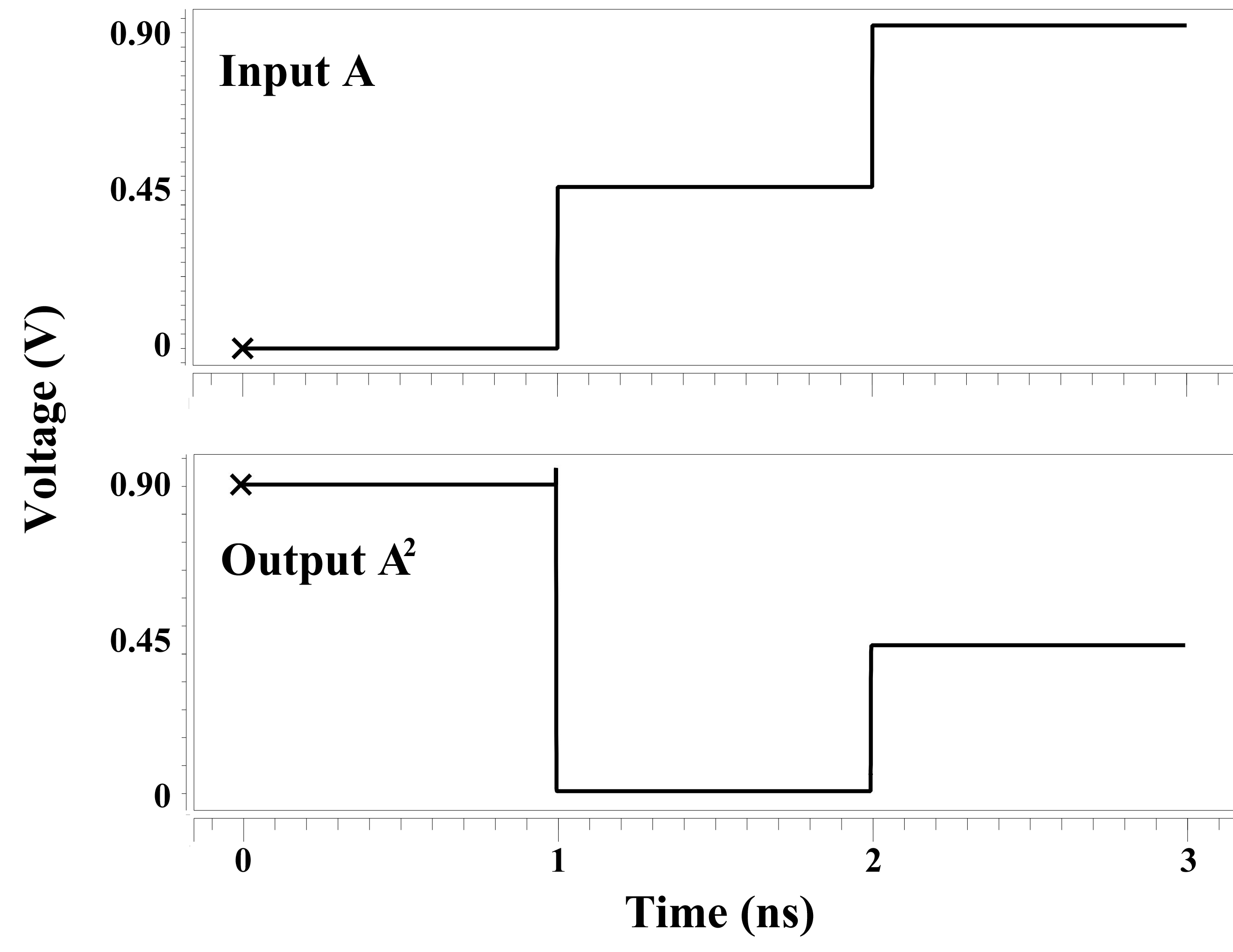}
\caption{Transient analysis of the proposed $A^2$.}
\label{simA2}
\end{figure}

\begin{figure}[!tb]
\centering
\includegraphics[width=10cm]{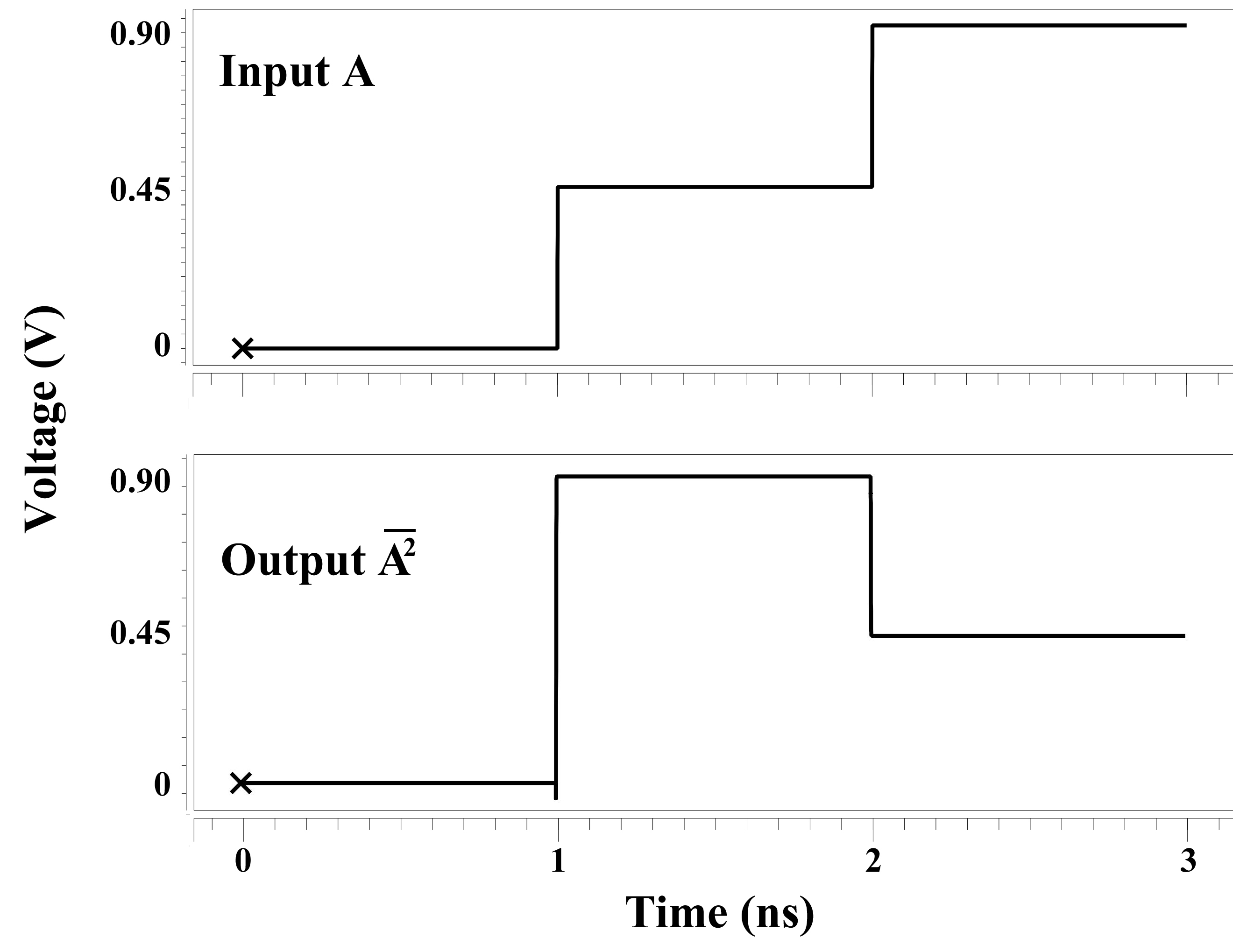}
\caption{Transient analysis of the proposed $\bar{A^2}$.}
\label{simA2b}
\end{figure}

\begin{figure}[!tb]
\centering
\includegraphics[width=10cm]{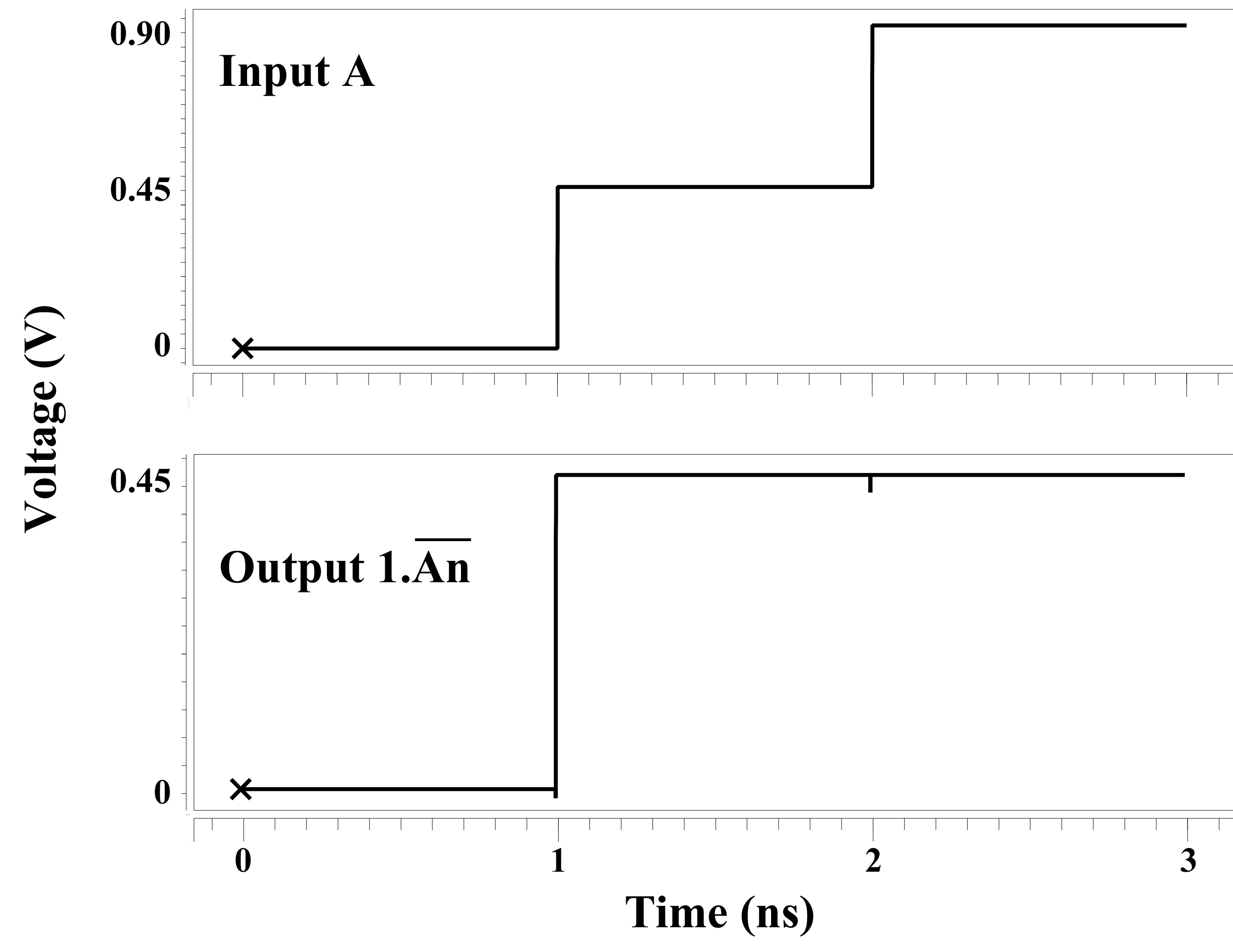}
\caption{Transient analysis of the proposed $1\cdot \bar{A_n}$.}
\label{sim1anb}
\end{figure}

\begin{figure}[!tb]
\centering
\includegraphics[width=10cm]{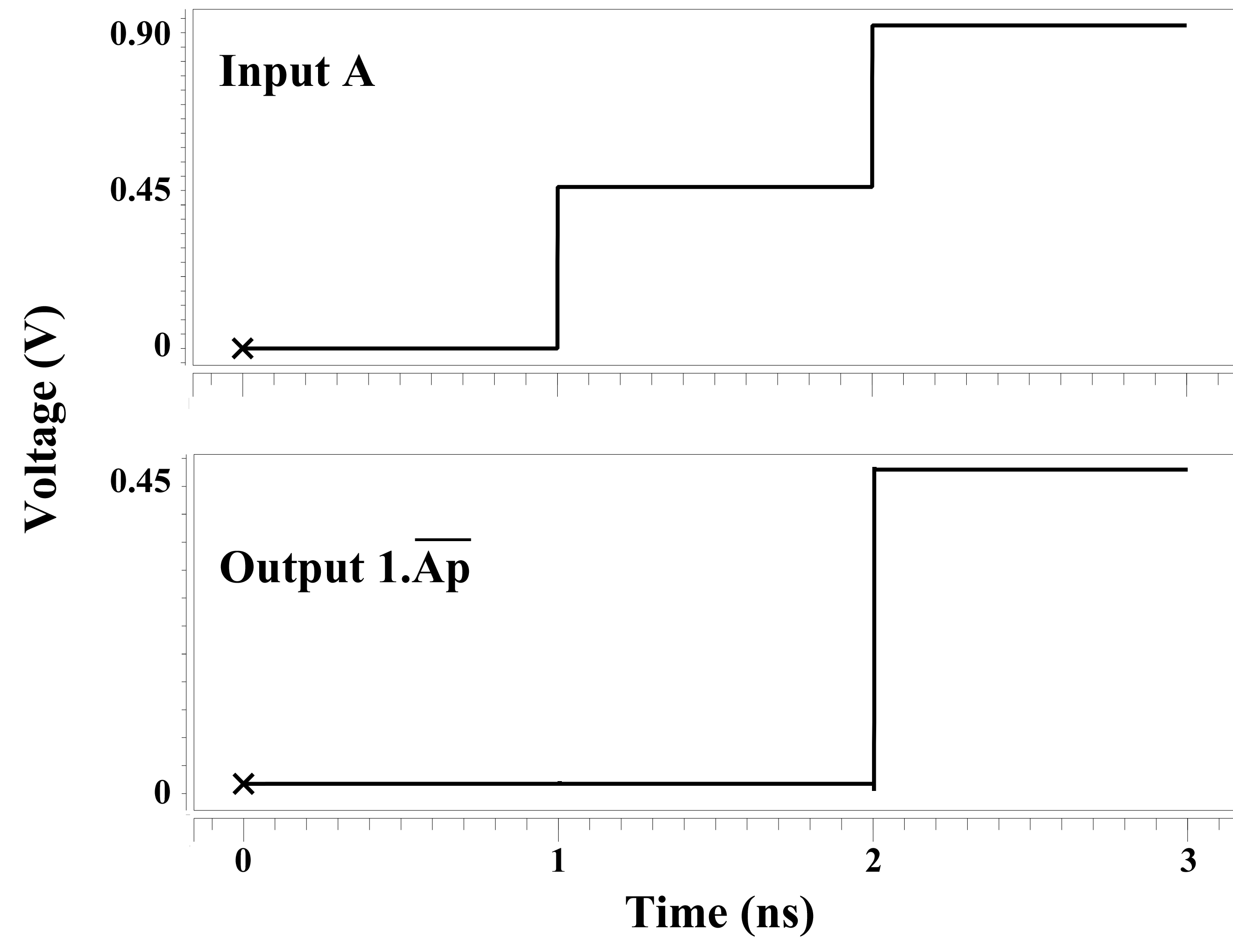}
\caption{Transient analysis of the proposed $1\cdot \bar{A_p}$.}
\label{sim1apb}
\end{figure}

\begin{figure}[!tb]
\centering
\includegraphics[width = 10cm]{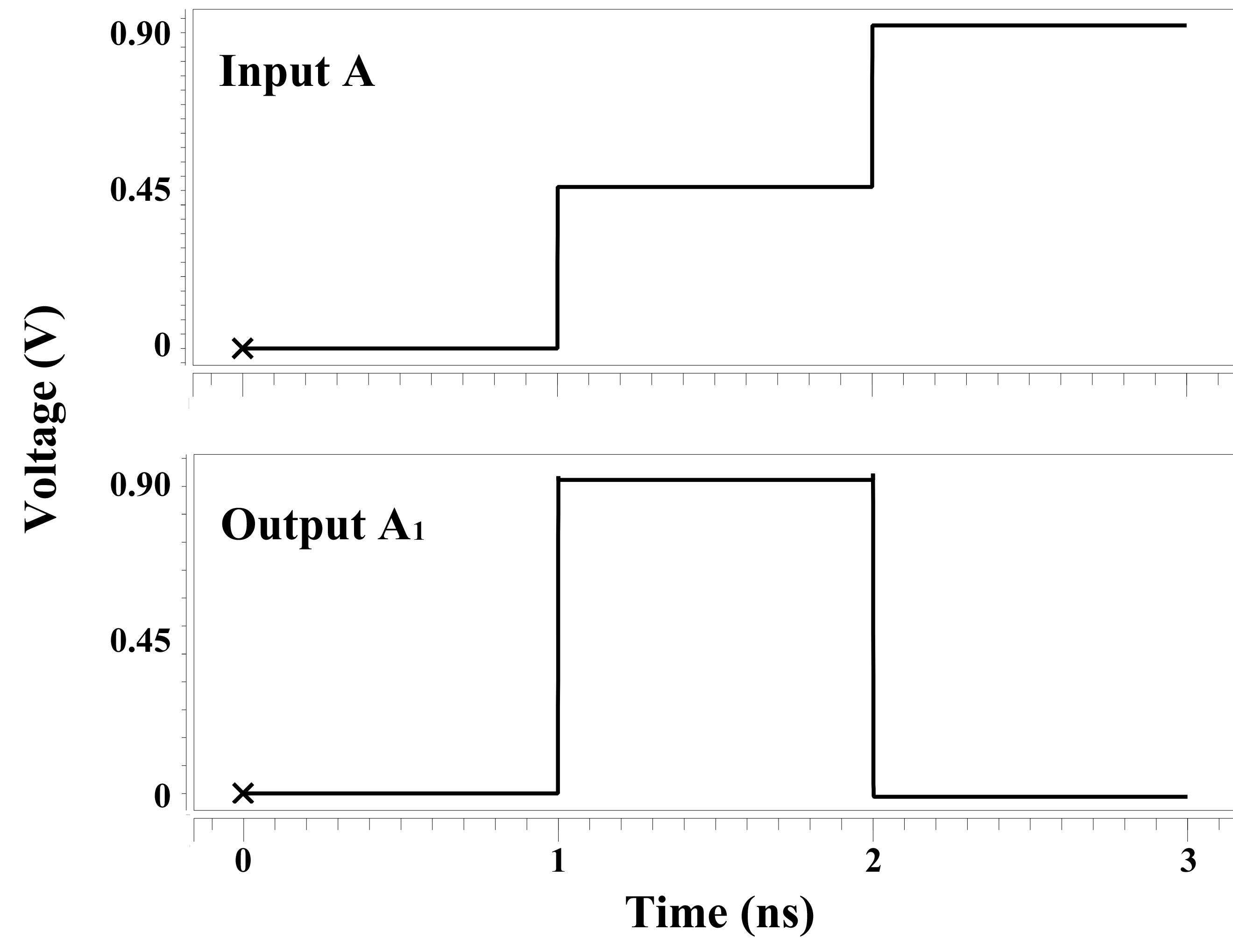}
\caption{Transient analysis of the proposed $A_1$.}
\label{simA11}
\end{figure}

\begin{figure}[!tb]
\centering
\includegraphics[width=10cm, height=5cm]{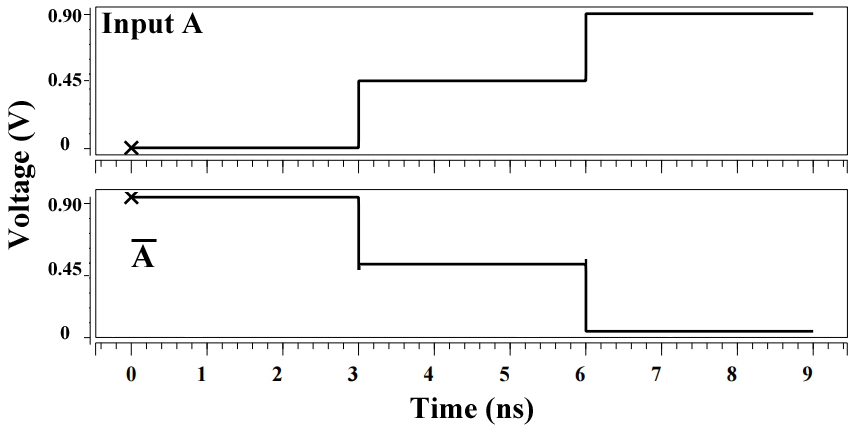}
\caption{Transient analysis of the proposed STI.}
\label{simSTI}
\end{figure}

\begin{figure}[!tb]
\centering
\includegraphics[width=10cm, height=7cm]{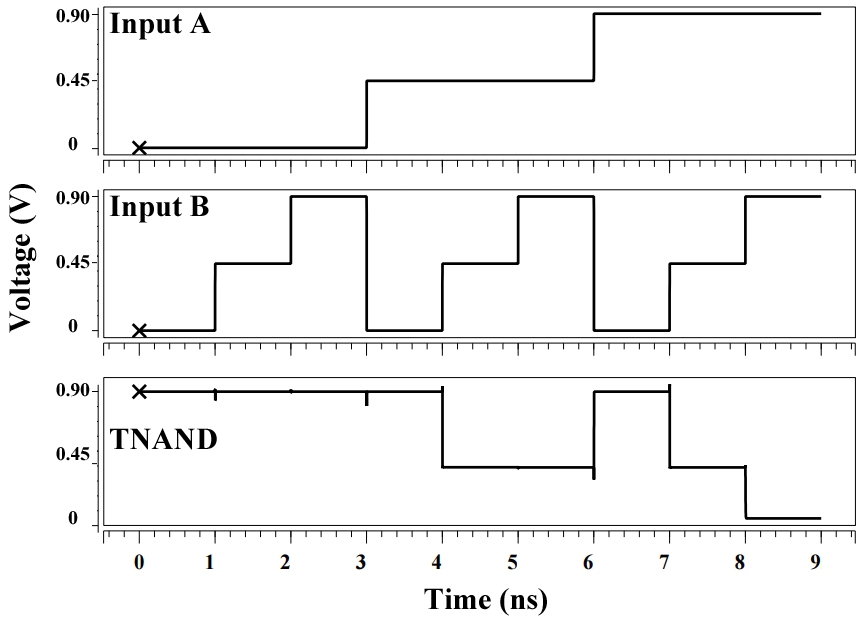}
\caption{Transient analysis of the proposed TNAND.}
\label{simnand}
\end{figure}

Figures from \ref{simA1} to \ref{simnand} illustrate the transient analysis of the proposed six Unary Operators, STI, and TNAND.

\clearpage
\subsection{Comparison of Transistors Count Of All Circuits}

The minimization of Transistors Count is not the only factor that affects the performance of the proposed circuits, but it is a important factor.

Table \ref{t2210} shows the comparison of Transistors Count for the 6 unary operators, STI, and TNAND compared to~\cite{205,123,t16,t18}. This comparison of the proposed circuits demonstrates a notable reduction in Transistors Count.
 
\renewcommand{\arraystretch}{1.5}
\begin{table}[b!]
\caption{Comparison of Transistors Count of All Circuits}
\label{t2210}
\setlength{\tabcolsep}{7pt}
\centering
\begin{tabular}{l|cccc|c|c}
\hline\hline

 								&\textbf{\cite{205}}&\textbf{\cite{123}}&\textbf{\cite{t16}}&\textbf{\cite{t18}} & \textbf{Proposed} & \textbf{Reduction}   \\
\hline
$\mathbf{A^1}$				& 7&-&-&-&\textbf{4}				& 42.85 \% \\
$\mathbf{A^2}$				& 7&-&-&-&\textbf{3}				& 57.14 \% \\
$\mathbf{\bar{A^2}}$			& 13*&-&-&-&\textbf{4}		& 69.2 \% \\
$\mathbf{1\cdot \overline{A_n}}$ 	& 5&-&-&-&\textbf{2}		& 60 \% \\
$\mathbf{1\cdot \overline{A_p}}$ 	& 5&-&-&-&\textbf{2}		& 60 \% \\
$\mathbf{A_1}$				&-&10&-& -&\textbf{3}			& 70 \% \\
\textbf{STI}					&-&6	&6			&6	&\textbf{5} & 16.66 \% 	\\
\textbf{TNAND}				&-&10  &10			&10&\textbf{10} &0 \% \\

\hline\hline
\multicolumn{6}{l}{*$\bar{A^2}$ = $A^2$ (7 T) + $STI$ (6 T) = 13 Transistors}\\
\end{tabular}
\end{table}
\renewcommand{\arraystretch}{1}

$A^1$, $A^2$, $\bar{A^2}$, $1\cdot \bar{A_n}$, and $1\cdot \bar{A_p}$ are compared to \cite{205}, and the reduction in transistors count are 42.9\%, 57.1\%, 69.2\%, 60\%, and 60\%, respectively.

For $A_1$, 70\% compared to $A_1$ in \cite{123}. For STI, 16.67\% compared to STI in \cite{123,t16}, and \cite{t18}. For TNAND, no reduction in Transistors Count but the results gives better performance.
\clearpage
\subsection{Comparison of the Unary Operators Circuits}

\renewcommand{\arraystretch}{1.3}
\begin{table}[!b]
\caption{Comparison between the six unary operators to \cite{205,123}~in terms of~the~average~power~consumption, maximum propagation delay, and maximum PDP at Vdd~(0.9V), temperature (27~\textdegree{}~C), and frequency (1 GHz)}
\label{t25}
\setlength{\tabcolsep}{5pt}
\centering
\begin{tabular}{c|ccc}
\hline \hline
						&Power 		& Delay		& PDP\\
						&(nw)		& (ps)			&(x$10^{-20}$ J)	\\ \hline
						
 $\mathbf{A^1}$ in \cite{205}		&190&4.42&83.98\\
									\textbf{Proposed $\mathbf{A^1}$}&50&1.73&8.65\\ \hline
									 Improvement			&73.68\%&60.86\%&89.7\%\\ \hline \multicolumn{4}{c}{} \\ \hline

$\mathbf{A^2}$ in \cite{205}		&180&4.86&87.48\\
									 \textbf{Proposed $\mathbf{A^2}$}&20&2.10&4.2\\ \hline
									 Improvement			&88.89\%&56.79\%&95.2\%\\ \hline \multicolumn{4}{c}{} \\ \hline

$\mathbf{\bar{A^2}}$ in \cite{205}	&580&7.38&428\\
									 \textbf{Proposed $\mathbf{\bar{A^2}}$}&4&2.32&0.93\\ \hline
									 Improvement			&99.31\%&68.56\%&99.78\%\\ \hline \multicolumn{4}{c}{} \\ \hline

$\mathbf{1\cdot \bar{A_n}}$ in \cite{205}&560&3.72&208.32\\
									 \textbf{Proposed $\mathbf{1\cdot \bar{A_n}}$}&6&4.5&2.7\\ \hline
									 Improvement			&98.93\%&-21.62\%&98.7\%\\ \hline \multicolumn{4}{c}{} \\  \hline

$\mathbf{1\cdot \bar{A_p}}$ in \cite{205}&200&3.10&62\\
									 \textbf{Proposed $\mathbf{1\cdot \bar{A_p}}$}&10&4.14&4.14\\ \hline
									 Improvement&95\%&-33.55\%&93.32\%\\ \hline \multicolumn{4}{c}{} \\ \hline
									 
$\mathbf{A_1}$ in \cite{123}		&572&6.12&350\\
									\textbf{Proposed $\mathbf{A_1}$}&5&1.95&1\\ \hline
									 Improvement			&99.12\%&68.13\%&99.72\%\\ 									 
\hline \hline
\end{tabular}
\end{table}
\renewcommand{\arraystretch}{1}

Table \ref{t25} shows the comparison between the six unary operators to \cite{205,123} in terms of the~average power consumption, maximum propagation delay, and maximum power delay product (PDP).

The comparison of the proposed six unary operators demonstrates a notable reduction in energy consumption (PDP) of 89.7\%, 95.2\%, 99.78\%, 98.7\%, and 93.32\% compared to $A^1$, $A^2$, $\bar{A^2}$, $1\cdot \bar{A_n}$, and $1\cdot \bar{A_p}$ in \cite{205},~respectively. And 99.72\% compared to $A_1$ in \cite{123}.

\subsection{Comparison of Different STI Circuits}
\begin{figure}[!b]
\centering
\includegraphics[width=10cm]{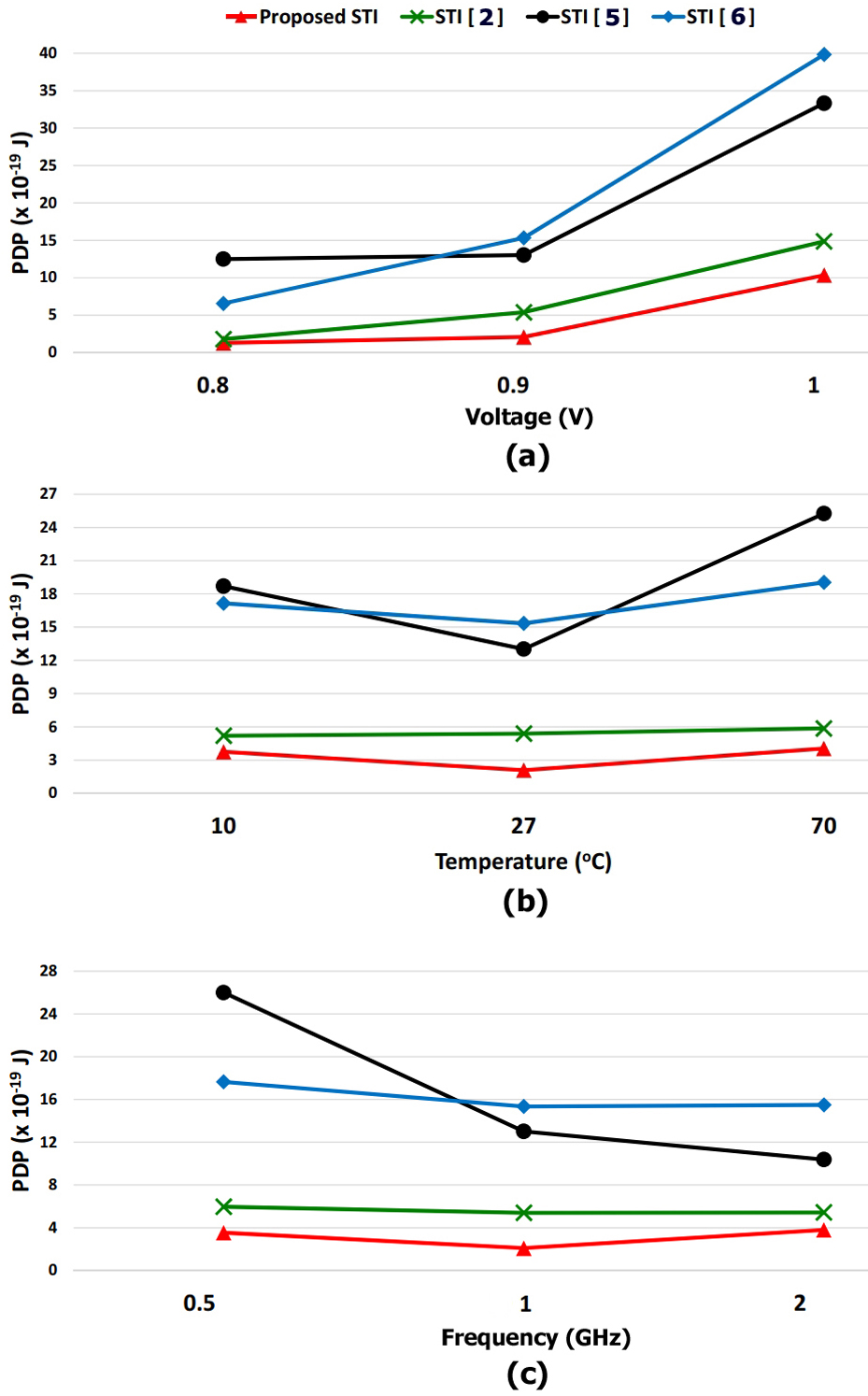}
\caption{PDP Comparison of the investigated STI for: (a) Different Power Supplies, (b) Different Temperatures, and (c) Different Frequencies.}
\label{ChartSTI}
\end{figure}

Fig. \ref{ChartSTI} shows the PDP Comparison of the investigated STI at (a) Different Power Supplies, (b) Different Temperatures, and (c) Different Frequencies.

\subsubsection{Impact of Different Power Supplies}
The effect of different power supplies (0.8V, 0.9V, 1V) on the performance metrics of all proposed circuits is studied. 

Simulation is done at 1 GHz operating frequency and room temperature at 27\textdegree{}C, as illustrated in Fig. \ref{ChartSTI} (a).
 
The comparison of the proposed STI demonstrates a notable reduction in PDP, as shown in Fig. \ref{ChartSTI} (a).\\
At Vdd=0.8V, 28.33\%, 88.68\%, and 80.37\% compared to \cite{123, t16}, and \cite{t18},~respectively.\\ 
At Vdd=0.9V, 61.22\%, 83.95\%, and 86.38\% compared to \cite{123, t16}, and \cite{t18},~respectively.\\
At Vdd=1V, 30.42\%, 68.98\%, and 74.05\% compared to \cite{123, t16}, and \cite{t18},~respectively.\\

\subsubsection{Impact of Different Temperatures}
Temperature noise is one of the most critical issues which negatively affect the performance of the circuit.

The effect of different temperatures (10\textdegree{}C, 27\textdegree{}C, 70\textdegree{}C) on the performance metrics of all proposed circuits is studied. In simulation, the operating frequency is 1 GHz, and the power supply (Vdd) is 0.9V, as illustrated in Fig. \ref{ChartSTI} (b).

The comparison of the proposed STI demonstrates a notable reduction in PDP, as shown in Fig. \ref{ChartSTI} (b).\\
At Temperature=10\textdegree{}C, 27.88\%, 79.95\%, and 78.15\% compared to \cite{123, t16}, and \cite{t18},~respectively. \\
At Temperature=27\textdegree{}C, 61.22\%, 83.95\%, and 86.38\% compared to \cite{123, t16}, and \cite{t18},~respectively.\\
At Temperature=70\textdegree{}C, 30.89\%, 83.96\%, and 78.73\% compared to \cite{123, t16}, and \cite{t18},~respectively.\\

\subsubsection{The Impact of Different Frequencies}
Electronic circuits behave very differently at high frequencies due to a change in the behavior of passive components (resistors, inductors, and capacitors) and parasitic effects on active components, PCB tracks.

Recently, high-frequency operation is in demand for electronic devices. 

The effect of different frequencies (0.5 GHz, 1 GHz, 2 GHz)  on the performance metrics of all proposed circuits is studied. Simulation is done at power supply Vdd equals 0.9 V, and at 27\textdegree{}C temperature, as illustrated in Fig. \ref{ChartSTI} (c).

The comparison of the proposed STI demonstrates a notable reduction in PDP, as shown in Fig. \ref{ChartSTI} (c).\\
At frequency=0.5 GHz, 40.60\%, 86.38\%, and 79.93\% compared to \cite{123, t16}, and \cite{t18},~respectively.\\
At frequency=1 GHz, 61.22\%, 83.95\%, and 86.38\% compared to \cite{123, t16}, and \cite{t18},~respectively. \\
At frequency=2 GHz, 30.20\%, 63.45\%, and 75.52\% compared to \cite{123, t16}, and \cite{t18},~respectively. \\

\subsection{Comparison of Different TNAND Circuits}
\begin{figure}[!t]
\centering
\includegraphics[width=10cm]{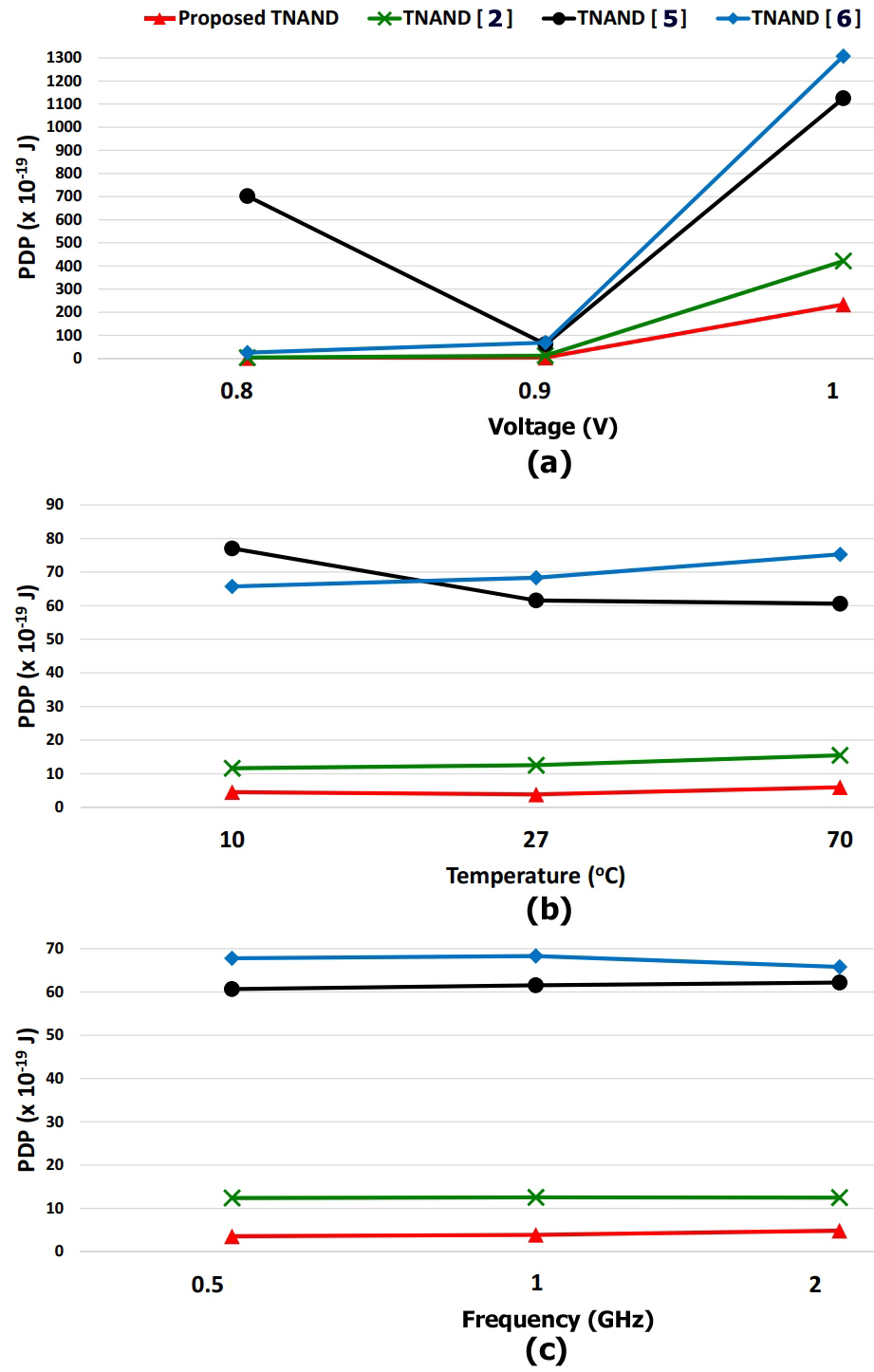}
\caption{PDP Comparison of the investigated TNAND at: (a) Different Power Supplies, (b) Different Temperatures, and (c) Different Frequencies.}
\label{ChartNAND}
\end{figure}

Figure \ref{ChartNAND} shows the PDP Comparison of the investigated TNAND for (a) Different Power Supplies, (b) Different Temperatures, and (c) Different Frequencies.

\subsubsection{Impact of Different Power Supplies}
Simulation is done at 1 GHz operating frequency and room temperature at 27\textdegree{}C, as~illustrated in Fig. \ref{ChartNAND} (a).

The comparison of the proposed TNAND demonstrates a notable reduction in PDP, as shown in Fig. \ref{ChartNAND} (a).\\
At Vdd=0.8 V, 32.53\%, 99.64\%, and 90.07\% compared to \cite{123, t16} and \cite{t18},~respectively. \\
At Vdd=0.9 V, 69.22\%, 93.73\%, and 94.35\% compared to \cite{123, t16} and \cite{t18},~respectively.\\
At Vdd=1 V, 44.6\%, 79.27\%, and 98.3\% compared to \cite{123, t16}, and \cite{t18},~respectively.\\

\subsubsection{Impact of Different Temperatures}
The effect of different temperatures (10\textdegree{}C, 27\textdegree{}C, 70\textdegree{}C) on the performance metrics of all proposed circuits is studied. 

Simulation is done at 1 GHz operating frequency, and at 0.9V power supply Vdd, as illustrated in Fig. \ref{ChartNAND} (b).

The comparison of the proposed TNAND demonstrates a notable reduction in PDP, as shown in Fig. \ref{ChartNAND} (b).\\
At Temperature=10\textdegree{}C, 60.88\%, 94.09\%, and 93.08\% compared to \cite{123, t16} and \cite{t18},~respectively. \\
At Temperature=27\textdegree{}C, 69.22\%, 93.73\%, and 94.35\% compared to \cite{123, t16} and \cite{t18},~respectively.\\
At Temperature=70\textdegree{}C, 61.34\%, 90.13\%, and 92.06\% compared to \cite{123, t16} and \cite{t18},~respectively.\\

\subsubsection{Impact of Different Frequencies}
The effect of different frequencies (0.5 GHz, 1 GHz, 2 GHz)  on the performance metrics of all proposed circuits is studied. 

Simulation is done at at 0.9V power supply Vdd, and at  27\textdegree{}C temperature equals, as illustrated in Fig. \ref{ChartNAND} (c).

The comparison of the proposed TNAND demonstrates a notable reduction in PDP, as shown in Fig. \ref{ChartNAND} (c).\\
At frequency=0.5 GHz, 71.46\%, 94.18\%, and 94.79\% compared to \cite{123, t16}, and \cite{t18},~respectively. \\
At frequency=1 GHz, 69.22\%, 93.73\%, and 94.35\% compared to \cite{123, t16}, and \cite{t18},~respectively.\\
At frequency=2 GHz, 61.27\%, 92.23\%, and 92.66\% compared to \cite{123, t16}, and \cite{t18},~respectively.\\

\section{Conclusion}
This chapter has proposed new designs of seven Unary Operators ($A^1$, $A^2$, $\bar{A^2}$, $A_1$, $1.\bar{A_n}$, $1.\bar{A_p}$, and the Standard Ternary Inverter (STI) $\bar{A}$), and TNAND that aims to optimize the trade-off between performance and energy efficiency.

The design process tried to optimize several circuit techniques such as reducing the number of used transistors, utilizing energy-efficient transistor arrangements, and applying the dual supply voltages (Vdd and Vdd/2).

The proposed ternary circuits are compared to the latest ternary circuits, simulated and tested using HSPICE simulator under various operating conditions with different supply voltages, different temperatures, and different frequencies.

The results prove the merits of the approach in terms of reduced energy consumption (PDP) compared to other existing designs. 

Therefore, the proposed circuits can be implemented in low-power portable electronics and embedded systems to save battery consumption.


\chapter{Ternary Combinational Logic Circuits} 

\label{Chapter3} 

\lhead{Chapter 3. \emph{Ternary Combinational Logic Circuits}} 

\section{Introduction and Literature Review}
The combinational logic circuits are time-independent, and they are only determined by the logical function of their current input state, whereas, sequential logic circuits whose outputs depend on both their present inputs and their previous outputs state giving them some form of memory \cite{301}.

The combinational logic circuits have two or more inputs and provide two or more outputs. They have no memory or feedback loops like sequential circuits.

In general, combinational logic circuits like Ternary Decoder (TDecoder) or Ternary Multiplexer (TMUX) are made up of basic logic gates like NAND, or NOR gates that are combined or connected.

In \cite{123}, the (1:3) TDecoder with 16 CNFETs was proposed, whereas, in \cite{t14} and \cite{t17}, a (1:3) TDecoder with 10 and 11 CNFETs was described in replacing  the TNOR of \cite{123} with a binary NOR.

However, in \cite{302}, the (3:1) TMUX was designed based on the TDecoder of \cite{123}, resulting in a circuit with 28 CNFETs.

In \cite{205}, the authors presented new unary circuits which are subsequently used to design a ``decoder-less'' (3:1) TMUX with 18 CNFETs.

This chapter proposes two different Ternary Decoders (a TDecoder1 using CNFET-based proposed unary operators with 9 CNFETs and a TDecoder2 using DPL binary gates with 12 CMOS transistors), and a (3:1) Ternary multiplexer using CNFET-based proposed unary operators with 15 CNFETs, which are published in (\cite{paper1}, \cite{paper2}, \cite{conf1}) and can be found in Appendices (A, B, C).

\section{The Proposed Ternary Decoders}
The ternary decoder converts $n trits$ information inputs to a maximum $3^n$ unique outputs. It is used in many applications such as the ternary adder, ternary multiplier, ternary memory, and others. 

Ternary decoder with one ternary input ($X$) and three binary outputs ($X_0$, $X_1$, $X_2$) are described in equation \eqref{3eq3} and its truth table is shown in Table~\ref{3t3}.

\begin{equation}
\label{3eq3}
  X_{k, k \in \{0, 1, 2\}} =
 \begin{cases}
2,  &   if  x  = k\\
0,  &   if  x \neq k
\end{cases} 
   \end{equation}

\renewcommand{\arraystretch}{1}
\begin{table}[h!]
\caption{Ternary Decoder truth table}
\label{3t3}
\centering
\begin{tabular}{cccc}
\hline \hline
Input &\multicolumn{3}{c}{Outputs}\\
\hline
X&  $X_2$ &	 $X_1$ & $X_0$ \\
\hline
0		 &0&	0&	2\\
1 	 &0&	2&	0\\
2	&2&	0&	0\\
\hline \hline
\end{tabular}
\end{table}
\renewcommand{\arraystretch}{1}

\subsection{The Proposed TDecoder1 using CNFET}

\begin{figure}[!t]
\centering
\includegraphics[height=20cm ]{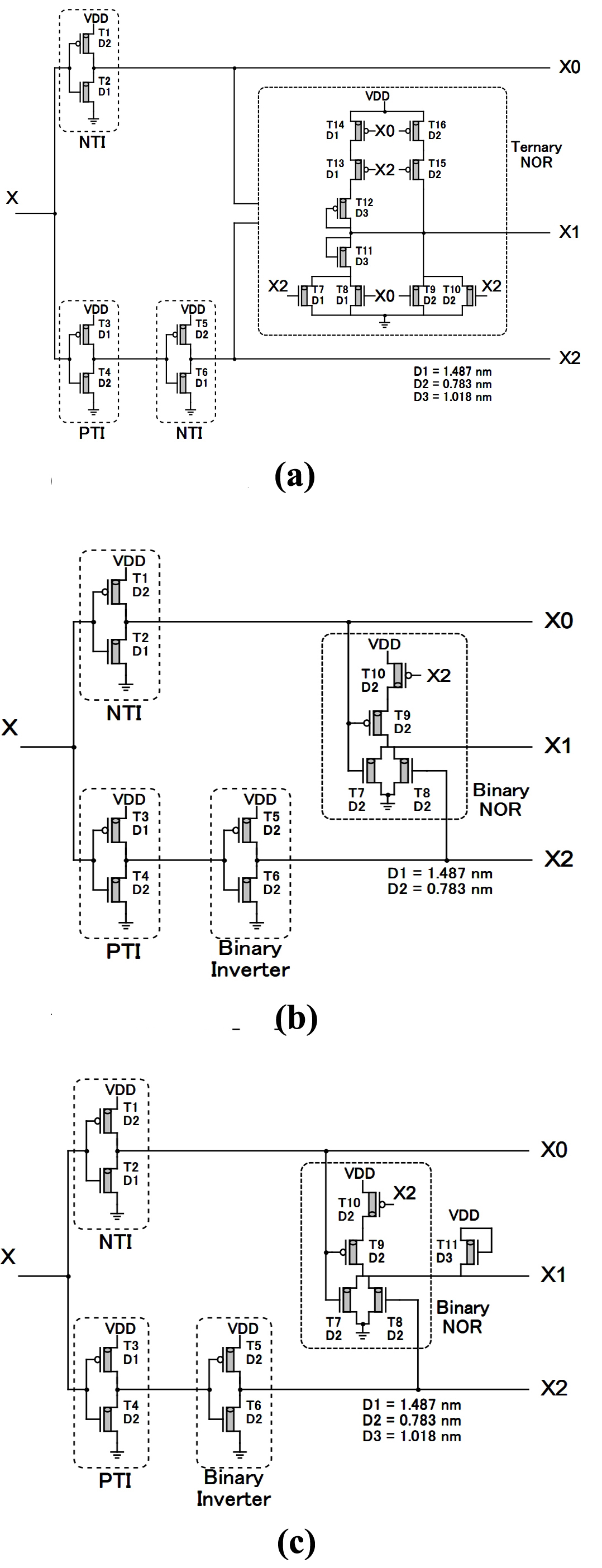}
\caption{The Existing TDecoder in (a) \cite{123} with 16 CNFETs, (b) \cite{t14} with 10 CNFETs, and (c) \cite{t17} with 11 CNFETs.}
\label{figdecoderExisting}
\end{figure}

The existing TDecoders in \cite{123,t14}, and \cite{t17} are shown in Fig. \ref{figdecoderExisting} with 16, 10, and 11 transistors respectively.

\begin{figure}[t!]
\centering
\includegraphics[width=12cm]{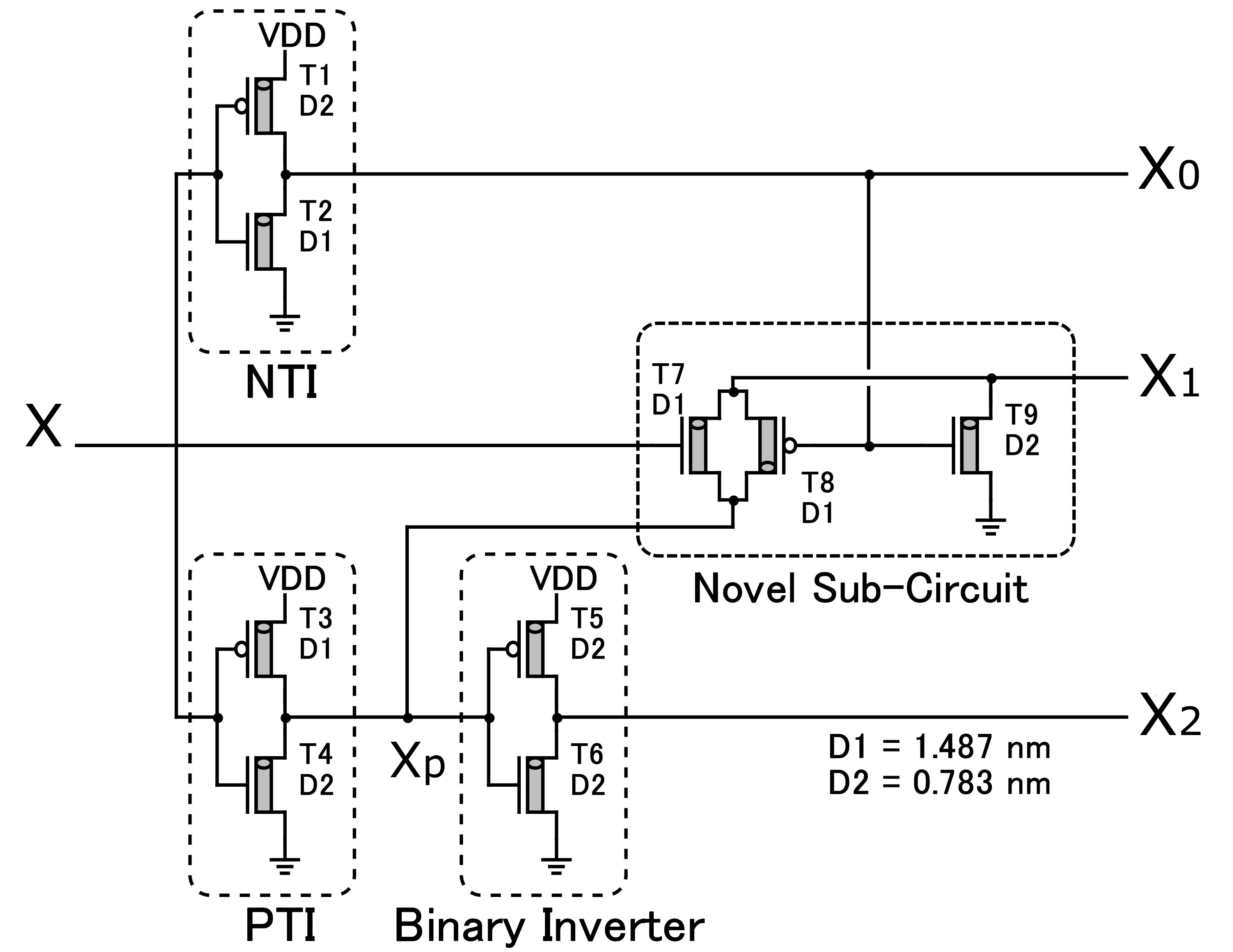}
\caption{Transistor level of the proposed TDecoder1 with 9 CNFETs.}
\label{figdecoder}
\end{figure}

This chapter proposes a new design of TDecoder1 with 9 CNFETs through replacing the TNOR and the second NTI with a novel sub-circuit and a binary inverter respectively, as represented in Fig.~\ref{figdecoder}.  

Figure~\ref{figdecoder} shows the transistor level design of the proposed TDecoder1. It consists of one negative ternary inverter (NTI), one positive ternary inverter (PTI), one binary inverter, and a novel sub-circuit.

The chirality, diameter, and threshold voltage (Vth) of the CNFETs used are shown in Table~\ref{3t4}, while the detailed operation of the proposed circuit is described in Table~\ref{3t5}.

\renewcommand{\arraystretch}{1.3}
\begin{table}[t!]
\caption{The chirality, diameter, and threshold voltage of the CNTs used in the proposed TDecoder1}
\label{3t4}
\centering
\begin{tabular}{lccr}
\hline
CNFET Type&	Chirality&	Diameter&	Vth \\
\hline
\hline
P-CNFET (T1, T5)&	(10, 0)&	0.783nm&	- 0.559V\\
P-CNFET (T3, T8)&	(19, 0)&	1.487nm&	- 0.289V\\
N-CNFET (T4, T6, T9)&	(10, 0)&	0.783nm&	 0.559V\\
N-CNFET (T2, T7)&	(19, 0)&	1.487nm&	 0.289V\\
\hline
\end{tabular}
\end{table}

\begin{table}[t!]
\caption{The detailed operation of the proposed TDecoder1}
\label{3t5}
\centering
\begin{tabular}{lccc}
\hline
\textbf{Ternary Input $X$}&	\textbf{0 (0 V)}&\textbf{1 (0.45 V)}&\textbf{2 (0.9 V)} \\
\hline
\hline
P-CNFET T1&	ON	&OFF	&OFF\\
N-CNFET T2&	OFF&	ON&	ON\\
\hline
\textbf{Output $X_0$}&\textbf{2}&\textbf{0}&	\textbf{0}\\
\hline
\hline
P-CNFET T3&	ON&	ON&	OFF\\
N-CNFET T4&	OFF&	OFF&	ON\\
\hline
\textbf{Intermediate $X_p$}&\textbf{2}&\textbf{2}	&\textbf{0}\\
\hline
\hline
P-CNFET T5&	OFF&	OFF	&ON\\
N-CNFET T6&	ON&	ON	&OFF\\
\hline
\textbf{Output $X_2$}&\textbf{0}&\textbf{0}&\textbf{2}\\
\hline
\hline
N-CNFET T7&	OFF&	ON&	ON\\
P-CNFET T8&	OFF&	ON	&ON\\
N-CNFET T9&	ON	&OFF	&OFF\\
\hline
\textbf{Output $X_1$}&\textbf{0}	&\textbf{$X_p$=2} &\textbf{$X_p$=0}\\
\hline
\end{tabular}
\end{table}
\renewcommand{\arraystretch}{1}

When the input $X$ is logic 0, then transistors (T1 and T3) are turned ON and (T2, T4, and T7) are turned OFF. The output $X_0$ and the intermediate $X_p$ are equal to logic 2. Then (T5 and T8) are turned OFF and (T6 and T9) are turned ON. Therefore, the outputs $X_1$ and $X_2$ are equal to logic 0.
 
When the input $X$ is logic 1, then transistors (T2, T3, and T7) are turned ON and (T1, and T4) are turned OFF. The output $X_0$ is equal to logic 0, and the intermediate $X_p$ is equal to logic 2. Then (T5, and T9) are turned OFF and (T6, and T8) are turned ON. Therefore, the output $X_1$ is equal to the value of $X_p$, which is logic 2, and the output $X_2$ is equal to logic 0.

Finally, when the input $X$ is logic 2, then transistors (T2, T4, and T7) are turned ON and (T1 and T3) are turned OFF. The output $X_0$ and the intermediate $X_p$ are equal to logic 0. Then (T6 and T9) are turned OFF and (T5 and T8) are turned ON. Therefore, the output $X_1$ is equal to the value of $X_p$, which is logic 0, and the output $X_2$ is equal to logic 2. 

Table \ref{3t350} shows the advantage of the proposed TDecoder using CNFET.

\renewcommand{\arraystretch}{1.3}
\begin{table}[thb!]
\caption{The advantage of the proposed TDecoder1}
\label{3t350}
\setlength{\tabcolsep}{5pt}
\begin{center}
\begin{tabular}{p{5pt}p{220pt}|p{200pt}}
\hline
 &\multicolumn {1}{c|}{Disadvantage of the existing TDecoder in \cite{123, t14, t17} }	& \multicolumn {1}{c}{Advantage of the proposed TDecoder1} \\ \hline \hline
&\multicolumn {1}{c|}{Shown in in Fig. \ref{figdecoderExisting}}&\multicolumn {1}{c}{Shown in Fig. \ref{figdecoder}}\\
 &\multicolumn {1}{c|}{The existing design suffers from } 	  	& \multicolumn {1}{c}{The proposed design provides }\\
 &\multicolumn {1}{c|}{High power consumption due to:} 	  	& \multicolumn {1}{c}{ Low power consumption due to:}\\
\hline
1- &The Transistors Count of \cite{123, t14, t17} equal to 16, 10, 11 respectively. &Transistors Count = 9.\\ \hline
2- &To get $X_1$, \cite{123} uses the Ternary $NOR$ (10 transistors). &  \multirow{3}{200pt}{To get $X_1$, the design uses the novel sub-circuit (3 transistors) including the Transmission Gate (T7, T8) which provides low power consumption and propagation delay.}  \\
3- &To get $X_1$, \cite{t14, t17} use the Binary $NOR$ (4 transistors). &  \\ 
&&\\
\hline
4- &Six transistors (T11, T12, T13, T14, T15, T16) must be active to get $X_1$ equals to logic 2 \cite{123}. & \multirow{2}{200pt}{Only the Transmission Gate (T7, T8) must be active to get logic $X_1$ equals to logic 2. }  \\
5- &Two transistors (T9, T10) in series must be active to get logic $X_1$ equals to logic 2 \cite{t14, t17}. & \\ \hline
\hline
\end{tabular}
\end{center}
\end{table}
\renewcommand{\arraystretch}{1}

\subsection{The Proposed TDecoder2 using DPL Binary Gates}

\begin{figure}[t!]
\centering
\includegraphics[width=12cm]{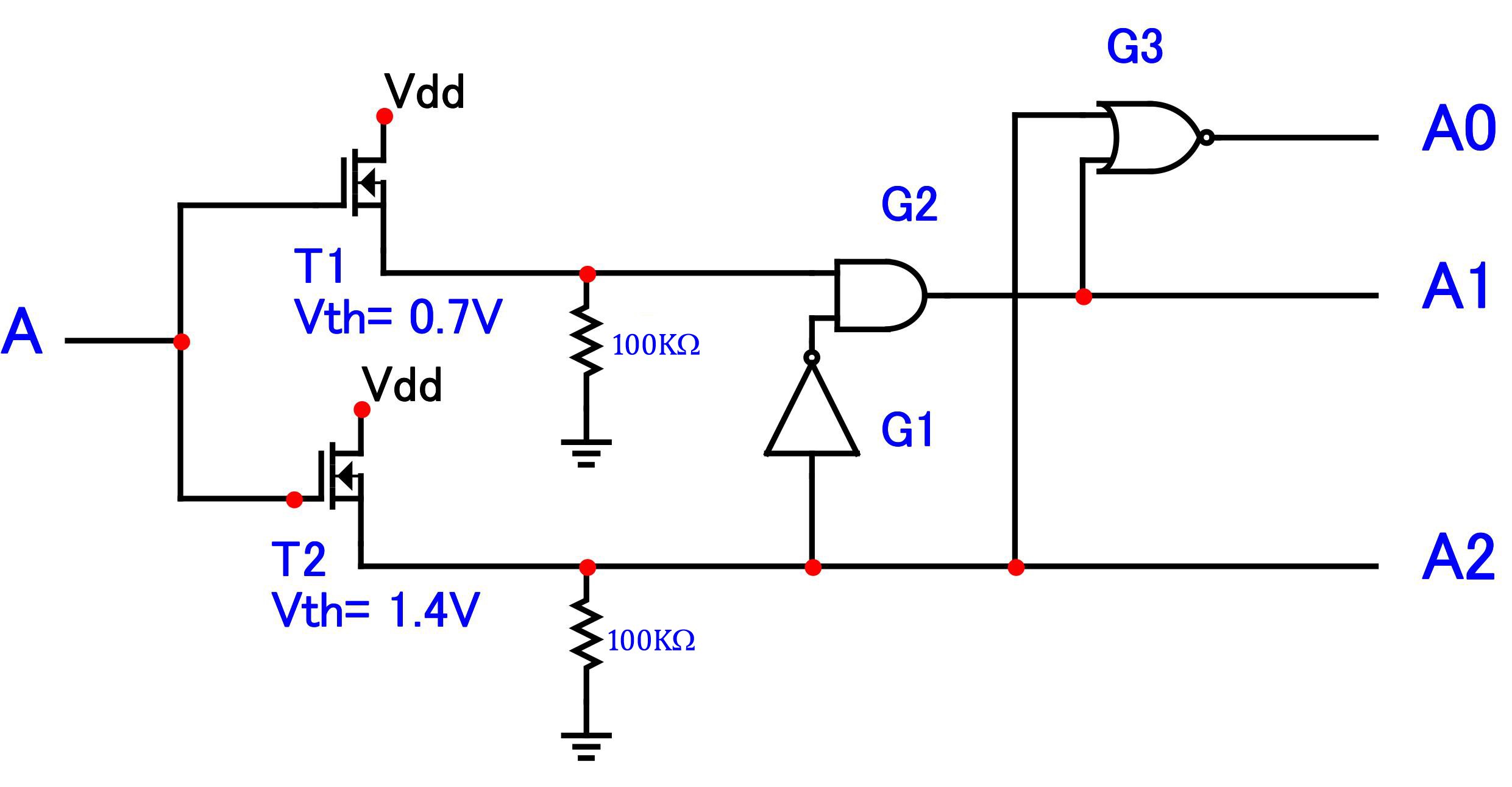}
\caption{Transistor and Gate level of the proposed TDecoder2 with 12 transistors.}
\label{FigDecoder_DPL}
\end{figure}

This chapter proposes TDecoder2 using CMOS DPL Binary Logic Gates with 12 transistors and use 0.18 $\mu$m CMOS DPL at power Vdd (1.8V), as shown in Fig. \ref{FigDecoder_DPL}.

Figure~\ref{FigDecoder_DPL} shows the transistor and gate-level design of the proposed TDecoder2. It consists of two N-channels MOSFET (T1, T2) with threshold voltage equal 0.7V and 1.4V, and three CMOS DPL Binary logic gates, G1 is Inverter, G2 is AND and G3 is NOR logic gate and two resistors (R) 100 K$\Omega$.

The detailed operation of the proposed TDecoder2 is described in Table~\ref{3t59}.

\renewcommand{\arraystretch}{1.3}
\begin{table}[t!]
\caption{The detailed operation of the proposed TDecoder2}
\label{3t59}
\centering
\begin{tabular}{c|cc|ccc}
\hline \hline
Ternary Input &T1&T2&\multicolumn{3}{c}{Binary Outputs}\\
$A$    &	$V_{th}= 0.7V$ &	$V_{th}= 1.4V$ &	$A_2$ & $A_1$ & $A_0$  \\
\hline
0 (0V) 		&OFF&OFF	&0		&	0&	2\\
1 (0.9V)  	&ON&OFF	&0		&	2&	0\\
2 (1.8V)   	&ON&ON	&2		&	0&	0\\
\hline \hline
\end{tabular}
\end{table}

When the input $A$ is logic 0, then transistors (T1 and T2) are turned OFF. The output $A_2$ (from T2 output) is equal to logic 0. The Inputs of the G2 gate are logic 0 (from T1 output) and $\bar{A_2}$ is equal to logic 2 (from G1 gate output). Then the output of (G2) $A_1$ is equal to logic 0. The inputs of the G3 gate are $A_1$ and $A_2$ equal to (0,0). Therefore, the output $A_0$ is equal to logic 2. 

When the input $A$ is logic 1, then transistor (T1) is turned ON and (T2) is turned OFF. The output $A_2$ (from T2 output) is equal to logic 0. The Inputs of the G2 gate are logic 2 (from T1 output) and $\bar{A_2}$ is equal to logic 2 (from G1 gate output). Then the output of (G2) $A_1$ is equal to logic 2. The inputs of the G3 gate are $A_1$ and $A_2$ equal to (2,0). Therefore, the output $A_0$ is equal to logic 0.  

Finally, when the input $A$ is logic 2, then transistors (T1 and T2) are turned ON. The output $A_2$ (from T2 output) is equal to logic 2. The Inputs of the G2 gate are logic 2 (from T1 output) and $\bar{A_2}$ is equal to logic 0 (from G1 gate output). Then the output of (G2) $A_1$ is equal to logic 0. The inputs of  the G3 gate are $A_1$ and $A_2$ equal to (0,2). Therefore, The output $A_0$ is equal to logic 0. 

\section{The Proposed Ternary Multiplexer}

A Multiplexer (MUX) can select between several analog or digital input signals and forward it to a single output line.

\begin{figure}[h!]
\centering
\includegraphics[width = 6 cm]{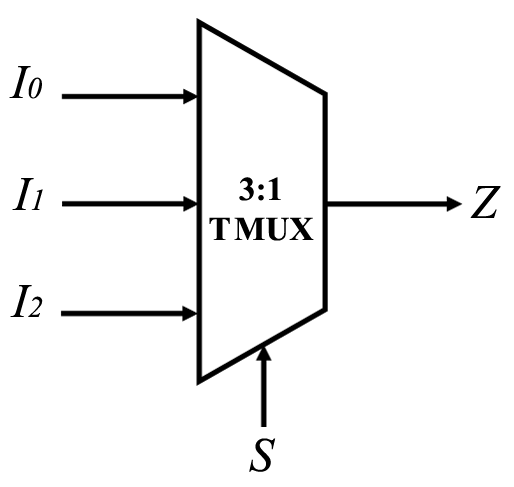}
\caption{The model of 3:1 TMUX}
\label{fig1}
\end{figure}

A 3:1 Ternary Multiplexer (TMUX) is introduced with the general model, as represented in Fig.\ref{fig1}, it has three inputs ($I_0$, $I_1$, $I_2$), one selection ($S$), and one output ($Z$) which depends on ($S$), as described in equation \eqref{3eqif} and Table \ref{33t3}. 

\begin{equation}
\label{3eqif}
  Z =
 \begin{cases}
I_0,  &   if  S  = 0\\
I_1,  &   if  S  = 1\\
I_2,  &   if  S  = 2
\end{cases} 
   \end{equation}

\begin{figure}[t!]
\centering
\includegraphics[width = 16cm, height=21cm]{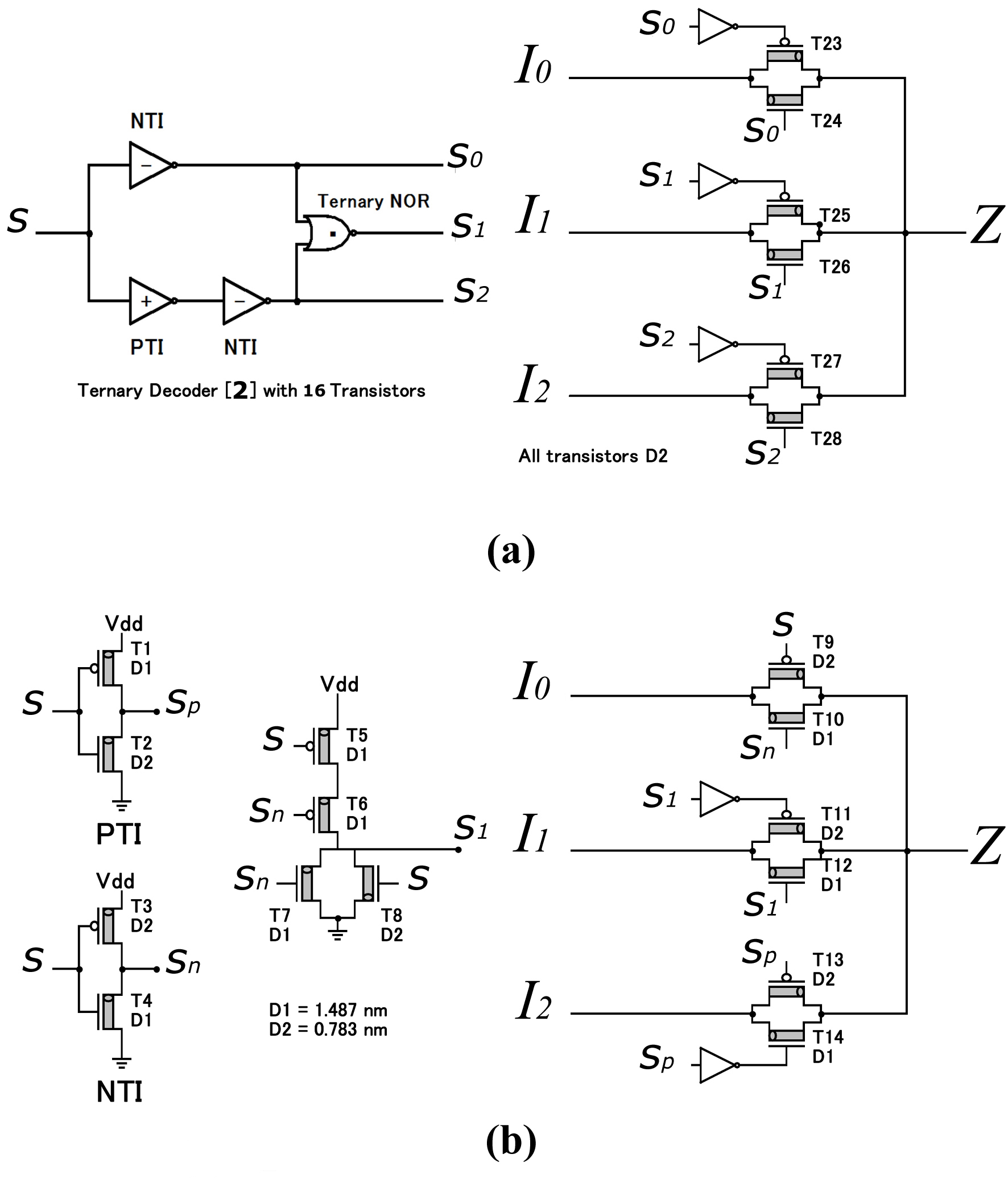}
\caption{The Existing 3:1 TMUX in (a) \cite{302} with 28 transistors, (b) \cite{205} with 18 transistors.}
\label{figExistingMUXboth}
\end{figure}

The existing (3:1) TMUX in \cite{302} and \cite{205} are shown in Fig. \ref{figExistingMUXboth}; \\
Fig. \ref{figExistingMUXboth} (a) shows the TMUX of \cite{302} with 28 CNFETs using the Ternary Decoder of \cite{123}.\\
Fig. \ref{figExistingMUXboth} (b) shows the TMUX of \cite{205} with 18 transistors without using the Ternary Decoder.
 
\begin{figure}[t!]
\centering
\includegraphics[width = 12cm]{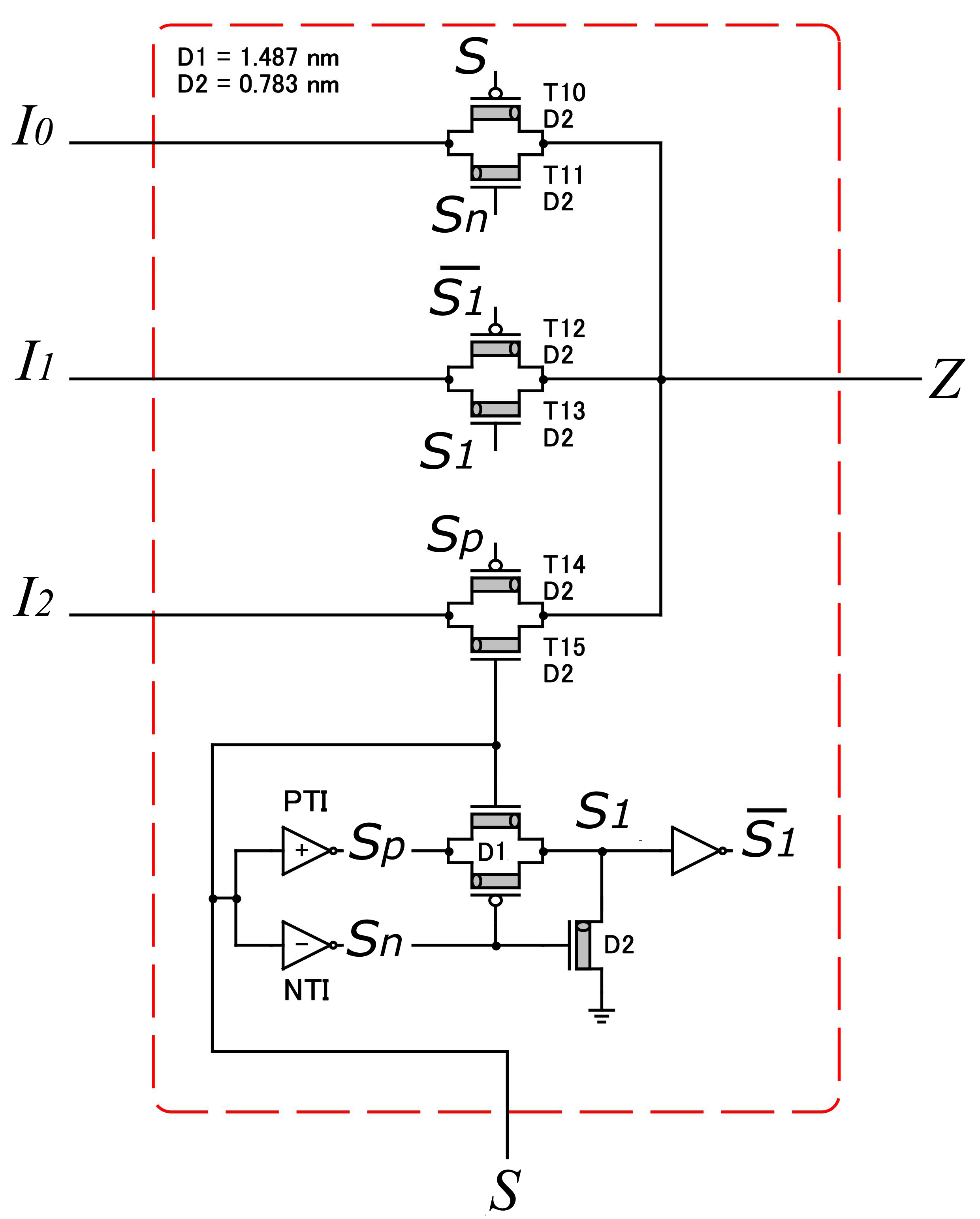}
\caption{Transistor level of the proposed TMUX with 15 CNFETs.}
\label{figmux}
\end{figure}

This chapter proposes a ``decoder-less'' (3:1) TMUX, as shown in Fig.\ref{figmux} with 15 transistors using unary operators, $S_n$, $S_p$, $S_1$, and the complement of $S_1$, $\bar{S_1}$. 
 
The diameter, and threshold voltage of the CNTs used in the proposed (3:1) TMUX are shown in Table \ref{33t444} and the detailed operation of the proposed (3:1) TMUX is shown in Table \ref{33t3}. 

\renewcommand{\arraystretch}{1.3}

\begin{table}[t!]
\caption{The diameter, and threshold voltage of the CNTs used in the proposed TMUX}
\label{33t444}
\centering
\begin{tabular}{lcr}
\hline\hline
CNFET Type &	Diameter&	Vth \\
\hline
P-CNFET (T10,T12,T14)&	0.783nm&	- 0.559V\\
N-CNFET (T11,T13,T15)&	0.783nm&	 0.559V\\
\hline \hline
\end{tabular}
\end{table}

The first nine transistors are: $S_n$ (2 transistors), $S_p$ (2 transistors), $S_1$ (3 transistors), and $\bar{S_1}$ (2 transistors) are described in Chapter \ref{ch2.2}.

When the selection $S$ is logic 0 (0 V), then transistors (T10, T11) are turned ON and (T12, T13, T14, T15) are turned OFF. Therefore, the output $Z$ is equal to the value of the input $I_0$.
 
When the selection $S$ is logic 1 (0.45 V), then transistors (T12, T13) are turned ON and (T10, T11, T14, T15) are turned OFF. Therefore, the output $Z$ is equal to the value of the input $I_1$.
 
Finally, when the selection $S$ is logic 2 (0.9 V), then transistors (T14, T15) are turned ON and (T10, T11, T12, T13) are turned OFF. Therefore, the output $Z$ is equal to the value of the input $I_2$.

\begin{table}[t!]
\caption{The operation of the proposed TMUX}
\label{33t3}
\centering
\begin{tabular}{l|ccc}
\hline \hline
\textbf{Selection ($S$)}		&\textbf{0}	&\textbf{1}&	\textbf{2}\\
\hline
$S_p$ \multirow{4}{*}{ \Bigg\} See Ch 2.2 \href{ch2.2} }				&2	&2&	0\\
$S_n$ 			&2	&0&	0\\
$S_1$ 				&0	&2&	0\\
$\bar{S_1}$			&2	&0&	2\\
\hline
P-CNFET T10		&ON&	OFF&	OFF\\
N-CNFET T11		&ON&	OFF&	OFF\\
\hline
P-CNFET T12		&OFF&	ON&	OFF\\
N-CNFET T13		&OFF&	ON&	OFF\\
\hline
P-CNFET T14	&OFF&	OFF&	ON\\
N-CNFET T15	&OFF&	OFF&	ON\\
\hline
\textbf{Output ($Z$)}	     &\textbf{$I_0$} &\textbf{$I_1$} &	\textbf{$I_2$}\\
\hline\hline
\end{tabular}
\end{table}
\renewcommand{\arraystretch}{1}


\clearpage
\section{Simulation Results and Comparisons}

The proposed TDecoder1, and TMUX are simulated, tested, and compared to \cite{123,t14,t17, 302, 205} using the HSPICE simulator and CNFET with the 32-nm channel length at different power supplies~(0.8V, 0.9V, 1V), different temperatures~(10\textdegree{}C, 27\textdegree{}C, 70\textdegree{}C ), and different frequencies~(0.5GHz, 1GHz, 2GHz).

While the proposed TDecoder2 is simulated, tested using MicroCap10 simulator and CMOS DPL Binary gates with 0.18 $\mu$m at power supply Vdd (1.8V), a temperature (25\textdegree{}C), and a frequency (0.1 GHz). The simulated circuits utilize an implementation of the AND and NOR DPL gates, as shown in chapter I with the average propagation delay is  42.56 ps, and the average power is 4.08 $\mu$w \cite{303}.

\begin{figure}[!t]
\centering
\includegraphics[width=12 cm, height=13cm]{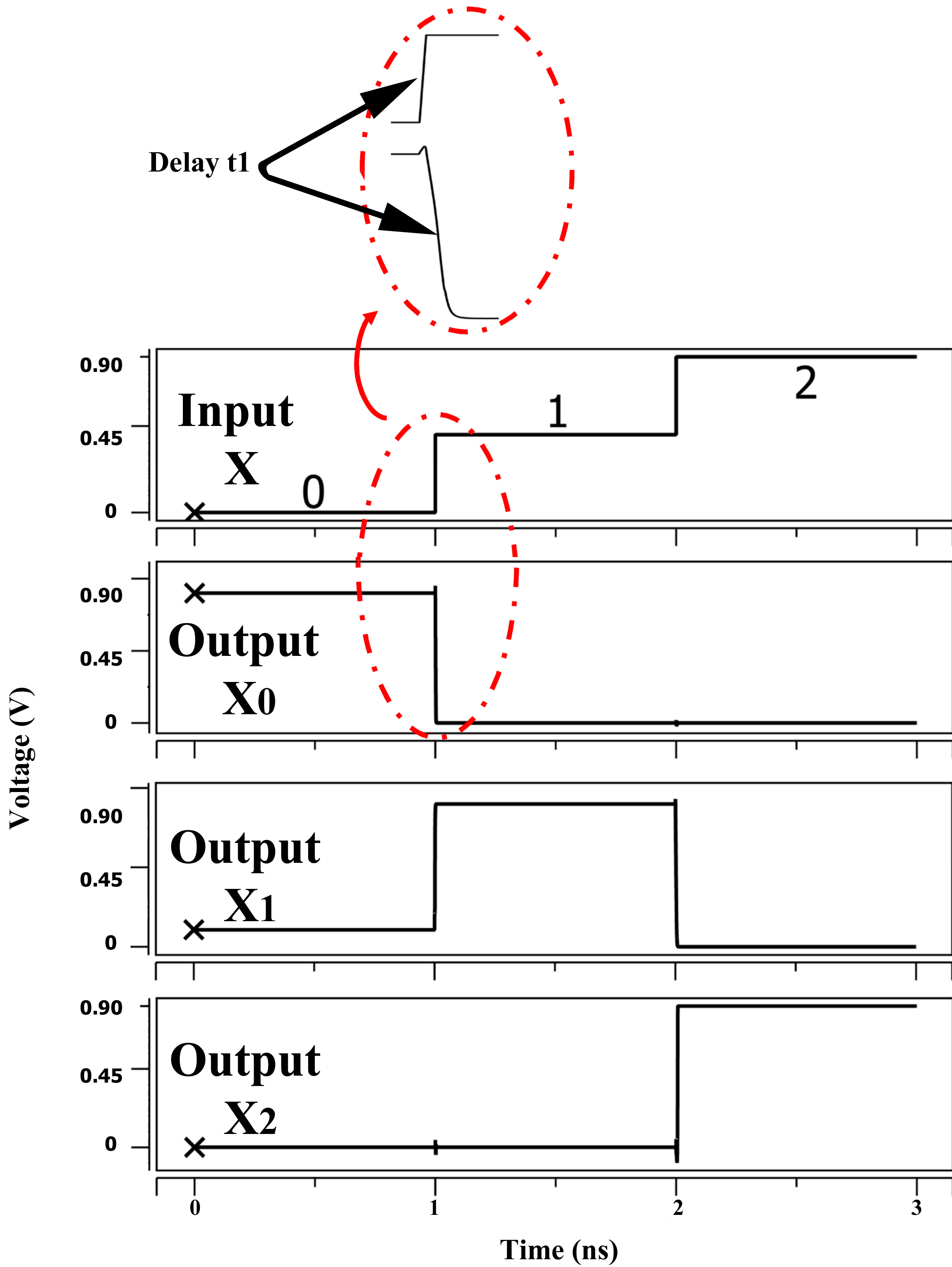}
\caption{Transient analysis of the proposed TDecoder1.}
\label{simdecoder}
\end{figure}

All input signals have a fall and rise time of 20 ps. The propagation delay of the circuit is measured, for example $t_1$, as shown in Fig. \ref{simdecoder}, when the input ($X$) is rising from 0 to 1, and the output ($X_0$) is falling from 2 to 0. A similar procedure is done to get all the possible rising and falling propagation delays for outputs of all studied circuits and find the maximum propagation delay for each circuit.\\
Then the average power consumption, maximum propagation delay, and maximum PDP are obtained for all circuits.

The performance of the proposed circuits will be compared to other designs for the PDP (energy consumption).

\begin{figure}[!t]
\centering
\includegraphics[width=12 cm, height=7cm]{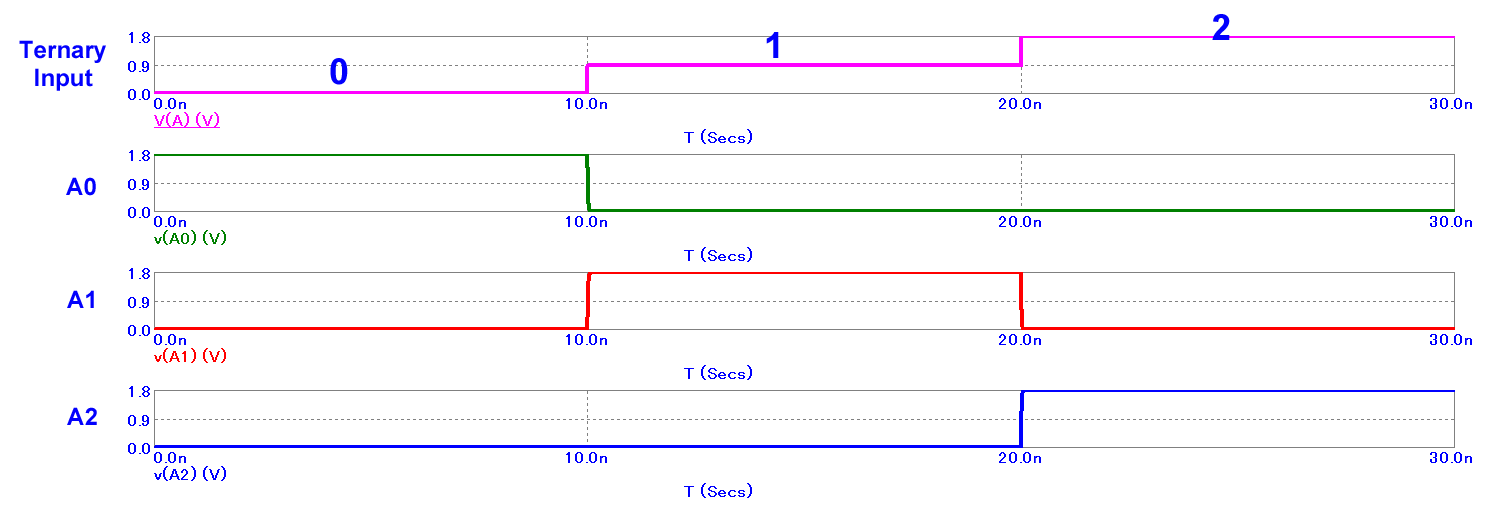}
\caption{Transient analysis of the proposed TDecoder2.}
\label{simdecoder2}
\end{figure}

Figures \ref{simdecoder} and \ref{simdecoder2} illustrate the transient analysis of the proposed TDecoder1 and TDecoder2.

\clearpage
\subsection{Comparison of Transistors Count of All Circuits}

Table \ref{tc3} shows the comparison of Transistors Count for the TDecoders and TMUXs compared to~\cite{123,t14,t17, 205, 302}. 
 
\renewcommand{\arraystretch}{1.5}
\begin{table}[t!]
\caption{Comparison of Transistors Count of All Circuits}
\label{tc3}
\setlength{\tabcolsep}{5pt}
\centering
\begin{tabular}{l|ccccc|cc}
\hline\hline

 	&\textbf{\cite{123}}&\textbf{\cite{t14}}&\textbf{\cite{t17}} &\textbf{\cite{205}} &\textbf{\cite{302}} & \textbf{Proposed1}  & \textbf{Proposed2} \\
\hline
TDecoder &16&10&11&-&-&9  & 12 \\ \hline
TMUX 		&-&-&-&18&28&15 &-\\

\hline\hline
\end{tabular}
\end{table}
\renewcommand{\arraystretch}{1}

This comparison of the proposed circuits demonstrates a notable reduction in Transistors Count.\\
For the TDecoder1, 43.75\%, 10\%, 18.18\%, and 35.71\% compared to TDecoder in \cite{123,t14,t17}, and proposed TDecoder2, respectively.
For the TMUX, 46.43\%, and 16.66\% compared to TMUX in \cite{302,205}, respectively. 


\subsection{Comparison of Different TDecoder Circuits}
\subsubsection{The Impact of Different Power Supplies}
Table~\ref{3t720} shows the comparison to the existing Ternary Decoder in \cite{123, t14}, and \cite{t17} with the proposed TDecoder1 using CNFET in terms of the average power consumption, maximum propagation delay, and maximum energy (PDP) at different supply voltages (0.8 V, 0.9 V, 1 V), same temperature (27\textdegree{}C), and same frequency (1 GHz). The boldface values are the best values, among others.

\renewcommand{\arraystretch}{1.3} 
\begin{table}[!htb]
\caption{Comparison of average power ($\mu$w), maximum delay (ps), and maximum PDP  (x$10^{-17}$ J) of 4 TDecoders at T=27\textdegree{}C, F=1 GHz, and for different supply voltages}
\label{3t720}
\setlength{\tabcolsep}{2pt}
\centering
\begin{tabular}{c|ccc|ccc|ccc}
\hline \hline
	& \multicolumn{3}{c|}{Vdd=0.8 V}& \multicolumn{3}{c|}{Vdd=0.9 V} & \multicolumn{3}{c}{Vdd=1 V}\\
	& Power &Delay 	&PDP 	& Power &Delay 	&PDP & Power &Delay 	&PDP \\
\hline	
TDecoder in \cite{123}  			&2.6	&10.4	&2.74			&3.5&8.91&3.11		&11.8	&8.24 &9.72\\
TDecoder in \cite{t14}			&1.9	&8.50	&1.65			&2.7&7.50&2.02		&10.4	&7.06 &7.37\\
TDecoder in \cite{t17}			&126	&8.67	&109.24		&185&7.68&142.08	&250	&7.18 &168.73\\
\hline
\textbf{Prop.TDecoder1} 	&\textbf{1.9}&\textbf{6.59}&\textbf{1.26}	&\textbf{2.6}	&\textbf{6.26}	&\textbf{1.62} &\textbf{9.1}	&\textbf{4.18}	&\textbf{3.82}\\
\hline\hline
\end{tabular}
\end{table}

\begin{table}[!htb]
\caption{Comparison of average power ($\mu$w), maximum delay (ps), and maximum PDP  (x$10^{-17}$ J) of 4 TDecoders at Vdd=0.9 V, F=1 GHz, and for different Temperatures}
\label{3t721}
\setlength{\tabcolsep}{2pt}
\centering
\begin{tabular}{c|ccc|ccc|ccc}
\hline \hline
	& \multicolumn{3}{c|}{Temp.=10\textdegree{}C}& \multicolumn{3}{c|}{Temp.=27\textdegree{}C} & \multicolumn{3}{c}{Temp.=70\textdegree{}C}\\
		& Power &Delay 	&PDP 	& Power &Delay 	&PDP & Power &Delay 	&PDP \\
\hline	
TDecoder in \cite{123}  	&5.1 	&9.18	&4.72		&3.5&8.91&3.11		&4.4	&8.31 &3.68\\
TDecoder in \cite{t14}	&4.3 	&7.7	&3.33		&2.7&7.50&2.02		&3.5	&7.05 &2.50\\
TDecoder in \cite{t17}	&184 	&7.89	&145.17	&185&7.68&142.08	&189	&7.20 &136.72\\
\hline
\textbf{Prop. TDecoder1} 	&\textbf{4}&\textbf{6.4}&\textbf{2.51}	  &\textbf{2.6}	&\textbf{6.26}	&\textbf{1.62}  &\textbf{3.5}&\textbf{5.85}	&\textbf{2.05}\\
\hline \hline
\end{tabular}
\end{table}

\begin{table}[!htb]
\caption{Comparison of average power ($\mu$w), maximum delay (ps), and maximum PDP  (x$10^{-17}$ J) of 4 TDecoders at T=27\textdegree{}C, Vdd=0.9 V, and for different frequencies}
\label{3t722}
\setlength{\tabcolsep}{2pt}
\centering
\begin{tabular}{c|ccc|ccc|ccc}
\hline
	& \multicolumn{3}{c|}{f= 2  GHz}& \multicolumn{3}{c|}{f= 1  GHz} & \multicolumn{3}{c}{f= 0.5  GHz}\\
		& Power &Delay 	&PDP 	& Power &Delay 	&PDP & Power &Delay 	&PDP \\
\hline	
\hline
TDecoder in \cite{123}  	&6.0		&8.93	&5.17		&3.5&8.91&3.11					&2.3	&8.91 &2.11\\
TDecoder in \cite{t14}	&4.1		&7.54	&3.12		&2.7&7.50&2.02					&2.0	&7.5 &1.46\\
TDecoder in \cite{t17}	&185 	&7.7	&142.24	&185&7.68&142.08			     &184	&7.68 &142.08\\
\hline
\textbf{Prop. TDecoder1} &\textbf{4}&\textbf{6.25}&\textbf{2.44}	  &\textbf{2.6}	&\textbf{6.26}	&\textbf{1.62}     &\textbf{1.8}&\textbf{6.26}	&\textbf{1.17}\\
\hline\hline
\end{tabular}
\end{table}
\renewcommand{\arraystretch}{1} 
\clearpage
The comparison of the proposed TDecoder1 demonstrates a notable reduction in PDP, as shown in Table~\ref{3t720}.\\
At Vdd=0.8 V, 54.01\%, 23.64\%, and 98.85\% compared to \cite{123, t14}, and \cite{t17} respectively. \\
At Vdd=0.9 V, 47.91\%, 19.8\%, and 98.86\% compared to \cite{123, t14}, and \cite{t17} respectively.\\
At Vdd=1 V, 60.7\%, 48.17\%, and 97.74\% compared to \cite{123, t14}, and \cite{t17} respectively.\\

\subsubsection{Impact of Different Temperatures}
Table~\ref{3t721} shows the comparison to the existing Ternary Decoder in \cite{123, t14}, and \cite{t17} with the proposed TDecoder1 in terms of the average power consumption, maximum propagation delay, and maximum energy (PDP) at different temperatures (10\textdegree{}C, 27\textdegree{}C, 70\textdegree{}C), same supply voltage Vdd (0.9 V), and same frequency (1 GHz).

The comparison of the proposed TDecoder1 demonstrates a notable reduction in PDP, as shown in Table~\ref{3t721}.\\
At Temp.=10\textdegree{}C, 46.82\%, 24.62\%, and 98.27\% compared to \cite{123, t14}, and \cite{t17} respectively. \\
At Temp.=27\textdegree{}C, 47.91\%, 19.8\%, and 98.86\% compared to \cite{123, t14}, and \cite{t17} respectively.\\
At Temp.=70\textdegree{}C, 44.29\%, 18\%, and 98.5\% compared to \cite{123, t14}, and \cite{t17} respectively.\\

\subsubsection{The Impact of Different Frequencies}
Table~\ref{3t722} shows the comparison to the existing Ternary Decoder in \cite{123, t14}, and \cite{t17} with the proposed TDecoder1 in terms of the average power consumption, maximum propagation delay, and maximum energy (PDP) at different frequencies (0.5 GHz, 1 GHz, 2 GHz), same temperatures (27\textdegree{}C), and same supply voltage Vdd (0.9 V).

The comparison of the proposed TDecoder1 demonstrates a notable reduction in PDP, as shown in Table~\ref{3t722}.\\
At frequency=2 GHz, 52.8\%, 21.79\%, and 98.28\% compared to \cite{123, t14}, and \cite{t17} respectively.\\ 
At frequency=1 GHz, 47.91\%, 19.8\%, and 98.86\% compared to \cite{123, t14}, and \cite{t17} respectively.\\
At frequency=0.5 GHz, 44.55\%, 19.86\%, and 99.18\% compared to \cite{123, t14}, and \cite{t17} respectively.\\

\subsection{Comparison between the Proposed TDecoders Circuits}

As mentioned in chapter I that CNFET provides better energy efficiency compared to CMOS, FinFET, and other transistor technologies \cite{119}.

Table~\ref{3t730} shows the comparison between the proposed TDecoder1 using CNFET and TDecoder2 using CMOS DPL Binary Gates in terms of the average power consumption, maximum propagation delay, and maximum energy (PDP) at a supply voltage (0.9 V), a room temperature (27\textdegree{}C), and a frequency (1 GHz).

\begin{table}[!htb]
\caption{Comparison of average power ($\mu$w), maximum delay (ps), and maximum PDP  (x$10^{-17}$ J) of the two proposed TDecoders}
\label{3t730}
\centering
\begin{tabular}{l|ccc}
\hline \hline

		& Power &Delay 	&PDP 	 \\
\hline	
\textbf{Proposed TDecoder2 using CMOS DPL Binary Gates}   &13	&140	&182   \\
\textbf{Proposed TDecoder1 using CNFET}   &\textbf{2.6}	&\textbf{6.26}	&\textbf{1.62}    \\
\textbf{TDecoder1 Improvement}   &\textbf{80\%}	&\textbf{95.5\%}	&\textbf{99.1\%}    \\
\hline \hline
\end{tabular}
\end{table}

The comparison of the proposed TDecoder1 demonstrates a notable reduction in Power (80\%), Delay (95.5\%), and PDP (99.1\%) compared to the proposed TDecoder2, as shown in Table~\ref{3t730}.

\subsection{Comparison of Different TMUX Circuits}

The proposed TMUX is compared to the TMUX in \cite{302,205}. The three circuits are simulated and tested using the HSPICE simulator at a frequency of 1 GHz, room temperature (27\textdegree{}C) and a supply voltage Vdd (0.9V).

For a fair comparison, the four circuits are simulated using the same software,~implementation techniques, CNFET model parameters, and the same values of ($I_0$, $I_1$, $I_2$, $S$), as follows:

$I_0$ =  $(210 021 002)_3$, $I_1$ =  $(102 201 001)_3$, $I_2$ =  $(012 012 012)_3$ and $S$ =  $(000 111 222)_3$.

\begin{figure}[!tb]
\centering
\includegraphics[height=20cm]{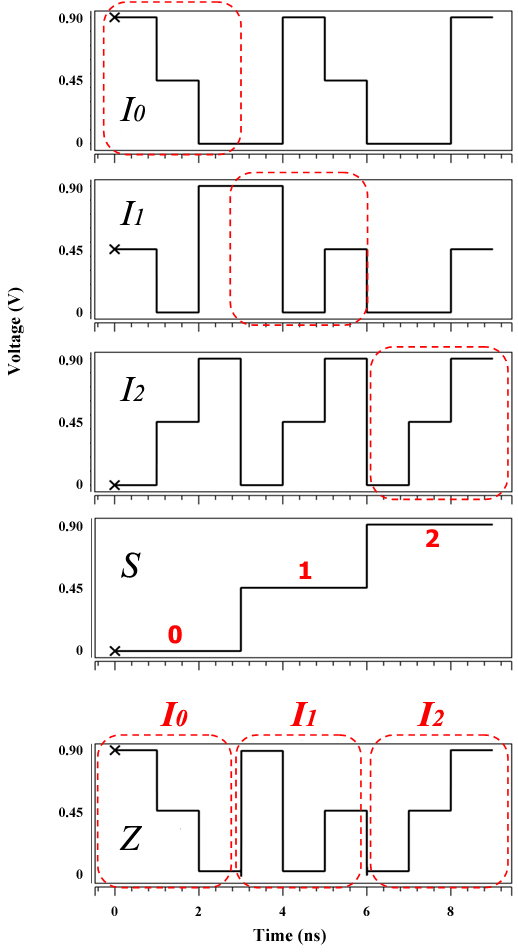}
\caption{Transient analysis of the proposed 3:1 TMUX (dotted rectangle corresponds to input selected according to S).}
\label{simmux}
\end{figure}

Figure \ref{simmux} illustrates the transient analysis of the proposed (3:1) TMUX. The dotted rectangle for each of the three inputs ($I_0$, $I_1$, $I_2$) is selected according to the ternary selection ($S$) to get the output ($Z$).

Table \ref{3t555} shows the comparison between the three TMUXs in terms of the average power consumption, maximum propagation delay, and maximum PDP.

\renewcommand{\arraystretch}{1.3}
\begin{table}[!t]
\caption{Comparison between three TMUXs at Vdd (0.9 V), temperature (27\textdegree{}C), and frequency (1 GHz)}
\label{3t555}
\setlength{\tabcolsep}{5pt}
\centering
\begin{tabular}{c|ccc}
\hline \hline
		&Power (nw)	 	& Delay (ps) & PDP (x$10^{-20}$ J)	 \\

\hline
TMUX in \cite{302}			&67.3				&14.3						&96.3	\\
TMUX in \cite{205}			&60.1				&9.64		 				&57.9	\\
\hline
\textbf{Proposed TMUX}&\textbf{48.4}		&\textbf{8.50}				&\textbf{41.1}	\\
\hline \hline
\end{tabular}
\end{table}
\renewcommand{\arraystretch}{1}

The comparison of the proposed “decoder-less” TMUX demonstrates a notable reduction in PDP of around 57.32\%, and 29.02\% compared to \cite{302}, and \cite{205}, respectively.

\clearpage
\section{Conclusion}

This chapter proposed three Ternary Combinational circuits (two different Ternary Decoders (a TDecoder1 using CNFET-based proposed unary operators of the ternary system and a TDecoder2 using DPL binary gates), and a Ternary multiplexer using CNFET-based proposed unary operators of the ternary system).

The design process tried to optimize several circuit techniques such as reducing the number of used transistors, utilizing energy-efficient transistor arrangements, and applying the dual supply voltages (Vdd and Vdd/2).

The proposed ternary circuits are compared to the latest ternary circuits, simulated and tested using HSPICE simulator under various operating conditions at different supply voltages, different temperatures, and different frequencies.

The results proved the merits of the approach in terms of reduced energy consumption (PDP) compared to other existing designs. 

Therefore, the proposed circuits can be implemented in low-power portable electronics and embedded systems to save battery consumption.


\chapter{Ternary Arithmetic Logic Units (TALU)} 

\label{Chapter4} 

\lhead{Chapter 4. \emph{Ternary Arithmetic Logic Units}} 

\section{Introduction and Literature Review}
A Ternary Arithmetic Logic Unit (TALU) is a combinational circuit that performs arithmetic operations like addition and multiplication.
A TALU is a fundamental building block of many types of computing circuits like the central processing unit (CPU) of computers, and graphics processing units (GPUs) \cite{401}.

In \cite{123}, the authors presented a Ternary Half Adder (THA) with 136 CNFETs, and a Ternary Multiplier (TMUL) with 100 CNFETs using their TDecoder (16 CNFETs), binary AND, binary OR, a ternary encoder (level shifter), and a ternary OR. Also, the authors in \cite{t15} the proposed a THA with 112 CNFETs, and a TMUL with 86 CNFETs using TDecoders of \cite{123}, binary NANDs, binary NORs, a ternary encoder (level shifter), a ternary OR, and a STI. Whereas, in \cite{t16} the authors used a TNAND instead of the ternary encoder to design a THA with 112 CNFETs, and a TMUL with 76 CNFETs using their TDecoder (10 transistors), binary NANDs, a binary NOT, a TNAND, and a STI.
 
However, in \cite{302} the authors the proposed a THA with 168 CNFETs, and a TMUL with 112 CNFETs using cascading their TMUXs (28 CNFETs).\\
While in \cite{205}, Unary Operators and their TMUX (18 CNFETs) were employed in the design of a THA with 64 CNFETs, and a TMUL with 58 CNFETs.

The designs mentioned above either suffer from a large number of transistors in the THA, and TMUL \cite{123,t15,t16,302} or even in the unary operators circuit \cite{205}.

Therefore, this thesis first proposes an efficient circuit implementation of six unary operators, a STI and TNANDs, as shown in Chapter 2, a novel (1:3) TDecoder1 with 9 CNFETs, and a novel “decoder-less” (3:1) TMUX with 15 CNFETs, as shown in Chapter 3, to be used in this Chapter to design three different models of THAs and TMULs, which are published in (\cite{paper1}, \cite{paper2}) and can be found in Appendices (A, B), as follow:
\setstretch{1}
\begin{enumerate}
 \item Proposes a THA1 with 85 CNFETs and a TMUL1 with 61 CNFETs using~the~proposed TDecoder1, a STI, TNANDs, and applying De Morgan's Law.
  \item Proposes a THA2 with 90 CNFETs and a TMUL2 with 60 CNFETs using~the~proposed cascading TMUX.
   \item Proposes a THA3 with 45 CNFETs and a TMUL3 with 40 CNFETs using~the~proposed Unary Operators combined with TMUX.
 \end{enumerate}
\setstretch{1.5}
\section{The Proposed Ternary Half Adders}
THA can add two ternary inputs and provides two outputs: the Sum and the Carry, as shown in Table~\ref{4t51}. 

The basic equations of the Sum and the Carry can be derived in Table~\ref{4t51} to lead three designs:
\begin{enumerate}
  \item Conventional design in \cite{123,t15,t16} which uses equation \eqref{4eq4}.
  \item Cascading TMUXs design in \cite{302} which uses equation \eqref{4eq42}.
  \item Unary operators-based design in \cite{205} which uses equation \eqref{4eq43}.
  \end{enumerate}
   
\renewcommand{\arraystretch}{1.3}
\begin{table}[!t]
\caption{Truth table of THA}
\label{4t51}
\centering
\begin{tabular}{|c||l l l|}
\multicolumn{4}{c}{Sum}\\ 
\hline
A/B&$B_0$(0)&$B_1$(1)&$B_2$(2)\\
 \hline\hline
$A_0$(0)&0 \multirow{3}{*}{ \Bigg\}	$\mathbf{A}$ }	&1 \multirow{3}{*}{ \Bigg\}	$\mathbf{A^1}$ } &2 \multirow{3}{*}{ \Bigg\} $\mathbf{A^2}$}\\
$A_1$(1)&	1						 				&2		&0\\
$A_2$(2)&	2										&0						&1\\
\hline
\multicolumn{4}{c}{}\\
\multicolumn{4}{c}{Carry}\\
\hline
A/B&$B_0$(0)&$B_1$(1)&$B_2$(2)\\ 
\hline\hline
$A_0$(0)&0 \multirow{3}{*}{ \Bigg\}	\textbf{0} }&0 \multirow{3}{*}{ \Bigg\}	$\mathbf{1\cdot \overline{A_p}}$ }    &0 \multirow{3}{*}{ \Bigg\} $\mathbf{1\cdot \overline{A_n}}$ }\\
$A_1$(1)&0&0&1\\
$A_2$(2)&0&1&1\\
\hline
\end{tabular}
\end{table}
\renewcommand{\arraystretch}{1}

\begin{equation}
\label{4eq4}
\begin{split}
Sum = 2 \bullet(A_0B_2+A_1B_1+A_2B_0)+ 1 \bullet(A_0B_1+A_1B_0+A_2B_2)
\\
Carry =  1 \bullet(A_1B_2+A_2B_1+A_2B_2)  
   \end{split}
     \end{equation}

\begin{equation}
\label{4eq42}
\begin{split}
Sum = A.B_0 + (1.A_0 + 2.A_1 +0.A_2).B_1+ (2.A_0 + 0.A_1 +1.A_2).B_2
\\
Carry = 0.B_0 + (0.A_0 + 0.A_1 +1.A_2).B_1+ (0.A_0 + 1.A_1 +1.A_2).B_2
\end{split}
  \end{equation}

\begin{equation}
\label{4eq43}
\begin{split}
Sum = A.B_0 + A^1.B_1 + A^2.B_2
\\
Carry = 0.B_0 + (1\cdot \bar{A_p}).B_1 + (1\cdot \bar{A_n}).B_2
 \end{split}
  \end{equation}

Where $A_k$ and $B_k$, $k$ $\in$ \{0,1,2\}, are the outputs of the TDecoder from inputs $A$ and $B$.

\begin{figure}[t!]
\centering
\includegraphics[width=\columnwidth]{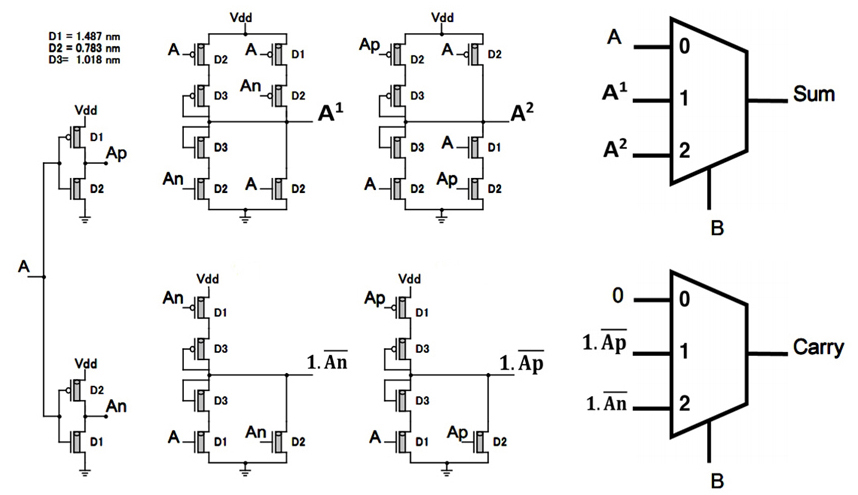}
\caption{The Existing THA in \cite{205} with 64 CNFETs.}
\label{figExistingHA2}
\end{figure}

\pagebreak
The existing THA in \cite{123, t15, t16} and \cite{205} are shown in Fig. \ref{figExistingHA2} and Fig. \ref{figExistingHA1};\\
Figure \ref{figExistingHA2} shows the THA of \cite{205} with 64 CNFETs that contains (3:1) TMUX (18 transistors) and Unary Operators.

\begin{figure}[t!]
\centering
\includegraphics[width=\columnwidth]{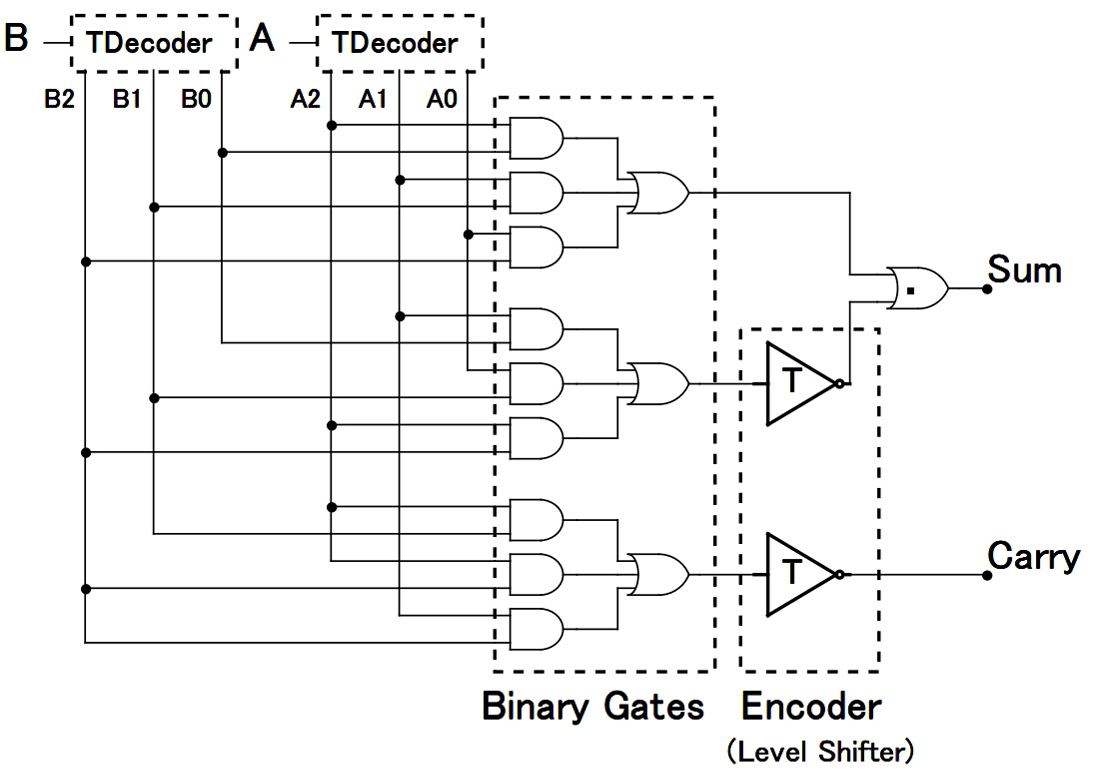}
\caption{The Existing THA in \cite{123} with 136 CNFETs.}
\label{figExistingHA1a}
\end{figure}

\begin{figure}[t!]
\centering
\includegraphics[width=\columnwidth]{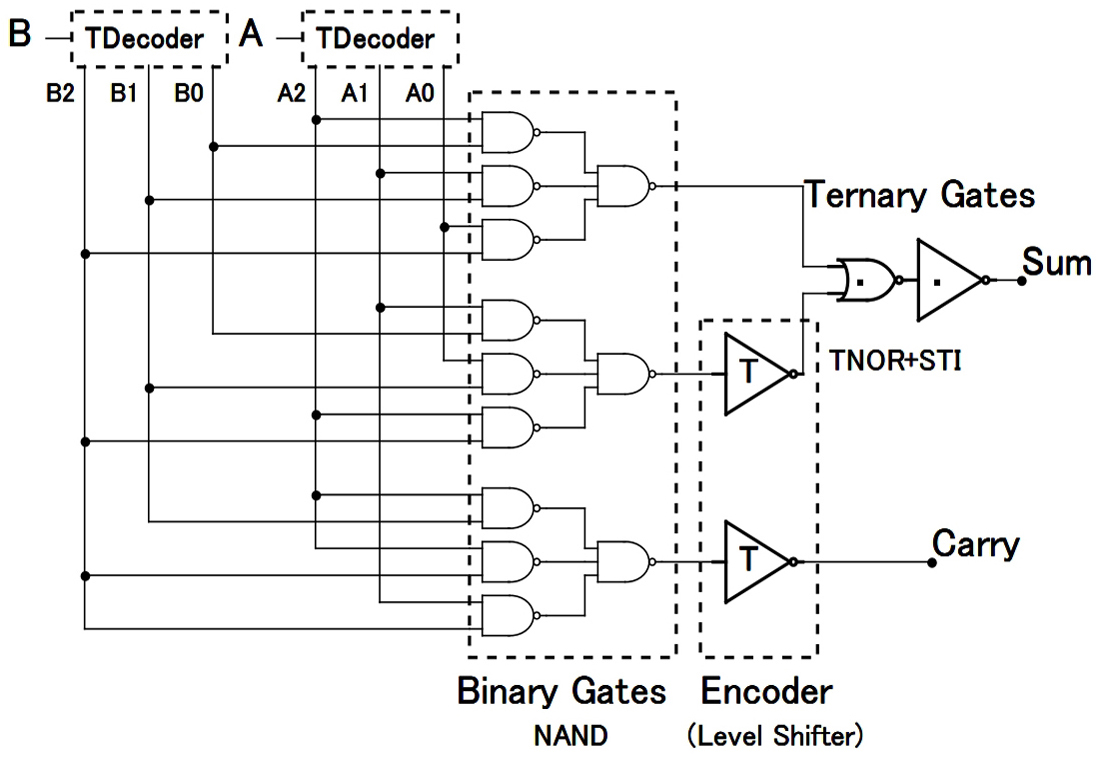}
\caption{The Existing THA in \cite{t15} with 112 CNFETs.}
\label{figExistingHA1b}
\end{figure}

\begin{figure}[t!]
\centering
\includegraphics[width=\columnwidth]{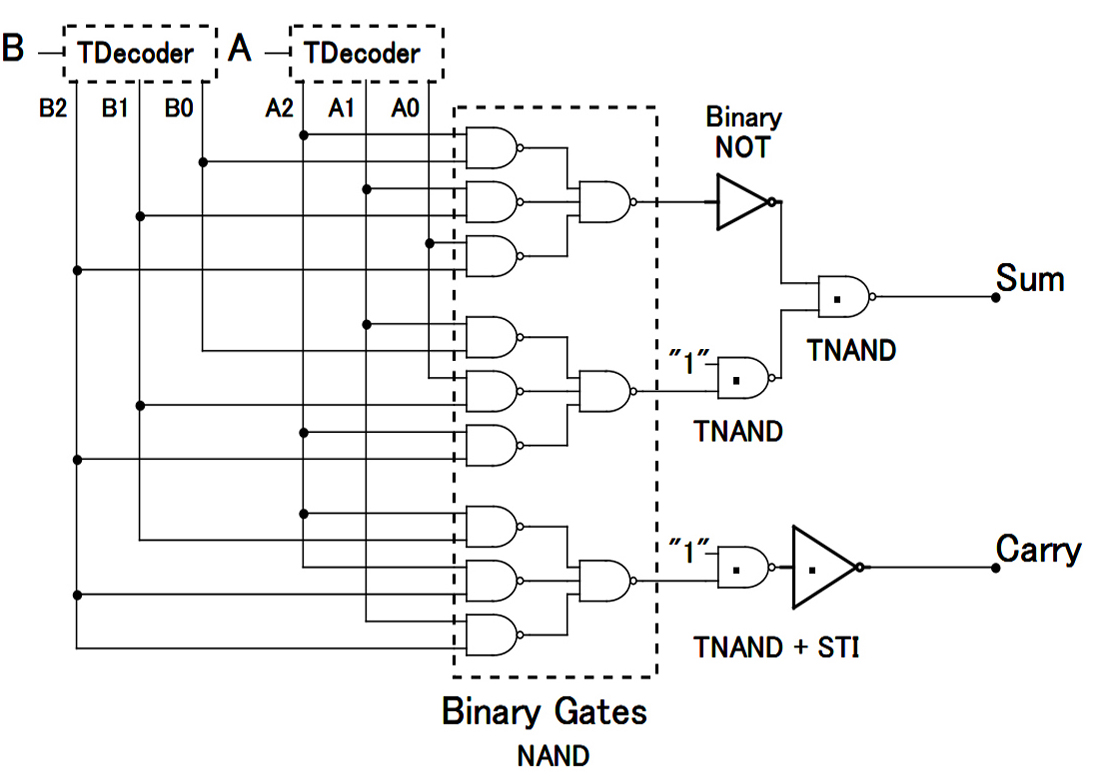}
\caption{The Existing THA in \cite{t16} with 112 CNFETs.}
\label{figExistingHA1c}
\end{figure}

Where Figure \ref{figExistingHA1a} shows the THA of \cite{123} with 136 CNFETs that contains their sTDecoder (16 transistors), binary ANDs, binary ORs, ternary encoders level shifter, and a ternary OR.\\
And Figure \ref{figExistingHA1b} shows the THA of \cite{t15} with 112 CNFETs that contains TDecoder of \cite{123}, binary NANDs, binary NORs, ternary encoders level shifter, a ternary OR, and a STI.\\
Finally, Figure \ref{figExistingHA1c} shows the THA of \cite{t16} with 112 CNFETs that contains their TDecoder (10 transistors), binary NANDs, a binary NOT, TNANDs, and a STI.

\clearpage
\subsection{The proposed THA1}

This Chapter proposes a new conventional design, which uses equation \eqref{4eq4}, of THA1 with 85 CNFETs using De Morgan's Law and dual power supply (Vdd and Vdd/2) to eliminate the encoder level shifter in \cite{123, t15}.

\begin{figure}[!b]
\centering
\includegraphics[width=15cm]{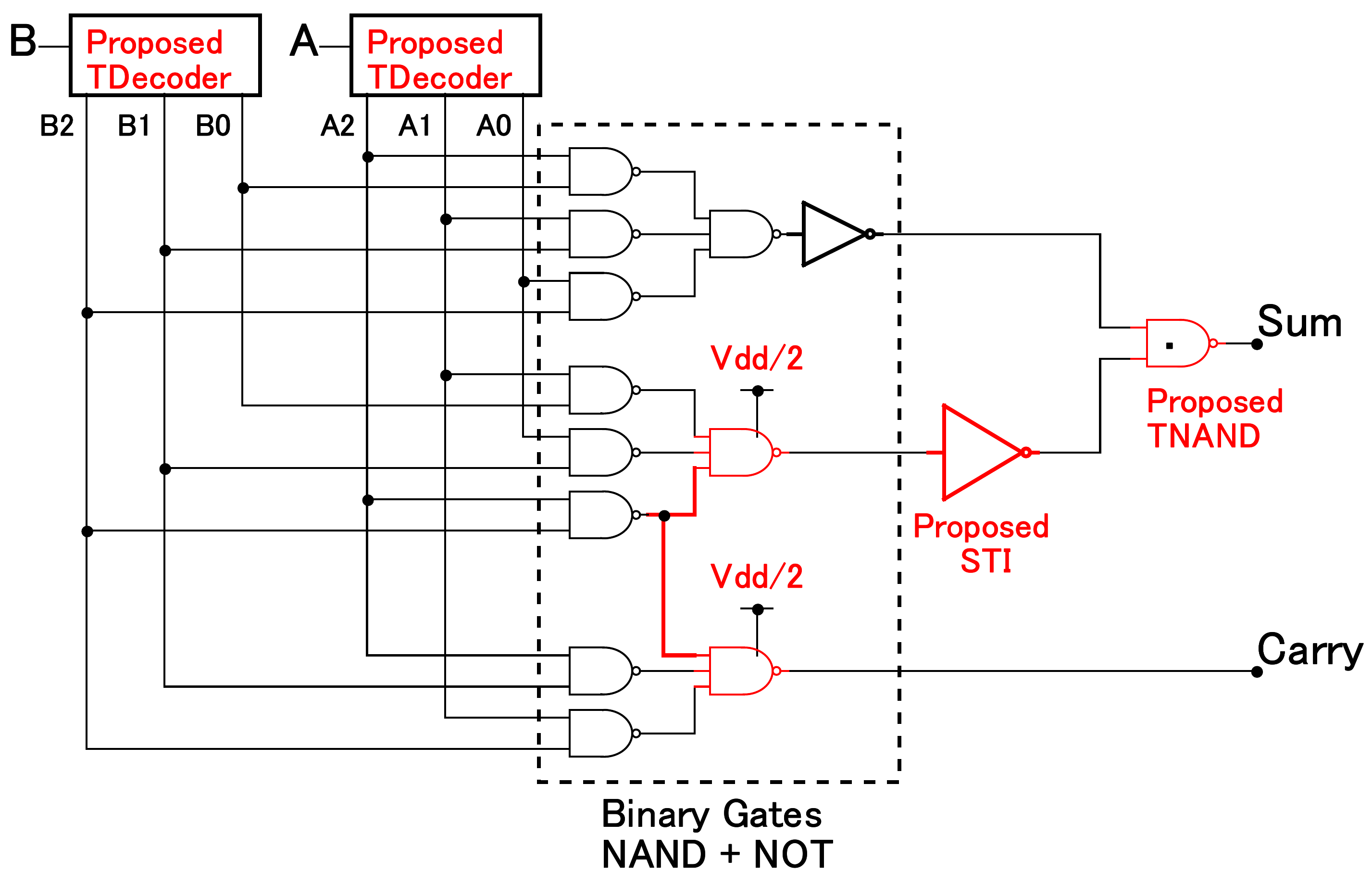}
\caption{The proposed THA1 with 85 CNFETs using the proposed TDecoder1, STI, and TNAND.}
\label{figha1}
\end{figure}

Figure~\ref{figha1} shows the proposed THA1. It consists of two the proposed TDecoder1 (9 CNFETs each), eight 2-inputs binary NAND, three 3-inputs binary NAND, one binary inverter, one the proposed TNAND (10 CNFETs), and one the proposed STI (5 CNFETs).

Table~\ref{4t75} shows the total Transistors Count of the proposed THA1.

 The operation of the proposed THA1: Two ternary inputs $A$ and $B$ fed to the proposed TDecoder1 and produce outputs as ($A_2$, $A_1$, $A_0$) and ($B_2$, $B_1$, $B_0$). These outputs will drive the eight 2-inputs binary NAND, the three 3-inputs binary NAND logic gates, the proposed STI, binary inverter, and the proposed TNAND to get the Sum and the Carry, as final outputs.

\renewcommand{\arraystretch}{1.3}
\begin{table}[!t]
\caption{Total transistors count of the proposed THA1}
\label{4t75}
\setlength{\tabcolsep}{5pt}
\centering
\begin{tabular}{l c c c}
\hline\hline
&No. of&No. of & Subtotal*  \\ 
&Devices&Transistors&\\ \hline
Proposed TDecoder1&2&9&18\\
Binary 2-NAND&8&4&32\\
Binary 3-NAND&3&6&18\\
Binary Inverter&1&2&2\\
Proposed STI&1&5&5\\
Proposed TNAND&1&10&10\\
\hline
Total&&&85\\
\hline\hline
\multicolumn{4}{l}{*Subtotal = No. of Devices x No. of Transistors}\\
\end{tabular}
\end{table}
\renewcommand{\arraystretch}{1}
\clearpage
\subsection{The proposed THA2}
\begin{figure}[b!]
\centering
\includegraphics[width = 15cm]{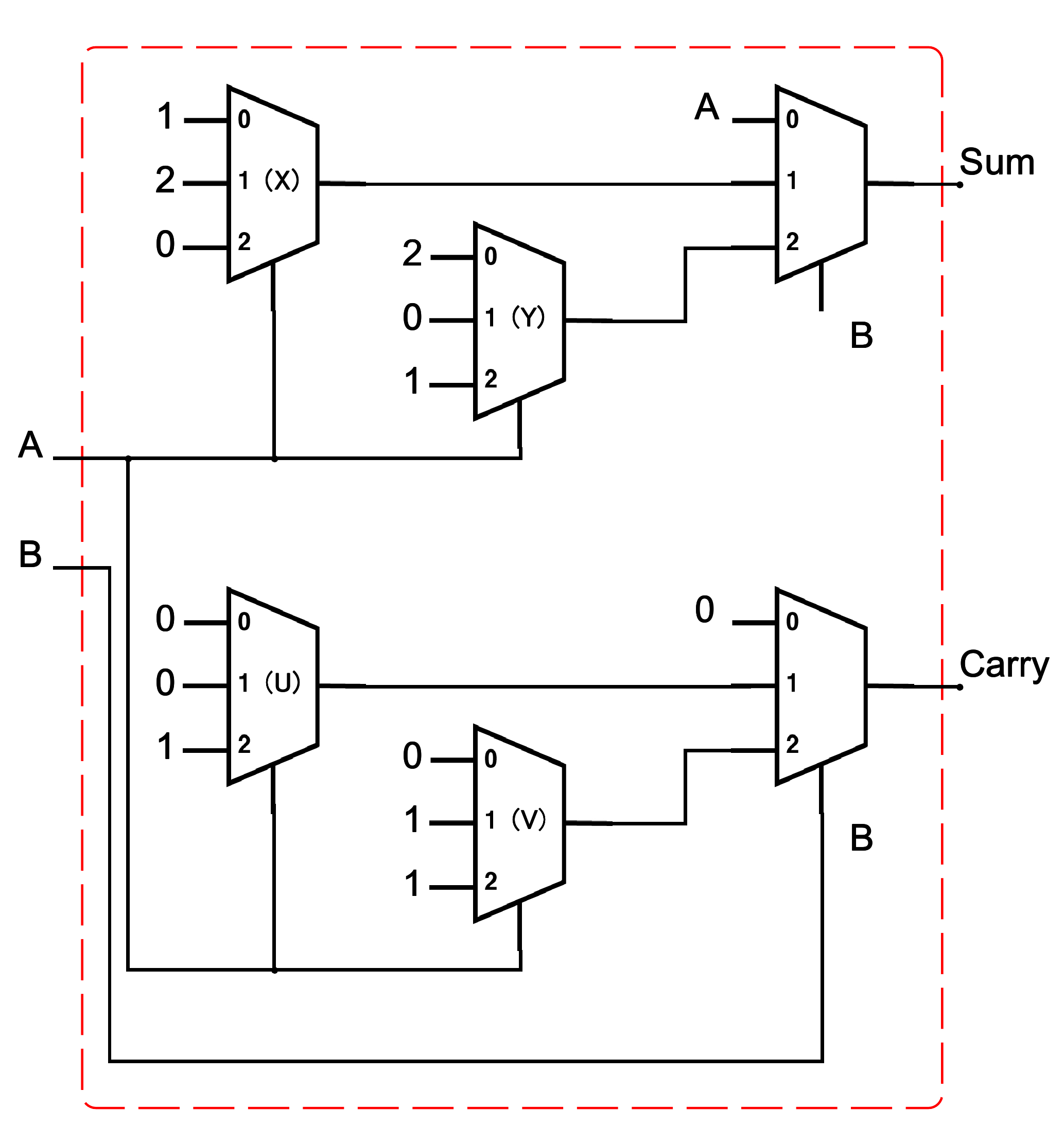}
\caption{The proposed THA2 with 90 CNFETs using cascading the proposed TMUX.}
\label{figha2}
\end{figure}

This Chapter proposes  another THA design using cascading TMUXs, which uses equation \eqref{4eq42}, of THA2 with 90 CNFETs (6*15 transistors), as shown in Fig.\ref{figha2}. 

The Existing THA design in \cite{302} with 168 (6*28 transistors) transistors is the same as Fig.\ref{figha2} but with TMUX equals to 28 transistors;

\pagebreak
The operation of the proposed THA2: Two ternary inputs $A$ and $B$ used as selection for the proposed TMUX.\\
When the selection $B$ is logic 0, then the Sum-TMUX output will be equal to the values of $A$ (0, 1, 2) and the Carry-TMUX will be equal to logic 0 for $A$ (0, 1, 2), respectively. \\
When the selection $B$ is logic 1, then the Sum-TMUX output will be equal to the output of X-TMUX (1, 2, 0) for $A$ (0, 1, 2), respectively.\\
The Carry-TMUX output will be equal to the output of U-TMUX (0, 0, 1) for $A$ (0, 1, 2), respectively.\\
Finally, when the selection $B$ is logic 2, then the Sum-TMUX output will be equal to the output of Y-TMUX (2, 0, 1) for $A$ (0, 1, 2), respectively.\\
The Carry-TMUX output will be equal to the output of V-TMUX (0, 1, 1) for $A$ (0, 1, 2), respectively.

\clearpage
\subsection{The proposed THA3}
\begin{figure}[b!]
\centering
\includegraphics[width = 15cm]{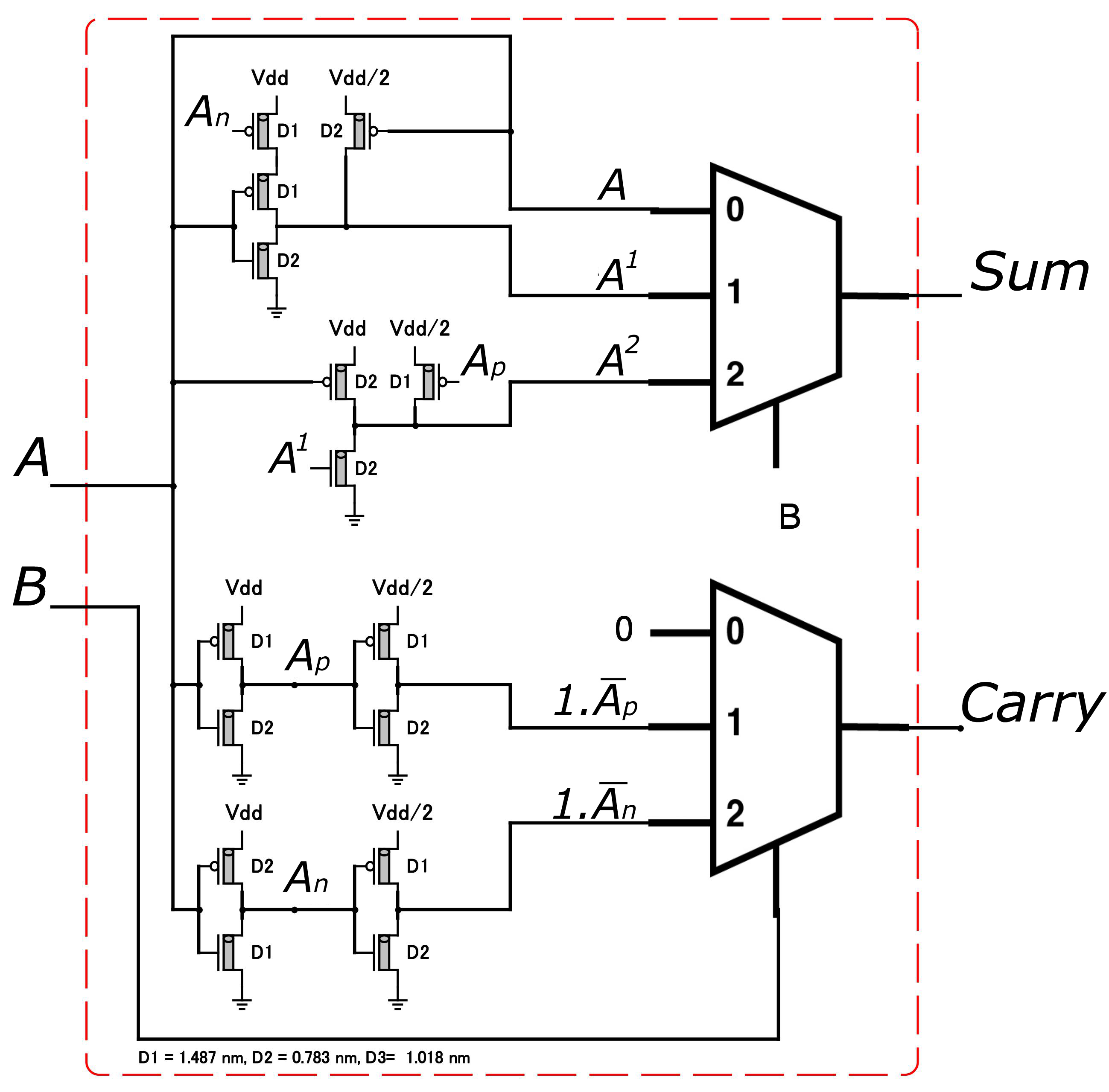}
\caption{The proposed THA3 with 45 CNFETs using the proposed TMUX and Unary Operators.}
\label{figha3}
\end{figure}

This Chapter proposes a third THA design based on new Unary operators combined with the proposed TMUX, which uses equation \eqref{4eq43}, of THA3 with 45 CNFETs, as shown in Fig.\ref{figha3}. 

Table \ref{4t475} shows total transistors count of the proposed THA3 which is equal to 45~transistors, and its operation is described in the following paragraphs where $A$ and $B$ are two ternary inputs.

The input $A$ fed to NTI, PTI, and the proposed Unary Operators produce the outputs as ($A^1$, $A^2$, $1\cdot \bar{A_p}$, $1\cdot \bar{A_n}$ ).\\
The outputs ($A^1$, $A^2$) and the input $A$ are fed to the first the proposed (3:1) TMUX to produce the Sum.\\
The outputs ($1\cdot \bar{A_p}$, $1\cdot \bar{A_n}$) are fed to the second the proposed (3:1) TMUX to produce the Carry.\\
The input $B$ is the selection for both the proposed TMUX.

\begin{table}[!t]
\caption{Total transistors count of the proposed THA3}
\label{4t475}
\setlength{\tabcolsep}{5pt}
\centering
\begin{tabular}{l c c c}
\hline \hline
&No. of&No. of & Subtotal  \\ 
&Devices&Transistors&\\ \hline
Proposed TMUX&2&15&30\\
NTI ($A_n$)&1&2&2\\
PTI ($A_p$)&1&2&2\\
Proposed $A^1$ & 1& 4&4\\
Proposed $A^2$ & 1& 3&3\\
Proposed $1\cdot \bar{A_p}$ & 1& 2&2\\
Proposed $1\cdot \bar{A_n}$ & 1& 2&2\\
\hline
Total&&&45\\
\hline\hline
\end{tabular}
\end{table}

The advantages of the proposed THA3 are: 
\begin{enumerate}
  \item Not using TDecoder compared to \cite{123} and the proposed THA1
  \item Not using basic logic gates like ``AND'', ``OR'', and ``NAND'' compared to \cite{123} and the proposed THA1
  \item Not using the cascading the proposed TMUX compared to \cite{302} and the proposed THA2
  \item Using the proposed (3:1) TMUX
  \item Using the proposed Unary Operators and the dual power supplies Vdd and Vdd/2
  \end{enumerate}

All these advantages can reduce the Transistors Count and energy consumption. 

\clearpage
\section{The proposed Ternary Multipliers}
A TMUL can multiply two ternary inputs and provides two outputs: the Product and the Carry, as shown in Table~\ref{4t55}

The basic equations of the Product and the Carry are shown in Table~\ref{4t55} to lead three designs:
\begin{enumerate}
  \item Conventional design in \cite{123,t15, t16} which uses equation \eqref{4eq51}.
  \item Cascading TMUXs design in \cite{302} which uses equation \eqref{4eq52}.
  \item Unary operators-based design in \cite{205} which uses equation \eqref{4eq53}.
  \end{enumerate}

\renewcommand{\arraystretch}{1.3}
\begin{table}[!h]
\caption{Truth table of TMUL}
\label{4t55}
\centering
\begin{tabular}{|c||l l l|}
\multicolumn{4}{c}{Product}\\ 
\hline
A/B&$B_0$(0)&$B_1$(1)&$B_2$(2)\\
 \hline\hline
$A_0$(0)&0 \multirow{3}{*}{ \Bigg\}	\textbf{0} }	&0 \multirow{3}{*}{ \Bigg\}	$\mathbf{A}$ } &0 \multirow{3}{*}{ \Bigg\} $\mathbf{\bar{A^2}}$}\\
$A_1$(1)&	0						 				&1														&2\\
$A_2$(2)&	0										&2														&1\\
\hline
\multicolumn{4}{c}{}\\
\multicolumn{4}{c}{Carry}\\
\hline
A/B&$B_0$(0)&$B_1$(1)&$B_2$(2)\\ 
\hline\hline
$A_0$(0)&0 \multirow{3}{*}{ \Bigg\}	\textbf{0} }&0 \multirow{3}{*}{ \Bigg\}	\textbf{0} }    &0 \multirow{3}{*}{ \Bigg\} $\mathbf{1\cdot \overline{A_p}}$ }\\
$A_1$(1)&0								&0							&0\\
$A_2$(2)&0								&0							&1\\
\hline
\end{tabular}
\end{table}
\renewcommand{\arraystretch}{1}

\begin{equation}
\label{4eq51}
\begin{split}
Product=2 \bullet(A_1B_2+A_2B_1)+ 1 \bullet(A_1B_1+A_2B_2)
\\
 Carry= 1 \bullet A_2B_2
   \end{split}
     \end{equation}

\begin{equation}
\label{4eq52}
\begin{split}
Product=0.B_0 + A.B_1+ (0.A_0 + 2.A_1 +1.A_2).B_2
\\
 Carry=0.B_0 + 0.B_1+ (0.A_0 + 0.A_1 +1.A_2).B_2
\end{split}
  \end{equation}

\begin{equation}
\label{4eq53}
\begin{split}
Product=0\cdot B_0 + A\cdot B_1A + \bar{A^2}B_2
\\
 Carry=0\cdot B_0 + 0\cdot B_1 + (1\cdot \bar{A_p}) B_2
 \end{split}
  \end{equation}

Where $A_k$ and $B_k$, $k$ $\in$ \{0,1,2\}, are the outputs of the TDecoder from inputs $A$ and $B$.

\begin{figure}[tb!]
\centering
\includegraphics[height=20cm]{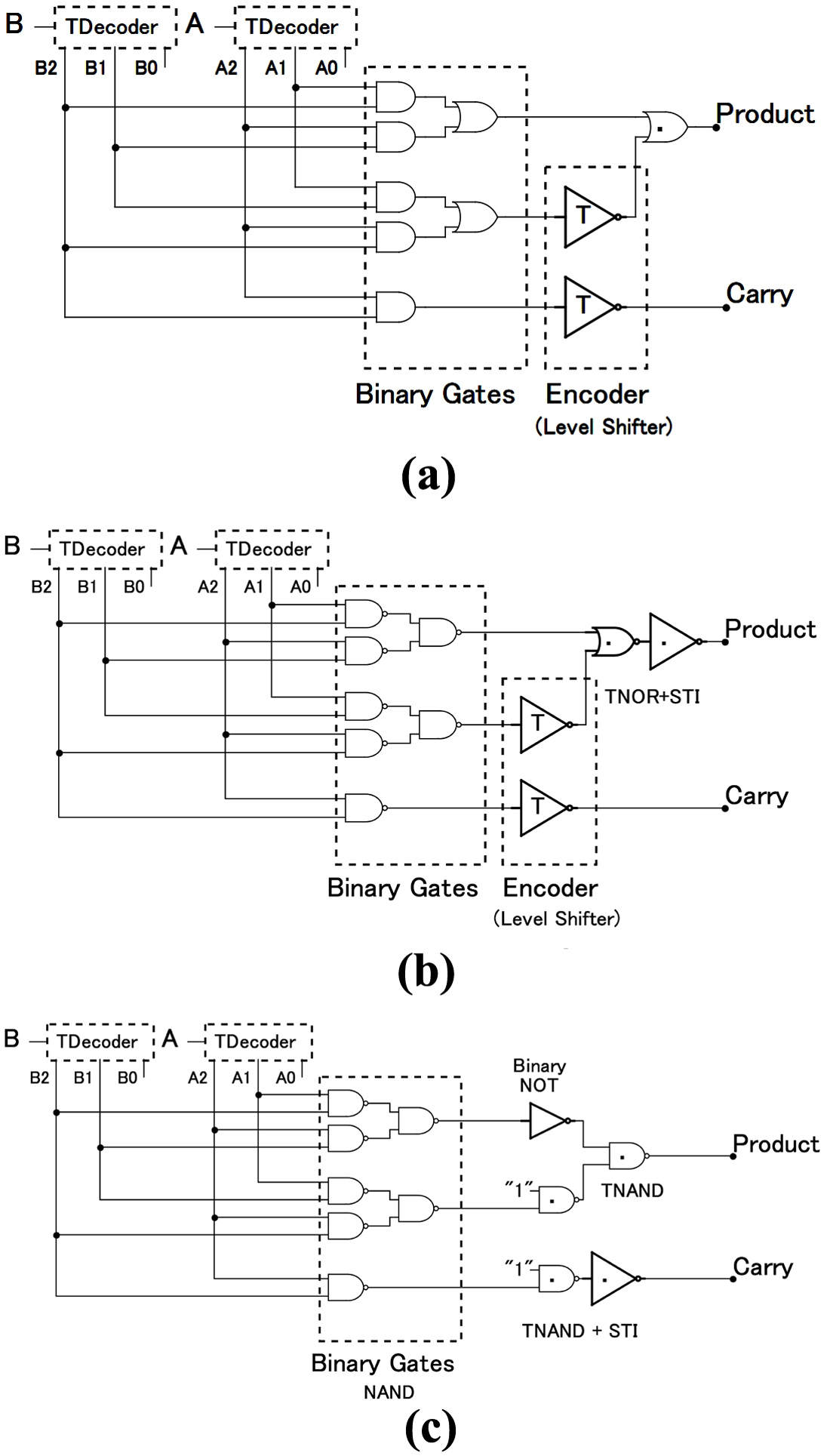}
\caption{The Existing TMUL in (a) \cite{123} with 100 CNFETs, (b)\cite{t15} with 86 CNFETs, and (b) \cite{t16} with 76 CNFETs.}
\label{figTMULExisting1}
\end{figure}

\begin{figure}[tb!]
\centering
\includegraphics[width=14cm]{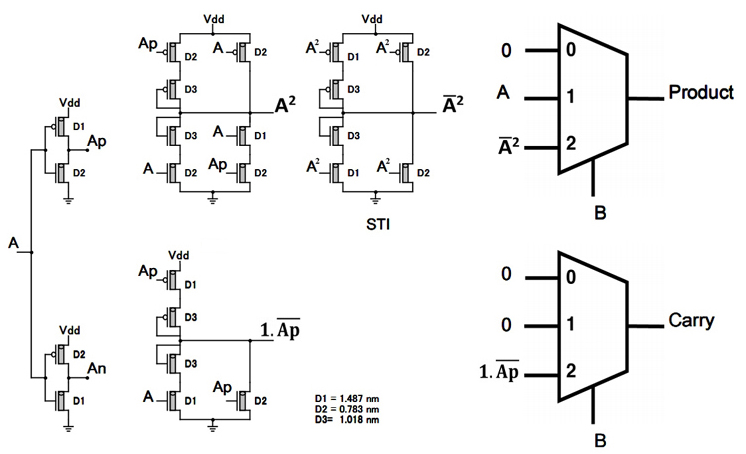}
\caption{The Existing TMUL in \cite{205} with 58 CNFETs.}
\label{figTMULExisting2}
\end{figure}

The existing TMUL in \cite{123} and \cite{205} are shown in Fig. \ref{figTMULExisting1} and Fig. \ref{figTMULExisting2};\\ 
Figure \ref{figTMULExisting1} (a) shows the TMUL of \cite{123} with 100 CNFETs that contains TDecoder (16~transistors), binary ANDs, binary ORs, a ternary encoder level shifter, and a ternary OR.\\
Figure \ref{figTMULExisting1} (b) shows the TMUL of \cite{t15} with 86 CNFETs that contains TDecoder of \cite{123}, binary NANDs, binary NORs, a ternary encoder level shifter, a ternary OR, and a STI.\\
Figure \ref{figTMULExisting1} (c) shows the TMUL of \cite{t16} with 76 CNFETs that contains their TDecoder (10 transistors), binary NANDs, a binary NOT, TNANDs, and a STI.

Figure \ref{figTMULExisting2} shows the TMUL of \cite{205} with 58 CNFETs that contains (3:1) TMUX (18 transistors) and Unary Operators.

\clearpage
\subsection{The proposed TMUL1}

This Chapter proposes a new conventional design of TMUL, which uses equation \eqref{4eq51}, with 61 CNFETs using De Morgan's Law and the dual power supply (Vdd and Vdd/2) to eliminate the encoder level shifter in \cite{123, t15}.

\begin{figure}[!b]
\centering
\includegraphics[width=15cm]{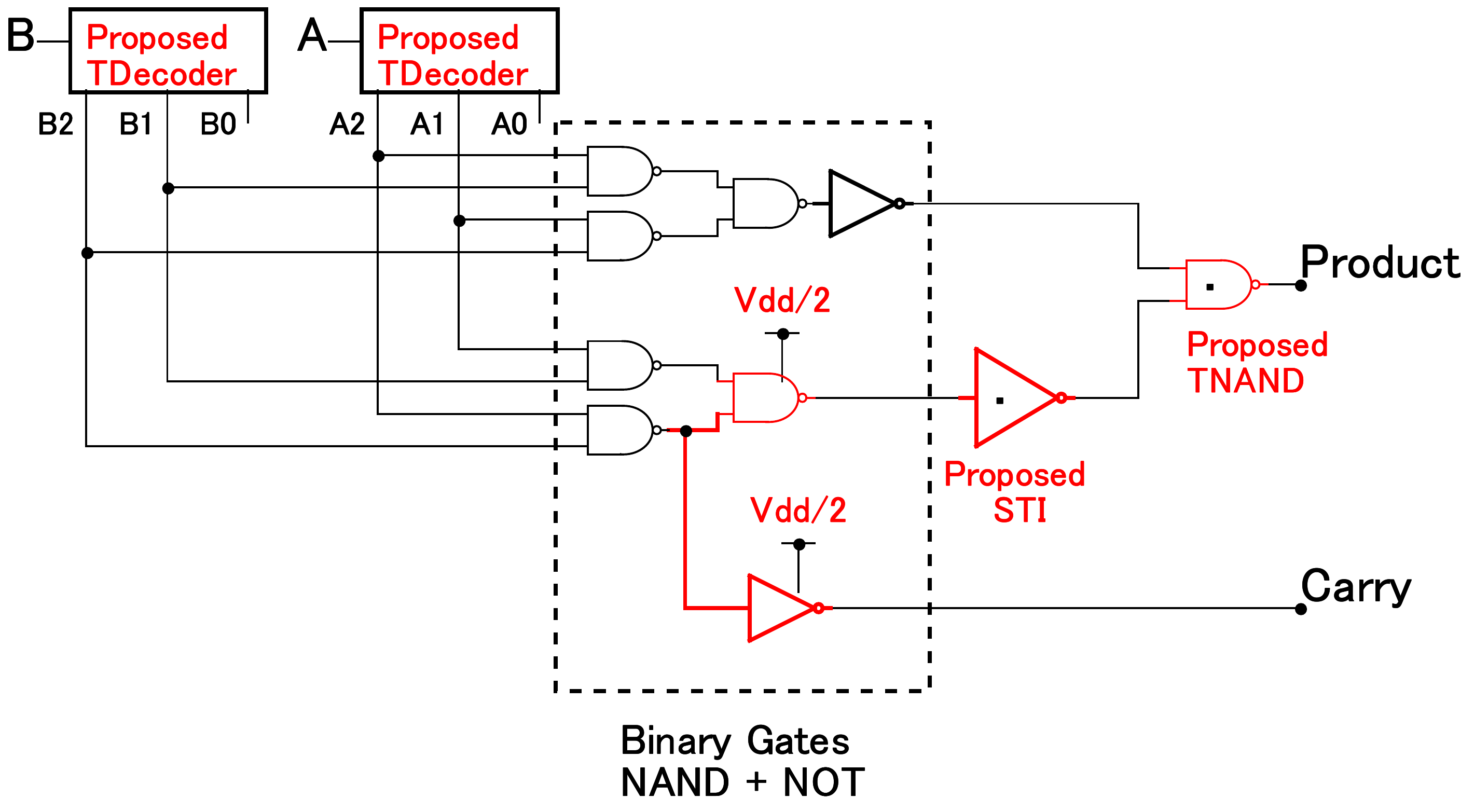}
\caption{The proposed TMUL1 with 61 transistors using the proposed TDecoder1, a STI, and TNANDs.}
\label{figtm1}
\end{figure}

Figure~\ref{figtm1} shows the proposed TMUL1. It consists of two the proposed TDecoders1 (9 CNFETs each), six binary NANDs, two binary inverters, one the proposed STI (5 CNFETs), and one the proposed TNAND (10 CNFETs).

Table~\ref{4t76} shows the total Transistors count of the proposed TMUL1.

The operation of the proposed TMUL1: Two ternary inputs $A$ and $B$ are fed to the two the proposed TDecoders1 and produce the outputs as ($A_2$, $A_1$, $A_0$) and ($B_2$, $B_1$, $B_0$), respectively.\\
These outputs will drive the four 2-inputs binary NAND, the two 2-inputs binary NAND logic gates, the proposed STI, the two binary inverters, and the proposed TNAND to get the Product and the Carry, as final outputs.

\renewcommand{\arraystretch}{1.3}
\begin{table}[!t]
\caption{Total transistors count of the proposed TMUL1}
\label{4t76}
\setlength{\tabcolsep}{5pt}
\centering
\begin{tabular}{l c c c}
\hline\hline
&No. of&No. of & Subtotal  \\ 
&Devices&Transistors&\\ \hline
Proposed TDecoder1&2&9&18\\
Binary NAND&6&4&24\\
Binary Inverter&2&2&4\\
Proposed STI&1&5&5\\
Proposed TNAND&1&10&10\\
\hline
Total&&&61\\
\hline\hline
\end{tabular}
\end{table}
\renewcommand{\arraystretch}{1}
\clearpage
\subsection{The proposed TMUL2}
\begin{figure}[b!]
\centering
\includegraphics[width=15cm]{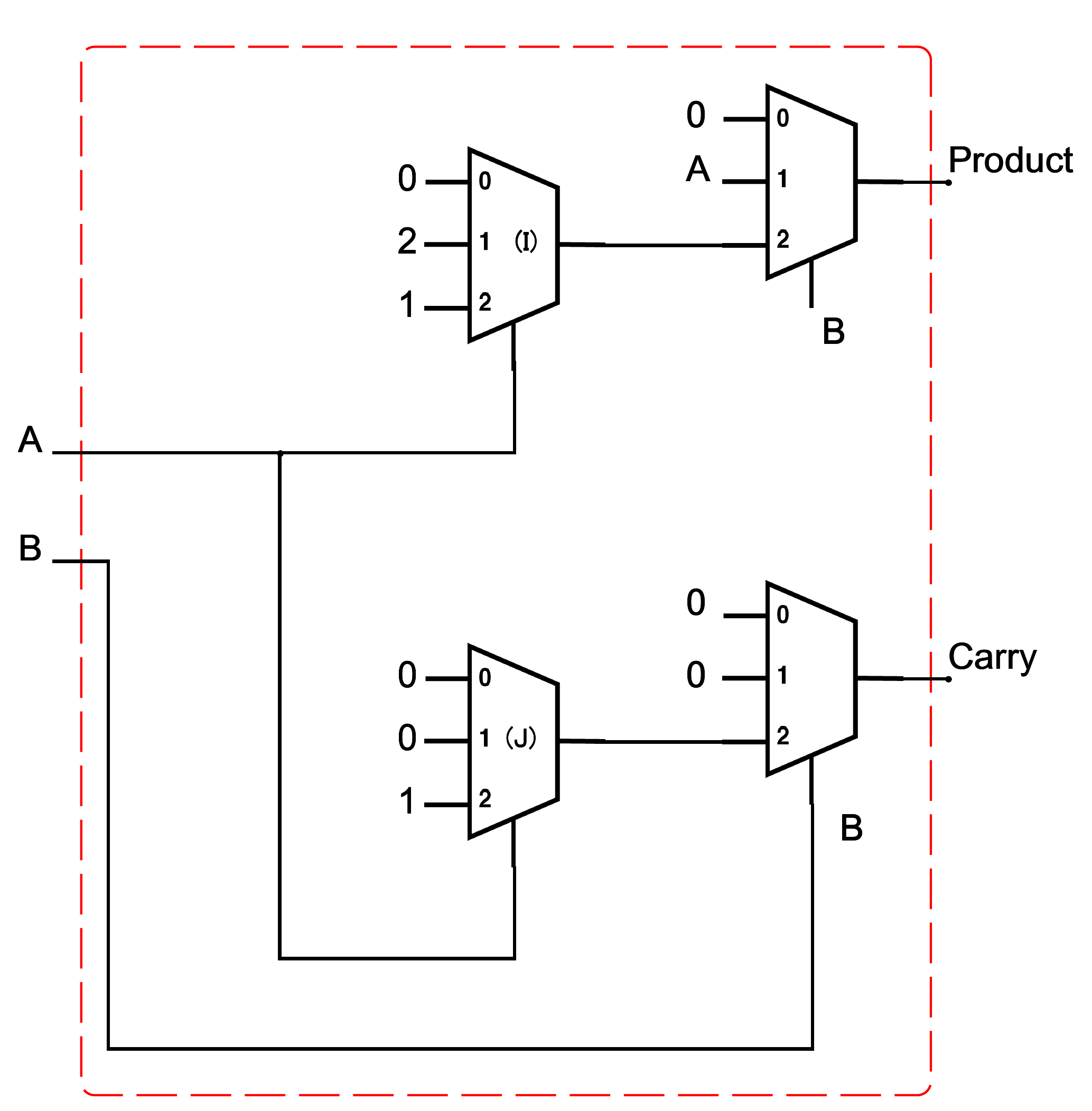}
\caption{The proposed TMUL2 with 60 CNFETs using cascading the proposed TMUXs.}
\label{figMUL2}
\end{figure}

This Chapter proposes another design of TMUL based on cascading the proposed TMUXs, which uses equation \eqref{4eq52}, of TMUL2 with 60 CNFETs (4*15 transistors), as shown in Fig.\ref{figMUL2}.

The Existing TMUL design in \cite{302} with 112 CNFETs (4*28 transistors) is the same as Fig.\ref{figMUL2} but with TMUX equals to 28 transistors.

The operation of the proposed TMUL2: Two ternary inputs $A$ and $B$ are used as selection for the proposed TMUX. \\
When the selection $B$ is logic 0, the Product-TMUX output and the Carry-TMUX output will be equal to the logic 0 for $A$ (0, 1, 2). \\ 
When the selection $B$ is logic 1, the Product-TMUX output will be equal to the values of $A$ (0, 1, 2) and the Carry-TMUX output will be equal to the logic 0 for $A$ (0, 1, 2), respectively. \\
Finally, when the selection $B$ is logic 2, the Product-TMUX output will be equal to the output of I-TMUX (0, 2, 1) for $A$ (0, 1, 2), respectively.
The Carry-TMUX output will be equal to the output of J-TMUX (0, 0, 1) for $A$ (0, 1, 2), respectively.
\clearpage
\subsection{The proposed TMUL3}
\begin{figure}[b!]
\centering
\includegraphics[width = 15cm]{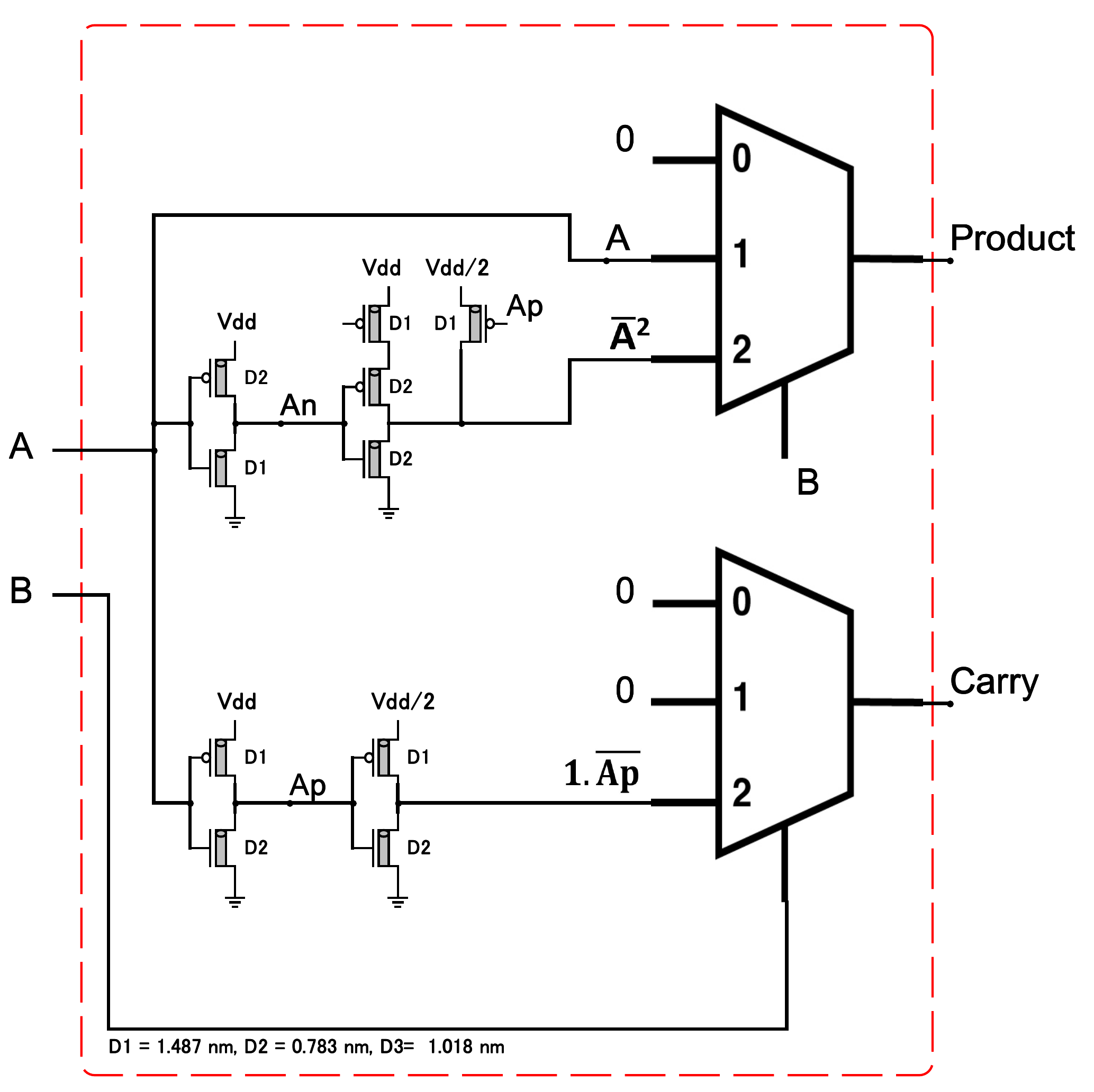}
\caption{The proposed TMUL3 with 40 CNFETs using the proposed TMUX and Unary Operators.}
\label{figMUL3}
\end{figure}

This Chapter proposes a third TMUL design based on Unary operators combined with the proposed TMUX, which uses equation \eqref{4eq53}, with 40 CNFETs, as shown in Fig.\ref{figMUL3}. 

Table~\ref{4t766} shows the total Transistors count of the proposed TMUL3.

\begin{table}[!t]
\caption{Total transistors count of the proposed TMUL3}
\label{4t766}
\setlength{\tabcolsep}{5pt}
\centering
\begin{tabular}{l c c c}
\hline
&No. of&No. of & Subtotal  \\ 
&Devices&Transistors&\\ \hline
Proposed TMUX&2&15&30\\
NTI ($A_n$)&1&2&2\\
PTI ($A_p$)&1&2&2\\
Proposed $\bar{A^2}$ & 1& 4&4\\
Proposed $1\cdot \bar{A_p}$ & 1& 2&2\\
\hline\hline
Total&&&40\\
\hline
\end{tabular}
\end{table}
\renewcommand{\arraystretch}{1}

The operation of the proposed TMUL3: Two ternary inputs $A$ and $B$ are used.\\ 
The input $A$ is fed to NTI ($A_n$), PTI ($A_p$), and to the proposed Unary Operators and produce outputs ($\bar{A^2}$,~$1\cdot\bar{A_p}$). \\
The output ($\bar{A^2}$) with the input $A$ and logic 0 are fed to the first the proposed (3:1) TMUX to get the Product.\\
The output ($1\cdot\bar{A_p}$) with two logic 0s are fed to the second the proposed (3:1) TMUX to get the Carry.\\
Whereas, the input $B$ is the selection input for both the proposed TMUX.

The advantages of the proposed TMUL3 are: 
\begin{enumerate}
  \item Not using TDecoders compared to \cite{123} and like the proposed TMUL1
  \item Not using basic logic gates like ``AND'', ``OR'', and ``NAND'' compared to \cite{123} and like the proposed TMUL1
   \item Not using the cascading the proposed TMUX compared to \cite{302} and the proposed TMUL2
  \item Using the proposed (3:1) TMUX
  \item Using the proposed Unary Operators and the dual power supplies Vdd and Vdd/2
  \end{enumerate}

All these advantages can reduce the Transistors count and energy consumption. 

\clearpage
\section{Simulation Results and Comparisons}
The proposed three THAs and three TMULs are simulated, tested, and compared to \cite{123,t15,t16,302, 205} using the HSPICE simulator and CNFET with the 32-nm channel length at different power supplies~(0.8V, 0.9V, 1V), different temperatures~(10\textdegree{}C, 27\textdegree{}C, 70\textdegree{}C ), and different frequencies~(0.5GHz, 1GHz, 2GHz).

All input signals have a fall and rise time of 15 ps. The average power consumption,~maximum propagation delay, and maximum PDP are obtained for all circuits.

The performance of the proposed circuits will be compared to other designs for the energy consumption (PDP).

\begin{figure}[!b]
\centering
\includegraphics[width=15cm]{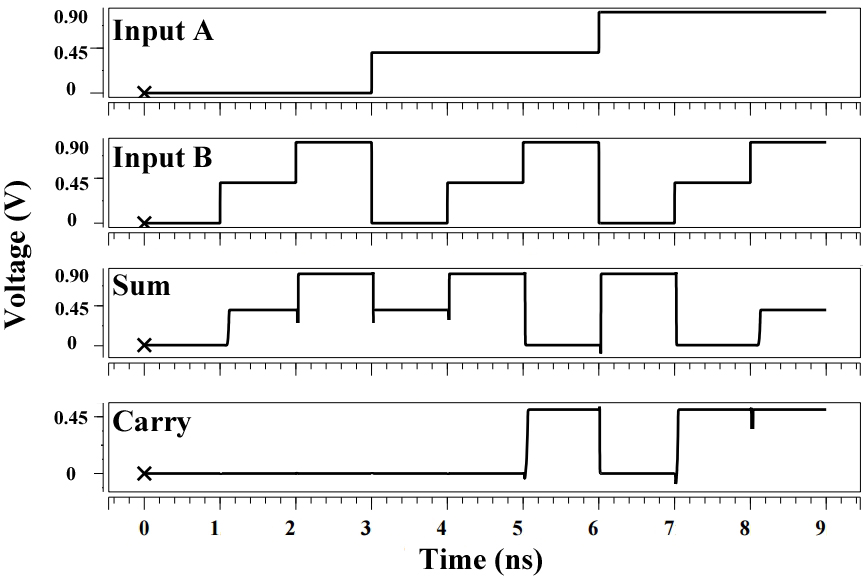}
\caption{Transient analysis of the proposed THA1.}
\label{simha}
\end{figure}

\begin{figure}[!t]
\centering
\includegraphics[width=15cm]{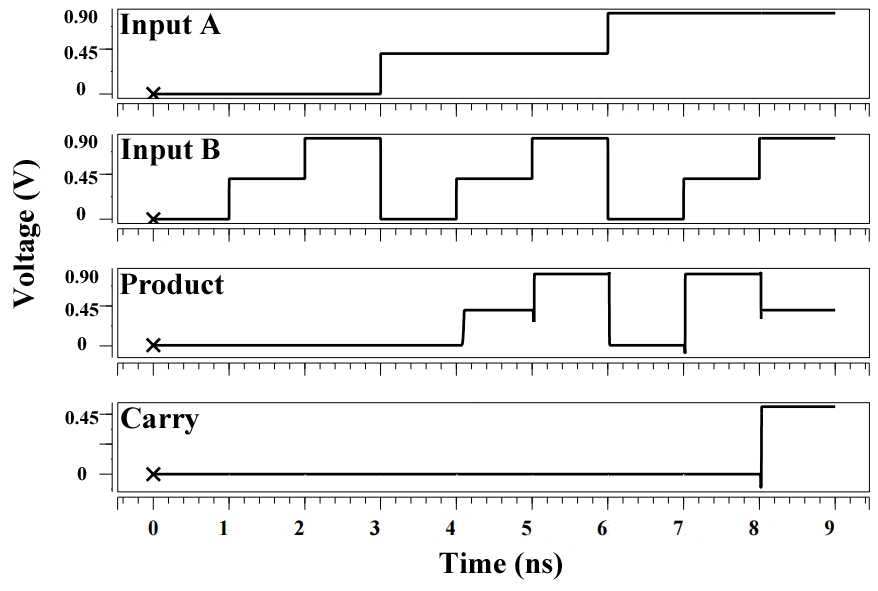}
\caption{Transient analysis of the proposed TMUL1.}
\label{simtm}
\end{figure}

Figures \ref{simha} and \ref{simtm} illustrate the transient analysis of the proposed THA1 and TMUL1.

\clearpage
\subsection{Comparison of Transistors Count}

Table \ref{4t210} shows the comparison of Transistors count for the THAs and TMULs of~\cite{123,t15,t16,302,205} and of the proposed ones. 
 
\renewcommand{\arraystretch}{1.5}
\begin{table}[b!]
\caption{Comparison of Transistors Count}
\label{4t210}
\setlength{\tabcolsep}{3pt}
\centering
\begin{tabular}{l|cc}
\hline\hline
																&\multicolumn{1}{c}{\textbf{THA}}&\multicolumn{1}{c}{\textbf{TMUL}}\\
\hline
\textbf{\cite{123}} Using TDecoder (16T*), AND, OR, TEncoder, and TOR 	  &\multicolumn{1}{c}{136} &\multicolumn{1}{c}{100}\\ 
\textbf{\cite{t15}} Using TDecoder of \cite{123}, NAND, NOR, TEncoder, and TNOR 	  &\multicolumn{1}{c}{112} &\multicolumn{1}{c}{86}\\ 
\textbf{\cite{t16}} Using TDecoder (10T), NAND, NOT, and TNAND 	  &\multicolumn{1}{c}{112} &\multicolumn{1}{c}{76}\\ 

\textbf{\cite{302}} Using cascading TMUXs (28T) 				 & \multicolumn{1}{c}{168 } &\multicolumn{1}{c}{112}\\ 
\textbf{\cite{205}} Using TMUX (18T), and Unary Operators 	 &\multicolumn{1}{c}{64} &\multicolumn{1}{c}{58}\\ 
\hline
\textbf{Proposed 1:} Using the proposed TDecoder1 (9T), STI, and TNAND &\multicolumn{1}{c}{85} &\multicolumn{1}{c}{61}\\ 
\textbf{Proposed 2:} Using cascading the proposed TMUXs (15T)				&\multicolumn{1}{c}{\textbf{90}}&\multicolumn{1}{c}{\textbf{60}}\\
\textbf{Proposed 3:} Using the proposed TMUX, and Unary Operators	&\multicolumn{1}{c}{\textbf{45}}&\multicolumn{1}{c}{\textbf{40}}\\
\hline\hline
\multicolumn{3}{l}{* T = Transistors}\\
\end{tabular}
\end{table}
\renewcommand{\arraystretch}{1}

The proposed THA1 and TMUL1, using conventional design, have a notable reduction in Transistors Count.\\
Around 37.5\%, 24.1\%, and 24.1\% compared to THA in \cite{123, t15} and \cite{t16}, respectively.\\
Around 39\%, 29\%, and 19.7\% compared to TMUL in \cite{123, t15} and \cite{t16}, respectively.

The proposed THA2 and TMUL2, using cascading TMUXs, have a notable reduction in Transistors count around 46.43\% compared to THA, and TMUL in \cite{302}. 

The proposed THA3 and TMUL3, using TMUX with Unary Operators, have the lowest Transistors count among all the investigated circuits:  
Around 66.91\%, 59.82\%, 59.82\%, 73.21\%, and 29.69\% compared to THA in \cite{123,t15,t16,302} and \cite{205}, respectively.
Also, around 47.06\%, and 50\% compared to the proposed THA1, and THA2.\\
Around 60\%, 53.49\%, 47.37\%, 64.23\%, and 31.03\% compared to TMUL in \cite{123,t15,t16,302} and \cite{205}, respectively. Also, around 34.43\%, and 33.33\% compared to the proposed TMUL1, and TMUL2.

\subsection{Comparison of Different THA Circuits}

\subsubsection{Comparison at Different Power Supplies}
Table~\ref{4t730} shows the comparison to the existing Ternary Half Adder in \cite{123, t15,t16,302,205} in terms of the average power consumption, maximum propagation delay, and maximum energy (PDP) at different supply voltages (0.8 V, 0.9 V, 1 V), same temperature (27\textdegree{}C), and same frequency (1 GHz). The boldface values are the best values, among others.

\renewcommand{\arraystretch}{1.3} 
\begin{table}[!htb]
\caption{Comparison of average power ($\mu$W), maximum delay (ps), and maximum PDP  (x$10^{-18}$ J) for 8 THAs at T=27\textdegree{}C, F=1 GHz, and for different supply voltages}
\label{4t730}
\setlength{\tabcolsep}{2pt}
\centering
\begin{tabular}{c|ccc|ccc|ccc}
\hline\hline
	& \multicolumn{3}{c|}{Vdd=0.8 V}& \multicolumn{3}{c|}{Vdd=0.9 V} & \multicolumn{3}{c}{Vdd=1 V}\\
	& Power &Delay 	&PDP 	& Power &Delay 	&PDP & Power &Delay 	&PDP \\
	
\hline
THA in \cite{123}  			&1.78	&112.10 &199.53			&2.54&56.51&143.53						&4.12	&55.78 	&229.81\\
THA in \cite{t15}				&1.34	&75.42	&101.1				&1.93&53.14&102.56						&3.62	&54.36 	&196.78\\
THA in \cite{t16}				&1.11	&70.4	&78.14				&1.84&43.25&79.58						&3.32	&43.45 	&144.25\\
THA in \cite{302}				&1.10	&27.31	&30.04				&1.97&21.15&41.66						&3.46	&19.15 	&66.26\\
THA in \cite{205}				&0.37	&\textbf{10.35}&3.82		&0.99&\textbf{8.52}&8.43					&2.97	&\textbf{9.10} &27.03\\
\hline
\textbf{Proposed THA1} 	&0.26&105.32&27.38				&0.53	&74.63	&39.55	 					&0.72		&73.89		&53.2\\
\textbf{Proposed THA2} 	&0.08&11.87&0.95					&0.14	&10.66&1.49 				  &\textbf{0.4}	&13.06	&5.22\\
\textbf{Proposed THA3} 	&\textbf{0.06}&12.65&\textbf{0.76}	&\textbf{0.13}	&9.50&\textbf{1.23}&\textbf{0.4}	&9.67	&\textbf{3.87}\\
\hline\hline
\end{tabular}
\end{table}

As shown in Table~\ref{4t730}, the comparison of the proposed THA1 demonstrates a notable~reduction in PDP.\\
At Vdd=0.8 V, around 86.28\%, 72.92\%, and 64.96\% compared to \cite{123, t15, t16}, respectively.\\ 
At Vdd=0.9 V, around 72.44\%, 61.44\%, and 50.3\% compared to \cite{123, t15, t16}, respectively.\\ 
At Vdd=1 V, around 76.85\%, 72.96\%, and 63.12\% compared to \cite{123, t15, t16}, respectively.

The comparison of the proposed THA2 demonstrates a notable reduction in PDP compared to \cite{302}, around 96.84\%, 96.42\%, and 92.12\% for Vdd (0.8V, 0.9V, and 1V),~respectively.

Whereas, the comparison of the proposed THA3 has the lowest PDP compared to all the investigated models and the proposed THA1 and THA2.\\
At Vdd=0.8 V, around 99.62\%, 99.25\%, 99.03\%, 97.47\%, 80.1\%, 97.22\%, and 20\%~compared to \cite{123, t15, t16,302,205}, the proposed THA1, and the proposed THA2, respectively.\\ 
At Vdd=0.9 V, around 99.14\%, 98.8\%, 98.45\%, 97.05\%, 85.41\%, 96.89\%, and 17.45\% compared to \cite{123, t15, t16,302,205}, the proposed THA1, and the proposed THA2, respectively.\\  
At Vdd=1 V, around 98.32\%, 98.03\%, 97.32\%, 94.16\%, 85.68\%, 92.73\%, and 25.86\% compared to \cite{123, t15, t16,302,205}, the proposed THA1, and the proposed THA2, respectively.

\subsubsection{Comparison at Different Temperatures}
Table~\ref{4t731} shows the comparison of the propsed THAs to the existing THAs in \cite{123, t15, t16,302,205} in terms of the average power consumption, maximum propagation delay, and maximum energy (PDP) at  different temperatures (10\textdegree{}C, 27\textdegree{}C, 70\textdegree{}C), same voltage supply Vdd (0.9 V), and same frequency~(~1~GHz).

\begin{table}[!htb]
\caption{Comparison of average power ($\mu$W), maximum delay (ps), and maximum PDP  (x$10^{-18}$ J) for 8 THAs at Vdd=0.9 V, F=1 GHz, and for different temperatures}
\label{4t731}
\setlength{\tabcolsep}{2pt}
\centering
\begin{tabular}{c|ccc|ccc|ccc}
\hline\hline
	& \multicolumn{3}{c|}{Temp.=10\textdegree{}C}& \multicolumn{3}{c|}{Temp.=27\textdegree{}C} & \multicolumn{3}{c}{Temp.=70\textdegree{}C}\\
		& Power &Delay 	&PDP 	& Power &Delay 	&PDP & Power &Delay 	&PDP \\
\hline	
THA in \cite{123}  	&2.67 	&57.68	&154				&2.54&56.51&143.53									&2.79	&54.25 &151.35\\
THA in \cite{t15}		&2.06 	&55.36	&114.04			&1.93&53.14&102.56									&2.21	&51.47 &113.75\\
THA in \cite{t16}		&1.91	&44	&84.04				&1.84&43.25&79.58									&2.15	&39.9	 &85.78\\
THA in \cite{302} 	&2.00	&21.86	&43.72				&1.97&21.15&41.66									&2.12	&19.73 &41.83\\
THA in \cite{205}		&0.92	&\textbf{8.69}	&7.99		&0.99&\textbf{8.52}&8.43								&1.28	&\textbf{8.11} &10.38\\
\hline
\textbf{Proposed THA1} 	&0.63&76.31&48.07	 	&0.53	&74.63	&39.55  								&0.65	&71.34	&46.37\\
\textbf{Proposed THA2} 	&0.15&11.40&1.71						&0.14	&10.66&1.49						 &0.20			&9.89	&1.98\\
\textbf{Proposed THA3} 	&\textbf{0.13}&9.96&\textbf{1.29}	&\textbf{0.13}	&9.50&\textbf{1.23} 		&\textbf{0.19}	&8.68	&\textbf{1.65}\\
\hline\hline
\end{tabular}
\end{table}

As shown in Table~\ref{4t731}, the comparison of the proposed THA1 demonstrates a notable~reduction in PDP.\\
At Temp.=10\textdegree{}C, around 68.79\%, 57.83\%, and 42.8\% compared to \cite{123, t15, t16}, respectively.\\
At Temp.=27\textdegree{}C, around 72.44\%, 61.44\%, and 50.3\% compared to \cite{123, t15, t16}, respectively.\\ 
At Temp.=70\textdegree{}C, around 69.36\%, 59.24\%, and 45.94\% compared to \cite{123, t15, t16}, respectively.

The comparison of the proposed THA2 demonstrates a notable reduction in PDP compared to \cite{302}, around 96.09\%, 96.42\%, and 95.27\% at temperature (10\textdegree{}C, 27\textdegree{}C, and 70\textdegree{}C), respectively.

Whereas, the comparison of the proposed THA3 has the lowest PDP compared to all the investigated models and the proposed THA1, and THA2.\\
At Temp.=10\textdegree{}C, around 99.16\%, 98.87\%, 98.47\%, 97.05\%, 83.85\%, 97.32\%, and 24.56\% compared to \cite{123, t15, t16,302,205}, the proposed THA1, and the proposed THA2, respectively.\\ 
At Temp.=27\textdegree{}C, around 99.14\%, 98.8\%, 98.45\%, 97.05\%, 85.41\%, 96.89\%, and 17.45\% compared to \cite{123, t15, t16,302,205}, the proposed THA1, and the proposed THA2, respectively.\\  
At Temp.=70\textdegree{}C, around 98.91\%, 98.55\%, 98.08\%, 96.06\%, 84.1\%, 96.44\%, and 16.67\% compared to \cite{123, t15, t16,302,205}, the proposed THA1, and the proposed THA2, respectively.

\subsubsection{Comparison at Different Frequencies}
Table~\ref{4t732} shows the comparison of the proposed THAs to the existing THAs in \cite{123, t15, t16,302,205} in terms of the average power consumption, maximum propagation delay, and maximum energy (PDP) at different frequencies (2 GHz, 1 GHz, 0.5 GHz), same temperatures (27\textdegree{}C), and same voltage supply Vdd (0.9 V).

\begin{table}[!htb]
\caption{Comparison of average power ($\mu$W), maximum delay (ps), and maximum PDP  (x$10^{-18}$ J) for 8 THAs at T=27\textdegree{}C, Vdd=0.9 V, and for different frequencies}
\label{4t732}
\setlength{\tabcolsep}{2pt}
\centering
\begin{tabular}{c|ccc|ccc|ccc}
\hline\hline
	& \multicolumn{3}{c|}{f= 2  GHz}& \multicolumn{3}{c|}{f= 1  GHz} & \multicolumn{3}{c}{f= 0.5  GHz}\\
		& Power &Delay 	&PDP 	& Power &Delay 	&PDP & Power &Delay 	&PDP \\
	
\hline
THA in \cite{123}  		&2.73	&55.9	&152.6			&2.54&56.51&143.53							&2.35	&56.81 &133.5\\
THA in \cite{t15}			&2.05	&52.3	&107.21		&1.93&53.14&102.56							&1.88	&53.61 &100.78\\
THA in \cite{t16}			&1.94 	&42.7	&82.84			&1.84&43.25&79.58			     				&1.79	&43.3 &77.5\\
THA in \cite{302}			&2.14	&21.20 &45.37			&1.97&21.15&41.66							&1.89	&21.20	&40.07						\\
THA in \cite{205}			&1.25	&\textbf{8.46}&10.58	&0.99&\textbf{8.52}&8.43						&1.17	&\textbf{8.56}	&10.02				\\
\hline
\textbf{Proposed THA1} 	&0.58&73.2&42.45			 			&0.53	&74.63	&39.55      						&0.48&71.3&34.22\\
\textbf{Proposed THA2} 	&0.21&10.90	&2.29						&0.14	&10.66&1.49          				&\textbf{0.11}&10.89&1.19     \\
\textbf{Proposed THA3} 	&\textbf{0.17}	&9.54&\textbf{1.62}	&\textbf{0.13}	&9.50&\textbf{1.23}  		&\textbf{0.11}&9.47&\textbf{1.04}\\
\hline\hline
\end{tabular}
\end{table}

As shown in Table~\ref{4t732}, the comparison of the proposed THA1 demonstrates a considerable reduction in PDP.\\
At freq.=2 GHz, around 72.18\%, 60.4\%, and 48.76\% compared to \cite{123, t15, t16}, respectively.\\ 
At freq.=1 GHz, around 72.44\%, 61.44\%, and 50.3\% compared to \cite{123, t15, t16}, respectively.\\ 
At freq.=0.5 GHz, around 74.37\%, 66.04\%, and 55.85\% compared to \cite{123, t15, t16}, respectively.

The comparison of the proposed THA2 demonstrates a notable reduction in PDP compared to \cite{302}, around 94.95\%, 96.42\%, and 97.03\% for frequency (2 GHz, 1 GHz, and 0.5 GHz), respectively.

Whereas, the comparison of the proposed THA3 has the lowest PDP compared to all the investigate models and the proposed THA1, and THA2.\\
At frequency=2 GHz, around 98.94\%, 98.49\%, 98.04\%, 96.43\%, 84.69\%, 96.18\%, and 29.26\% compared to \cite{123, t15, t16,302,205}, the proposed THA1, and the proposed THA2, respectively.\\ 
At frequency=1 GHz, around 99.14\%, 98.8\%, 98.45\%, 97.05\%, 85.41\%, 96.89\%, and 17.45\% compared to \cite{123, t15, t16,302,205}, the proposed THA1, and the proposed THA2, respectively.\\  
At frequency=0.5 GHz, around 99.22\%, 98.97\%, 98.66\%, 97.4\%, 89.62\%, 96.96\%, and 12.61\% compared to \cite{123, t15, t16,302,205}, the proposed THA1, and the proposed THA2, respectively.

\subsection{Comparison at Different TMUL Circuits}

\subsubsection{Comparison to Different Power Supplies}
Table~\ref{44t730} shows the comparison of the proposed TMULs to the existing Ternary Multipliers in \cite{123, t15,t16,302,205} in terms of the average power consumption, maximum propagation delay, and maximum energy (PDP) at different supply voltages (0.8 V, 0.9 V, 1 V), same temperature (27\textdegree{}C), and same frequency (1 GHz). The boldface values are the best values, among others.

\renewcommand{\arraystretch}{1.3} 
\begin{table}[!htb]
\caption{Comparison of average power ($\mu$W), maximum delay (ps), and maximum PDP  (x$10^{-18}$ J) for 8 TMULs at T=27\textdegree{}C, F=1 GHz, and for different supply voltages}
\label{44t730}
\setlength{\tabcolsep}{2pt}
\centering
\begin{tabular}{c|ccc|ccc|ccc}
\hline\hline
	& \multicolumn{3}{c|}{Vdd=0.8 V}& \multicolumn{3}{c|}{Vdd=0.9 V} & \multicolumn{3}{c}{Vdd=1 V}\\
	& Power &Delay 	&PDP 	& Power &Delay 	&PDP & Power &Delay 	&PDP \\
	
\hline
TMUL in \cite{123}  			   &1.32	&68.72	&90.71				&1.88&46.32&87.08				&3.07	&48.46 &148.77\\
TMUL in \cite{t15}				&1	&57.32	&57.32				&1.45&43.05&62.42					&2.48	&45.25 &112.22\\
TMUL in \cite{t16}				&0.8&48.75&39				&1.32&31.26&41.26					&2.24	&32.9 &73.69\\
TMUL in \cite{302}				&1.07	&23.75	&25.41				&1.93&18.21&35.15				&3.37	&16.32 &54.99\\
TMUL in \cite{205}				&0.27	&18.09&4.88				&0.64&16.63&10.64				&1.40	&16.82 &23.55\\
\hline
\textbf{Proposed TMUL1} 	&0.30	&67.80	&20.34				&0.42&54.82&	23.02			     		&0.72	&56.34 &40.56\\
\textbf{Proposed TMUL2} 	&0.08&\textbf{13.83}&1.10			&0.17	&\textbf{9.65}&1.64 			 &0.53			&\textbf{9.18}	&4.86\\
\textbf{Proposed TMUL3} 	&\textbf{0.06}&16.96&\textbf{1.02}	&\textbf{0.07}	&11.61&\textbf{0.81}	&\textbf{0.15}	&10.02	&\textbf{1.50}\\

\hline\hline
\end{tabular}
\end{table}

As shown in Table~\ref{44t730}, the comparison of the proposed TMUL1 demonstrates a notable reduction in PDP.\\
At Vdd=0.8 V, around 77.58\%, 64.52\%, and 47.85\% compared to \cite{123, t15, t16}, respectively.\\ 
At Vdd=0.9 V, around 73.56\%, 63.12\%, and 44.21\% compared to \cite{123, t15, t16}, respectively.\\ 
At Vdd=1 V, around 72.74\%, 63.86\%, and 44.96\% compared to \cite{123, t15, t16}, respectively.

The comparison of the proposed TMUL2 demonstrates a notable reduction in PDP~compared to \cite{302}: around 95.67\%, 95.33\%, and 91.16\% at Vdd (0.8V, 0.9V, and 1V),~respectively.

Whereas, the comparison of the proposed TMUL3 has the lowest PDP compared to all the investigated models and the proposed TMUL1, and TMUL2.\\
At Vdd=0.8 V, around 98.88\%, 98.22\%, 97.38\%, 95.99\%, 79.1\%, 94.99\%, and 7.27\% compared to \cite{123, t15, t16,302,205}, the proposed TMUL1, and the proposed TMUL2, respectively.\\ 
At Vdd=0.9 V, around 99.07\%, 98.7\%, 98.04\%, 97.7\%, 92.39\%, 96.48\%, and 50.61\% compared to \cite{123, t15, t16,302,205}, the proposed TMUL1, and the proposed TMUL2, respectively.\\  
At Vdd=1 V, around 98.99\%, 98.66\%, 97.96\%, 97.27\%, 93.63\%, 96.3\%, and 69.14\% compared to \cite{123, t15, t16,302,205}, the proposed TMUL1, and the proposed TMUL2, respectively.

\subsubsection{Comparison at Different Temperatures}
Table~\ref{44t731} shows the comparison to the existing TMULs in \cite{123, t15, t16,302,205} in terms of the average power consumption, maximum propagation delay, and maximum energy (PDP) at different temperatures (10\textdegree{}C, 27\textdegree{}C, 70\textdegree{}C), same voltage supply Vdd (0.9 V), and same frequency (1 GHz).

\begin{table}[!htb]
\caption{Comparison of average power ($\mu$W), maximum delay (ps), and maximum PDP  (x$10^{-18}$ J) for 8 TMULs at Vdd=0.9 V, F=1 GHz, and at different temperatures}
\label{44t731}
\setlength{\tabcolsep}{2pt}
\centering
\begin{tabular}{c|ccc|ccc|ccc}
\hline\hline
	& \multicolumn{3}{c|}{Temp.=10\textdegree{}C}& \multicolumn{3}{c|}{Temp.=27\textdegree{}C} & \multicolumn{3}{c}{Temp.=70\textdegree{}C}\\
		& Power &Delay 	&PDP 	& Power &Delay 	&PDP & Power &Delay 	&PDP \\
\hline	
TMUL in \cite{123}  	&1.98 	&48.23	&95.49				&1.88&46.32&87.08				&2.06	&44.9 &92.50\\
TMUL in \cite{t15}		&1.58 	&45.92	&72.55			&1.45&43.05&62.42									&1.65	&42.12 &69.49\\
TMUL in \cite{t16}		&1.37	&33.45&45.82			&1.32&31.26&41.26									&1.57	&29.83&46.83\\
TMUL in \cite{302} 	&1.95	&18.74	&36.54				&1.93&18.21&35.15				&2.08	&17.05 &35.46\\
TMUL in \cite{205}		&0.60	&17.25&10.35				&0.64&16.63&10.64				&0.83	&15.34 &12.73\\
\hline
\textbf{Proposed TMUL1} 	&0.50	&56.15	&28.07				&0.42&54.82&	23.02			     &0.52	&52.89 &27.50\\
\textbf{Proposed TMUL2} 	&0.16&\textbf{10.03}&1.60				&0.17	&\textbf{9.65}&1.64 			 &0.23		&\textbf{8.82}	&2.03\\
\textbf{Proposed TMUL3} 	&\textbf{0.09}&11.97&\textbf{1.07}	&\textbf{0.07}	&11.61&\textbf{0.81}	&\textbf{0.10}	&10.75	&\textbf{1.07}\\
\hline\hline
\end{tabular}
\end{table}

As shown in Table~\ref{44t731}, the comparison of the proposed TMUL1 demonstrates a notable reduction in PDP.\\
At Temp.=10\textdegree{}C, around 70.6\%, 61.31\%, and 38.74\% compared to \cite{123, t15, t16}, respectively.\\
At Temp.=27\textdegree{}C, around 73.56\%, 63.12\%, and 44.21\% compared to \cite{123, t15, t16}, respectively.\\
At Temp.=70\textdegree{}C, around 70.27\%, 60.43\%, and 41.28\% compared to \cite{123, t15, t16}, respectively.

The comparison of the proposed TMUL2 demonstrates a notable reduction in PDP~compared to \cite{302}: around 95.62\%, 95.33\%, and 94.28\% at Temperature (10\textdegree{}C, 27\textdegree{}C, and 70\textdegree{}C), respectively.

Whereas, the comparison of the proposed TMUL3 has the lowest PDP compared to all the investigated models and the proposed TMUL1, and TMUL2.\\
At Temperature=10\textdegree{}C, around 98.88\%, 98.53\%, 97.66\%, 97.07\%, 89.66\%, 96.19\%, and 33.13\% compared to \cite{123, t15, t16,302,205}, the proposed TMUL1, and the proposed TMUL2, respectively.\\ 
At Temperature=27\textdegree{}C, around 99.07\%, 98.7\%, 98.04\%, 97.7\%, 92.39\%, 96.48\%, and 50.61\% compared to \cite{123, t15, t16,302,205}, the proposed TMUL1, and the proposed TMUL2, respectively.\\   
At Temperature=70\textdegree{}C, around 98.84\%, 98.46\%, 97.72\%, 96.98\%, 91.59\%, 96.11\%, and 47.29\% compared to \cite{123, t15, t16,302,205}, the proposed TMUL1, and the proposed TMUL2, respectively.

\subsubsection{Comparison at Different Frequencies}
Table~\ref{44t732} shows the comparison to the existing TMUL in \cite{123, t15, t16,302,205} in terms of the average power consumption, maximum propagation delay, and maximum energy (PDP) at different frequencies (2 GHz, 1 GHz, 0.5 GHz), same temperatures (27\textdegree{}C), and same voltage supply Vdd (0.9 V).

\begin{table}[!htb]
\caption{Comparison of average power ($\mu$W), maximum delay (ps), and maximum PDP  (x$10^{-18}$ J) for 8 TMULs at T=27\textdegree{}C, Vdd=0.9 V, and at different frequencies}
\label{44t732}
\setlength{\tabcolsep}{2pt}
\centering
\begin{tabular}{c|ccc|ccc|ccc}
\hline\hline
	& \multicolumn{3}{c|}{f= 0.5  GHz}& \multicolumn{3}{c|}{f= 1  GHz} & \multicolumn{3}{c}{f= 2  GHz}\\
		& Power &Delay 	&PDP 	& Power &Delay 	&PDP & Power &Delay 	&PDP \\
	
\hline
TMUL in \cite{123}  			&1.74	&46.11 &80.23				&1.88&46.32&87.08				&2.03	&45.85	&93.08\\
TMUL in \cite{t15}			&1.41	&43.15 &60.84		&1.45&43.05&62.42							&1.54		&42.21	&65\\
TMUL in \cite{t16}			&1.28	&31.42&40.21		&1.32&31.26&41.26			     				&1.39 	&30.11 &41.85\\
TMUL in \cite{302}			&1.87	&18.18	&33.99				&1.93&18.21&35.15				&2.05	&18.20 &37.31\\
TMUL in \cite{205}			&0.62	&16.67&10.34				&0.64&16.63&10.64				&0.89	&6.55 &5.83\\
\hline
\textbf{Proposed TMUL1} 	&0.38	&51.89	&19.72				&0.42&54.82&	23.02			     &0.46	&53.15 &24.45\\
\textbf{Proposed TMUL2} 	&0.15&\textbf{9.63}&1.44			&0.17	&\textbf{9.65}&1.64 			 &0.20		&\textbf{9.64}	&1.93\\
\textbf{Proposed TMUL3} 	&\textbf{0.06}&11.60&\textbf{0.69}	&\textbf{0.07}	&11.61&\textbf{0.81}	&\textbf{0.09}	&11.62	&\textbf{1.05}\\
\hline\hline
\end{tabular}
\end{table}

As shown in Table~\ref{44t732}, the comparison of the proposed TMUL1 demonstrates a considerable reduction in PDP.\\
At freq.=2 GHz, around 73.73\%, 62.38\%, and 41.58\% compared to \cite{123, t15, t16}, respectively.\\ 
At freq.=1 GHz, around 73.56\%, 63.12\%, and 44.21\% compared to \cite{123, t15, t16}, respectively.\\
At freq.=0.5 GHz, around 75.42\%, 67.59\%, and 50.96\% compared to \cite{123, t15, t16}, respectively.

The comparison of the proposed TMUL2 demonstrates a notable reduction in PDP compared to \cite{302}: around 94.83\%, 95.33\%, and 95.76\% at frequency (2 GHz, 1 GHz, and 0.5 GHz), respectively.

Whereas, the comparison of the proposed TMUL3 has the lowest PDP compared to all the investigated models and the proposed TMUL1, and TMUL2.\\
At frequency=2 GHz, around 98.87\%, 98.38\%, 97.49\%, 97.19\%, 81.99\%, 95.71\%, and 45.6\% compared to \cite{123, t15, t16,302,205}, the proposed TMUL1, and the proposed TMUL2, respectively.\\ 
At frequency=1 GHz, around 99.07\%, 98.7\%, 98.04\%, 97.7\%, 92.39\%, 96.48\%, and 50.61\% compared to \cite{123, t15, t16,302,205}, the proposed TMUL1, and the proposed TMUL2, respectively.\\ 
At frequency=0.5 GHz, around 99.14\%, 98.87\%, 98.28\%, 97.97\%, 93.33\%, 96.5\%, and 52.08\% compared to \cite{123, t15, t16,302,205}, the proposed TMUL1, and the proposed TMUL2,~respectively.


\begin{figure}[!b]
\centering
\includegraphics[width = 7 cm]{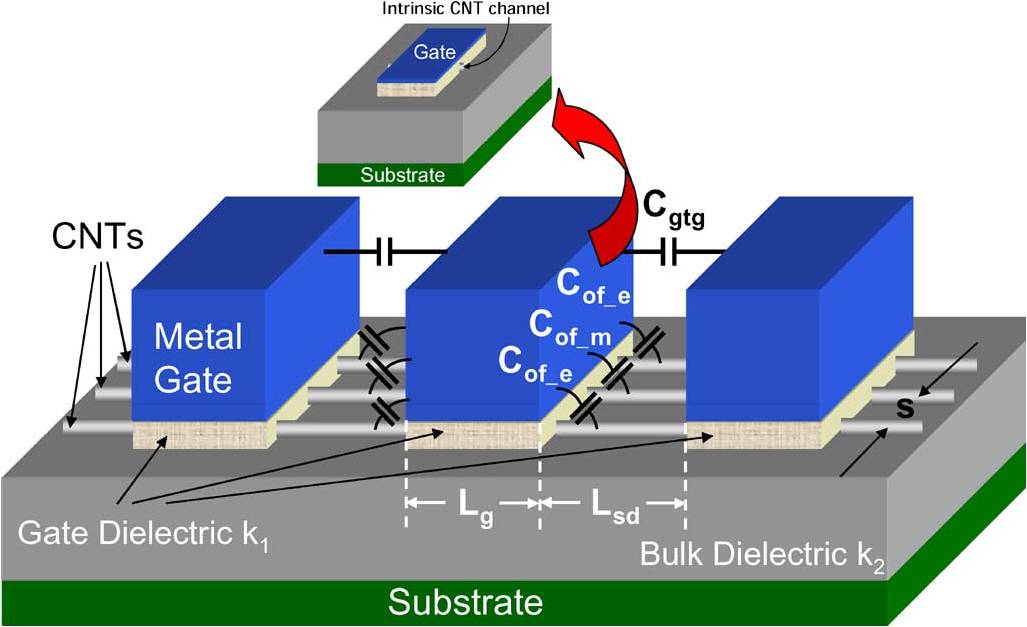}
\caption{Stanford CNFET Model.}
\label{figcnfet}
\end{figure}

\subsection{Process Variations}
Process variations have a great impact on the performance and robustness of nanoscale devices and circuits.
Hence, all circuits are tested in the presence of major process variations: TOX, CNT diameter, CNT's count, and channel length \cite{refA}.\\
\textbf{TOX}\\
In the Stanford CNFET model, TOX is the oxide thickness of the Gate Dielectric $k_1$ material ($HfO_2$), as shown in \ref{figcnfet}. The variation in TOX will lead to leakage current.
By decreasing TOX, the leakage current will increase, then the PDP will also increase and Vise Versa.\\
\textbf{CNT Diameter}\\
The CNT diameter variation is one of the multiple problems of CNFET imperfections caused due to nonidealities in the CNT synthesis process, which negatively affects the performance of CNFET circuits because this variation will lead to variation in threshold voltage. This problem has more impact on MVL designs where transistors with different threshold voltages are applied.\\
\textbf{CNT's Count}\\
Variation in the count of CNTs leads to a change in circuit parameters because it changes the output current of the transistor, which may cause a problem in the circuit functionality. \\
\textbf{Channel Length}\\
Variation in channel length will lead to the variation of the channel between drain and source, which negatively affects the performance of CNFET circuits.

Therefore, this paper uses Monte Carlo analysis based on the Gaussian distributions with 5\%, 10\%, and 15\% variations at the ±3 $\sigma$ level with the number of simulation running is equal to 1000.

Monte Carlo analysis is based on statistical distributions. It simulates mismatching and process variations. In each simulation run, it calculates every parameter randomly according to a statistical distribution model.

\begin{figure}[!t]
\centering
\includegraphics[width = \columnwidth]{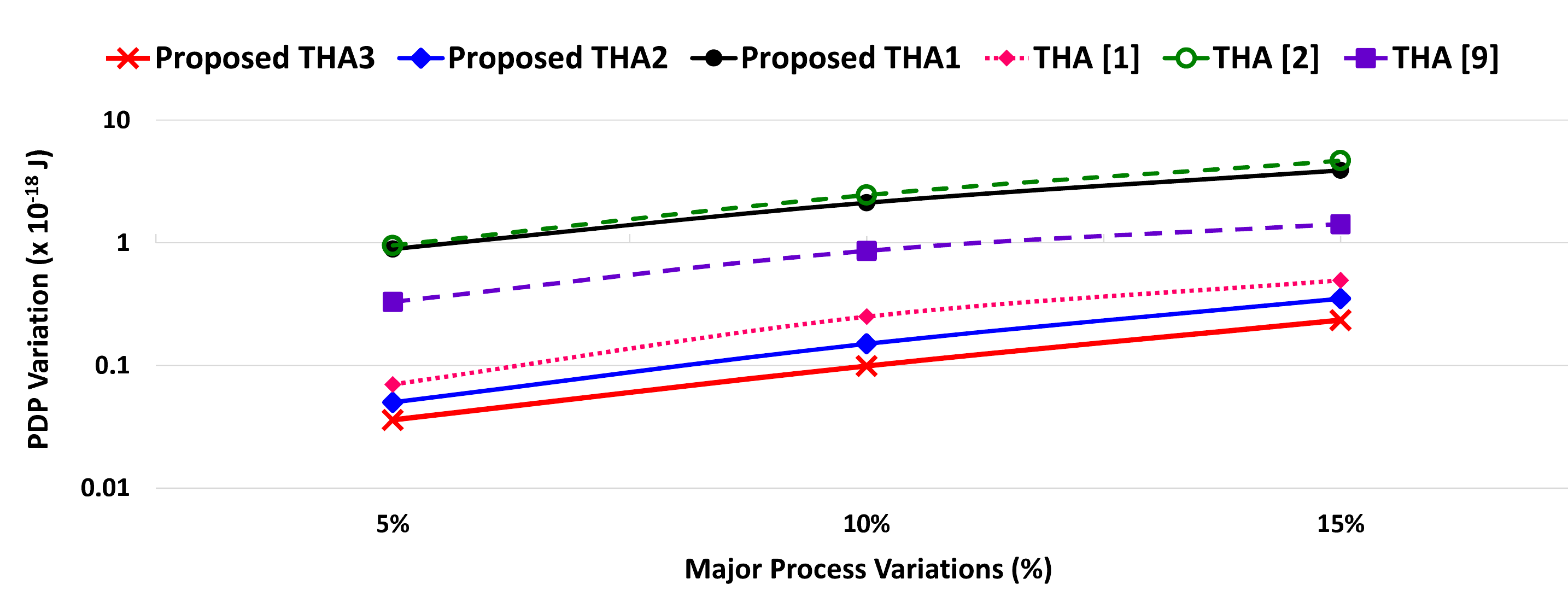}
\caption{THAs Major Process Variations: TOX, CNT Diameter, CNT's Count, and Channel length.}
\label{chartProcess1}
\end{figure}

\begin{figure}[!t]
\centering
\includegraphics[width = \columnwidth]{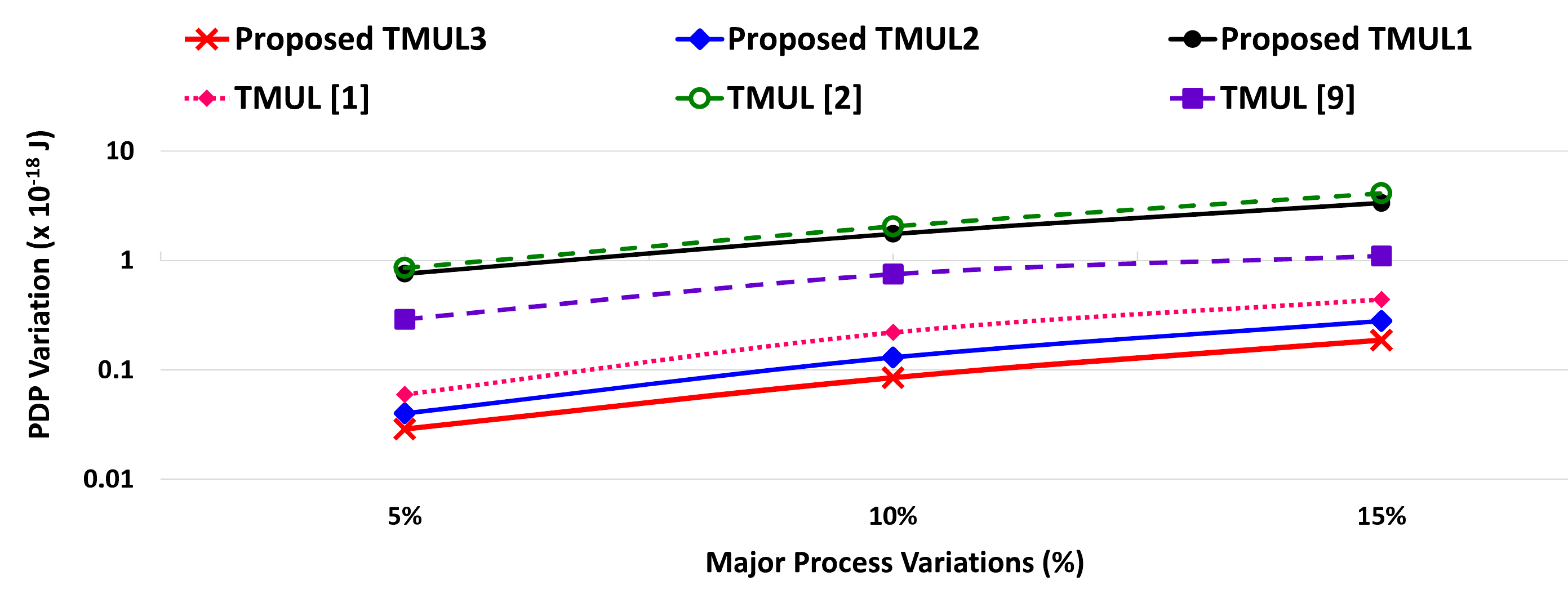}
\caption{TMULs Major Process Variations: TOX, CNT Diameter, CNT's Count, and Channel length.}
\label{chartProcess2}
\end{figure}

The energy variations of all THA and TMUL circuits in the presence of the major process variations are shown in Fig. \ref{chartProcess1} and \ref{chartProcess2}.

As shown in Fig. \ref{chartProcess1} and \ref{chartProcess2}, The proposed THA3 and TMUL3 have a lower sensitivity to process variations and are more robustness compared to the other designs because their PDP variations are the smallest among all the investigated circuits.

\subsection{Noise Effect}
Digital circuits are inherently noise-tolerant, and they are only affected by noises with high amplitude and wide width.

The Noise Immunity Curve (NIC) is used to determine the impact of noisy inputs on all THA and TMUL circuits.

\begin{figure}[!t]
\centering
\includegraphics[width = \columnwidth]{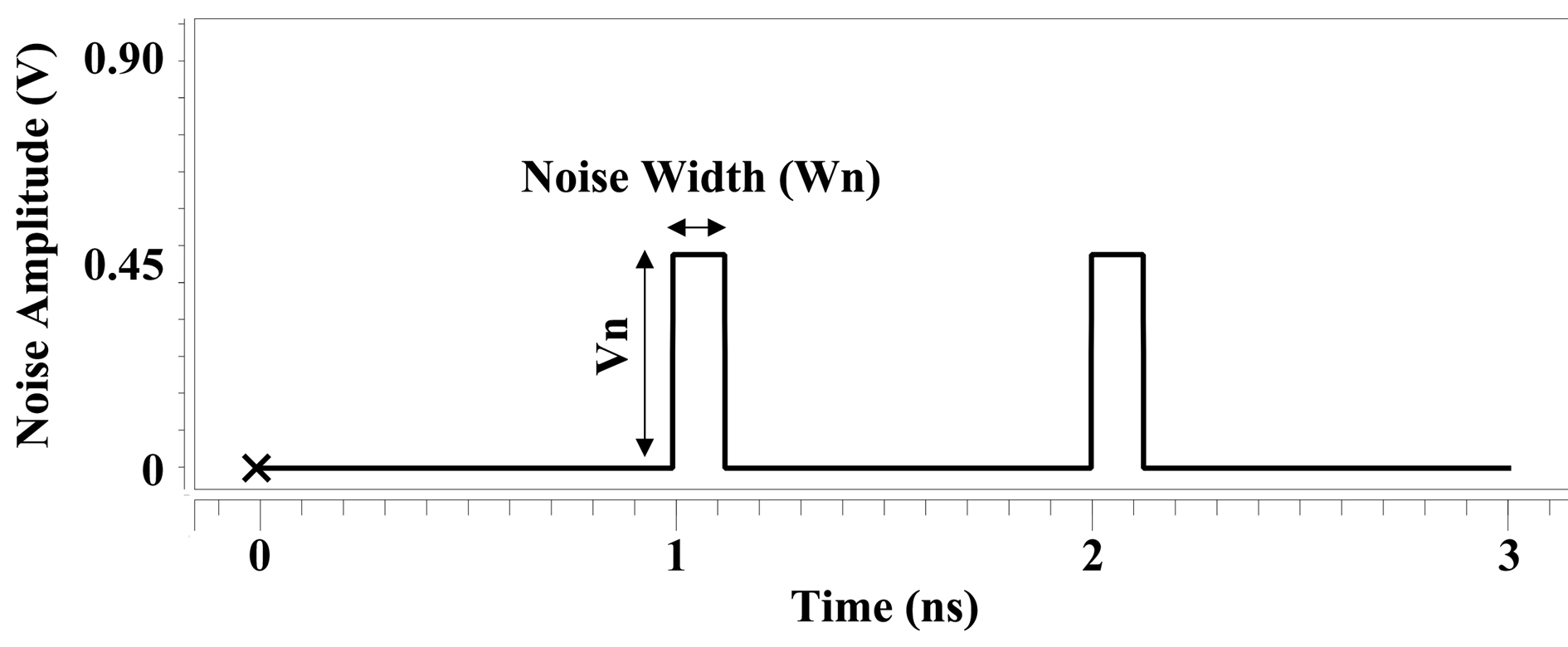}
\caption{Noise Signal.}
\label{fignoise}
\end{figure}

The noise signal, shown in Fig.\ref{fignoise}, is injected into inputs of all THAs and TMULs.

As shown in Fig.\ref{fignoise}, the noise signal has pulse width ($W_n$) and pulse amplitude ($V_n$).

Each point on the NIC curve is a pair of ($W_n$, $V_n$). Above that point, the circuit will produce an error on the output.
The region above the NIC curve is an unsafe zone, whereas the region below the NIC curve is a safe zone against noise pulses.

Therefore, any circuit with higher NIC demonstrates a more noise-tolerant circuit \cite{refB}.

\begin{figure}[!t]
\centering
\includegraphics[width = \columnwidth]{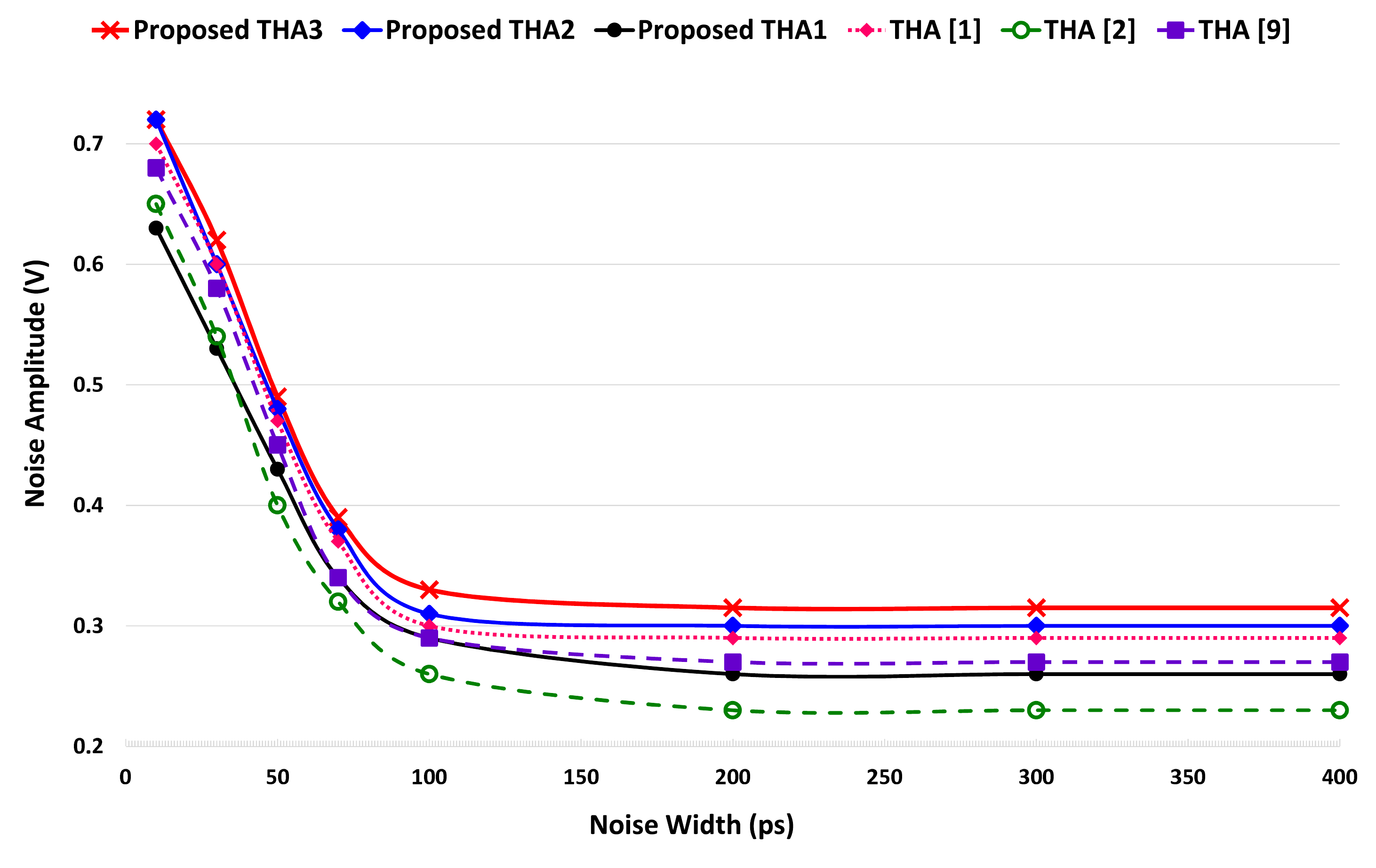}
\caption{THAs Noise Immunity Curve (NIC).}
\label{chartNIC1}
\end{figure}

\begin{figure}[!t]
\centering
\includegraphics[width = \columnwidth]{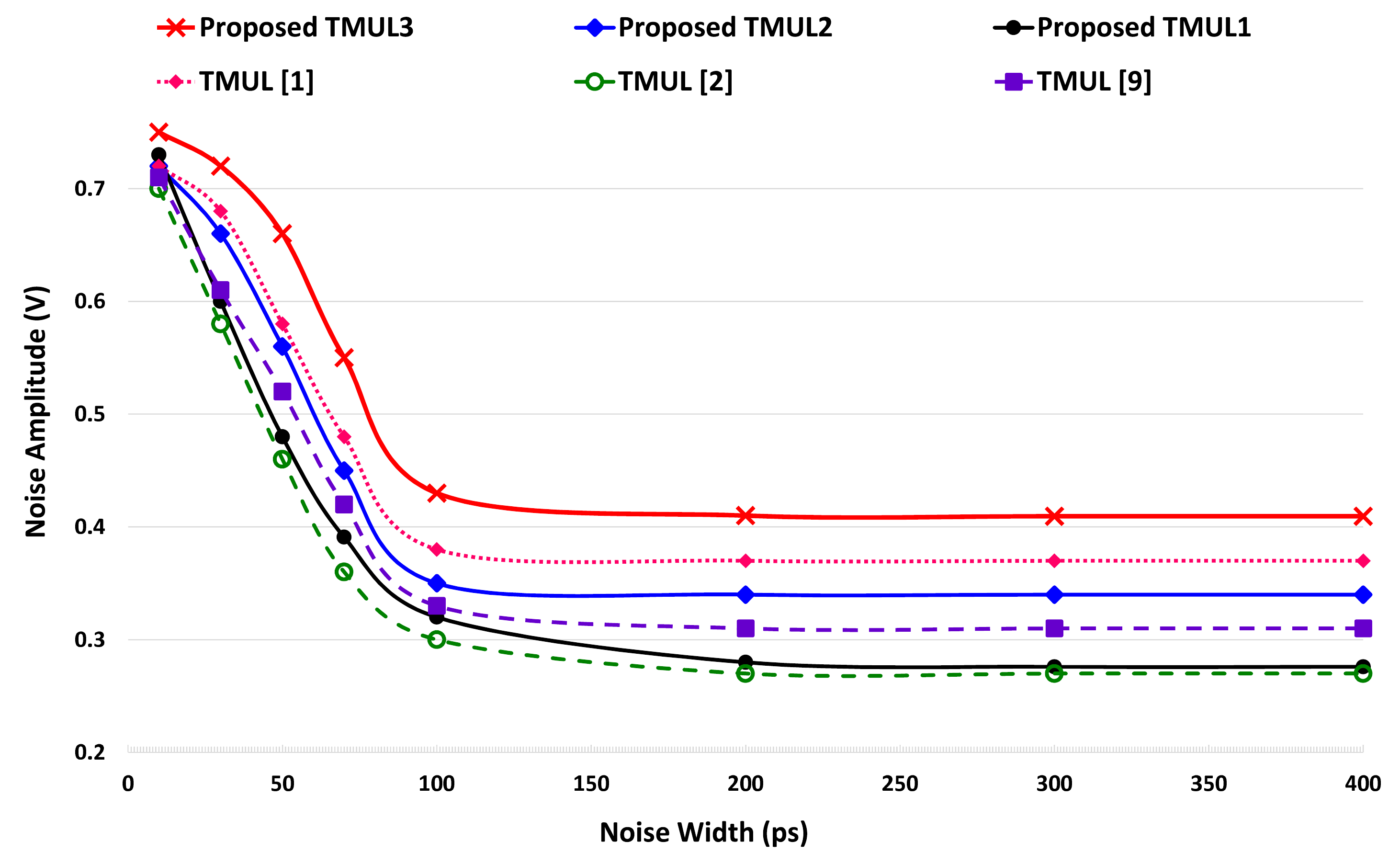}
\caption{TMULs Noise Immunity Curve (NIC).}
\label{chartNIC2}
\end{figure}

The proposed THA3 and TMUL3 show higher noise immunity compared to all the investigated circuits, as shown in Fig. \ref{chartNIC1} and \ref{chartNIC2}.

\clearpage
\section{Conclusion}
This Chapter has proposed three different designs of THA and TMUL. The first design uses the proposed TDecoder1, binary NANDs, the proposed STI, and the proposed TNAND. The second design uses the proposed cascading TMUX. The third design uses the proposed unary operators and the proposed TMUX.

The design process tried to optimize several circuit techniques such as reducing the number of the used transistors, utilizing energy-efficient transistor arrangements, and applying the dual supply voltages (Vdd and Vdd/2).

The proposed ternary circuits are compared to the latest exixting ternary circuits, simulated and tested using HSPICE simulator under various operating conditions at different supply voltages, different temperatures, and different frequencies.

The results prove the merits of the approach in terms of reduced energy consumption (PDP) compared to other existing designs. 

Moreover, the noise immunity curve (NIC) and Monte Carlo analysis for major process variations (TOX, CNT Diameter, CNT's Count, and Channel length) were studied. The results confirmed that the third proposed THA3 and TMUL3 had higher robustness and higher noise tolerance, among other designs.

Therefore, the proposed circuits can be implemented in low-power portable electronics and embedded systems to save battery consumption.


\chapter{Ternary Data Transmission} 

\label{Chapter5} 

\lhead{Chapter 5. \emph{Ternary Data Transmission}} 

\section{Introduction and Literature Review}
In general, data transmission transfer data over a communication medium to one or more computers, smartphones, embedded systems, or any electronic devices \cite{501, 504}. 

Besides external communication, data transmission also may be internal between the computer components, for example, between the CPU and the RAM or HDD \cite{502}.

Transmission media usually used in communication channels are a copper wire (coaxial cable, twisted pair), glass fiber, microwave, radio, or infrared \cite{503}.

There are limitations on binary data transmission at the physical layer in cable connections between hosts (computers, hubs, switches, and routers), such as bandwidth, delay, and size of data. 

Therefore, this chapter will propose a novel bi-directional circuit that contains two converters (Binary-to-Ternary and Ternary-to-Binary).
When sending data, the circuit will convert from the binary system to the ternary system, whereas, when receiving data, the circuit will convert from the ternary system to the binary system.

\begin{figure}[h!]
\centering
\includegraphics[width = 15 cm]{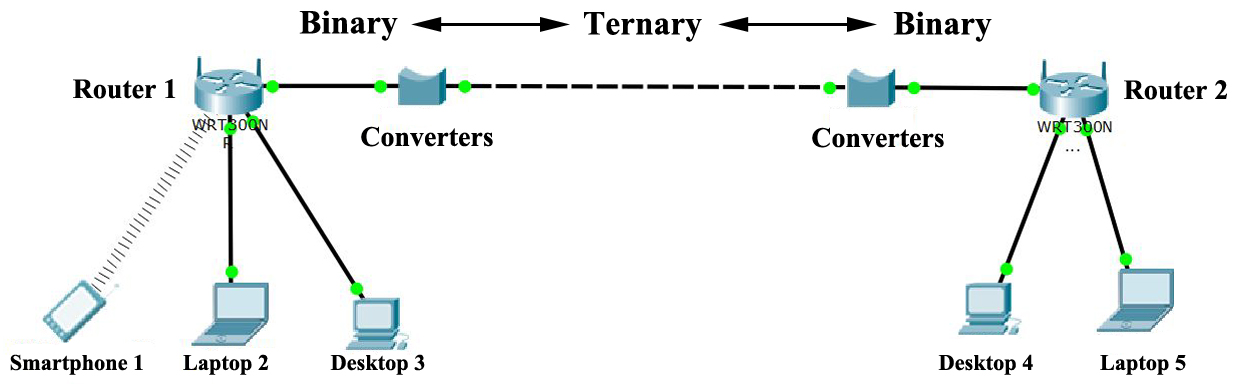}
\caption{The proposed Data Transmission model between routers. }
\label{FigTrans}
\end{figure}

The proposed Data Transmission model between routers is introduced with the general model, as represented in Fig.\ref{FigTrans}. It has two bi-directional converters between the routers. 

The design of binary-to-ternary converters and ternary-to-binary converters has recently constituted an active area of research, and valued papers are selected.

In \cite{misc01, misc02}, a circuit that converts two binary bits to one ternary digit (trit) was proposed. In \cite{article05}, the authors proposed a binary-to-ternary converter based on the Josephson junction (JJ). In \cite{proceedings08}, the designed converter circuit is characterized by a dual supply operation. Finally, in \cite{article06}, the authors proposed a novel binary-to-ternary converter based on DPL, which gives the best performance than all mentioned papers above.

In \cite{t4}, CMOS circuit is proposed to convert one trit into two bits, while the circuits in \cite{t5, t6, t61, t62, t63} represent ternary, quaternary, quinary, octal and hexadecimal to binary converters that use multiple input floating gate MOSFETs. Also, in \cite{proceedings09}, a novel Ternary-to-Binary Converter used in Quantum-dot Cellular Automata proposed, while in \cite{t7}, three circuits are used to convert ternary data to bits.

The above mentioned converters suffer from high power consumption \cite{article05, proceedings08, t4,t5,t6}, large propagation delay \cite{proceedings08, article06}, complex circuit \cite{proceedings08, proceedings09, t7} or even errors in the output when two binary inputs are set as high level voltage \cite{misc01, misc02}.

Therefore, this chapter proposes novel bi-directional converters, which are published in (\cite{conf2}, \cite{conf3}) and can be found in Appendices (D, E).

Binary-to-ternary converter using CMOS DPL Gates and ternary-to-binary converter using CNFETs, which provide a drastic reduction in power consumption, lower the propagation delay, and provide a small, simple and error-free circuit~implementation (for all possible inputs) compared to others circuits.

\section{The Proposed Binary-to-Ternary Converter using CMOS DPL Gates}

The proposed circuit converts four binary bits to three ternary trits using Double~Pass-Transistor Logic (DPL) CMOS binary logic gates. It is simulated using Micro-Cap V10 PSPICE simulator and it demonstrates a notable reduction in devices count, propagation delay, and energy consumption.

\begin{figure}[h!]
\centering
\includegraphics[width=12cm]{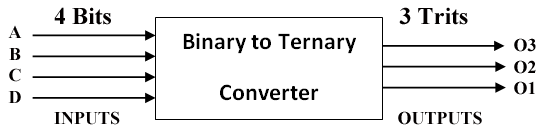}
\caption{Binary-to-Ternary model}
\label{fig52}
\end{figure}

A novel four to three binary-to-ternary converter model is shown in Fig.\ref{fig52}. 

As shown in Fig.\ref{fig52}, four binary bits are provided as input (A, B, C, D). The input “A” is the most significant bit (MSB), and the input “D” is the least significant bit (LSB). Three trits are provided as output (O3, O2, O1). The output “O3” is the most significant trit (MST), and “O1” is the least significant trit (LST), using simple ternary logic (0, 1, 2).

Among the 27 trits combinations, only 16 are used to represent all the four-bit combinations while the remaining 11 could be used for additional features such as error detection.
 
The 16 combinations that are produced in the converter circuit are shown in Table \ref{table51}.
The voltage supply used in the proposed circuit is 1.8V to decrease power consumption. Therefore, DPL is used to eliminate noise, as described in chapter 1.

Binary digits are (0: Low, and 1: High) and Ternary trits are (0: Low, 1: Medium, and 2: High). This chapter will use, “0” to represents low-level voltage (0V), “1” to represents medium-level voltage (0.9V), and “2” to represents high-level voltage (1.8V) for the both binary and ternary system.

\begin{table}[h!]
\centering
\caption{Binary-to-Ternary truth table}
\label{table51}
 \begin{tabular}{c|cccc|ccc } 
\hline\hline
 & \multicolumn{4}{c|}{\textbf{Binary Inputs}} & \multicolumn{3}{c}{\textbf{Ternary Outputs}}\\
\hline
 \textbf{Decimal} &\textbf{A}&\textbf{B}& \textbf{C}& \textbf{D} & \textbf{O3} & \textbf{O2} & \textbf{O1}\\
\hline
0&	0&	0&	0&	0&	0&	0&	0\\
1&	0&	0&	0&	2&	0&	0&	1 \\
2&	0&	0&	2&	0&	0&	0&	2 \\
3&	0&	0&	2&	2&	0&	1&	0\\
4&	0&	2&	0&	0&	0&	1&	1 \\
5&	0&	2&	0&	2&	0&	1&	2 \\
6&	0&	2&	2&	0&	0&	2&	0 \\
7&	0&	2&	2&	2&	0&	2&	1 \\
8&	2&	0&	0&	0&	0&	2&	2 \\
9&	2&	0&	0&	2&	1&	0&	0 \\
10&2&	0&	2&	0&	1&	0&	1 \\
11&2&	0&	2&	2&	1&	0&	2 \\
12&2&	2&	0&	0&	1&	1&	0 \\
13&2&	2&	0&	2&	1&	1&	1 \\
14&2&	2&	2&	0&	1&	1&	2 \\
15&2&	2&	2&	2&	1&	2&	0 \\
\hline\hline
\end{tabular}
\end{table}

The procedure to build the proposed circuit starts with a novel technique by decompositioning each ternary output in Table \ref{table51} to two intermediate binary outputs, as shown in Table \ref{table52} and Table \ref{table53}.

\begin{table}[h!]
\centering
\caption{Decomposition of O1 to two intermediate Binary Outputs }
\label{table52}
 \begin{tabular}{c|cc} 
\hline\hline
 \textbf{Ternary Output} & \multicolumn{2}{c}{\textbf{Intermediate Binary Outputs}}\\
\hline
 \textbf{O1} &\textbf{O1a}&\textbf{O1b}\\
\hline
Logic 0 (0V)    &	0&	0	\\
Logic 1 (0.9V)&	0& 2	\\
Logic 2 (1.8V)&	2&	0	\\
\hline\hline
\end{tabular}
\end{table}

\begin{table}[h!]
\centering
\caption{Decomposition of the three ternary outputs }
\label{table53}
 \begin{tabular}{cccc|c|cc|cc} 
\hline\hline
 \multicolumn{4}{c|}{\textbf{Binary Inputs}} & \multicolumn{5}{c}{\textbf{Ternary Outputs as Intermediate Binary}}\\
\hline
 &&&&\textbf{O3} & \multicolumn{2}{c|}{\textbf{O2}} & \multicolumn{2}{c}{\textbf{O1}}\\
\hline
 \textbf{A} &\textbf{B} & \textbf{C} & \textbf{D} & \textbf{O3b} & \textbf{O2a} & \textbf{O2b} & \textbf{O1a} & \textbf{O1b}\\
\hline
0&	0&	0&	0&	0&	0&	0&	0&	0\\
0&	0&	0&	2&	0&	0&	0&	0&	2\\
0&	0&	2&	0&	0&	0&	0&	2&	0\\
0&	0&	2&	2&	0&	0&	2&	0&	0\\
0&	2&	0&	0&	0&	0&	2&	0&	2\\
0&	2&	0&	2&	0&	0&	2&	2&	0\\
0&	2&	2&	0&	0&	2&	0&	0&	0\\
0&	2&	2&	2&	0&	2&	0&	0&	2\\
2&	0&	0&	0&	0&	2&	0&	2&	0\\
2&	0&	0&	2&	2&	0&	0&	0&	0\\
2&	0&	2&	0&	2&	0&	0&	0&	2\\
2&	0&	2&	2&	2&	0&	0&	2&	0\\
2&	2&	0&	0&	2&	0&	2&	0&	0\\
2&	2&	0&	2&	2&	0&	2&	0&	2\\
2&	2&	2&	0&	2&	0&	2&	2&	0\\
2&	2&	2&	2&	2&	2&	0&	0&	0\\

\hline\hline
\end{tabular}
\end{table}

\begin{figure}[h!]
\centering
\includegraphics[width=12cm]{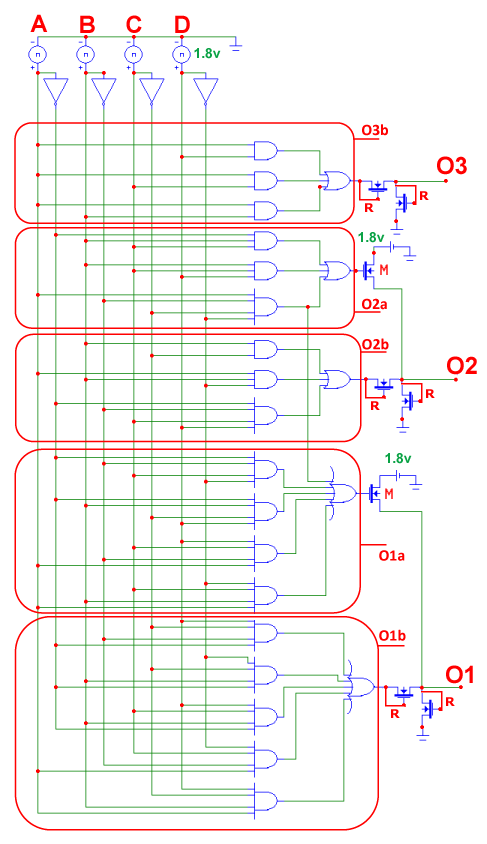}
\caption{The Proposed binary-to-ternary Converter Circuit}
\label{fig53}
\end{figure}

\clearpage
The second step involves using Karnaugh maps to get an optimized circuit for each intermediate output in Table \ref{table53}. The minimization process results in equations \ref{eq53}- \ref{eq57}.\\
\begin{equation} \label{eq53}
\small O1a = \xbar{A}\, \xbar{B}\, C \,\xbar{D} + \xbar{A}\, B \xbar{C}\, D + A \,\xbar{B} \,\xbar{C}\, \xbar{D} + A \,\xbar{B}\, C\, D + A\, B\, C\, \xbar{D}
\end{equation}
\begin{equation} \label{eq54}
\small O1b= \xbar{A} \, \xbar{B} \, \xbar{C} \, D +\xbar{A} \, B \xbar{C} \, \xbar{D}+ \xbar{A} B\, C \, D + A \xbar{B} \, C \, \xbar{D} + A \, B \xbar{C} \, D
\end{equation}
\begin{equation} \label{eq55}
\small O2a = \xbar{A} \,B \,C + \xbar{A} \, B \, \xbar{C} \,D + B\,C\,D + A \xbar{B} \,\xbar{C} \, \xbar{D}
\end{equation}
\begin{equation} \label{eq56}
\small O2b = B \xbar{C}+ A\,B\xbar{D} + \xbar{A} \, \xbar{B} \,C\,D
\end{equation}
\begin{equation} \label{eq57}
\small O3b = AB +AC +AD
\end{equation}
The resulting equations are implemented in the binary-to-ternary converter circuit shown in Fig. \ref{fig53}.
 
Figure \ref{fig53} represents the binary-to-ternary converter with four binary inputs (A, B, C, D) and three ternary outputs (O1, O2, O3). The intermediate binary outputs labeled as O1a, O1b, O2a, O2b, and O3b.

An enhancement MOSFET (M) is used as a switch after each of the intermediate outputs O1a and O2a. When O1a or O2a is at high-level voltage, then the transistor M will be turned “ON” to pass the 1.8 volts to the final output O1, and it will be  “OFF” when the O1a or O2a is at low-level voltage.
 
Similarity, two diode-connected transistors (R) that act like resistors are added after the intermediate outputs (O1b, O2b, O3b), these transistors act as a voltage divider. For example, when O1b is equal to 1.8V,  then the output O1 is equal to 0.9V.
When O1b is equal to 0V and O1a is equal to 1.8V, then the two transistors (Rs) will become two parallel resistors with equivalent resistance (Req) which will drop the output voltage O1 to 1.7V, as described in Table \ref{table54}.

\clearpage
\begin{table}[t!]
\centering
\caption{Combination of Ternary Output O1}
\label{table54}
 \begin{tabular}{cc|cc|c} 
\hline\hline
\textbf{O1a} &\textbf{O1b} &\textbf{Transistor (M)} & \textbf{(2 Rs)} & \textbf{O1}\\
\hline
 0V&	   0V& 	OFF & Req &	0V		(Logic 0)\\
0V&	 1.8V&	OFF& Divider&	0.9V	(Logic 1)\\
1.8V     &0V&	 ON& Req &	1.7V	(Logic 2)\\
\hline
\multirow{2}{*}{1.8V}&\multirow{2}{*}{1.8V}&\multicolumn{3}{c}{\textbf{This Case Does Not Exist}}\\
&&\multicolumn{3}{c}{\textbf{Whereas, in \cite{misc01} and \cite{misc02} have errors }}\\
\hline\hline
\end{tabular}
\end{table}

The same procedure will be done for the outputs O2 and O3.

The proposed design has a notable advantage over previous converter designs in \cite{misc01} and \cite{misc02} since no error occurs for any input combination. As demonstrated in Tables \ref{table53} and \ref{table54}, the intermediate binary outputs Oia, and Oib (i=1, 2) cannot be high at the same time, which offers a significant advantage in simplifying the circuit.

The device count in the proposed circuit is calculated in Table \ref{table56}.

\begin{table}[b!]
\centering
\caption{Device count in the proposed circuit (Fig.\ref{fig53})}
\label{table56}
 \begin{tabular}{c|c|c|c} 
\hline\hline
\textbf{Device Name} &\textbf{No. of devices} &\textbf{No. of Transistors} & \textbf{Subtotal}\\
\hline
Inverter Gate&		4&	1 input *2 =2&	8\\
2 inputs AND Gate&		2&	2 inputs *2 = 4&	8\\
3 inputs AND Gate&		3&	3 inputs *2 = 6&	18\\
4 inputs AND Gate&		11&	4 inputs *2 = 8&	88\\
3 inputs OR Gate&		3&	3 inputs *2 = 6&	18\\
5 inputs OR Gate&		2&	5 inputs *2 = 10&	20\\
MOSFET&			8&	1&			8\\
\hline
\multicolumn{3}{c|}{Internal Resistor (one per each gate)} &25		\\
\multicolumn{3}{c|}{Internal Capacitor (one per each gate)}&25		\\ \hline
\multicolumn{3}{c|}{\textbf{Total No. of Devices}}					&\textbf{218}	\\
\hline\hline
\end{tabular}
\end{table}

\clearpage
\subsection{Simulation Results and Comparisons}

In \cite{article06}, the authors proposed a novel binary-to-ternary converter based on DPL, which gives the best performance than binary-to-ternary converters in \cite{misc01, misc02, article05, proceedings08, proceedings09}.

Therefore, the proposed converter will be compared to the latest converter circuit designs described in \cite{article06}.

For fair comparisons, both circuits are implemented, simulated, and analyzed using the same software Micro-Cap V10 PSPICE Simulator with CMOS technology 0.18 $\mu$m.

The simulated circuits utilize an implementation of the AND and OR DPL gates as previously described in \cite{124}. At a temperature of 25 \textdegree{}C, average propagation delay (42.56 ps), and average power (4.08 $\mu$w), as stated in \cite{article07}.

\begin{figure}[h!]
\centering
\includegraphics[width=15cm]{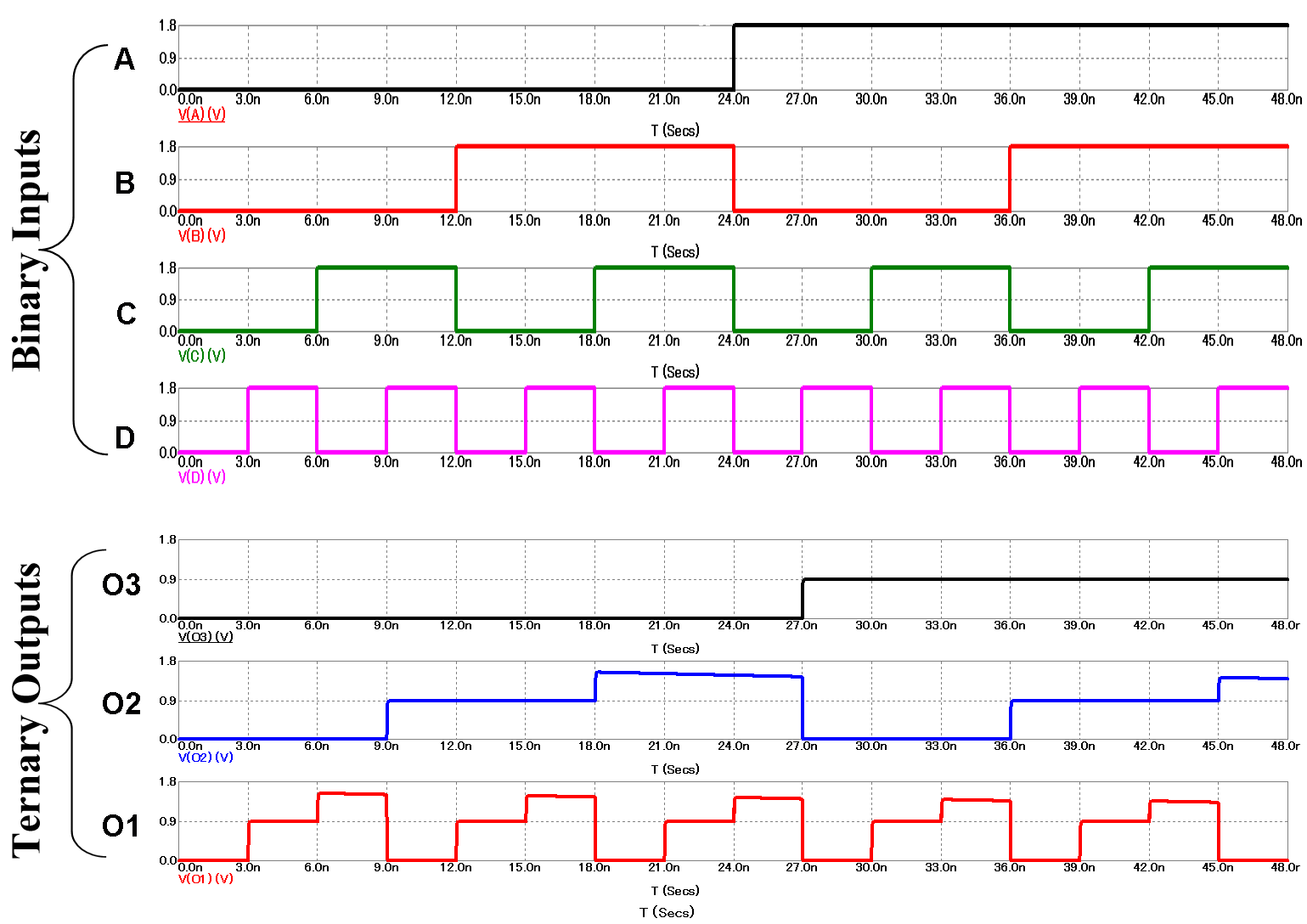}
\caption{The simulation transient analysis of the proposed circuit}
\label{fig54}
\end{figure}

Figure \ref{fig54} shows the transient simulation analysis of the proposed circuit at a temperature of 25 \textdegree{}C, power supply (1.8V), and frequency (0.33 GHz).

The following performance metrics will be used to compare to \cite{article06}: the frequency, the device count inside the circuit, the maximum propagation delay, the average power consumption, the maximum energy consumption, and the type of used logic gates.

The propagation delays in the proposed circuit are: From ternary “0” to “1” is 0.07 ns, from “1” to “2” is 0.09 ns and from ”2” to “0” is 0.06 ns.

The simulation of the proposed converter shows that the average power consumption is 349.9 $\mu$w.

Table \ref{table57} summarizes the comparison to the latest novel converter in \cite{article06}.

\renewcommand{\arraystretch}{1.3}
\begin{table}[t!]
\centering
\caption{Comparison to the latest novel converter in \cite{article06}}
\label{table57}
\begin{tabular}{l|c|c|c} 
\hline\hline
 &\textbf{Converter \cite{article06}} &\textbf{Proposed} &\textbf{Improvement}\\
\hline
Frequency	&		0.33 GHz				&	0.33 GHz			&		 \\
Device Count&	426							&	218				&		48.83\%\\
MAX. Delay &		0.91 ns					&	0.09 ns			&		90.1\%\\
Avg. Power	&270.42 µw 					&263.5 µw 		&         2.56\% \\
 MAX. Energy&246 $10^{-18}$J			&23.7 $10^{-18}$J&		90.36\%\\
  &&&\\
Type of used gates &	Ternary  			&	Binary gates&				Faster \cite{article09} \\
&\& binary gates&&\\
\hline\hline
\end{tabular}
\end{table}
\renewcommand{\arraystretch}{1}

The proposed (4:3) converter is bigger than the latest (3:2) converter in \cite{article06}. Despite that, the proposed converter gets better results compared to \cite{article06}, as shown in Table \ref{table57}.

\newpage
\section{The Proposed Ternary-to-Binary Converter using~CNFET}

\begin{figure}[h!]
\centering
\includegraphics[width=12cm]{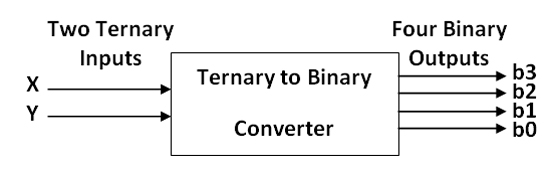}
\caption{Ternary-to-Binary model}
\label{FigTBC1}
\end{figure}

This chapter proposes two to three CNFET-Based Ternary-to-Binary converter model, as shown in Fig.\ref{FigTBC1}. 

Figure \ref{FigTBC1} shows two ternary inputs (X, Y): Where the input “X” is the most significant trit (MST), and the input “Y” is the least significant trit (LST). Four binary outputs (b3, b2, b1, b0): Where the output “b3” is the most significant bit (MSB), and “b0” is the least significant bit (LSB). 
Among the 16 ($2^4$) binary combinations, it will cover the 9 ($3^2$) ternary combinations while the remaining 7 combinations could use for additional features such as error detection.

\begin{figure}[h!]
\centering
\includegraphics[width=12cm]{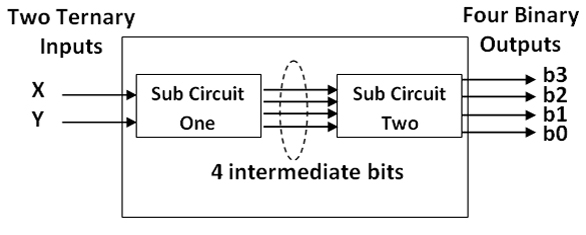}
\caption{Ternary-to-Binary model}
\label{FigTBC2}
\end{figure}

The “decoder-less” circuit of Fig. \ref{FigTBC2} converts 2-trits to 4-bits via 2-subcircuits.

The novel sub-circuit-one converts 2-trits to 4 intermediate bits while the sub-circuit-two converts the intermediate bits to four final bit outputs, as described in Table \ref{table58}.
A novel technique used: each trit transformed into the corresponding two bits, i.e., ternary logic “0” transformed into two bits ”00”, ternary logic “1” into binary ”02” and ternary logic “2” into ”20”.

\begin{table}[h!]
\centering
\caption{Convert 2 trits to 4 intermediate bits then final 4 bits}
\label{table58}
 \begin{tabular}{cc|cccc|cccc} 
\hline\hline
 \multicolumn{2}{c|}{\textbf{Ternary Inputs}} & \multicolumn{4}{|c}{\textbf{Intermediate bits}} & \multicolumn{4}{|c}{\textbf{Binary Outputs}} \\
\hline
\textbf{$X$} &\textbf{$Y$} & \textbf{$X_2$} & \textbf{$X_1$} & \textbf{$Y_2$} & \textbf{$Y_1$}  & \textbf{b3} & \textbf{b2} & \textbf{b1} & \textbf{b0}\\
\hline
0&	0&	0&	0&	0&	0&	0&	0&	0&	0 \\
0&	1&	0&	0&	0&	2&	0&	0&	0&	2\\
0&	2&	0&	0&	2&	0&	0&	0&	2&	0\\
1&	0&	0&	2&	0&	0&	0&	0&	2&	2\\
1&	1&	0&	2&  0&	2&	0&	2&	0&	0\\
1&	2&	0&	2&	2&	0&	0&	2&	0&	2\\
2&	0&	2&	0&	0&	0&	0&	2&	2&	0\\
2&	1&	2&	0&	0&	2&	0&	2&	2&	2\\
2&	2&	2&	0&	2&	0&	2&	0&	0&	0\\
\hline\hline
\end{tabular}
\end{table}

\subsection{The Novel Sub-Circuit-One Design}
\begin{figure}[h!]
\centering
\includegraphics[width=12cm]{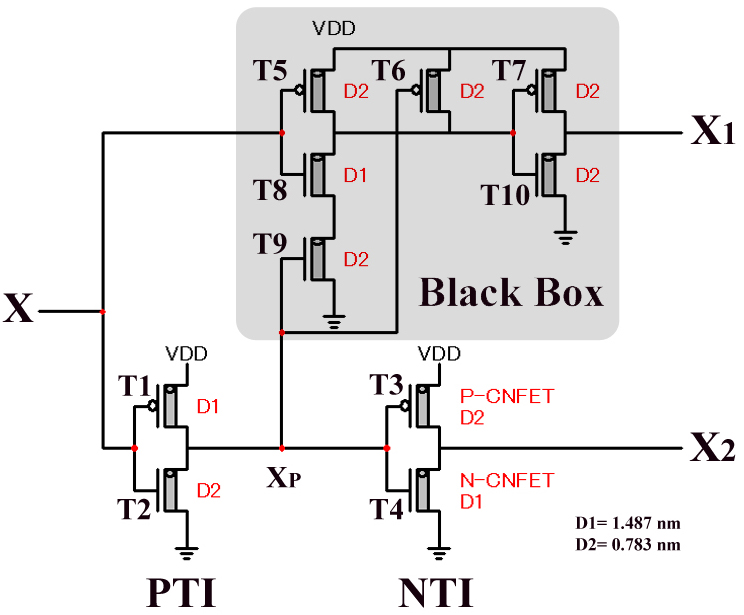}
\caption{Novel transistor level for 1-trit input to 2-intermediate bits outputs}
\label{FigSUB1}
\end{figure}

This novel sub-circuit-one made of two similar circuits, each one will convert 1-trit ($X$) to 2-intermediate bits ($X_1$ and $X_2$), as shown in Fig. \ref{FigSUB1}. It is containing a PTI inverter, an NTI inverter, and a key component “black box” , as described in Table \ref{table60}.

The ternary input $X$ is fed to the inverter (PTI) which has two transistors (T1, T2) to get $X_p$ as output which is fed to the inverter (NTI) which has two transistors (T3, T4) to get the binary intermediate target $X_2$. Also, $X$ and $X_p$ are fed to the black box which has 6 transistors (T5-T10) to get the binary~intermediate target $X_1$, as shown in Fig. \ref{FigSUB1}.

Two CNFET-diameters are used in the novel design: D1= 1.487nm which corresponds to the voltage thresholds Vth=0.289V for N-CNFET and -0.289V for P-CNFET, and D2=0.783 nm which corresponds to Vth=0.559V for N-CNFET and -0.559V for P-CNFET.

Table \ref{table60} summarizes all the detailed operations of the sub-circuit-one functionslity of Fig. \ref{FigSUB1}.
 
\begin{table}[t!]
\caption{The Sub-Circuit-One detailed truth table}
\label{table60}
\centering
\begin{tabular}{l|ccc}
\hline \hline
\textbf{$X$}		&\textbf{0}	&\textbf{1}&	\textbf{2}\\
\hline
$X_p$             &2	&2&	0\\
\hline
P-CNFET T3 (-0.559V)		&OFF&	OFF&	ON\\
N-CNFET T4 (0.289V)		&ON&	ON&	OFF\\
\hline
Output $X_2$             &0	&0&	2\\
\hline
P-CNFET T5 (-0.559V)		&ON&	OFF&	OFF\\
P-CNFET T6 (-0.559V)		&OFF&	OFF&	ON\\
P-CNFET T7 (-0.559V)		&OFF&	ON&	OFF\\
N-CNFET T8 (0.289V)		&OFF&	ON&	ON\\
N-CNFET T9 (0.559V)		&ON&	ON&	OFF\\
N-CNFET T10 (0.559V)		&ON&	OFF&	ON\\
\hline
Output $X_1$             &0	&2&	0\\
\hline\hline
\end{tabular}
\end{table}

\subsection{The Sub-Circuit-Two Design}

All the 4-bit intermediate output combinations, detailed in Table \ref{table58}, are now processed using the sub-circuit-two of the ternary to binary converter.
It will be optimized using a Karnaugh map to get the following Boolean equations:
\begin{equation} \label{6eq53}
\small b_0 = \xbar{X_1} \, Y_1+ X_1 \, \xbar{Y_1}
\end{equation}
\begin{equation} \label{6eq54}
\small b_1= X_2 \, \xbar{Y_2}+\xbar{X_2} \, \xbar{X_1} \, Y_2+ X_1 \, \xbar{Y_2} \, \xbar{Y_1}
\end{equation}
\begin{equation} \label{6eq55}
\small b_2= X_1 \, Y_1+X_1 \, Y_2 + X_2 \, \xbar{Y_2}
\end{equation}
\begin{equation} \label{6eq56}
\small b_3= X_2 \, Y_2
\end{equation}

\begin{figure}[h!]
\centering
\includegraphics[width=12cm]{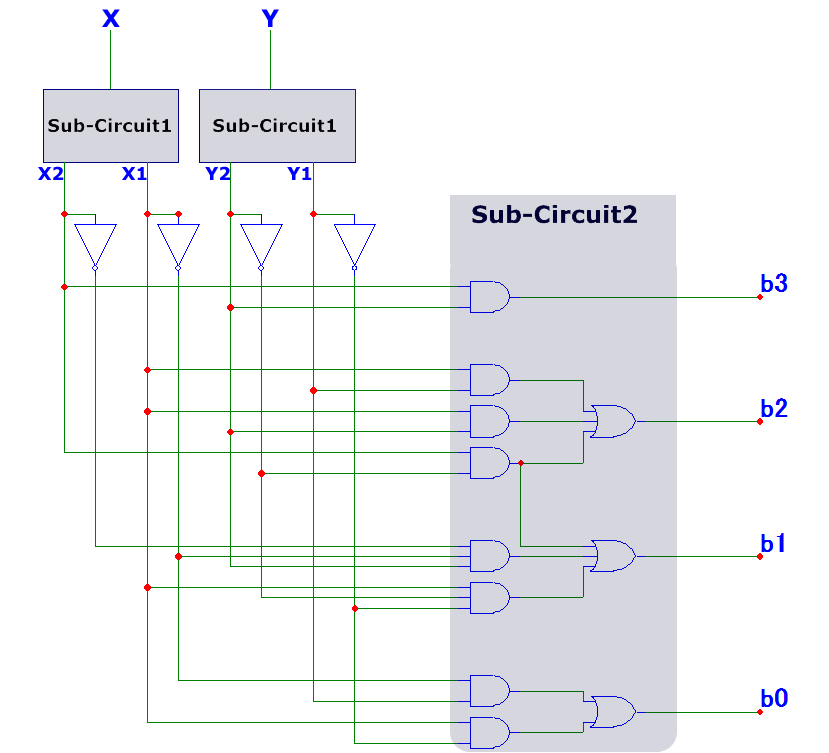}
\caption{The proposed schematic Ternary-to-Binary Converter.}
\label{FigTBC3}
\end{figure}

Figure \ref{FigTBC3} shows the complete schematic Ternary-to-Binary circuit based on the previous four equations \eqref{6eq53} till \eqref{6eq56}.
As a result, the Ternary-to-Binary converter has 102 transistors, as detailed in Table \ref{table61}.

\begin{table}[t!]
\caption{Total transistors count in the Ternary-to-Binary converter circuit}
\label{table61}
\centering
\begin{tabular}{l|ccc}
\hline \hline
			&\textbf{No. of devices}	&\textbf{No. Transistors}&	\textbf{Subtotal}\\
\hline
1-trit to 2 bits circuit             &2	&10&	20\\
Inverter             &4	&2&	8\\
2 inputs AND             &6	&6&	36\\
3 inputs AND             &2	&8&	16\\
2 inputs OR             &1	&6&	6\\
3 inputs OR             &2	&8&	16\\
\hline
\multicolumn{3}{l}{Total}&102\\
\hline\hline
\end{tabular}
\end{table}

Two ternary inputs X and Y are fed to the novel two sub-circuit-one and output the intermediate binary (X2, X1) and (Y2, Y1). After that, the outputs are passed to eight binary “AND” logic gates followed by three binary “OR” logic gates to get the 4 binary outputs (b3, b2, b1, b0), as shown in Fig. \ref{FigTBC3}.

\subsection{Simulation Results and Comparisons}

\begin{figure}[t!]
\centering
\includegraphics[width=16cm]{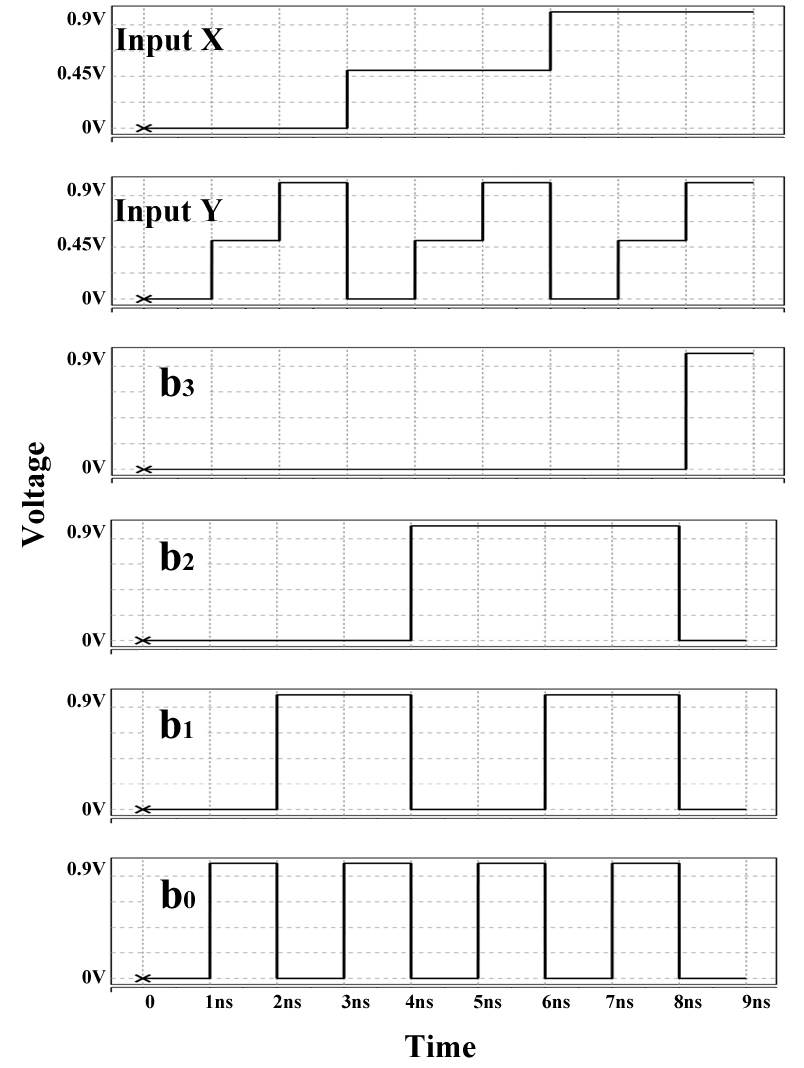}
\caption{Transient analysis of the proposed Ternary-to-Binary converter.}
\label{simTBC}
\end{figure}

The proposed Ternary-to-Binary converter is simulated and tested using HSPICE simulator with a power supply of 0.9V as Vdd.
Fig. \ref{simTBC} illustrates the transient analysis of the proposed converter: the X-axis represents the time with a step of 1 ns, and the Y-axis represents the voltage with a step of 0.2 volts.

Table \ref{table62} shows the comparisons between the CNFET and CMOS converters concerning the number of CNFET, the average power consumption, the maximum propagation delay, and the maximum Power Delay Product (PDP).

\begin{table}[t!]
\caption{Comparison between the three converter circuits}
\label{table62}
\centering
\begin{tabular}{l|cccc}
\hline \hline
			&\textbf{CNFETs No.}	&\textbf{Power}&	\textbf{Delay} &	\textbf{PDP}\\
			&\textbf{No.}	&\textbf{$\mu$W}&	\textbf{ps} &	\textbf{x $10^{-17}$J}\\
\hline
\cite{t4} Using CMOS						&127	&8.7 x $10^6$ & 8000 &69.6 x $10^9$\\
\cite{123} Using CNFET and TDecoder            &110	&0.425& 16.55 &0.703\\
\hline
The Proposed converter 	 &102	&0.143 & 16.19 &0.232\\
using CNFET and ``Decoder-less''&&&&\\
\hline\hline
\end{tabular}
\end{table}

The performance of the CNFET-based circuit design is better than the CMOS-based design, as shown in Table~\ref{table62}.\\
The comparison of the proposed Ternary-to-Binary converter demonstrates a~notable~reduction in Transistors count of around 19.68\%, and 72.72\% compared to \cite{t4} and \cite{123}, respectively.

Also, a notable reduction in PDP of around 99.99\%, and 66.99\% compared to \cite{t4} and \cite{123}, respectively.

\section{Conclusion}
This chapter proposed a bi-directional circuit that contains two converters. A binary-to-ternary converter and ternary-to-binary converter.

The first part proposed a binary-to-ternary converter with novel techniques that takes four bits and produces three trits using Double Pass-Transistor Logic (DPL).

The proposed binary-to-ternary converter is designed and simulated using Micro-Cap V10 PSPICE simulator. Finally, logical analysis and simulation results demonstrate the merits of the approach in terms of reduced latency and energy consumption compared to other converters.

The second part has proposed a novel ternary-to-binary converter "decoder-less" using CNFET that takes two trits as input and provides four bits as output. 

The proposed ternary-to-binary converter is then simulated and tested by the HSPICE simulator. Logical analysis and simulation results prove the merits of the implementation compared to the existing designs in terms of transistor count, reduced latency, and energy efficiency.


\chapter{Conclusion and Perspectives} 

\label{Chapter6} 

\lhead{Chapter 6. \emph{Conclusion and Perspectives}} 

\section{Conclusion}
This thesis proposes new designs include eight ternary logic gates, three ternary combinational circuits, and six Ternary Arithmetic Logic Units (TALU).  
The ternary logic gates are seven unary operators of the ternary system ($A^1$, $A^2$, $\bar{A^2}$, $A_1$, $1.\bar{A_n}$, $1.\bar{A_p}$, and the Standard Ternary Inverter (STI) $\bar{A}$), and Ternary NAND based on Carbon Nanotube Field-Effect Transistor (CNFET). Ternary combinational circuits, two different designs for Ternary Decoders (TDecoder) and Ternary Multiplexer (TMUX). TDecoder1 using CNFET-based proposed unary operators and TDecoder2 using Double-Pass Logic (DPL) binary gates. TMUX using CNFET-based proposed unary operators. 
And Ternary Arithmetic Logic Units are three different designs for Ternary Half-Adders (THA) and Ternary Multipliers (TMUL): (1) The first design uses the proposed TDecoder1, STI, and TNAND. (2) While the second design uses the cascading proposed TMUX. (3) As for the third design, it uses the proposed unary operators and TMUX.

All proposed circuits aim to optimize the trade-off between high-performance and energy-efficiency in order to implement them in low-power portable electronics and embedded systems to preserve battery consumption.\\
The design process tried to optimize several circuit techniques such as reducing the number of the used transistors, utilizing energy-efficient transistor arrangements, and applying the dual supply voltages (Vdd and Vdd/2) to achieve its aim.\\
The proposed ternary circuits are then compared to the latest ternary circuits, simulated and tested using HSPICE simulator under various operating conditions at different supply voltages, different temperatures, and different frequencies.\\
The results prove the merits of the approach in terms of reduced energy consumption (PDP) compared to other existing designs. 

Moreover, the noise immunity curve (NIC) and Monte Carlo analysis for major process variations (TOX, CNT Diameter, CNT's Count, and Channel length) were studied. The results confirmed that the proposed THA had higher robustness and higher noise tolerance, among other designs.

In addition, the second aim is using ternary data transmission to improve data~communications between hosts. Also, this thesis proposes a bi-directional circuit that contains two converters: (1) a binary-to-ternary converter and (2) a ternary-to-binary converter.\\
The proposed binary-to-ternary converter with novel techniques takes four bits as inputs and produces three trits as outputs using Double Pass-Transistor (DPL). The proposed circuit is then simulated and tested using the Micro-Cap V10 PSPICE simulator. 

While the proposed ternary-to-binary converter "decoder-less" using CNFET takes two trits as inputs and provide four bits as outputs. The proposed circuit is then simulated and tested using the HSPICE simulator. 

Finally, logical analysis and simulation results prove the merits of the approaches compared to existing designs in terms of transistor count, reduced latency, and energy efficiency.
\clearpage
\section{Perspectives}

Our methodology can be used to design Ternary Sequential Circuits like:
\begin{enumerate}
  \item Ternary Flip Flops
  \item Ternary Memory
  \item Ternary CPU
\end{enumerate}
Moreover, Application to different Engineering areas, as follow:

\begin{enumerate}
  \item Biomedical \& healthcare like:
  		\begin{itemize}
     		 \item{ElectroCardioGram (ECG)}
      	\item{ElectroEncephaloGram (EEG)}
      	\item{Implantable Medical Devices (IMD)}
      	\item{Magnetic Resonance Imaging (MRI)}
      	\item{Sensors}
    		\end{itemize}
  
  \item Mechanical Engineering like:
  		\begin{itemize}
     		 \item{Computer box or Electronic Control Unit (ECU)}
      	\item{Cloud Vehicular Networks (CVN)}
      	\item{Electronic Fuel Injection (EFI)}
      	\end{itemize}
  \item Almost in all digital and control circuits
   \end{enumerate}

\addtocontents{toc}{\vspace{2em}} 

\begin{appendices}



\chapter{High-Performance and Energy-Efficient CNFET-Based Designs for Ternary Logic Circuits} 

\label{AppendixA} 

\lhead{Appendix A. \emph{High-Performance and Energy-Efficient CNFET-Based Designs for Ternary Logic Circuits}} 

\setboolean{@twoside}{false}
\includepdf[pages=-, offset=75 -75]{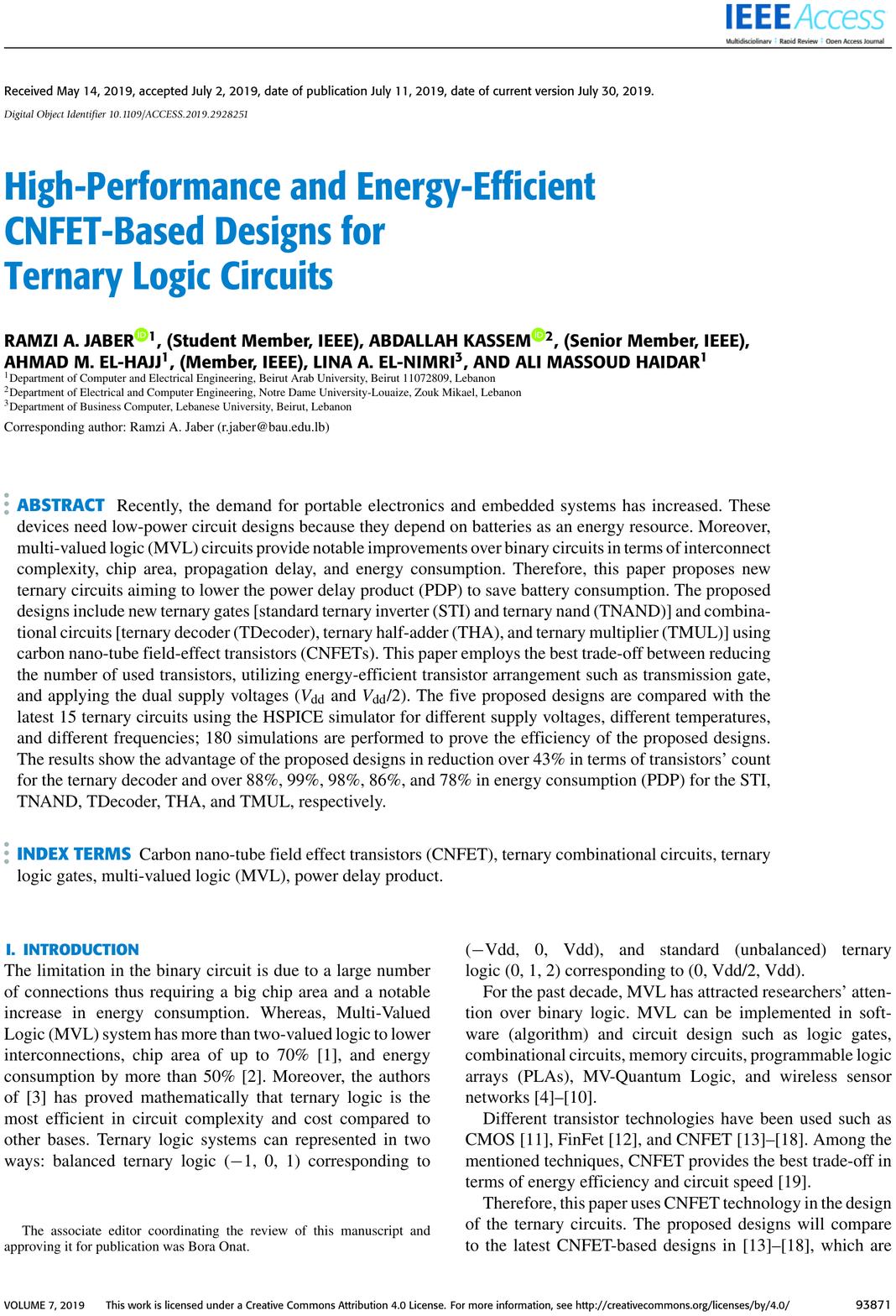}


\chapter{CNFET-Based Designs of Ternary Half-Adder using a Novel ``Decoder-less" Ternary Multiplexer based on Unary Operators} 

\label{AppendixB} 

\lhead{Appendix B. \emph{CNFET-Based Designs of Ternary Half-Adder using a Novel ``Decoder-less" Ternary Multiplexer based on Unary Operators}} 

\setboolean{@twoside}{false}
\includepdf[pages=-, offset=75 -75]{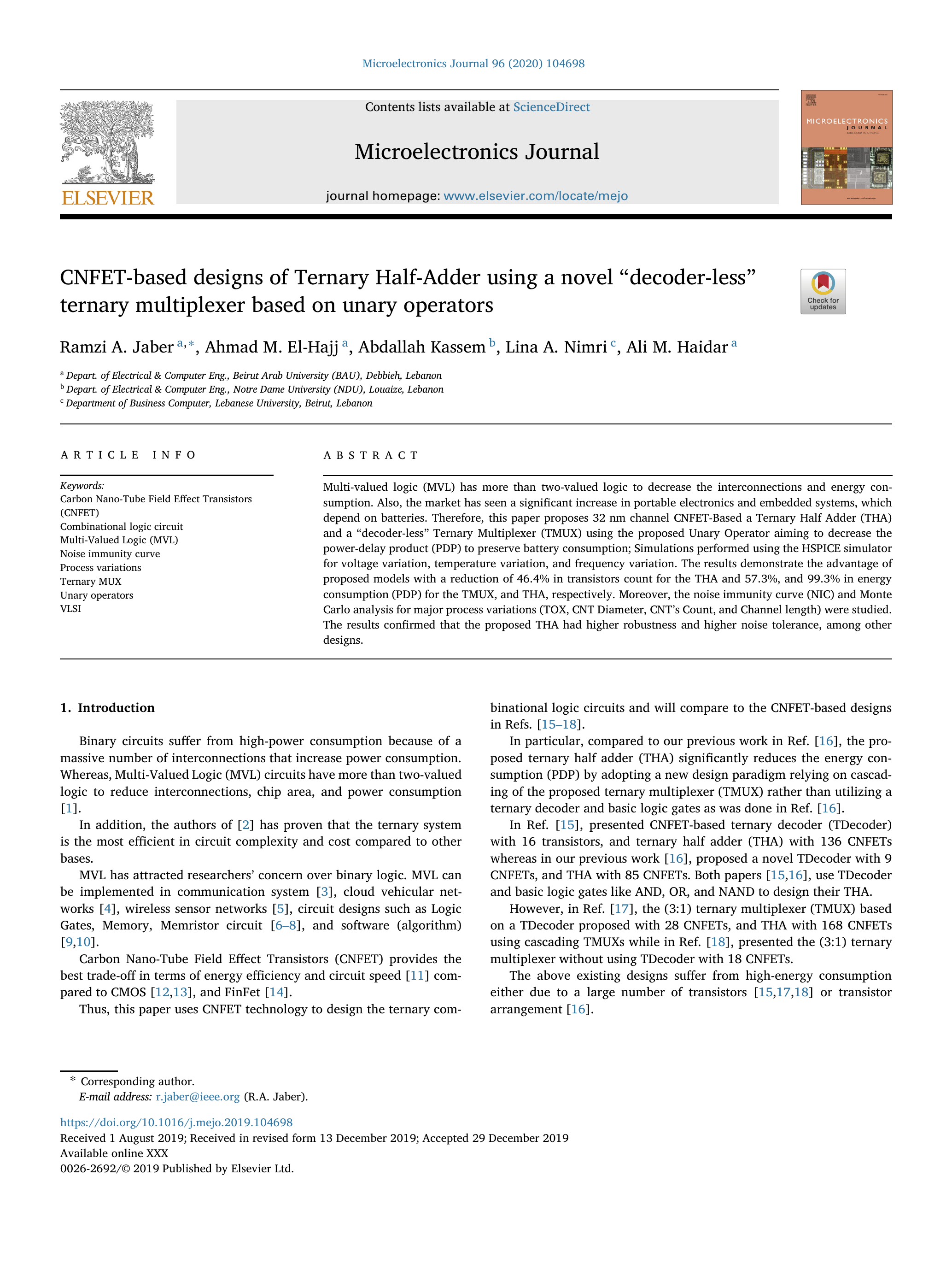}


\chapter{A Novel Implementation of Ternary Decoder Using CMOS DPL Binary Gates} 

\label{AppendixC} 

\lhead{Appendix C. \emph{A Novel Implementation of Ternary Decoder Using CMOS DPL Binary Gates}} 

\setboolean{@twoside}{false}
\includepdf[pages=-, offset=75 -75]{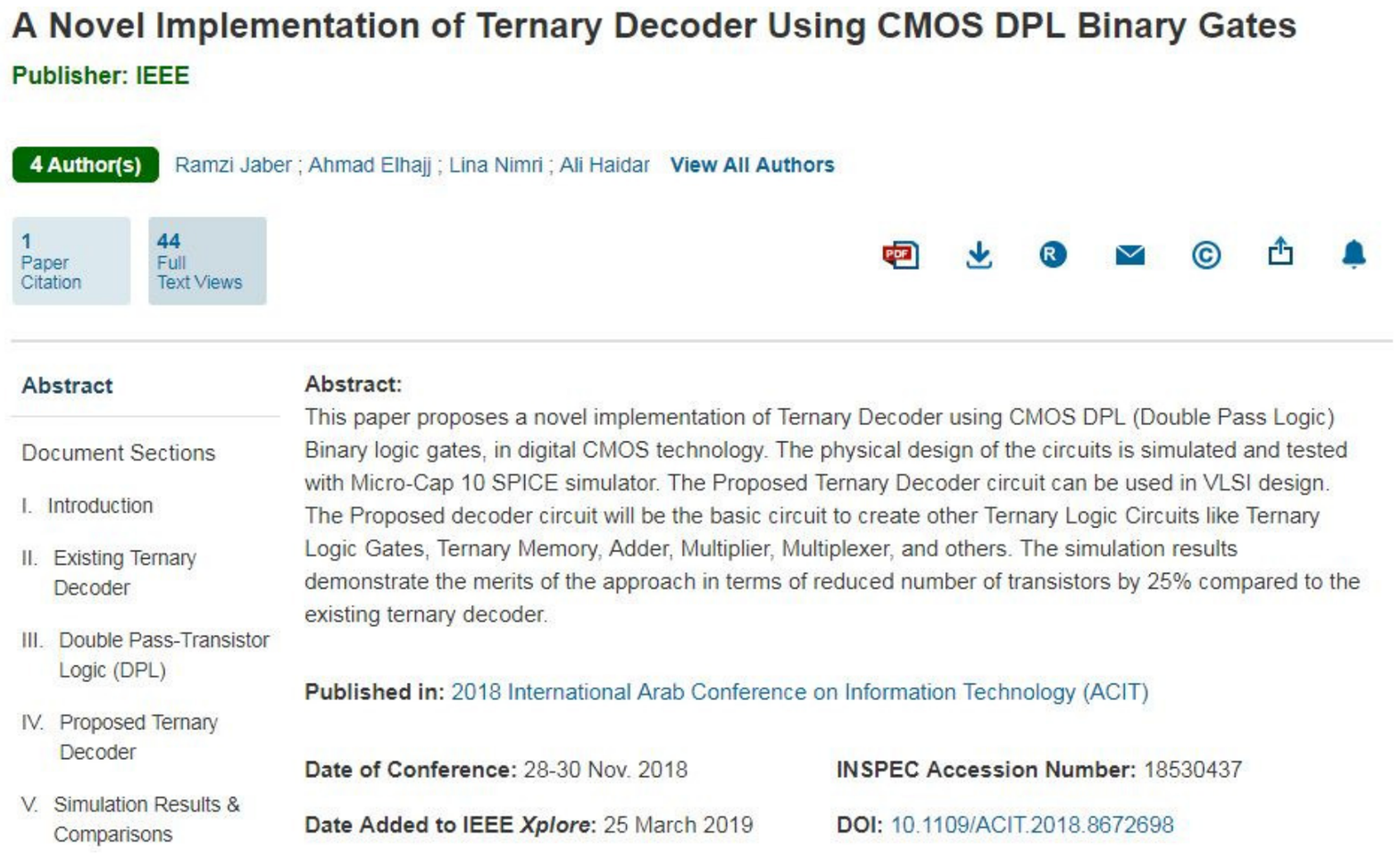}


\chapter{A Novel Binary to Ternary Converter using Double Pass-Transistor Logic} 

\label{AppendixD} 

\lhead{Appendix D. \emph{A Novel Binary to Ternary Converter using Double Pass-Transistor Logic}} 

\setboolean{@twoside}{false}
\includepdf[pages=-, offset=75 -75]{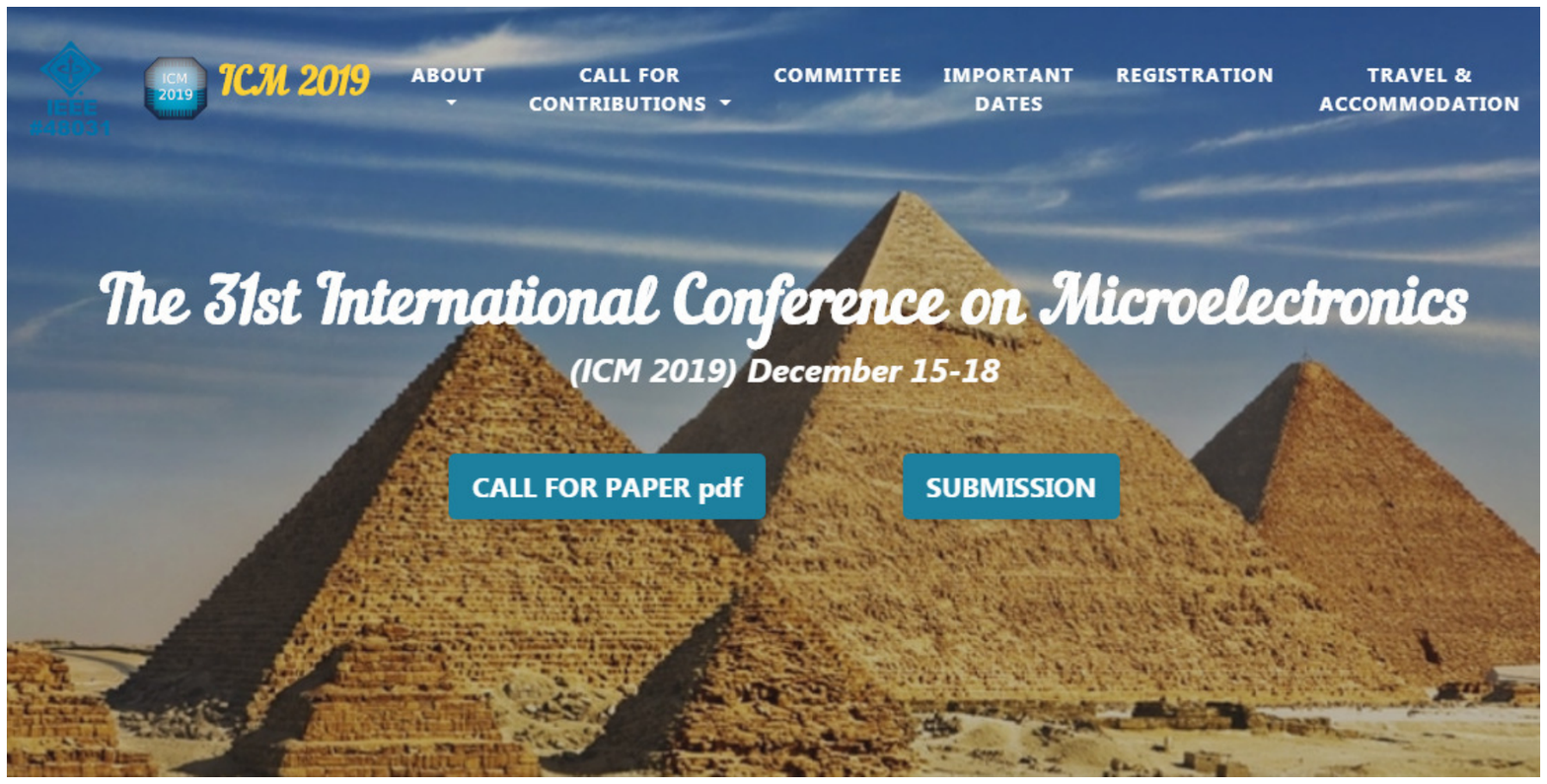}


\chapter{Ternary Data Transmission Between Hosts} 

\label{AppendixE} 

\lhead{Appendix E. \emph{Ternary Data Transmission Between Hosts}} 

\setboolean{@twoside}{false}
\includepdf[pages=-, offset=75 -75]{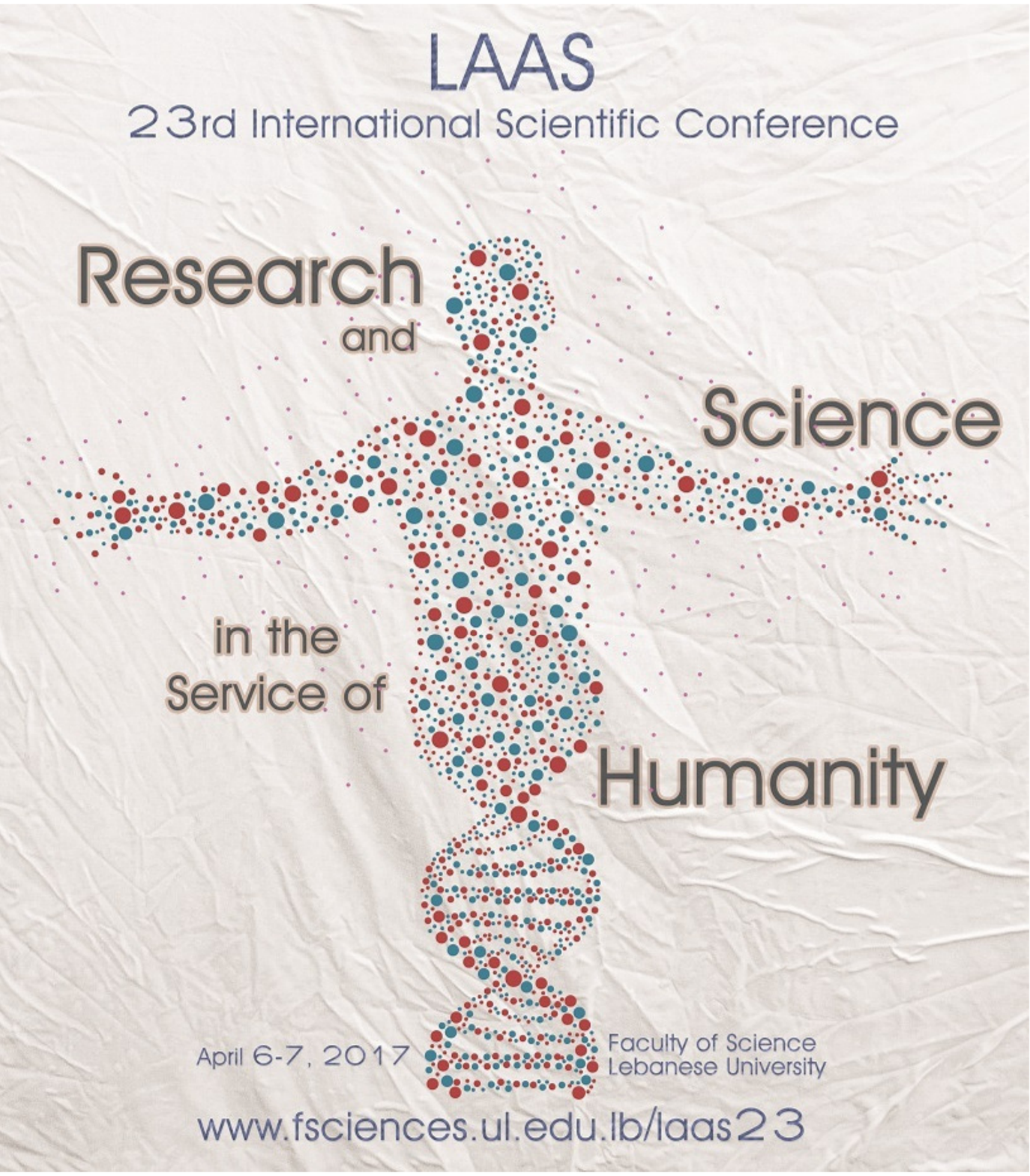}
\end{appendices}

\addtocontents{toc}{\vspace{2em}} 

\backmatter


\label{Bibliography}

\lhead{\emph{Bibliography}} 



\begin{thebibliography}{69}
\providecommand{\natexlab}[1]{#1}
\providecommand{\url}[1]{\texttt{#1}}
\expandafter\ifx\csname urlstyle\endcsname\relax
  \providecommand{\doi}[1]{doi: #1}\else
  \providecommand{\doi}{doi: \begingroup \urlstyle{rm}\Url}\fi

\bibitem[Knuth(1981)]{100}
D.~E. Knuth.
\newblock \emph{The Art of Computer Programming: Seminumerical Algorithms},
  volume~2.
\newblock Addison-Wesley Pub (Sd), 2nd edition, 1981.
\newblock ISBN 0201038226,9780201038224.
\newblock pp. 190-193.

\bibitem[Smith(1981)]{101}
K.~C. Smith.
\newblock The prospects for multivalued logic: A technology and applications
  view.
\newblock \emph{IEEE Transactions on Computers}, 30\penalty0 (9):\penalty0
  619--634, 1981.
\newblock URL \url{https://doi.org/10.1109/TC.1981.1675860}.

\bibitem[Hosseini and Etezadi(2019)]{102}
S.A. Hosseini and S.~Etezadi.
\newblock A novel very low-complexity multi-valued logic comparator in
  nanoelectronics.
\newblock \emph{Springer Circuits, Systems, and Signal Processing}, 2:\penalty0
  1--22, 2019.
\newblock URL \url{https://doi.org/10.1007/s00034-019-01158-2}.

\bibitem[Ifrah(2000)]{103}
G.~Ifrah.
\newblock \emph{The Universal History of Numbers. From Prehistory to the
  Invention of the Computer}.
\newblock John Wiley \&Sons, New York, 2000.
\newblock ISBN 0-471-39340-1.

\bibitem[Post(1921)]{104}
E.L. Post.
\newblock Introduction to a general theory of elementary propositions.
\newblock \emph{Amer. J. Math}, 43:\penalty0 163–185, 1921.
\newblock URL \url{https://doi.org/10.2307/2370324}.

\bibitem[Brusentsov and Alvarez(2006)]{105}
N.~Brusentsov and J.R. Alvarez.
\newblock Ternary computers: The setun and the setun 70.
\newblock pages 74--80, Berlin, Heidelberg, 2006. Springer, Perspectives on
  Soviet and Russian Computing (SoRuCom).

\bibitem[Moore(2006)]{106}
G.~Moore.
\newblock \emph{Understanding Moore's Law: Four Decades of Innovation (Chapter
  7)}.
\newblock Brock, David (ed.), New York, 2006.
\newblock ISBN 978-0-941901-41-3.

\bibitem[Miller and Soeken(2018)]{107}
D.~Miller and M.~Soeken.
\newblock A spectral algorithm for ternary function classification.
\newblock Linz, Austria, 2018. IEEE 48th Int. Symp. on Multiple-Valued Logic
  (ISMVL).

\bibitem[Chen et~al.(2019)Chen, Chen, Feng, Huang, and Chen]{108}
J.~Chen, Y.~Chen, C.~Feng, Y.~Huang, and R.~Chen.
\newblock Some new classes of entanglement-assisted quantum mds codes derived
  from constacyclic codes.
\newblock \emph{IEEE Access}, 7:\penalty0 91679 -- 91695, 2019.
\newblock URL \url{https://doi.org/10.1109/ACCESS.2019.2927294}.

\bibitem[Saleh et~al.(2018)Saleh, Kassem, and Haidar]{109}
N.~Saleh, A.~Kassem, and A.M. Haidar.
\newblock Energy-efficient architecture for wireless sensor networks in
  healthcare applications.
\newblock \emph{IEEE Access}, 6:\penalty0 6478 -- 6486, 2018.
\newblock URL \url{https://doi.org/10.1109/ACCESS.2018.2789918}.

\bibitem[Yang et~al.(2016)Yang, Zhu, and Wu]{110}
Q.~Yang, B.~Zhu, and S.~Wu.
\newblock An architecture of cloud-assisted information dissemination in
  vehicular networks.
\newblock \emph{IEEE Access}, 4:\penalty0 2764--2770, 2016.
\newblock URL \url{https://doi.org/10.1109/ACCESS.2016.2572206}.

\bibitem[Abdelaziz and Gulliver(2019)]{111}
M.~Abdelaziz and T.A. Gulliver.
\newblock Ternary trellis coded modulation.
\newblock \emph{IEEE Access}, 7:\penalty0 49027 -- 49038, 2019.
\newblock URL \url{https://doi.org/10.1109/ACCESS.2019.2909707}.

\bibitem[Toulabinejad et~al.(2019)Toulabinejad, Taheri, Navi, and
  Bagherzadeh]{112}
M.~Toulabinejad, M.R. Taheri, K.~Navi, and N.~Bagherzadeh.
\newblock Toward efficient implementation of basic balanced ternary arithmetic
  operations in cnfet technology.
\newblock \emph{Microelectronics Journal}, 90:\penalty0 267--277, 2019.
\newblock URL \url{https://doi.org/10.1016/j.mejo.2019.05.010}.

\bibitem[Abdelaziz and Gulliver(2017)]{113}
M.~Abdelaziz and T.A. Gulliver.
\newblock G-aetcam: gate-based area-efficient ternary content-addressable
  memory on fpga.
\newblock \emph{IEEE Access}, 5:\penalty0 20785 -- 20790, 2017.
\newblock URL \url{https://doi.org/10.1109/ACCESS.2017.2756702}.

\bibitem[Soliman et~al.(2019)Soliman, Fouda, Alhurbi, Said, Madian, and
  Radwan]{114}
N.~Soliman, M.E. Fouda, A.G. Alhurbi, L.A. Said, A.H. Madian, and A.G. Radwan.
\newblock Ternary functions design using memristive threshold logic.
\newblock \emph{IEEE Access}, 7:\penalty0 48371--48381, 2019.
\newblock URL \url{https://doi.org/10.1109/ACCESS.2019.2909500}.

\bibitem[Hurst(1984)]{115}
S.L. Hurst.
\newblock Multiple-valued logic its status and its future.
\newblock \emph{IEEE Transactions on Computers}, 133\penalty0 (1):\penalty0
  1160–1179, 1984.
\newblock URL \url{https://doi.org/10.1109/TC.1984.1676392}.

\bibitem[Gordon(2003)]{118}
J.R. Gordon.
\newblock \emph{Transistor circuit techniques: discrete and integrated}.
\newblock CRC Press, 2003.
\newblock ISBN 0748740759,9780748740758.

\bibitem[Valizadeh(2016)]{118a}
P.~Valizadeh.
\newblock \emph{FET A COMPREHENSIVE OVERVIEW: From Basic Concepts to Novel
  Technologies}.
\newblock John Wiley \& Sons, Inc., Canada, 2016.
\newblock ISBN 978-1-119-15549-2.

\bibitem[Rasit and Philip(2019)]{118c}
O.T. Rasit and H.-S. Philip.
\newblock \emph{Beyond-CMOS Technologies for Next Generation Computer Design}.
\newblock Springer International Publishing, 2019.
\newblock ISBN 978-3-319-90384-2;978-3-319-90385-9.

\bibitem[Sudeb et~al.(2017)Sudeb, Brajesh, and Pankaj]{118f}
D.~Sudeb, K.~Brajesh, and K.P. Pankaj.
\newblock \emph{Spacer engineered FinFET architectures : high-performance
  digital circuit applications}.
\newblock CRC Press, 2017.
\newblock ISBN 9781351751049,1351751042,1498783597,978-1-4987-8359-0.

\bibitem[Hills et~al.(2018)Hills, Bardon, Doornbos, Yakimets, Schuddinck,
  Baert, Jang, Mattii, Sherazi, Rodopoulos, Ritzenthaler, Lee, Thean, Radu,
  Spessot, Debacker, Catthoor, Raghavan, Shulaker, Philip~Wong, and Mitra]{119}
G.~Hills, M.G. Bardon, G.~Doornbos, D.~Yakimets, P.~Schuddinck, R.~Baert,
  D.~Jang, L.~Mattii, S.Y. Sherazi, D.~Rodopoulos, R.~Ritzenthaler, C.-S. Lee,
  A.~Thean, I.~Radu, A.~Spessot, P.~Debacker, F.~Catthoor, P.~Raghavan,
  M.~Shulaker, H.-S. Philip~Wong, and S.~Mitra.
\newblock Understanding energy efficiency benefits of carbon nanotube
  field-effect transistors for digital {VLSI}.
\newblock \emph{IEEE Transactions on Nanotechnology}, 17\penalty0 (6):\penalty0
  1259--1269, 2018.
\newblock URL \url{https://doi.org/10.1109/TNANO.2018.2871841}.

\bibitem[120(Accessed: August, 16 2019)]{120}
Stanford university cnfet model website [online], Accessed: August, 16 2019.
\newblock URL \url{http://nano.stanford.edu/model.php?id = 23}.

\bibitem[Deng and Wong(2007{\natexlab{a}})]{121}
J.~Deng and H.-S.~P. Wong.
\newblock A compact spice model for carbon-nanotube field-effect transistors
  including nonidealities and its application - part i: model of the intrinsic
  channel region.
\newblock \emph{IEEE Transactions on Electron Devices}, 54\penalty0
  (12):\penalty0 3186--3194, 2007{\natexlab{a}}.
\newblock URL \url{https://doi.org/10.1109/TED.2007.909030}.

\bibitem[Deng and Wong(2007{\natexlab{b}})]{122}
J.~Deng and H.-S.~P. Wong.
\newblock A compact spice model for carbon-nanotube field-effect transistors
  including nonidealities and its application - part ii: full device model and
  circuit performance benchmarking.
\newblock \emph{IEEE Transactions on Electron Devices}, 54\penalty0
  (12):\penalty0 3195--3205, 2007{\natexlab{b}}.
\newblock URL \url{https://doi.org/10.1109/TED.2007.909043}.

\bibitem[Deschamps et~al.(2017)Deschamps, Valderrama, and Terés]{201}
J.P. Deschamps, E.~Valderrama, and L.~Terés.
\newblock \emph{Digital Systems: From Logic Gates to Processors}.
\newblock Springer International Publishing, 2017.
\newblock ISBN 978-3-319-41198-9,978-3-319-41197-2.

\bibitem[Miller and Thornton(2008)]{117}
D.~M. Miller and M.~A. Thornton.
\newblock \emph{Multiple Valued Logic: Concepts and Representations}.
\newblock Morgan \& Claypool, San Rafael, CA, USA, 2008.
\newblock URL \url{https://doi.org/10.2200/S00065ED1V01Y200709DCS012}.

\bibitem[Lin et~al.(2011)Lin, Kim, and Lombardi]{123}
S.~Lin, Y.B. Kim, and F.~Lombardi.
\newblock Cntfet-based design of ternary logic gates and arithmetic circuits.
\newblock \emph{IEEE Transactions on Nanotechnology}, 10:\penalty0 217–225,
  2011.
\newblock URL \url{https://doi.org/10.1109/TNANO.2009.2036845}.

\bibitem[Suzuki et~al.(1993)Suzuki, Ohkubo, Shinbo, Yamanaka, Shimizu, and
  Nakagome]{124}
M.~Suzuki, N.~Ohkubo, T.~Shinbo, T.~Yamanaka, K.~Shimizu, and Y.~Nakagome,
  editors.
\newblock \emph{A 1.5 ns 32 b CMOS ALU in double pass-transistor logic},
  volume~1 of \emph{1}, San Francisco, CA, USA, 24-26 Feb. 1993. IEEE
  International Solid-State Circuits Conference Digest of Technical Papers,
  IEEE.

\bibitem[Shahrom and Hosseini(2018)]{206}
E.~Shahrom and S.~A. Hosseini.
\newblock A new low power multiplexer based ternary multiplier using cntfets.
\newblock \emph{AEU - International Journal of Electronics and Communications},
  93\penalty0 (1):\penalty0 191--207, 2018.
\newblock URL \url{https://doi.org/10.1016/j.aeue.2018.06.011}.

\bibitem[Zimpeck et~al.(2018)Zimpeck, Meinhardt, Artola, Hubert, Kastensmidt,
  and Reisa]{tg}
A.L. Zimpeck, C.~Meinhardt, L.~Artola, G.~Hubert, F.L. Kastensmidt, and R.A.L.
  Reisa.
\newblock Impact of different transistor arrangements on gate variability.
\newblock \emph{Microelectronics Reliability}, 88:\penalty0 111--115, Sept.
  2018.
\newblock https://doi.org/10.1016/j.microrel.2018.06.090.

\bibitem[Jaber et~al.(2019{\natexlab{a}})Jaber, Kassem, El-Hajj, Nimri, and
  Haidar]{paper1}
R.A. Jaber, A.~Kassem, A.M. El-Hajj, L.A. Nimri, and A.M. Haidar.
\newblock High-performance and energy-efficient cnfet-based designs for ternary
  logic circuits.
\newblock \emph{IEEE Access}, 7:\penalty0 93871 -- 93886, July
  2019{\natexlab{a}}.
\newblock https://doi.org/10.1109/ACCESS.2019.2928251.

\bibitem[Jaber et~al.(2020)Jaber, Kassem, El-Hajj, Nimri, and Haidar]{paper2}
R.A. Jaber, A.~Kassem, A.M. El-Hajj, L.A. Nimri, and A.M. Haidar.
\newblock Cnfet-based designs of ternary half-adder using a novel
  "decoder-less" ternary multiplexer based on unary operators.
\newblock \emph{Microelectronics Journal}, 36, Feb. 2020.
\newblock https://doi.org/10.1016/j.mejo.2019.104698.

\bibitem[Mouftah(1976)]{202}
H.T. Mouftah.
\newblock A study on the implementation of three-valued logic.
\newblock Logan, Utah, USA, 1976. MVL '76 Proceedings of the sixth
  international symposium on Multiple-valued logic.

\bibitem[Lloris et~al.(1980)Lloris, Prieto, and Velasco]{203}
A.~Lloris, A.~Prieto, and J.~Velasco.
\newblock C.m.o.s. circuit for implementation of unary operators in ternary
  logic.
\newblock \emph{IET-Electronics Letters}, 16\penalty0 (5):\penalty0 161--162,
  1980.
\newblock URL \url{https://doi.org/10.1049/el:19800115}.

\bibitem[LLoris et~al.(1982)LLoris, Cobo, and Prieto]{204}
A.~LLoris, A.~Cobo, and A.~Prieto.
\newblock Implementation of the unary operators in ternary logic : A universal
  cmos circuit.
\newblock \emph{International Journal of Electronics}, 52\penalty0
  (4):\penalty0 307--311, 1982.
\newblock URL \url{https://doi.org/10.1080/00207218208901433}.

\bibitem[Srinivasu and Sridharan(2017)]{205}
B.~Srinivasu and K.~Sridharan.
\newblock A synthesis methodology for ternary logic circuits in emerging device
  technologies.
\newblock \emph{IEEE Transactions on Circuits and Systems I: Regular Papers},
  64\penalty0 (8):\penalty0 2146–2159, 2017.
\newblock URL \url{https://doi.org/10.1109/TCSI.2017.2686446}.

\bibitem[Samadi et~al.(2017)Samadi, Shahhoseini, and Aghaei-liavali]{t16}
H.~Samadi, A.~Shahhoseini, and F.~Aghaei-liavali.
\newblock A new method on designing and simulating cntfet-based ternary gates
  and arithmetic circuits.
\newblock \emph{Microelectronics Journal, Elsevier}, 63:\penalty0 41--48, 2017.
\newblock URL \url{https://doi.org/10.1016/j.mejo.2017.02.018}.

\bibitem[Tabrizchi et~al.(2019)Tabrizchi, Taheri, Navi, and Bagherzadeh]{t18}
S.~Tabrizchi, M.R. Taheri, K.~Navi, and N.~Bagherzadeh.
\newblock Novel cnfet ternary circuit techniques for high-performance and
  energy-efficient design.
\newblock \emph{IET Circuits, Devices \& Systems}, 13\penalty0 (2):\penalty0
  193--202, 2019.
\newblock URL \url{https://doi.org/10.1049/iet-cds.2018.5036}.

\bibitem[Ndjountche(2016{\natexlab{a}})]{301}
T.~Ndjountche.
\newblock \emph{Digital Electronics, Volume 1: Combinational Logic Circuits}.
\newblock Electronics Engineering. Wiley-ISTE, 1 edition, 2016{\natexlab{a}}.
\newblock ISBN 1848219849,9781848219847.

\bibitem[Sridharan et~al.(2013)Sridharan, Gurindagunta, and Pudi]{t14}
K.~Sridharan, S.~Gurindagunta, and V.~Pudi.
\newblock Efficient multi-ternary digit adder design in cntfet technology.
\newblock \emph{IEEE Transactions on Nanotechnology}, 12\penalty0 (3):\penalty0
  283–287, 2013.
\newblock URL \url{https://doi.org/10.1109/TNANO.2013.2251350}.

\bibitem[Wang et~al.(2018)Wang, Gong, Zhang, and Kang]{t17}
P.~Wang, D.~Gong, Y.~Zhang, and Y.~Kang.
\newblock Ternary 2 - 9 line address decoder realized by cnfet, 2018.
\newblock U.S. Patent 20180182450.

\bibitem[Das et~al.(2018)Das, Banerjee, and Prasad]{302}
D.~Das, A.~Banerjee, and V.~Prasad.
\newblock Design of ternary logic circuits using cntfet.
\newblock Howrah, India, 2018. Int. Symp. On Devices, Circuits and Systems
  (ISDCS).
\newblock URL \url{https://doi.org/10.1109/ISDCS.2018.8379661}.
\newblock pp. 1-6.

\bibitem[Jaber et~al.(2018)Jaber, El-Hajj, Nimri, and Haidar]{conf1}
R.A. Jaber, A.M. El-Hajj, L.A. Nimri, and A.M. Haidar, editors.
\newblock \emph{A Novel implementation of ternary decoder using CMOS DPL binary
  gates}, Werdanye, Lebanon, Nov. 2018. 2018 Int. Arab Conf. on Information
  Technology (ACIT), IEEE.
\newblock https://doi.org/10.1109/ACIT.2018.8672698.

\bibitem[Aloke et~al.(2017)Aloke, Dipankar, and Mahesh]{303}
S.~Aloke, P.~Dipankar, and C.~Mahesh.
\newblock Benchmarking of dpl based 8b×8b novel wave-pipelined multiplier.
\newblock \emph{Int. J. of Electronics Letters (IJEL)}, 5\penalty0
  (1):\penalty0 115--128, 2017.
\newblock URL \url{https://doi.org/10.1080/21681724.2016.1175031}.

\bibitem[Ndjountche(2016{\natexlab{b}})]{401}
T.~Ndjountche.
\newblock \emph{Digital Electronics, Volume 2: Sequential and Arithmetic Logic
  Circuits}.
\newblock Wiley-ISTE, 1st edition edition, 2016{\natexlab{b}}.
\newblock ISBN 1119329779,9781119329770.

\bibitem[Muglikar et~al.(2016)Muglikar, Sahoo, and Sahoo]{t15}
M.~Muglikar, R.~Sahoo, and S.~K. Sahoo.
\newblock High-performance ternary adder using cntfet.
\newblock Coimbatore, India, 2016. 3rd Int. Symp. On Devices, Circuits and
  Systems (ISDCS).
\newblock URL \url{https://doi.org/10.1109/ICDCSyst.2016.7570599}.

\bibitem[Ebrahimi et~al.(2019)Ebrahimi, Reshadinezhad, and Bohlooli]{refA}
S.A. Ebrahimi, M.R. Reshadinezhad, and A.~Bohlooli.
\newblock A new design method for imperfection-immune cnfet-based circuit
  design.
\newblock \emph{Microelectronics Journal}, Jan. 2019.
\newblock https:// doi.org/10.1016/j.mejo.2019.01.013.

\bibitem[Balamurugan and Shanbhag(2001)]{refB}
G.~Balamurugan and N.R. Shanbhag.
\newblock The twin-transistor noise-tolerant dynamic circuit technique.
\newblock \emph{IEEE Journal of Solid-State Circuits}, 36\penalty0
  (2):\penalty0 273--280, 2001.
\newblock https://doi.org/10.1109/4.902768.

\bibitem[Doelz et~al.(1957)Doelz, Heald, and Martin]{501}
M.L. Doelz, E.T. Heald, and D.L. Martin.
\newblock Binary data transmission techniques for linear systems.
\newblock \emph{Proceedings of the IRE}, 45:\penalty0 656--661, 1957.
\newblock ISSN 0096-8390.
\newblock \doi{10.1109/jrproc.1957.278415}.
\newblock URL \url{http://doi.org/10.1109/jrproc.1957.278415}.

\bibitem[Simmonds(1997)]{504}
A.~Simmonds.
\newblock \emph{Data Communications and Transmission Principles: An
  Introduction}.
\newblock Macmillan New Electronics. Macmillan Education UK, 1997.
\newblock ISBN 978-0-333-64689-2,978-1-349-13900-2.

\bibitem[Yan(2019)]{502}
W.Q. Yan.
\newblock \emph{Introduction to Intelligent Surveillance: Surveillance Data
  Capture, Transmission, and Analytics}.
\newblock Texts in Computer Science. Springer International Publishing, 3rd ed.
  edition, 2019.
\newblock ISBN 978-3-030-10712-3,978-3-030-10713-0.

\bibitem[Bidgoli(2007)]{503}
H.~Bidgoli.
\newblock \emph{The Handbook of Computer Networks, Key Concepts, Data
  Transmission, and Digital and Optical Networks (Volume 1)}, volume~1.
\newblock Wiley, 2007.
\newblock ISBN 0471784583,9780471784586.

\bibitem[Trampel(1961)]{misc01}
M.~Trampel.
\newblock Binary to ternary converter.
\newblock US Patent 3,217,316, 12 1961.
\newblock https://patents.google.com/patent/US3217316A/en.

\bibitem[Nakanishi et~al.(2016)Nakanishi, Kawaguchi, and Sekiya]{misc02}
Y.~Nakanishi, T.~Kawaguchi, and M.~Sekiya.
\newblock Pulse synthesizing circuit.
\newblock US Patent 9,287,867, 2016.
\newblock https://patents.google.com/patent/US9287867B2/en.

\bibitem[Li et~al.(1995)Li, Morisue, and Ogata]{article05}
F.Q Li, M.~Morisue, and T.~Ogata.
\newblock A proposal of josephson binary-to-ternary converter.
\newblock \emph{IEEE Transactions on Applied Superconductivity}, 5\penalty0
  (2):\penalty0 2632--2635, 1995.

\bibitem[Zlatko et~al.(2010)Zlatko, Bundalo, Softic, and
  Kostadinovic]{proceedings08}
B.~Zlatko, D.~Bundalo, F.~Softic, and M.~Kostadinovic, editors.
\newblock \emph{Interconnection of binary and ternary {CMOS} digital circuits
  and systems}, volume~1 of \emph{1}, Opatija, Croatia, 24-28 May 2010. The
  33rd International Convention MIPRO, IEEE.

\bibitem[Aloke and Dipankar(2018)]{article06}
S.~Aloke and P.~Dipankar.
\newblock Dpl-based novel binary-to-ternary converter on {CMOS} technology.
\newblock \emph{AEU - International Journal of Electronics and Communications},
  92\penalty0 (1):\penalty0 69--73, August 2018.

\bibitem[Bowles(1996)]{t4}
G.~Bowles.
\newblock Ternary/binary converter circuit.
\newblock US Patent 5,49,898, 1996.
\newblock https://patents.google.com/patent/US5498980A/en.

\bibitem[Harish(2002)]{t5}
V.~Harish.
\newblock Ternary and quaternary logic to binary bit conversion cmos integrated
  circuit design using multiple input floating gate mosfets, 2002.
\newblock LSU Master's Theses.
  https://digitalcommons.lsu.edu/gradschooltheses/2548.

\bibitem[Ratna(2009)]{t6}
S.~Ratna.
\newblock Ternary to binary converter design in cmos using multiple input
  floating gate mosfets, 2009.
\newblock LSU Master's Theses. digitalcommons.lsu.edu/gradschooltheses/3963.

\bibitem[Haidar et~al.(2005)Haidar, Oseily, Nassar, and Shirahama]{t61}
A.~M. Haidar, H.~Oseily, E.~Nassar, and H.~Shirahama, editors.
\newblock \emph{Quirnary Coded Decimal Conversion Techniques}, Bruges, Belgium,
  October 2005. The Proceeding of the NOLTA 2005.

\bibitem[Oseily and Haidar(2011)]{t62}
H.~Oseily and A.~M. Haidar, editors.
\newblock \emph{Octal to Binary Conversion Using Multi-Input Floating Gate
  CMOS}, Iasi, Romania, June 30 – July 1 2011. The Proceeding of the 10-th
  International Symposium on Signals, Circuits and Systems. ISSCS 2011.

\bibitem[Oseily and Haidar(2015)]{t63}
H.~Oseily and A.~M. Haidar, editors.
\newblock \emph{Hexadecimal to Binary Conversion Using Multi-Input Floating
  Gate Complementary Metal Oxide Semiconductors}, 8 – 9 October 2015. The
  proceeding of International Conference on Applied Research in Computer
  Science and Engineering (ICAR).

\bibitem[Mohsen et~al.(2012)Mohsen, Soryani, Navi, and Tehrani]{proceedings09}
A.~Mohsen, M.~Soryani, K.~Navi, and MA. Tehrani, editors.
\newblock \emph{A novel ternary-to-binary converter in quantum-dot cellular
  automata}, volume~1 of \emph{1}, Amherst, MA, USA, 19-21 Aug. 2012. 2012 IEEE
  Computer Society Annual Symposium on VLSI, IEEE.

\bibitem[Ito(2011)]{t7}
H.~Ito.
\newblock Ternary valve input circuit, 2011.
\newblock https://www.google.com/patents/US8013630.

\bibitem[Jaber et~al.(2019{\natexlab{b}})Jaber, Kassem, El-Hajj, Nimri, and
  Haidar]{conf2}
R.A. Jaber, A.~Kassem, A.M. El-Hajj, L.A. Nimri, and A.M. Haidar, editors.
\newblock \emph{A Novel Binary to Ternary Converter using Double
  Pass-Transistor Logic}, Cairo, Egypt, 15-19 Dec. 2019{\natexlab{b}}. 31st
  International Conference on Microelectronics (ICM 2019), IEEE.
\newblock https://doi.org/10.1109/ICM48031.2019.9021886.

\bibitem[Jaber et~al.(2017)Jaber, Nimri, and Haidar]{conf3}
R.A. Jaber, L.A. Nimri, and A.M. Haidar, editors.
\newblock \emph{Ternary Data Transmission between Hosts}, Beirut, Lebanon,
  April 2017. 23rd International Scientific Conference LAAS.

\bibitem[Saha et~al.(2017)Saha, Dipanka, and Mahesh]{article07}
A.~Saha, P.~Dipanka, and C.~Mahesh.
\newblock Benchmarking of dpl-based 8b × 8b novel wave-pipelined multiplier.
\newblock \emph{International Journal of Electronics Letters}, 5\penalty0
  (1):\penalty0 115--128, Feb 2017.

\bibitem[Danial and Michel(1977)]{article09}
E.~Danial and I.~Michel.
\newblock Implementation of ternary circuits with binary integrated circuits.
\newblock \emph{IEEE Transactions on Computers}, 26\penalty0 (12):\penalty0 291
  -- 300, Dec. 1977.

\end{thebibliography}


\end{document}